\documentclass[epj,final,numbook]{svjour}

\usepackage{amsmath}
\usepackage{amssymb}
\usepackage{epsfig}
\usepackage{graphicx}
\usepackage{color}
\usepackage[colorlinks,allcolors=blue]{hyperref}
\usepackage{cite}

\newcommand{\ofR}{(\mathbf{r})}

\newcommand{\si}{\sigma}
\newcommand{\al}{\alpha}
\newcommand{\tet}{\theta}
\newcommand{\az}{\varphi}
\newcommand{\ro}{\rho}

\newcommand{\be}{\beta}
\newcommand{\ga}{\gamma}

\newcommand{\tro}{\tilde{\rho}}
\newcommand{\na}{\nabla}
\newcommand{\vsi}{\mbox{\boldmath{$\sigma$}}}
\newcommand{\vna}{\mbox{\boldmath{$\na$}}}
\newcommand{\Del}{\Delta}

\newcommand{\oeq}{\begin{equation}}
\newcommand{\ceq}{\end{equation}}
\newcommand{\oeqn}{\begin{eqnarray}}
\newcommand{\ceqn}{\end{eqnarray}}

\renewcommand{\>}{\rangle}
\newcommand{\<}{\langle}
\renewcommand{\(}{\left(}
\renewcommand{\)}{\right)}
\renewcommand{\[}{\left[}
\renewcommand{\]}{\right]}

\newcommand{\stf}{\,\,\,}
\newcommand{\sdf}{\,\,}
\newcommand{\stb}{\!\!\!}
\newcommand{\sdb}{\!\!}

\newcommand{\kfi}{|\phi \>}

\newcommand{\ksp}{|s'\>}
\newcommand{\kvac}{|-\>}

\newcommand{\bfi}{\<\phi |}

\newcommand{\bs}{\< s |}

\newcommand{\oQ}{\hat{Q}}

\newcommand{\oP}{\hat{P}}
\newcommand{\oM}{\hat{M}}
\newcommand{\oH}{\hat{H}}

\newcommand{\oV}{\hat{V}}

\newcommand{\oD}{\hat{D}}

\newcommand{\oro}{\hat{\rho}}
\newcommand{\oh}{\hat{h}}

\newcommand{\osi}{\hat{\sigma}}
\newcommand{\ovr}{\hat{\bf r}}
\newcommand{\ovp}{\hat{\bf p}}
\newcommand{\ovP}{\hat{\bf P}}
\newcommand{\ovR}{\hat{\bf R}}
\newcommand{\oT}{\hat{T}}

\newcommand{\op}{\hat{p}}

\newcommand{\of}{\hat{f}}

\newcommand{\oad}{\hat{a}^\dagger}
\newcommand{\oa}{\hat{a}}
\newcommand{\oA}{\hat{A}}
\newcommand{\oB}{\hat{B}}
\newcommand{\oF}{\hat{F}}
\newcommand{\oN}{\hat{N}}
\newcommand{\oX}{\hat{X}}
\newcommand{\oY}{\hat{Y}}
\newcommand{\oZ}{\hat{Z}}

\newcommand{\oL}{\hat{L}}
\newcommand{\ovsi}{\hat{\boldsymbol{\sigma}}}
\newcommand{\odel}{\hat{\delta}}
\newcommand{\ovk}{\hat{\bf k}}

\newcommand{\ov}{\hat{v}}

\newcommand{\del}{\delta\!}

\renewcommand{\d}{{\mbox d}}

\newcommand{\hb}{\hbar}

\newcommand{\vr}{{\bf r}}

\newcommand{\vj}{{\bf j}}
\newcommand{\vT}{{\bf T}}
\newcommand{\vJ}{{\bf J}}

\newcommand{\vR}{{\bf R}}

\newcommand{\vV}{{\bf V}}
\newcommand{\vW}{{\bf W}}
\newcommand{\vC}{{\bf C}}

\newcommand{\vS}{{\bf S}}

\newcommand{\ve}{{\bf e}}

\newcommand{\mH}{{\mathcal{H}}}

\newcommand{\mN}{{\mathcal{N}}}
\newcommand{\mD}{{\mathcal{D}}}
\newcommand{\mP}{{\mathcal{P}}}

\newcommand{\mL}{{\mathcal{L}}}

\newcommand{\Tr}{\mathrm{Tr}}
\newcommand{\tr}{\mathrm{tr}}

\definecolor{gris}{gray}{0.3}

\begin{document}

\title{Nuclear Quantum Many-Body Dynamics ($2^{nd}$ edition)}
\subtitle{From Collective Vibrations to Heavy-Ion Collisions}
\headnote{Review article}
\author{C\'edric Simenel}
\institute{Department of Fundamental and Theoretical Physics and Department of Nuclear Physics and Accelerator Applications, Research School of Physics, The Australian National University, Canberra ACT 2601, Australia\\
\email{cedric.simenel@anu.edu.au}
}
\date{Received: date / Revised version: date}
\dedication{This review article is dedicated to the memory of Paul Bonche who pioneered the application of the time-dependent Hartree-Fock theory to nuclear systems.}
\abstract{\normalfont\small
This article is a new edition of the review  [Eur. Phys. J. A 48, (2012) 152]. The increase in computational power has naturally led to new applications of mean-field (and beyond) methods. This is particularly the case of quasi-fission reactions. Since the first edition, significant progress has also been made in treating pairing correlations dynamically, leading to realistic applications in multi-nucleon transfer, fusion and fission reactions. A new section has been added on fission dynamics. 
A summary of recent researches on nuclear dynamics with realistic microscopic quantum approaches is presented. 
The Balian-V\'en\'eroni variational principle is used to derive the time-dependent Hartree-Fock (TDHF) equation describing the dynamics at the mean-field level, as well as an extension including small-amplitude quantum fluctuations which is equivalent to the time-dependent random-phase approximation (TDRPA). 
Such formalisms as well as their practical implementation in the nuclear physics framework with modern three-dimensional codes are discussed.
Recent applications to nuclear dynamics, from collective vibrations to heavy-ion collisions are presented. 
A particular attention is devoted to the interplay between collective motions and internal degrees of freedom.
For instance, the harmonic nature of collective vibrations is questioned.
Large amplitude collective motions are investigated in the framework of heavy-ion collisions and fission. 
How fusion is affected by the internal structure of the collision partners, such as their deformation, is discussed. 
Other mechanisms in competition with fusion, and responsible for the formation of  fragments which differ from the entrance channel (transfer reactions, deep-inelastic collisions, and quasi-fission)  are investigated. 
Finally, studies of actinide collisions forming, during very short times of few zeptoseconds, the heaviest nuclear systems available on Earth, are presented. 
}
\maketitle
\tableofcontents

\section{Introduction}

Nuclei are ideal to investigate fundamental  aspects of the quantum many-body problem. 
They exhibit collective motions built from coherent superpositions of the states of their constituents.
Examples range from collective vibrations to the formation of a compound system in collisions.
These features are common to other composite systems (atomic clusters, molecules, Bose-Einstein condensates...).
Their study in nuclear systems is obviously part of a wider physics field. 

An interesting feature of the dynamics of quantum many-body systems is the interplay between collective motion and single-particle degrees of freedom.
The latter is a source of complex and fascinating behaviours.
For instance, giant resonances are characterised by a collective vibration of many particles, but their  decay may occur  by the emission of a single nucleon.
Another example could be taken from the collision of composite systems where the transfer of few particles may have a strong impact on the final outcome of the reaction.

To describe these complex systems, one needs to solve the quantum many-body problem. 
The description of the dynamics of composite systems can be very challenging, especially when two such systems interact. 
An important goal of nuclear physics is then to find a unified way to describe the dynamics of nuclear systems.
Ultimately, the same theoretical model should be able to describe  vibration, rotation, fission, all the possible outcomes of heavy-ion collisions (elastic and inelastic scattering, particle transfer, fusion, and multifragmentation), and even the dynamics of neutron star crusts.

This desire for a global approach to nuclear dynamics has strongly influenced past research activities.
Beside the quest for a unified model of nuclear dynamics, possible applications of heavy-ion collisions to the formation of nuclear systems in extreme conditions (new exotic nuclei, large deformation and angular momentum, super-heavy elements...) are also strong motivations for the study of reaction mechanisms to optimise their production cross-sections. 

The Balian-V\'en\'eroni variational principle is a starting point leading to well-know formalisms, such as the time-dependent Hartree-Fock (TDHF) theory and the time-dependent random-phase approximation (TDRPA). 
Thanks to the recent increase of computational power, these approaches have been applied to investigate several aspects of the nuclear dynamics with modern numerical codes.
Recent applications to collective vibrations, fission, heavy-ion collisions, and neutron stars, as well as the underlying formalism and numerical details are discussed in this review article. 

Several general reviews and more specific  ``mini-reviews'' have been published since the first edition of this article~\cite{simenel2012}. 
Vibrations and requantisation of mean-field trajectories were addressed in~\cite{nakatsukasa2016}.
Microscopic methods to evaluate nucleus-nucleus potentials, including Pauli repulsion and coupled-channels effects, were reviewed in  \cite{simenel2018} together with realistic applications of time-dependent BCS simulations for fission dynamics. 
Connections between microscopic and macroscopic aspects of low-energy reactions were discussed in \cite{washiyama2020}.
See also  \cite{lacroix2015} for a focus on astrophysics applications. 
Ref.~\cite{bulgac2020} adressed applications to fission. 
Some reviews were dedicated to beyond mean-field methods, e.g., stochastic mean-field approaches~\cite{lacroix2014} and the time-dependent density matrix theory~\cite{tohyama2020}.
Multi-nucleon transfer and quasi-fission reactions were covered in  \cite{nakatsukasa2016,simenel2018,sekizawa2019,godbey2020}.
The role of the Skyrme functional was studied in \cite{stevenson2019}, and more specifically, the impact of the tensor terms was the topic of \cite{sun2022c}.
Quantum vortices in fermionic superfluids such as neutron stars are reviewed in \cite{magierski2024}. 
Finally, transport models for intermediate and relativistic energies were investigated in \cite{colonna2020}.
Naturally, some of these new applications and developments will only briefly be discussed here, and the reader is referred to these reviews, as well as the original works for more details. 
Nevertheless, some recent applications to fission and quasi-fission that were not included in these reviews will be presented in more details here. 

The formalism and numerical developments are presented in section~\ref{chap:formalism}.
Studies of collective vibrations are then collected in section~\ref{chap:vib}.
Section~\ref{chap:fission} is dedicated to recent applications to fission. 
Finally, heavy-ion collisions are investigated in section~\ref{chap:HIC}.

\section{The time-dependent Hartree-Fock theory and its extensions\label{chap:formalism}}

\subsection{Introduction}

The quantum many-body problem is common to many theoretical fields~\cite{neg98,bla86}.
It aims at describing the structure and dynamics of interacting particles. 
Electrons, atoms, molecules and nucleons are usual constituents of quantum many-body systems. 

In the non-relativistic regime, these systems obey the Schr\"odinger equation. 
\oeq
i \frac{d|\Psi(t)\>}{dt} = \oH |\Psi(t)\>,
\label{eq:schrod}
\ceq
where $|\Psi(t)\>$ describes the state of the many-body system and $\oH$ is the microscopic Hamiltonian. 
We use the notation $\hb\equiv1$.
This equation can be solved exactly for simple cases only. 
Generally, one has to rely on some approximations.

Variational principles offer an elegant starting point to build such approximations.
Indeed, they ensure an optimization of the equations of motion under the approximation that the variational space is limited to a sub-space of the full Hilbert (or Fock) space. 
Of course, without any restriction of the variational space, it is required that the chosen variational principle allows to recover the Schr\"odinger equation. 
However, their usefulness appears when restricting the variational space. 
Then, the validity of the approximation relies entirely on the choice of the variational space. 
In one hand, the latter has to be small enough so that the problem is numerically tractable.
In the other hand, the variational space should contain the relevant degrees of freedom to allow for a realistic description of the physical processes. 

Although the Schr\"odinger equation is unique,  different variational principles have been developed in the past. 
One is based on the stationarity of the Dirac action
\oeq
S[t_0,t_1;\Psi(t)]=\int_{t_0}^{t_1}dt \stf \<\Psi(t)| \(i \frac{d}{dt}-\oH \)|\Psi(t)\>.
\label{eq:SDirac}
\ceq
The variational principle $\delta S=0$ is applied with the boundary conditions $\delta \Psi(t_0) = \delta \Psi(t_1) = 0$. 
If $\Psi$ is allowed to span the entire Hilbert space, one recovers the Schr\"odinger equation. 

In most practical applications, mean-field models are considered in a first approximation, 
and, eventually, serve as a basis for beyond-mean-field approaches~\cite{sim10a,lac04}. 
To construct such a mean-field theory from the above variational principle, one restricts the variational space by assuming that the $N$ particles (we consider fermions) are independent. 
In this case, they may be described by a Slater determinant
\oeq
\kfi=\prod_{i=1}^N\oad_i\kvac,
\ceq 
where $\oad_i$ creates a particle in the state $|\az_i\>$ when it acts on the particle vacuum~$\kvac$.
In such a state, all the information is contained in the one-body density-matrix $\ro$ 
associated with  the single-particle operator 
\oeq
\oro=\sum_{i=1}^N|\az_i\>\<\az_i|.
\ceq
Solving the variational principle where the action defined in Eq.~(\ref{eq:SDirac}) is required to be stationary in the subspace of Slater determinants $\kfi$ with fixed boundary conditions at times $t_0$ and $t_1$ leads to (see appendix~\ref{annexe:standardTDHF})
\oeq 
i\frac{\partial \ro}{\partial t}=\[h[\ro],\ro\],
\label{eq:tdhf}
\ceq
where $h[\ro]$ is the Hartree-Fock (HF) single-particle Hamiltonian with matrix elements 
\oeq 
h_{\al\be}=\frac{\delta \bfi \oH \kfi}{\delta \ro_{\be\al}},
\label{eq:hHF}
\ceq
and 
\oeq
\ro_{\al\be}=\<\az_\al|\oro|\az_\be\>=\bfi \oad_\be\oa_\al \kfi.
\label{eq:roalbe}
\ceq
Equation~(\ref{eq:tdhf}) is the  time-dependent Hartree-Fock (TDHF) equation. It was obtained by Dirac in 1930~\cite{dir30}.
It provides a self-consistent mean-field evolution where the interaction between the particles is replaced by a one-body mean-field potential generated by all the particles. 
It is, then, assumed that each particle evolves independently in this potential. 

We may question the validity of such a mean-field approximation in the nuclear physics context. 
This assumption could only give an approximation of the exact dynamics and we do expect, in 
general, that the system will deviate from the independent particle picture.
Indeed, the exact dynamics  is given by the time-dependent 
Schr\"odinger equation~[Eq.~(\ref{eq:schrod})], and, unless the Hamiltonian 
contains one-body operators only, the mean-field theory can only 
approximate the exact evolution of the system. 
The exact and mean-field Hamiltonians differ by the residual interaction.
The validity of the mean-field approximation depends, then, on the intensity of the residual interaction. 
The latter is a function of the state $\left| \Psi \right\rangle$ and, therefore, depends   
on the physical situation. 

Starting from simple arguments \cite{lic76}, 
the time $\tau$ over which the Slater determinant picture breaks down could be expressed as:
\begin{eqnarray}
\tau &=& \frac{\hbar}{2}  \Big(\frac{1}{N}
\sum_{\bar \alpha \bar \beta \alpha \beta} 
|\left\langle \bar \alpha \bar \beta \left| \bar v \right| \alpha \beta \right\rangle|^2 \Big)^{-1/2},
\end{eqnarray} 
where $N$ is the number of particles, and $\alpha$ and $\beta$ denote hole states while $\bar{\alpha}$ and $\bar{\beta}$ denote particle states. 
In the nuclear physics context, typical values of the residual interaction lead to $\tau \simeq 100-200$~fm/$c$.
This gives an estimate of the time during which one can safely consider that the independent particle approximation is valid. 
For longer times, as encountered, e.g., in heavy-ion collisions, the validity of the mean-field approximation can only be verified by comparison with experiment and/or with beyond mean-field calculations. 

One important aspect of the TDHF approach is the treatment of one-body dissipation mechanisms. 
In particular, one-body dissipation is crucial to properly describe low-energy heavy-ion collisions. 
Two kind of one-body dissipation mechanisms can easily be identified:
\begin{itemize}
\item {\it Coupling of collective motions with one-particle one-hole (1p1h) excitations.} In the case of giant resonances, this coupling leads to the so-called Landau damping (see sec.~\ref{chap:vib}). In heavy-ion collisions, the field of the  collision partner generates a distortion of the single-particle wave-functions. This is particularly true when part of the wave functions are transferred from one fragment to another, leading to an excitation of the fragments, and, then, to a dissipation of the translational kinetic energy. Another example is the case of fusion reactions. Multiple reflexions of single-particle wave-functions on the wall of the mean-field dissipate at least some of the original collective translational kinetic energy into single-particle excitations and collective vibrations of the compound system.
\item {\it Single-particle wave-function emission to the continuum.}  Emission of nucleons into the continuum is a natural cooling mechanism of excited nuclei. As an example, the direct decay by nucleon emission induces an escape width of giant resonances (see sec.~\ref{sec:decay}).
\end{itemize}

Naturally, extensions of TDHF including pairing and in-medium two-body correlations should be considered, at least to determine the validity of the mean-field approximation in terms of energy and simulation time. 
Larger variational spaces could then be considered in order to incorporate some effects of the residual interaction which are missing at the TDHF level. 
This is the case, for instance, with the inclusion of pairing correlations by taking a variational space of quasi-particle vacua, leading to the time-dependent Hartree-Fock-Bogoliubov (TDHFB) formalism~\cite{bla81}. 
Another possible approach to include beyond mean-field correlations is to consider different levels of truncation of the Bogoliubov-Born-Green-Kirwood-Yvon (BBGKY) hierarchy~\cite{bog46,bor46,kir46}. 
The two first equations of this hierarchy read
\begin{equation}
\left\{ 
\begin{array}{cl}
i\hbar \frac{\partial }{\partial t}\rho _{1}
=&\left[ t_1,\rho _{1}\right] + \frac{1}{2}{\rm \Tr}_{2}\left[ \bar v_{12},\rho_{12}\right]  \\ 
&  \\ 
i \hbar \frac{\partial }{\partial t}\rho_{12} =& [t_1 + t_2 + \frac{1}{2} \bar v_{12}, \rho_{12}] 
+ \frac{1}{2} \Tr_3 \left[ \bar v_{13} + \bar v_{23} , \rho_{123} \right], \\ 
\end{array}
\right.  \label{eq:BBGKY}
\end{equation}
where $\rho_1$, $\rho_{12}$ and $\rho_{123}$ are the one-, two-, and three-body density matrices, respectively.
We see that $\rho_1$ and $\rho_{12}$ are coupled. 
In fact, each equation of the BBGKY hierarchy couples $\rho_{1\cdots k}$ to $\rho_{1\cdots k+1}$, with $k<N$.
Put together, they form a closed system of coupled equations equivalent to the Schr\"odingier equation for a finite many-body system. 

The two-body  density matrix can be expressed as $\rho_{12}= \rho_1\rho_2 (1-P_{12})+C_{12}$ where $P_{12}$ is the permutation operator between particles 1 and 2, and $C_{12}$ is the two-body correlation matrix describing, e.g., pairing correlations and in-medium two-body collisions. 
Neglecting $C_{12}$ in the first equation of the BBGKY hierarchy leads to the TDHF equation [Eq.~(\ref{eq:tdhf})]. 
One could also include these two-body correlations by solving the coupled equations (\ref{eq:BBGKY})
and neglecting the three-body correlations.
The resulting set of coupled equations for the evolution of $\rho_1$ and $\rho_{12}$ is known as the time-dependent density-matrix (TDDM) formalism\footnote{Note that the TDDM formalism is not obtained by solving  a variational principle.}. 
It is obvious that solving the TDDM equations request much more computational efforts than standard TDHF calculations~\cite{cas90,bla92,luo99,toh01,toh02a,toh02b,ass09,tohyama2016,lackner2015,lackner2017,wen2018,tohyama2020,barton2021}. 

The difficulties of solving quantum many-body models have  sometimes been overcome by using their semi-classical limit.
The semi-classical limit of the TDHF approach is the Vlasov equation which
 is obtained by taking the Wigner transform of Eq.~(\ref{eq:tdhf}) and keeping only the first order in $\hbar$.
The in-medium two-body collision term can then be added, leading to the Landau-Vlasov equation which is the root of the Vlasov-Uehling-Uhlenbeck (VUU) \cite{kru85} and Boltzmann-Uehling-Uhlenbeck (BUU) \cite{aic85} transport theories. 
The latter approaches are common tools to interpret heavy-ion collision data at intermediate energy where the collision term is expected to play a significant role.  
However, at  energies closer to the barrier, the collision term is expected to be hindered by the Pauli blocking. 
In addition,  semi-classical approaches lead to a poor description of the ground state structure of the nuclei, such as their binding energies and deformations.
As one of the goal of the present review article is to study the interplay between nuclear structure and dynamics, it is a natural choice to focus on fully quantum approaches only and semi-classical models won't be further discussed here.

\begin{figure}[h]
\begin{center}
\includegraphics[width=9cm]{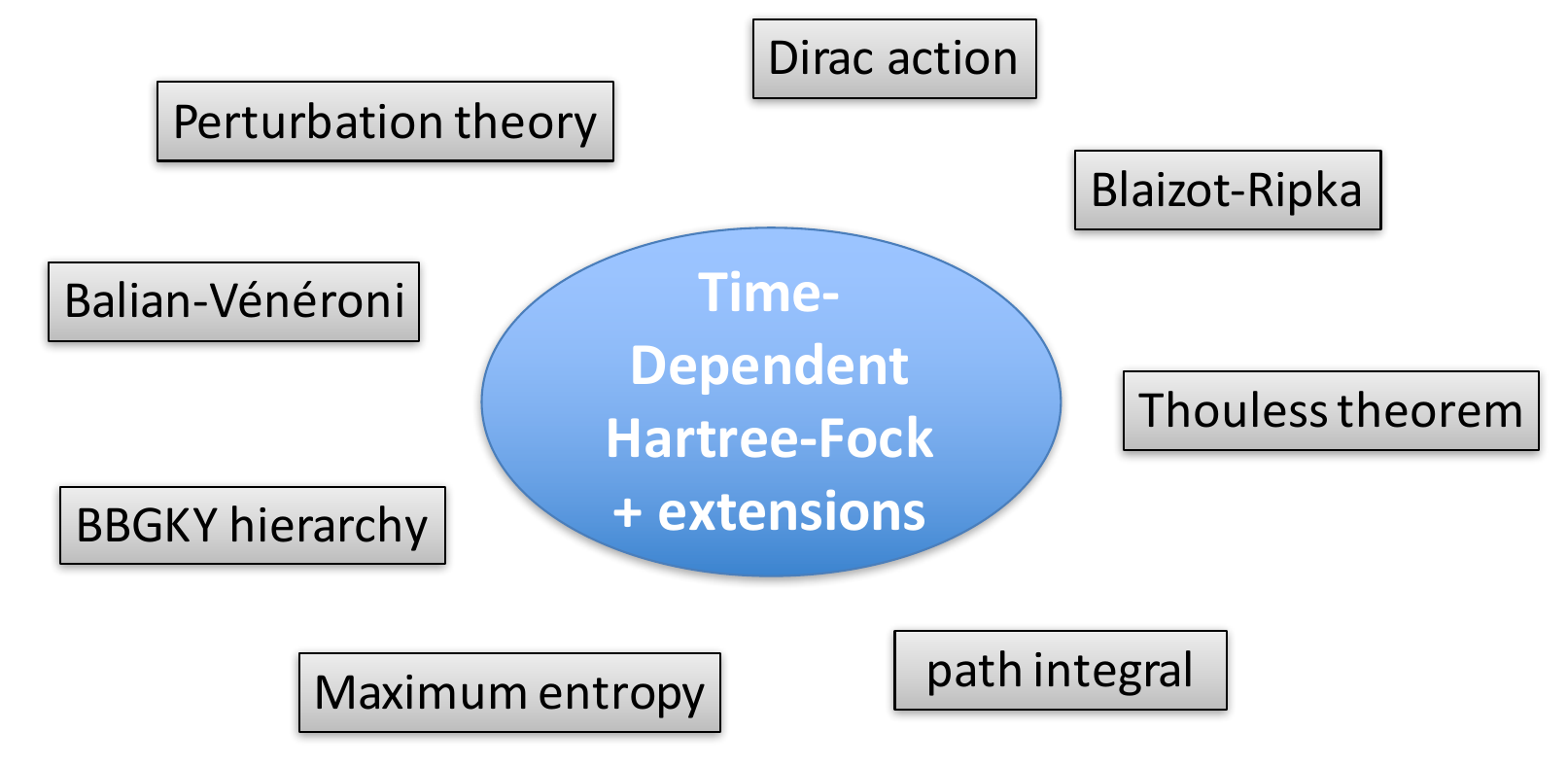}
\end{center}
\caption{Various approaches leading to the TDHF equation and to its extensions. From Ref. \cite{simenel2018}.
\label{fig:TDHFderiv}}
\end{figure}

Although several approaches are traditionally used to describe dynamics of quantum many-body systems, 
some being based on variational principles and others on perturbation theory, it is interesting to note that, at the lowest level of approximation, they usually lead to the time-dependent Hartree-Fock self-consistent mean-field theory \cite{dirac1930}  (see Fig.~\ref{fig:TDHFderiv} and Ref.~\cite{simenel2018} for a more detailed discussion of these approaches).
The choice of the approach depends on the problem to study. 
For instance, Blaizot and Ripka introduced in 1981 a variational principle appropriate to the calculation of transition amplitudes~\cite{bla81}. 
Inspired by this work, Balian and V\'en\'eroni (BV) presented the same year their variational principle for the expectation value of an observable~\cite{bal81}. 
In the latter, both the state of the system and the observable of interest are allowed to vary in their own variational space. 

In particular, the BV variational principle allows a more transparent interpretation of the TDHF theory~\cite{bal81}: 
it is shown that the TDHF equation~(\ref{eq:tdhf}) is optimised for the {\it expectation value of one-body observables}. 
TDHF calculations are indeed successful to predict such quantities (several examples are presented in the following sections). 
It also explains why TDHF sometimes fail to reproduce other quantities such as fluctuations of one-body operators~\cite{koo77,dav78,das79}, which are outside the variational space used to derive the TDHF equation.
In fact, to predict such fluctuations, Balian and V\'en\'eroni proposed a prescription also based on their variational principle, but with a different variational space for the evolution of the observable~\cite{bal84,bal92}.

In the following part of this section, we first describe the Balian-V\'en\'eroni variational principle.
The latter is  used to derive the TDHF equation as well as fluctuations and correlations of one-body observables. 
Then, the Skyrme energy density functional describing the strong interaction between the nucleons is introduced. 
Numerical aspects are also discussed.
Finally, perspectives for beyond TDHF calculations are presented. 

\subsection{The Balian-V\'en\'eroni variational principle}

The BV variational principle has been applied to different problems 
in nuclear physics~\cite{tro85,mar85,bon85,zie88,bro08,bro09,sim11,scamps2015a,williams2018,godbey2020b,gao2025}, 
hot Fermi gas~\cite{mar91}, $\phi^4$-theory~\cite{mar95,mar99}, and Boson systems~\cite{ben99,bou10}.
The first realistic application of the BV prescription (also referred to as time-dependent random phase approximation, TDRPA) to fragment mass and charge distributions in heavy-ion collisions are reported in Ref.~\cite{sim11} and will be presented in section~\ref{sec:DIC}.
The importance of the BV variational principle for the interpretation of the TDHF theory, which will be discussed and applied in the following sections, as well as the derivation of the BV prescription for fluctuations and correlations of one-body observables, justify the more detailed discussion in this subsection. 

Let us first define two variational quantities: $\oD(t)$, describing the state of the system, and $\oA(t)$, describing the evolution of the observable in the Heisenberg picture. 
The application of the BV variational principle requires two boundary conditions:
\oeq
\oD(t_0)=\oD_0,
\label{eq:BCD}
\ceq
where the initial state of the system $\oD_0$ is assumed to be known, and
\oeq
\oA(t_1) = \oA_1,
\label{eq:BCA}
\ceq
where $\<\oA_1\>$ is the final expectation value we want to compute at $t_1>t_0$. 

The action-like quantity defined by Balian and V\'en\'eroni reads~\cite{bal81}
\oeqn
&J& = \Tr\[ \oA(t_1)\oD(t_1) \] \nonumber\\
&-& \int_{t_0}^{t_1} \stb dt \,\Tr\!\[ \oA(t) \(\frac{d\oD(t)}{dt} + i[\oH(t),\oD(t)]\) \]\sdb . 
\label{eq:JA}
\ceqn
 We see that, imposing $\delta_A J=0$, where $\delta_A$ induces small variations of $\oA(t)$, leads to the Liouville-Von Neumann equation
\oeq
i\frac{d\oD(t)}{dt}=\[\oH,\oD(t)\]
\label{eq:Liouville}
\ceq
which is fully equivalent to the Schr\"odinger equation. 

{
To get Eq.~(\ref{eq:Liouville}), we first note that, according to the boundary condition in Eq.~(\ref{eq:BCA}), $\oA(t_1)$ is fixed and we get
$$\delta_AJ= - \int_{t_0}^{t_1} dt\stf \Tr\[ \delta_A\oA(t)\, \(\frac{d\oD(t)}{dt} + i[\oH(t),\oD(t)]\) \].$$
To get $\delta_AJ=0$ for any variation of $\oA$, the term inside the brackets must be zero, giving Eq.~(\ref{eq:Liouville}).
}

Variations of $\oD(t)$ should also be considered. 
It is easier to first rewrite Eq.~(\ref{eq:JA}) as
\oeqn
&J& = \Tr\[ \oA(t_0)\oD(t_0) \] \nonumber\\
&+& \int_{t_0}^{t_1} \stb dt\, \Tr\[ \oD(t) \(\frac{d\oA(t)}{dt} + i[\oH(t),\oA(t)]\) \]\sdb .
\label{eq:JD}
\ceqn
{
To get Eq.~(\ref{eq:JD}) we integrate by part the $\int dt \oA\partial_t\oD$ term in Eq.~(\ref{eq:JA}) and we use the relation $$\Tr(\oA[\oH,\oD])=-\Tr(\oD[\oH,\oA]).$$}

With the boundary condition in Eq.~(\ref{eq:BCD}), requiring $\delta_DJ=0$ leads to 
\oeq
i\frac{d\oA(t)}{dt}=\[\oH,\oA(t)\],
\label{eq:Ehrenfest}
\ceq
which is also equivalent to the Schr\"odinger equation, and is expressed in the Heisenberg picture. 

\subsection{Derivation of the time-dependent Hartree-Fock equation}

The TDHF theory is obtained under the approximation that $\oA(t)$ is constrained to be a one-body operator  and that $\oD(t)$ is an independent particle state for all $t$. 
As a result, TDHF is optimised for the expectation value of one-body operators. 
On the contrary, other quantities, such as expectation values of two-body operators, 
 are not guaranteed to be well predicted because they are outside the  variational space of $\oA(t)$. 

To get the TDHF equation, we impose that the variation  $\delta_A$  leaves $\oA$ in the space of one-body operators. 
As we consider arbitrary variations, we can choose 
\oeq
\delta_A\oA(t)\equiv\oad_\al\oa_\be \mbox{ for } t_0\le t<t_1
\ceq
and $\delta\oA(t_1)=0$ due to the boundary condition in Eq.~(\ref{eq:BCA}).
Requiring $\delta_AJ=0$, we get from Eq.~(\ref{eq:JA}) 
\oeq
\Tr\[\oad_\al\oa_\be\(\frac{d\oD}{dt}+i[\oH,\oD]\)\]=0.
\ceq
In addition, the state of the system is constrained to be an independent particle state.
The variational space for $\oD(t)$ is then defined by $\oD(t)=|\phi(t)\>\<\phi(t)|$ where $|\phi(t)\>$ is a Slater determinant. 
Using the one-body density matrix defined in Eq.~(\ref{eq:roalbe}), we get
\oeq
i\frac{d\ro_{\be\al}(t)}{dt}=\<\phi(t)|\[\oad_\al\oa_\be,\oH\]|\phi(t)\>.
\label{eq:idrodt}
\ceq

Consider a Hamiltonian of the form
\oeq
\oH = \sum_{ij} \sdf t_{ij}\sdf \oad_i \oa_j + \frac{1}{4} \sum_{ijkl} \sdf  \bar{v}_{ijkl} \sdf \oad_i \oad_j \oa_l \oa_k 
\label{eq:oH2}
\ceq
where matrix elements associated with the kinetic energy and with the anti-symmetric two-body interaction are given, respectively, by 
\oeqn
t_{ij} &=& \frac{1}{2m} \sdf \<i | \op^2 | j \>\stf \mbox{ and }  \label{eq:tij}\\
\bar{v}_{ijkl} &=& v_{ijkl} - v_{ijlk}. \label{eq:vbar}
\ceqn

Reporting the Hamiltonian expression  
[Eq.~(\ref{eq:oH2})] in Eq.~(\ref{eq:idrodt}), we get
\oeqn
i\frac{d\ro_{\be\al}}{dt} &=& \sum_{kl} t_{kl} \< [\oad_\al\, \oa_\be , \oad_k \,\oa_l ]\,\>\nonumber\\
&+& \frac{1}{4}  \sum_{klmn}  \bar{v}_{klmn}  \< \, [  \oad_\al\, \oa_\be ,   \oad_k \,\oad_l \,\oa_n \,\oa_m ]\,\>,
\label{eq:erhenfest_detail}
\ceqn
where the time variable has been omitted and $\<\cdots\>$ denotes the expectation value on $|\phi(t)\>$ to simplify the notation.
Equation~(\ref{eq:erhenfest_detail}) leads to the TDHF equation
 \oeq
i \frac{d{\ro}_{\be\al}}{dt} =\[ h[\ro], \ro \]_{\be\al} .
 \label{eq:TDHFbeal}
\ceq
The single-particle Hartree-Fock Hamiltonian reads
\oeq
h[\ro]=t+U[\ro]
\label{eq:h}
\ceq
with the self-consistent mean-field 
\oeq
U[\ro]_{ij} = \sum_{kl} \sdf\bar{v}_{ikjl} \sdf \ro_{{lk}}.
\label{eq:Uro}
\ceq

{
To show the equivalence between Eq.~(\ref{eq:erhenfest_detail}) and the TDHF equation, let us start with the  term associated with the kinetic energy:
\oeqn
\< \, [ \, \oad_i \, \oa_j \, , \, \oad_k \, \oa_l \, ] \, \>
&=& \del_{jk} \,  \<  \oad_i \, \oa_l \> -  \<  \oad_i \, \oad_k \, \oa_j  \,  \oa_l \>  \nonumber \\
&& - \del_{il} \,  \<  \oad_k \, \oa_j \>+   \<  \oad_k \, \oad_i \, \oa_l  \,  \oa_j \>  \nonumber \\
&=& \del_{jk} \,  {\ro}_{li} - \del_{il} \,  {\ro}_{jk} .\nonumber
\ceqn
The kinetic energy term reduces to 
$$
\sum_{kl} \sdf t_{kl} \< [\oad_i\, \oa_j\sdf ,\sdf \oad_k \,\oa_l ]\,\>
 = \sum_k \sdf \( t_{jk} \sdf {\ro}_{ki} - t_{ki} \sdf {\ro}_{jk} \)=\[t,\rho\]_{ji}.
\label{eq:terme_cinetique}
$$
For the two-body interaction, we need the expectation value of the commutator  

\oeqn
&&\stb\stb\stb\stb\stb\stb\stb\stb\<\, [ \, \oad_i \, \oa_j \, , \, \oad_k \, \oad_l  \, \oa_n \, \oa_m \, ] \, \>\nonumber\\
&=&\<  \oad_i \, \oad_l  \, \oa_n \, \oa_m  \> \sdf \del_{jk} -  \<  \oad_i \, \oad_k  \, \oa_n \, \oa_m  \> \sdf\del_{jl} 
\nonumber\\
&&+ \<  \oad_i \, \oad_k \, \oad_l  \, \oa_j \, \oa_n \, \oa_m  \>  
 - \<  \oad_k \, \oad_l  \, \oa_n \, \oa_j  \> \sdf \del_{mi} \nonumber\\
 && +  \<  \oad_k \, \oad_l  \, \oa_m \, \oa_j  \> \sdf \del_{ni} 
- \<  \oad_k \, \oad_l \, \oad_i  \, \oa_n \, \oa_m \, \oa_j  \>. \nonumber 
\label{eq:groscom}
\ceqn
The two terms with 6 annihilation/creation operators cancel out and we get
\oeqn
&&\stb\stb\stb\stb\stb\< \, [ \, \oad_i \, \oa_j \, , \, \oad_k \, \oad_l  \, \oa_n \, \oa_m \, ] \, \>\nonumber\\
&=& \( \ro_{{mi}}  \ro_{{nl}} -  \ro_{{ml}}  \ro_{{ni}} \) \del_{jk} + \( \ro_{{mk}}  \ro_{{ni}} -  \ro_{{mi}}  \ro_{{nk}}  \) \del_{jl} \nonumber \\
&&+ \( \ro_{{jl}}  \ro_{{nk}} -  \ro_{{jk}}  \ro_{{nl}} \) \del_{mi} + \( \ro_{{jk}}  \ro_{{ml}} -  \ro_{{jl}}  \ro_{{mk}}  \) \del_{ni} \nonumber 
\ceqn
Altogether, the two-body interaction contribution reduces to 
\oeqn
&&\stb\stb\stb\stb\stb\stb \stb \stb \frac{1}{4}\sum_{klmn}   \bar{v}_{klmn}  \< \, [  \oad_i\, \oa_j ,   \oad_k \,\oad_l \,\oa_n \,\oa_m ]\,\> \nonumber\\
&=& 
\frac{1}{2}  \sum_{klm}  \[ \bar{v}_{jklm} \( \ro_{{li}} \ro_{{mk}} - \ro_{{lk}}  \ro_{{mi}} \) \right.\nonumber \\
& &\left.\stf \stf \stf \stf \stf  +    \bar{v}_{klim} \( \ro_{{jl}} \ro_{{mk}} - \ro_{{jk}}  \ro_{{ml}}  \) \]\nonumber\\
&=& \sum_{klm} \[ \bar{v}_{jklm}  \(\ro_{{li}} \ro_{{mk}} \)- \bar{v}_{klim} \( \ro_{{jk}} \ro_{{ml}}\) \]\nonumber\\
&=&\sum_{k}  \( U[\ro]_{jk}  \ro_{{ki}} - U[\ro]_{ki}  \ro_{{jk}} \)=\[U[\rho],\rho\]_{ji}\nonumber
\ceqn
where we have used $\bar{v}_{klmn} = - \bar{v}_{klnm} = - \bar{v}_{lkmn}$.
Gathering the kinetic and interaction terms gives Eq.~(\ref{eq:TDHFbeal}).

We see that the TDHF equation is obtained by solving the BV variational principle with the variational spaces restricted to Slater determinants for the state of the system and to one-body operators for the observable. 
It is interesting to see that only the variation of $\oA$ is needed to get the TDHF equation. 
In addition, solving the TDHF equation allows to compute any one-body observable, and the equation does not depend on the final time $t_1$ entering the BV action. 
These properties are specific to the TDHF case. 
In a more general situation, the resulting equations of motion are obtained from both the variation of $\oA$ and $\oD$, and the results are valid for only one observable $\oA_1$ and one final time $t_1$.

We recall that the TDHF equation is  optimised for the expectation value of one-body operators and its predictive power for other purposes may be questionable. 
In particular, two-body operators and fluctuations of one-body operators are outside the variational space of the observables used to derive the TDHF equation.  
In the next section, we solve the BV variational principle in order to compute such fluctuations. (The case of more general two-body operators would be more complicated and the resulting equations of motions are not expected to be easily solvable numerically.) 

\subsection{Equivalence between Hamiltonian and EDF approaches \label{sec:equivalence}}

The BV derivation of the TDHF equation makes explicit use of an expression for the Hamiltonian. 
This is at variance with the derivation based on the stationarity of the Dirac action (see appendix~\ref{annexe:standardTDHF}) in which $\<\phi|\oH|\phi\>$ is replaced by the EDF $E[\rho]$. 
To demonstrate that both approaches are equivalent, it is sufficient to show that \cite{simenel2018}
\oeq\<\phi|\[\oad_\beta\oa_\al\,,\,\oH\]|\phi\>=\[h,\rho\]_{\al\beta}\label{eq:goal}\ceq
for any  $\oH$ and Slater $|\phi\>$, with 
\oeq h_{\al\be}=\frac{\delta \<\phi|\oH|\phi\>}{\delta \rho_{\be\al}}=\frac{\delta E[\rho]}{\delta \rho_{\be\al}}.\ceq
Let us use the eigenbasis  $\{|\nu\>\}$  of $\oH$ with eigenenergies $\{E_\nu\}$ to rewrite 
\oeqn
\<\phi|\[\oad_\beta\oa_\al\,,\,\oH\]|\phi\>
&=&\sum_\nu E_\nu\<\phi|\[\oad_\be\oa_\al\,,\,|\nu\>\<\nu|\]|\phi\>.\nonumber
\ceqn
We now expand  $|\nu\>$ in the $n-$particle $n-$hole basis 
\oeqn
|\nu\> &=& C^\nu_0|\phi\> + \sum_{ph} C^\nu_{ph} \oad_p\oa_h|\phi\>\nonumber\\
&&+ \sum_{pp'hh'} C^\nu_{pp'hh'}\oad_p\oad_{p'}\oa_h\oa_{h'}|\phi\>+\cdots,
\label{eq:nu}
\ceqn
with $C^\nu_{ph}=\<\phi|\oad_h\oa_p|\nu\>$ and $C^\nu_{hp}=0$, leading  to 
\oeq
\<\phi|\[\oad_\be\oa_\al,\oH\]|\phi\> = \sum_\nu E_\nu \(C^\nu_{\al\be} {C_0^\nu}^*-C_0^\nu {C^\nu_{\be\al}}^*\).
\label{eq:commoad}
\ceq
Using $\rho_{\al\be}=\<\phi|\oad_\be\oa_\al|\phi\>\equiv \<\oad_\be\oa_\al\>$, the $r.h.s.$ of Eq.~(\ref{eq:goal}) can be expressed  as 
\oeqn
\[h,\rho\]_{\al\be} 
&=& \sum_\gamma \(\frac{\delta \<\oH\>}{\delta\<\oad_\al\oa_\gamma\>}\<\oad_\be\oa_\gamma\> - \<\oad_\gamma\oa_\al\>\frac{\delta \<\oH\>}{\delta\<\oad_\gamma\oa_\beta\>}\). \nonumber\\\label{eq:comm}
\ceqn
Using $\<\oH\>=\sum_\nu E_\nu \<\phi|\nu\>\<\nu|\phi\>$ and  Eq.~(\ref{eq:nu}), we can show that the functional derivatives in Eq.~(\ref{eq:comm}) include terms like
\oeq
\frac{\delta ( \<\phi|\nu\>\<\nu|\phi\>)}{\delta\<\oad_\al\oa_\gamma\>} = C^\nu_{\al\gamma}\<\nu|\phi\> + {C^\nu_{\gamma\al}}^* \<\phi|\nu\>. 
\ceq
For a Slater determinant, we have $\rho_{hh'}=\delta_{hh'}$, the other terms being zero. 
Therefore, $C^\nu_{\gamma\al}\,\rho_{\gamma\be}=C^\nu_{\gamma\al}\,\rho_{\be\gamma}=0$ 
as $\gamma$ cannot be a particle and a hole at the same time. 
Equation~(\ref{eq:comm}) then becomes
\oeqn
\[h,\rho\]_{\al\be}&=&\sum_\nu E_\nu \sum_\gamma \(C_{\al{\gamma}}^\nu \<\nu|\phi\>\rho_{\gamma\be}-\rho_{\al{\gamma}} {C_{\be{\gamma}}^\nu}^*\<\phi|\nu\>\)\nonumber\\
&=& \sum_\nu E_\nu \(C_{\al{\be}}^\nu {C_0^\nu}^*- {C_{\be{\al}}^\nu}^*C_0^\nu\),
\ceqn
which, according to Eq.~(\ref{eq:commoad}), is equal to $\<\phi|\[\oad_\be\oa_\al,\oH\]|\phi\>$. 

\subsection{Fluctuations and correlations of one-body observables \label{sec:fluccor}}

Let $\oX$ and $\oY$ be two observables. Their correlation in the state $|\Psi\>$ is defined as
\oeq
\sigma_{XY}=\sqrt{\frac{1}{2}\(\<\oX\oY\>+\<\oY\oX\>\)-\<\oX\>\<\oY\>}.
\ceq
The case $\oX=\oY$ defines the fluctuations of $\oX$:
\oeq
\sigma_{XX}\equiv\sigma_X=\sqrt{\<\oX^2\>-\<\oX\>^2}.
\ceq

If $\oX$ and $\oY$ are one-body operators, we see that $\sigma_{XY}^2$ includes the expectation value of the {\it square} of a one-body operator, which contains a two-body contribution.
{
Indeed,
\oeqn
\oX\oY&=&\sum_{\al\be}X_{\al\be}\oad_\al\oa_\be\sum_{\mu\nu}Y_{\mu\nu}\oad_\mu\oa_\nu\nonumber \\
&=&\sum_{\al\be\mu\nu} X_{\al\be} Y_{\mu\nu} (\delta_{\beta\mu}\oad_\al\oa_\nu-\oad_\al\oad_\mu\oa_\be\oa_\nu).\nonumber
\ceqn
The last term is clearly of a two-body nature. 
}
The TDHF theory is then not optimised for the prediction of correlations and fluctuations of one-body operators.
 
A possible improvement would be to solve the BV variational principle with a variational space for the observable which includes square of one-body operators~\cite{flo82}. This approach, however, leads to complicated equations of motion due to an intricate coupling between the evolution of the observable $\oA(t)$ and of the state $\oD(t)$. 

Fluctuations and correlations of one-body operators $\oQ_i=\sum_{\al\be}Q_{i_{\al\be}}\oad_{\al}\oa_\be$ can also be computed from the expectation value of 
\oeq
\oA_1\equiv e^{-\sum_i\varepsilon_i\oQ_i}
\ceq
in the limit $\varepsilon_i\rightarrow0$.
Indeed, 
\oeq
\ln\<\oA_1\> = -\sum_i\varepsilon_i\<\oQ_i\>+\frac{1}{2}\sum_{ij}\varepsilon_i\varepsilon_jC_{ij} +O(\varepsilon^3),\label{eq:lnA}
\ceq
where $C_{ij} =\sigma_{Q_iQ_j}^2$.
The linear and quadratic dependences in $\varepsilon$ of  $\ln\<\oA_1\>$ then lead to the expectation values and fluctuations/correlations of the one-body observables $\oQ_i$, respectively.

In Ref.~\cite{bal84}, Balian and V\'en\'eroni applied their variational principle with a variational space for the observable parametrised by
\oeq
\oA(t)=e^{-\oL(t)}=e^{-\sum_{\al\be}L_{\al\be}(t)\oad_\al\oa_\be}, 
\label{eq:defA}
\ceq
while the density matrix is constrained to be that of an independent particle 
state. 
For a pure state, the latter takes the form
\oeq
\oD(t) = |\phi(t)\>\<\phi(t)|,
\ceq
where $|\phi(t)\>$ is a Slater determinant. 
In fact, the original derivation of Balian and V\'en\'eroni~\cite{bal92} involves more general mean-field states of the form
\oeq
\oD(t) = e^{-m(t)-\oM(t)}=e^{-m(t)-\sum_{\al\be}M_{\al\be}(t)\oad_{\al}\oa_{\be}}.
\label{eq:defD}
\ceq
{
The particular case of a Slater determinant can be obtained with this parametrisation by letting the eigenvalues of the matrix $M(t)$ tend to $\pm\infty$, with the normalisation factor $m(t)$ also tending to $+\infty$ in such a way that the norm $z(t)=\Tr(\oD(t))$ is equal to~1~\cite{bal85}.
In this case, the Slater determinant $|\phi(t)\>$ is built from the eigenvectors of $M(t)$ associated with the eigenvalues~$-\infty$. }
Comparing eqs.~(\ref{eq:defA}) and~(\ref{eq:defD}), we see that both variational spaces for the state and the observable are similar, being both composed of exponentials of one-body operators. 
The resolution of the BV variational principle with this choice of variational spaces and assuming the boundary conditions given in eqs.~(\ref{eq:BCD}) and~(\ref{eq:BCA}) can be found in Refs.~\cite{bal92,bro09} 
and in Appendix~\ref{annexe:BV}.

Expanding the one-body density matrix in terms of $\varepsilon$,
\oeq
\rho(t)=\rho^{(0)}(t)+\rho^{(1)}(t)+O(\varepsilon^2),
\ceq
where $\rho^{(1)}$ is of the order $\varepsilon$,
the main results are:
\begin{itemize}
\item The expectation value of $\oQ_i$ is given by
\oeq 
\<\oQ_i\>(t_1)=\tr \(\ro^{(0)}(t_1)Q_i\)
\ceq
where $\rho^{(0)}$ is given by the TDHF equation~(\ref{eq:tdhf}) with the boundary condition
\oeq
\rho^{(0)}_{\al\be}(t_0)=\Tr\(\oD_0\oad_\be\oa_\al\)
\ceq
and $\oD_0$ is the initial density matrix. 
\item The fluctuations/correlations $\sigma_{Q_iQ_j}=\sqrt{C_{ij}}$ obey
\oeq
\sum_j \varepsilon_jC_{ij}(t_1)
=\sum_j \varepsilon_j C_{ij}^{TDHF}(t_1)-\tr\(\ro^{(1)}(t_1)Q_i\)
\label{eq:BVprescription1}
\ceq
where 
\oeqn
\stb \stb \stb C_{ij}^{TDHF}(t_1)&=&\frac{1}{2}\tr\(\ro^{(0)}(t_1)[Q_i,Q_j]\)\nonumber\\
&+&\tr\[Q_i\rho^{(0)}(t_1)Q_j(1-\rho^{(0)}(t_1))\]
\label{eq:CiiTDHF}
\ceqn
are the (square of the) fluctuations/correlations obtained from the standard TDHF approach. 
\end{itemize}
Equation~(\ref{eq:BVprescription1}) gives fluctuations and correlations which differ from the standard TDHF result [Eq.~(\ref{eq:CiiTDHF})]. 
This is not surprising as only Eq.~(\ref{eq:BVprescription1}) is optimised for these fluctuations/correlations.
The result in Eq.~(\ref{eq:BVprescription1}) takes into account possible fluctuations around the TDHF mean-field evolution in the small amplitude limit, i.e., at the RPA level~\cite{bal84,bal92,ayi08}. 

The additional term in Eq.~(\ref{eq:BVprescription1}) involves $\rho^{(1)}$, i.e., the part of the one-body density matrix which is linear in~$\varepsilon$. However, the equation of motion for the latter is not trivial. Fortunately, for a Slater determinant, Eq.~(\ref{eq:BVprescription1}) can be re-written so that it is easier to implement. The final result reads (see  appendix~\ref{annexe:BV})
\oeqn
C_{ij}(t_1) = \lim_{\varepsilon_i,\varepsilon_j\rightarrow0} \frac{1}{2\varepsilon_i\varepsilon_j}
&\tr&\[\(\rho^{(0)}(t_0)-\eta_i(t_0,\varepsilon_i)\)\right.\nonumber\\
&&\stb \stb \stb\left.\(\rho^{(0)}(t_0)-\eta_j(t_0,\varepsilon_j)\)\],
\label{eq:CijBV}
\ceqn
where the single-particle matrices $\eta(t,\varepsilon)$  obey the TDHF equation~(\ref{eq:tdhf}) with a boundary condition defined at the final time $t_1$:
\oeq
\eta_j(t_1,\varepsilon_j)=e^{i\varepsilon_jQ_j}\rho^{(0)}(t_1)e^{-i\varepsilon_jQ_j}.
\label{eq:BCeta}
\ceq
The fluctuations $\sigma_{Q_i}=\sqrt{C_{ii}}$ are determined by taking $Q_i=Q_j$, leading to
\oeq
C_{ii}(t_1) = \lim_{\varepsilon_i\rightarrow0} \frac{1}{2\varepsilon_i^2}
\tr\[\(\rho^{(0)}(t_0)-\eta_i(t_0,\varepsilon_i)\)^2\],
\label{eq:CiiBV}
\ceq
with the boundary condition given in Eq.~(\ref{eq:BCeta}).
Hereafter, the BV prescription to fluctuations refers to Eqs.~(\ref{eq:BCeta}) and (\ref{eq:CiiBV}), which are also called the time-dependent random phase approximation (TDRPA) equations. 

The fluctuations are generated by the transformation in Eq.~(\ref{eq:BCeta}) and propagated in the backward Heisenberg picture from $t_1$ to $t_0$ according to the dual of the TDRPA  equation.
This is why $C_{ij}(t_1)$ is expressed as a function of density matrices at the initial time $t_0$. 
It is easy to show that, if the backward trajectories $\eta_i$ have the same mean-field as the forward evolution, then Eq.~(\ref{eq:CijBV}) leads to the TDHF expression in Eq.~(\ref{eq:CiiTDHF}).
If, however, deviations occur around the original mean-field, then additional terms appear and lead to a variation of $C_{ij}(t_1)$.

Equation~(\ref{eq:BCeta}) imposes to solve the TDHF equation first for $\rho^{(0)}(t)$ forward in time, and then for $\eta(t)$ backward in time. 
Numerical applications solving Eq.~(\ref{eq:CiiBV}) with the boundary condition given in Eq.~(\ref{eq:BCeta}) are detailed in~\cite{tro85,mar85,bon85,bro08,bro09,sim11,williams2018,godbey2020b,gao2025}.
In practice, several (typically $\sim5$ in~\cite{sim11}) backward TDHF trajectories with different but small values of $\varepsilon$ are performed to compute the limit in Eq.~(\ref{eq:CiiBV}). 
Figure~\ref{fig:BVnum} gives a schematic illustration of the numerical technique.
In Ref.~\cite{sim11}, Eq.~(\ref{eq:CijBV}) is also solved. It is used to compute the correlations between proton and neutron numbers in fragments following deep-inelastic collisions in addition to their fluctuations  (see section~\ref{sec:DIC}). 

In principle, TDRPA predictions of fluctuations of the particle number in the fragments can only be converted into a width of the associated distribution if the latter is Gaussian. 
In particular, any skewness in the distribution can lead to unphysical predictions in terms of width of fragment mass and charge distributions. See supplemental material of \cite{williams2018} for a detailed discussion. 

\begin{figure}
\begin{center}
\includegraphics[width=5cm]{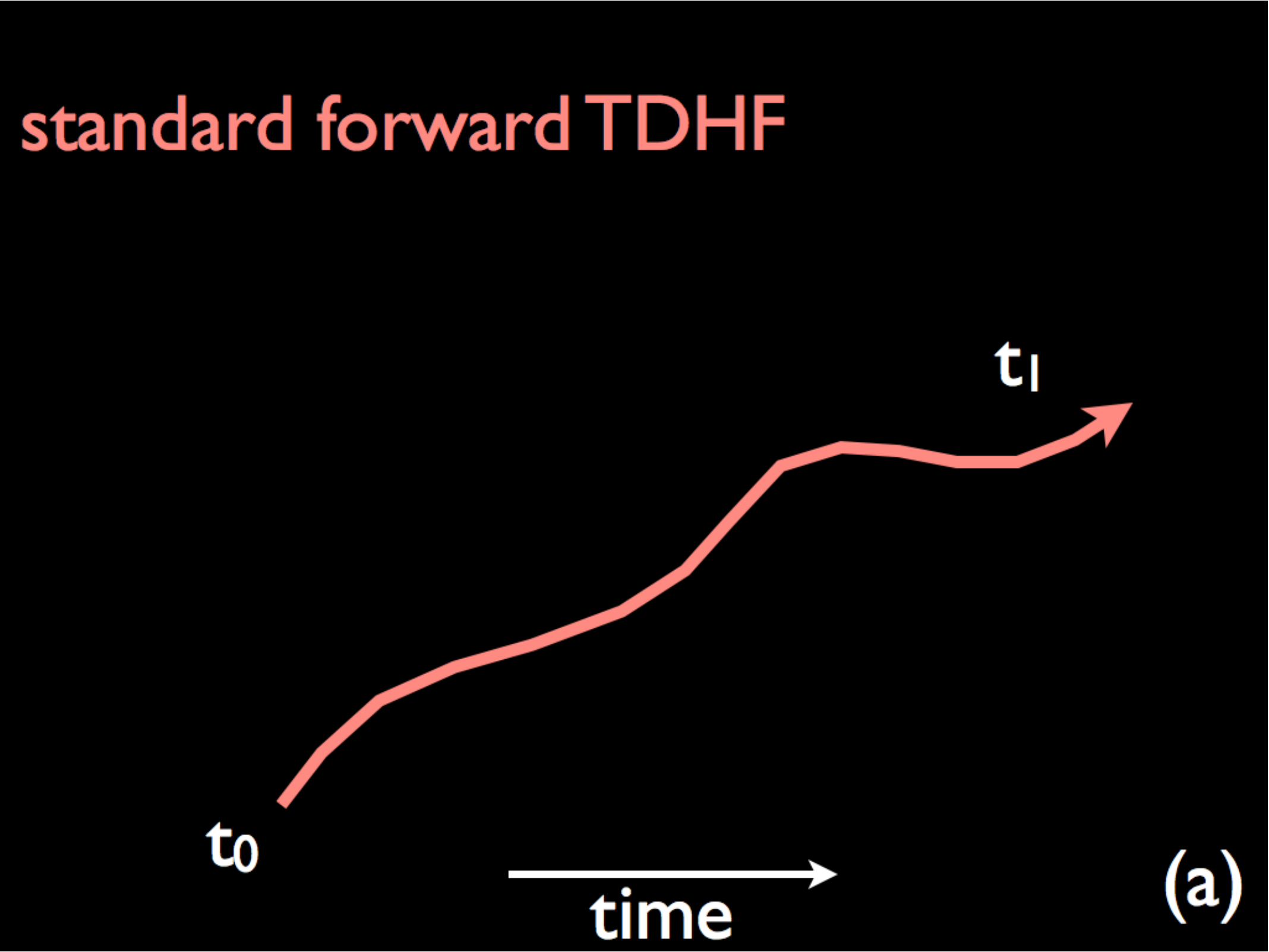}
\includegraphics[width=5cm]{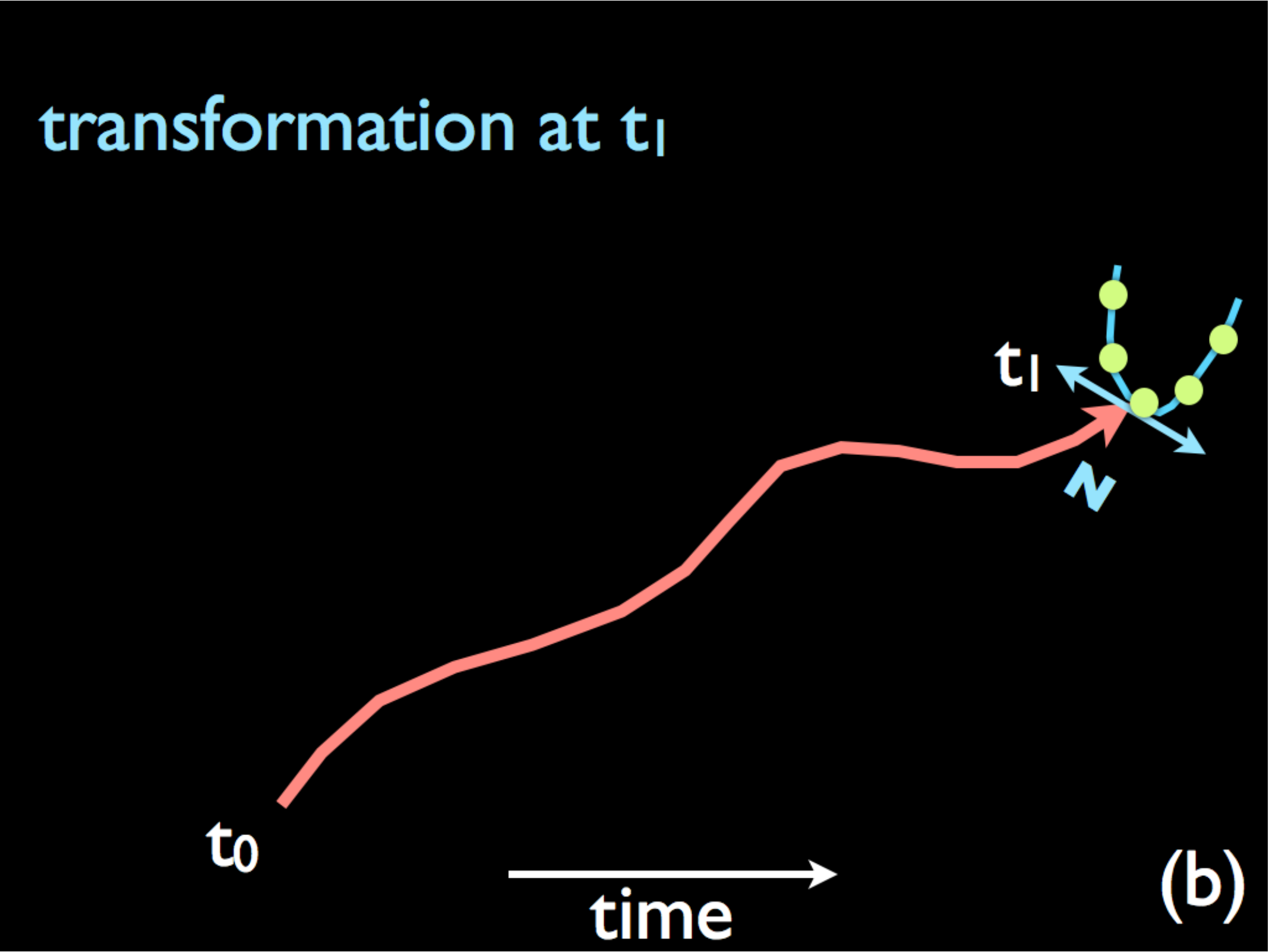}
\includegraphics[width=5cm]{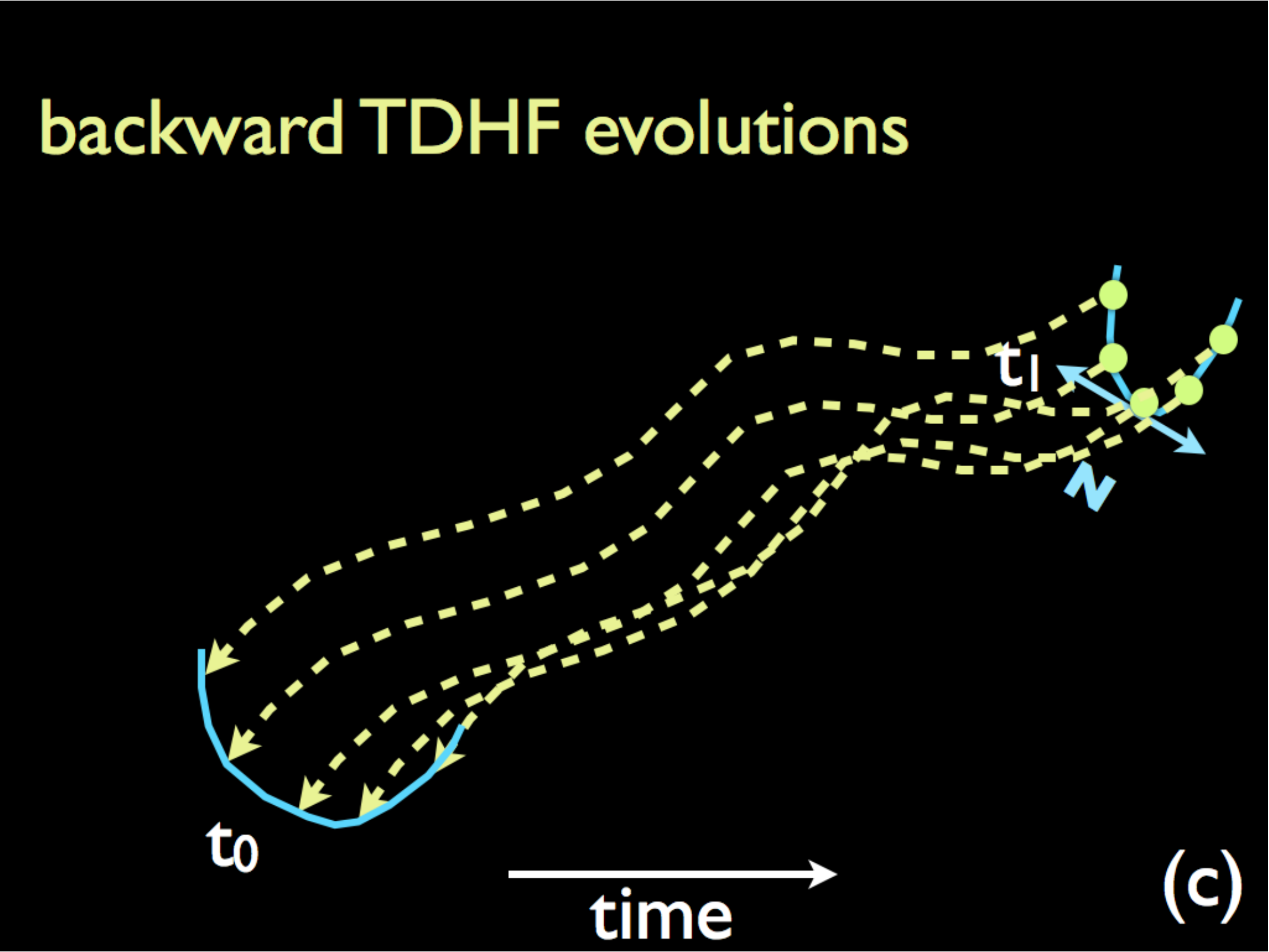}
\end{center}
\caption{Schematic illustration of the main numerical steps to compute the fluctuation of the one-body observable $\oN$ (e.g., neutron numbers in one fragment following a heavy-ion collision). (a) A standard TDHF evolution is performed from $t_0$ to $t_1$. (b) The transformation given in Eq.~(\ref{eq:BCeta}) is applied on the one-body density matrix at time $t_1$ with several (small) values of $\varepsilon$. (c) Backward TDHF evolutions from $t_1$ to $t_0$ are performed for each $\varepsilon$ and the resulting one-body density matrices at time $t_0$ are used to compute the (square of the) fluctuations from Eq.~(\ref{eq:CiiBV}).
\label{fig:BVnum}}
\end{figure}

It is important to note that, although eqs.~(\ref{eq:CijBV}) and (\ref{eq:CiiBV}) are rather easy to solve numerically with a slightly modified TDHF code, they only provide the fluctuations and correlations for a specific choice of $\oQ_i$ and $\oQ_j$ at the time $t_1$. 
Another choice of operator(s) and final time implies to compute numerically another set of backward TDHF evolutions. 
This is not the case for the calculation of expectation values of one-body observables which only requires one forward TDHF evolution. 
Indeed, solving the TDHF equation is an initial value problem and the resulting one-body density matrix can be used to compute any one-body observable (i.e., the TDHF equation depends neither on $t_1$ nor on $\oQ_{i}$). 
As a result, the determination of fluctuations and correlations of one-body observables from the BV variational principle is much more computational time consuming than standard TDHF calculations of their expectation values. 

Finally, it is worth mentioning an alternative derivation of Eq.~(\ref{eq:CijBV}) considering small fluctuations in the initial state at time $t_0$ and propagated within the stochastic mean-field (SMF) approach~\cite{ayi08}.  
Numerical applications of the SMF method can be found in Refs.~\cite{ayi09,was09b,yil11,lacroix2014,ayik2015,ayik2016,tanimura2017,sekizawa2020,ayik2020,ayik2020b,ayik2021,arik2023,ayik2023a,ayik2023b}, where fluctuations are assumed to be small and can then be evaluated from mean-field evolutions. 
Note that another implementation of SMF, called phase space approximation, has also been proposed. 
In this approach, initial correlations are propagated explicitly, allowing for large fluctuations as well as possible bifurcations in the many-body dynamics (see Refs. \cite{lacroix2012,lacroix2013,lacroix2014,lacroix2014b,yilmaz2014a,lacroix2016,ulgen2019,lacroix2022,lacroix2024} for applications in various contexts). 
See also Ref. \cite{czuba2020} that combines SMF with BBGKY.

\subsection{Pairing correlations\label{sec:pairingTh}}

Pairing correlations are essentially due to an attractive short range contribution of the residual interaction in the $^1S_0$ channel\footnote{This notation means that the two nucleons are coupled to produce a total isospin 1, a total orbital angular momentum $L=0$ ($S-$wave), and a total spin 0.}~\cite{may50,dea03}.
Pairing correlations affect then mostly (but not only) time-reversed states. 
The pairing residual interaction induces a scattering of a pair of nucleons across the Fermi surface.
As a result, the ground-state with pairing correlations is primarily a sum of $2p2h$ states where the $2p$ ($2h$) are essentially time-reversed states. 
Such a state is represented schematically in the upper part of Fig.~\ref{fig:pairing}.

\begin{figure}
\includegraphics[width=8.8cm]{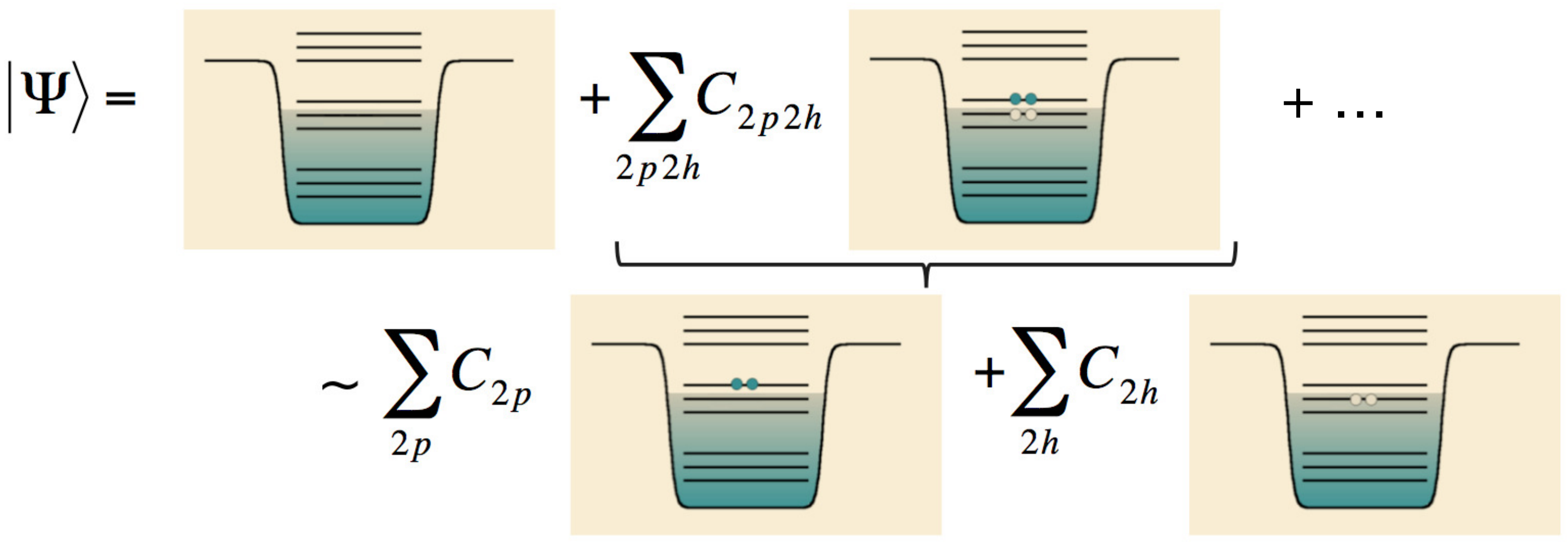}
\caption{Illustration of the configurations used to generate a ground state with pairing correlations. (top) Exact case. (bottom) BCS approximation. 
\label{fig:pairing}}
\end{figure}

The treatment of pairing correlations in finite nuclei is simplified with the Bardeen-Cooper-Schrieffer (BCS) approximation initially developed to interpret supraconductivity in metals~\cite{bar57}. 
In this approximation, the $2p2h$ states are replaced by a sum of $2p$ and $2h$ states (see bottom part of Fig.~\ref{fig:pairing}).
The resulting approximation of the ground-state can then be written as a vacuum of quasiparticles, allowing for the application of the Wick theorem~\cite{rin80}. 
As a result, the BCS approximation leads to a {\it generalised} mean-field theory. 

As we can see in Fig.~\ref{fig:pairing}, the ``price to pay'' is that the BCS ground-state is not an eigenstate of the particle number operator anymore. 
Pairing correlations are  included thanks to a gauge symmetry breaking. 
The Hartree-Fock-Bogoliubov (HFB) theory is more general but shares the same features with the BCS approach. 
It is an extension to the BCS theory where pairs are not limited to time-reversed states. 

In the (TD)HFB theory, all the information on the state of the system is contained in the {\it generalised} density matrix ${\cal R}$ defined as 
\begin{eqnarray}
{\cal R}  = \left( 
\begin{array} {cc}
\left( \langle \oad_j \oa_i \rangle \right) & \left(\langle \oa_j^{ } \oa_i \rangle \right)\\
&\\
\left(\langle \oad_j \oad_i \rangle \right) &  \left(\langle \oa_j \oad_i \rangle   \right)
\end{array} 
\right)
 = \left( 
\begin{array} {cc}
\rho & \kappa \\
- \kappa^* & 1-\rho^*  
\end{array} 
\right)\!,
\label{eq:Rmatrix}
\end{eqnarray}
where $\kappa$ is the so-called {\it pairing tensor}.
$\kappa$ and $\kappa^*$ contain the pairing correlations (at the HFB level). 

The time evolution of the generalised density matrix is given by the TDHFB equation 
\begin{eqnarray}
i \hbar \frac{d{\cal R}}{dt} = \left[{\cal H} , {\cal R} \right] 
\label{eq:tdhfbR}
\end{eqnarray}
which has the same form as the TDHF equation~(\ref{eq:tdhf}). 
The generalised HFB Hamiltonian reads
\begin{eqnarray}
{\cal H}  \equiv \left( 
\begin{array} {cc}
h & \Delta \\
- \Delta^* & - h^*
\end{array} 
\right),\label{eq:Hmatrix}
\end{eqnarray}
where 
\begin{eqnarray}
h_{\mu\nu}=\frac{\delta \mathcal{E}[\rho,\kappa,\kappa^*]}{\delta \rho_{\nu\mu}} \mbox{~~and~~} 
\Delta_{\mu\nu}=\frac{\delta \mathcal{E}[\rho,\kappa,\kappa^*]}{\delta
 \kappa^*_{\mu\nu}} 
 \label{eq:field}
\end{eqnarray}
are the HF Hamiltonian and the pairing field, respectively, and $ \mathcal{E}[\rho,\kappa,\kappa^*]$ is the EDF including pairing. 

Simplifications can be obtained by using the  
BCS approximation where the time-dependent state is written \cite{blocki1976,reinhard1997,eba10,sca12}
\begin{equation}
|\Psi(t)\rangle = \prod_{k>0} \left( u_k(t) + v_k(t) a^{\dagger}_k(t) a^{\dagger}_{\overline k}(t) \right) | 0 \rangle, \label{eq:tdstate}
\end{equation} 
where $\overline k$ denotes the time-reversed state of $k$. 
The TDHF+ BCS equations (also called TDBCS or canonical basis TDHFB in the literature) are written in terms of the occupation numbers $n_k (t)= v^2_k(t)$ of single-particle states~\cite{eba10,sca12}
and anomalous density components $\kappa_k(t) = u_k^*(t) v_k(t)$
\begin{eqnarray}
i\hbar\frac{d n_k}{dt} &=& \Delta^*_k \kappa_k - \Delta_k \kappa^*_k, \quad\label{eq:TDBCS1}\\
i\hbar\frac{d \kappa_k}{dt} &=&  \kappa_i(\epsilon_k-\epsilon_{\overline{k}}) + \Delta_k(2n_k-1)\;\label{eq:TDBCS2},
\end{eqnarray}
where the pairing field is now diagonal. 
As a result, the evolution of the single particle states closely resembles TDHF equations
\begin{equation}
i\hbar \frac{d}{dt}|\varphi_k\rangle = (h[\rho] - \epsilon_k(t)) |\varphi_k\rangle\;,
\end{equation}
where $\epsilon_k(t) = \langle \varphi_k |h[\rho]| \varphi_k \rangle $~\cite{sca12}.

The BCS approximation can introduce spurious effects such as an artificial low-density gas in the
numerical box when the pairing residual interaction couples to single-particle states
with positive energy. TDHF+BCS also violates the one-body continuity equation, potentially leading to spurious transfer between fragments \cite{sca12,scamps2013a}.
The latter limitation, however, does not seem to impact fission studies \cite{scamps2015a,tanimura2015,tanimura2017}.

A further simplification is obtained with the frozen occupation approximation (FOA) where the occupation numbers $n_k$ of the initial state are kept constant during the dynamics. 
Although the effect of pairing correlations on the initial density distribution are accounted for within the FOA, the approach neglects any subsequent pairing dynamics which could have induced a rearrangement of the occupation numbers. 

\subsection{Skyrme Energy-Density-Functional\label{sec:Skyrme}}

Early TDHF calculations~\cite{bon76,neg82} were based on simplified Skyrme effective interactions~\cite{sky56}.
Modern applications, however, use full Skyrme functionals including spin-orbit interactions to reach a realistic description of the strong interactions between the nucleons. 
Studies with the Gogny effective interaction~\cite{has12,hashimoto2013,hashimoto2016,scamps2017b,scamps2019b} and with covariant energy density functionals \cite{ren2020a,ren2020b,ren2022,ren2022b,li2023,zhang2024a,zhang2024b,li2024} have also been reported. 

Modern Skyrme energy density functionals (EDF) are usually derived from a Skyrme effective interaction of the form
\oeqn
\ov (1,2) &=& t_0 \sdf \( 1+x_0\, \oP_\si \) \sdf \odel \nonumber \\
&+& \frac{1}{2} \sdf t_1 \sdf \( 1+x_1\, \oP_\si \) 
\sdf \(\ovk'^2 \sdf  \odel + \odel \sdf \ovk^2 \) \nonumber \\
&+& t_2 \sdf \( 1+x_2\, \oP_\si \) 
\sdf \(\ovk' \cdot  \odel \sdf  \ovk \) \nonumber \\
 &+& \frac{1}{6} \sdf t_3 \sdf \( 1+x_3\, \oP_\si \) 
 \sdf \ro^\al\! (\ovR) \sdf \odel \nonumber \\
 &+& i\, W_0 \sdf \ovsi \cdot \(\ovk' \times \odel \, \ovk \) 
\label{eq:skyrme}
\ceqn
where $\odel = \del\(\ovr(1)-\ovr(2)\)$, $\ovk = \frac{1}{2i}\(\vna(1)-\vna(2)\)$ 
(relative momentum), $\ovk'$ is the complex conjugate of $\ovk$ acting on the left, and  $\ovR = \(\ovr(1)+\ovr(2)\)/2$.
The operators $\ovsi = \ovsi(1)+ \ovsi(2)$, with
$\ovsi(i) = \osi_x\!(i) \, \ve_x+ \osi_y\!(i) \, \ve_y + \osi_z\!(i) \, \ve_z$, 
are expressed in terms of the Pauli matrices $\osi_{x/y/z}(i)$ 
acting on the spin of the particle $i$.
 $\oP_\si = \[1+  \ovsi(1) \cdot \ovsi(2) \]/2$ exchanges the spins of the particles. 
 The particle density in~$\vr$
is noted $\ro(\vr) \equiv \sum_{sq} \ro(\vr s q ,\vr s q )\equiv\rho_q(\mathbf{r}s,\mathbf{r}s')$ where $\rho$ is the one-body density matrix, $s$ the spin and $q$ the isospin.
The ``$t_1$'' and ``$t_2$'' terms are non-local in space and simulate the short range part of the interaction.
Finally the last term accounts for the spin-orbit interaction.

The very interesting aspect of this interaction is its zero range nature, which greatly 
simplifies the mean-field expression in coordinate space (see below).
Parameters ($t_{0-3}$, $x_{0-3}$, $W_0$ and $\al$) are generally adjusted to 
reproduce nuclear properties like saturation and incompressibility of nuclear matter 
and selected properties of finite nuclei (see for instance \cite{cha98}).

In the following of this section, when there is no ambiguity, the spin and isospin indices, as well as the indices $\al$ denoting single-particle states  are omitted in the notation for clarity.
We also  assume that $\ro$ is diagonal in isospin. 
Let us define the usual densities entering the Skyrme EDF:
\oeqn
\ro(\vr) &=& \sum_{\al s}\sdf  \az_{\al }^*(\vr s)\,  \az_{\al }(\vr s) \equiv  \sum_{\al s}\sdf  \az^*\,  \az \label{eq:def_rho}  \\
\tau(\vr) &=&   \sum_{\al s}\sdf  |\vna \az|^2  \label{eq:def_tau}  \\
\vj(\vr) &=& \frac{1}{2\,i} \sum_{\al s} \az^* \sdf \vna \sdf \az \stf +c.c.  \label{eq:def_j}  \\
\vna . \vJ(\vr) &=& -i\sum_{\al ss'}   \vna\az^*(\vr s) \times \vna \az (\vr s')\cdot \bs \vsi \ksp \\
\vS(\vr) &=& \sum_{\al s} \sdf  \az^*(\vr s) \sdf \az (\vr s')\sdf \bs \vsi \ksp ,
\ceqn
where $c.c.$ means ``complex conjugate''. These are the local, kinetic, current, (gradient of) spin-orbit, and spin densities, respectively. 
The $\vj$ and $\vS$ densities are time-odd and vanish in time-reversal invariant systems. 
They are, however, important in time-dependent calculations to ensure Galilean invariance~\cite{eng75}. 

The total energy of an interacting system can be written as an integral of a local energy density
\oeq
E=\int d\vr \mH(\vr).
\ceq
Within the framework of the Skyrme EDF, we have~\cite{bon87}
\oeqn
\mH(\vr) &=& \frac{\hb^2}{2m}\tau + B_1 \rho^2 + B_2 \sum_q\rho_q^2 \nonumber \\
&&+B_3(\rho \tau -\vj^2)+B_4\sum_q(\rho_q \tau_q -\vj_q^2)\nonumber \\
&&+B_5\rho\Delta\rho+B_6\sum_q\rho_q\Delta\rho_q+B_7\rho^{2+\alpha}+B_8\rho^\al \sum_q\rho_q^2\nonumber\\
&&+B_9(\rho\vna\!\cdot\!\vJ + \vj\!\cdot\! \vna\!\times \!\vS+\sum_q\rho_q\vna\!\cdot\!\vJ_q+\vj_q\!\cdot\!\vna\!\times\!\vS_q)\nonumber\\
&&+B_{10} \vS^2+B_{11}\sum_q\vS_q^2 +B_{12}\rho^\al \vS^2 +B_{13} \rho^\al \sum_q\vS_q^2,\nonumber \\
&&\label{eq:Hr}
\ceqn
where the coefficients $B_i$ are the usual Skyrme functional coefficients:
\oeqn
B_1&=&\frac{t_0}{2}\(1+\frac{x_0}{2}\)\nonumber\\
B_2&=&-\frac{t_0}{2}\(x_0+\frac{1}{2}\)\nonumber\\
B_3&=&\frac{1}{4}\[t_1\(1+\frac{x_1}{2}\)+t_2\(1+\frac{x_2}{2}\)\]\nonumber\\
B_4&=&-\frac{1}{4}\[t_1\(x_1+\frac{1}{2}\)-t_2\(x_2+\frac{1}{2}\)\]\nonumber\\
B_5&=&-\frac{1}{16}\[3t_1\(1+\frac{x_1}{2}\)-t_2\(1+\frac{x_2}{2}\)\]\nonumber\\
B_6&=&\frac{1}{16}\[3t_1\(x_1+\frac{1}{2}\)+t_2\(x_2+\frac{1}{2}\)\]\nonumber\\
B_7&=&\frac{t_3}{12}\(1+\frac{x_3}{2}\)\nonumber\\
B_8&=&-\frac{t_3}{12}\(x_3+\frac{1}{2}\)\nonumber\\
B_9&=&-\frac{1}{2}W_0\nonumber\\
B_{10}&=&\frac{t_0x_0}{4}\nonumber\\
B_{11}&=&-\frac{t_0}{4}\nonumber\\
B_{12}&=&\frac{t_3x_3}{24}\nonumber\\
B_{13}&=&-\frac{t_3}{24}.\nonumber\\
\ceqn 

In fact, the complete Skyrme functional is more general and contains other terms of the form $\vS\cdot\Delta\vS$ and with other densities, i.e., the spin-current pseudo-tensor $\stackrel{\leftrightarrow}{J}$ and the spin kinetic energy density $\vT$~\cite{eng75,uma06a}. All or some of these additional terms are sometimes included in TDHF calculations~\cite{uma06a,mar06,loe12,fracasso2012,dai2014a,stevenson2016}. In Eq.~(\ref{eq:Hr}), only the anti-symmetric part of $\stackrel{\leftrightarrow}{J}$, which is the spin-orbit density $\vJ$, is  included. The spin-orbit energy is indeed expected to be more important (by about one order of magnitude) than the other spin-gradient terms~\cite{cha98}.

The Skyrme-HF mean-field is derived from Eq.~(\ref{eq:hHF}) by replacing the expectation value of the Hamiltonian on the Slater determinant by the expression of the Skyrme EDF. 
The action of this field on single-particle wave functions is then given by \cite{bon87}
\oeqn
&&\(h[\ro]  \az_\al \)(\vr, s ) = \nonumber\\
&&\sum_{s'} \!\[\! \(\!-\vna \frac{\hb^2}{2m^*_{q_\al}\!(\vr)} \vna \!+\! U_{q_\al}\!(\vr)\!  +\! i\vC_{q_\al}\!(\vr)\! \cdot\! \vna\! \)\!\delta_{ss'}\right. \nonumber \\
&&\left. + \vV_{q_\al}\!(\vr)\cdot  \bs \vsi \ksp 
 +i\vW_{q_\al}\!(\vr) \cdot \( \bs \vsi \ksp \times \vna \)\frac{}{} \] \az_\al\! (\vr,s').
 \nonumber \\
\label{eq:HFskyrme}
\ceqn
The derivatives act on each term sitting on their right, including the wave function.
The fields (functions of $\vr$) read
\oeqn
\frac{\hb^2}{2\,m_q^*} &=& \frac{\hb^2}{2\,m} + B_3 \,\ro + B_4\, \ro_q \\
U_q &=& 2B_1\ro+2B_2\ro_q+B_3(\tau+i\vna\cdot\vj)+B_4(\tau_q+i\vna\cdot\vj_q)\nonumber\\
&&+2B_5\Delta\ro+2B_6\Delta\ro_q+(2+\al)B_7\ro^{1+\al}\nonumber\\
&&+B_8[\al\ro^{\al-1}\sum_q\ro_q^2+2\ro^\al\ro_q]+B_9(\vna\cdot \vJ +\vna \cdot \vJ_q)\nonumber \\
&&+\al\ro^{\al-1}(B_{12}\vS^2+B_{13}\sum_q\vS_q^2)\\
\vV_q&=&B_9\vna\times (\vj+\vj_q)+2B_{10}\vS+2B_{11} \vS_q\nonumber\\
&&+2\rho^\al(B_{12}\vS+B_{13}\vS_q)\\
\vW_q &=& -B_9 \, \vna \, \(\ro+ \ro_q \)\label{eq:W}\\
\vC_q &=& 2 \, B_3 \, \vj + 2 \, B_4 \, \vj_q - B_9 \,  \vna \times \(\vS + \vS_q \) ,\label{eq:C} 
\ceqn
where the derivatives act on the first term sitting on their right only. 

In addition to this mean-field potential, protons are also affected by the Coulomb interaction.
The direct part of the Coulomb energy reads
\oeq
E_c^{dir}=\frac{ e^2}{2} \int d^3r \int d^3r' \frac{\rho_p(\vr)\rho_p(\vr')}{|\vr-\vr'|}.
\ceq
The latter is usually computed by solving first the Poisson equation to get the Coulomb potential $V_c(\vr)$, and, then, by evaluating the integral $\frac{1}{2}\int d^3r \rho_pV_c$.
The exchange part of the Coulomb energy is usually determined within the Slater approximation as
\oeq
E_c^{ex}=\frac{-3e^2}{4}\(\frac{3}{\pi}\)^{\frac{1}{3}} \int d^3r \rho_p(\vr)^\frac{4}{3}.
\ceq
As a result, the contribution of the Coulomb interaction to the proton mean-field reads 
\oeq
U_c=V_c-e^2\(\frac{3\rho_p}{\pi}\)^{\frac{1}{3}}.
\ceq

In order to describe superfluid systems with a general mean-field approach, the EDF should also be a function of the pairing tensor, or equivalently, the 
anomalous density 
$\tilde{\rho}_q(\mathbf{r}s,\mathbf{r}'s') = -2s'\kappa_q(\mathbf{r}s,\mathbf{r}'-s')$.
The pairing functional is usually assumed to be local and often written as (see, e.g.,~\cite{bender2003} and references therein)
 \begin{equation}
\mathcal{E}_{pair} = \int {d^3r}
\,\,\,  \frac{g}{4} \, 
\left[ 1 - \left(\frac{\rho\ofR}{\rho_c}\right)^\gamma \right]
 \sum_{q} \tilde{\rho}_q^*(\mathbf{r})\,  \tilde{\rho}_q(\mathbf{r}) 
\label{pairfunc}
\end{equation}
where $\tilde{\rho}_q(\mathbf{r})=\sum_{s} \tilde{\rho}_q(\mathbf{r}s,\mathbf{r}s)$, and $g$, $\rho_c$ and $\gamma$ are parameters. 
To avoid divergences, one should use a regularization scheme, such as introducing a cutoff in the quasiparticle spectrum~\cite{bulgac1999,bulgac2002b,dobaczewski1996}.

\subsection{Numerical solution to the TDHF(B) equation}


As we saw in section \ref{sec:fluccor},  the calculations of both expectation values and fluctuations/ correlations of one-body observables imply to determine the time evolution of the one-body density matrix with the TDHF equation~(\ref{eq:tdhf}). 
Few numerical codes solving the TDHF equation in three dimensions with a full Skyrme EDF including spin-orbit terms are now available. 
The \textsc{tdhf3d} solver initially written by Paul Bonche \cite{kim97} has been expanded to solve the TDHF+BCS  \cite{scamps2013a} and TDRPA \cite{sim11} equations.
Unlike \textsc{tdhf3d} which assumes a plane of symmetry, the \textsc{sky3d}  \cite{mar05,maruhn2014,schuetrumpf2018,abhishek2024} and \textsc{hit3d} \cite{shi2024} codes solve the TDHF equation in full 3-dimensional grids. 
\textsc{hit3d} and versions of \textsc{sky3d} \cite{dai2014a,stevenson2016} also include a tensor force.
A version of \textsc{sky3d} that uses the Barcelona-Catania-Paris-Madrid (BCPM) EDF\footnote{The BCPM functional is simpler than Skyrme, leading to a lower computational cost \cite{giuliani2023}.}
 is currently being developed \cite{solermiras2025}.
Full 3-dimensional TDHF solvers like the one developed at Vanderbilt University \cite{uma05} and the one written by the Tsukuba group \cite{nak05,sekizawa2013} have also been used to solve the TDRPA equation \cite{williams2018,godbey2020b,gao2025}. 
The TDHF+BCS equation is solved with Skyrme in \cite{eba10,scamps2013a} and with covariant energy-density functionals in \cite{ren2020a,ren2020b}.
The \textsc{lise} code solves 3-dimensional TDHFB equation with Skyrme functional \cite{jin2021}. 
TDHFB equation has also been solved with the Gogny interaction \cite{has12,hashimoto2013}.

TDHF solvers (and their extensions) were  used in  studies of giant resonance properties \cite{sim03,cho04,ste04,nak05,uma05,mar05,rei07,bro08,sim09,ste11,has12,fracasso2012,avez2013,scamps2013b,pardi2013,goddard2013,scamps2014a,stetcu2015,burrello2019,shi2020,aitbenmennana2020,aitbenmennana2021,li2023b,aitbenmennana2023,marevic2023,shi2023}, low-lying vibrations \cite{simenel2013a,simenel2013b,scamps2013b,simenel2016b,vophuoc2016,burrello2019}, fission \cite{simenel2014a,scamps2015a,tanimura2015,goddard2015,goddard2016,bulgac2016,tanimura2017,scamps2018,scamps2019,bulgac2019,bulgac2019b,pancic2020,bulgac2021,qiang2021a,qiang2021b,bulgac2022,scamps2022,ren2022,iwata2022,ren2022b,scamps2024a,tong2022b,abdurrahman2024,huang2024a,huang2024b}, neutron star crust properties \cite{seb09,seb11,schuetrumpf2013,schuetrumpf2014,schuetrumpf2015a,schuetrumpf2015b,wlazlowski2016,sekizawa2022,yoshimura2024} 
and heavy-ion collisions. Among the latter, fusion and nucleus-nucleus potentials~\cite{kim97,sim01,sim04,mar06,uma06a,uma06b,uma06c,uma06d,guo07,sim07,uma07,sim08,uma08b,was08,uma09a,ayi09,uma09b,uma09c,was09a,uma10b,obe10,loe11,iwa11,leb12,uma12a,obe12,kes12,uma12b,iwata2013,desouza2013,simenel2013a,simenel2013b,dai2014,umar2014a,steinbach2014,vophuoc2016,bourgin2016,hashimoto2016,reinhard2016a,stevenson2016,simenel2017,godbey2017,magierski2017,sekizawa2017b,guo2018,guo2018b,wen2018,sekizawa2019b,scamps2019b,godbey2019b,godbey2019c,li2019,ren2020b,umar2021,tong2022a,sun2022,sun2022b,godbey2022,magierski2022,gumbel2023,tong2023,sun2023,sun2023b,yao2024,desouza2024}, isospin equilibration \cite{sim01,sim07,uma07,iwa09,iwa10a,iwa10b,iwa12,obe12,sim11,sim12,umar2017,stone2017,godbey2017,simenel2020,gumbel2023}, Coulomb excitation \cite{sim04,uma06c,uma06d,uma07}, rotational properties \cite{guo08}, multi-nucleon transfer \cite{sim08,uma08a,was09b,sim10b,eve11,yil11,sim12,scamps2013a,sekizawa2013,sekizawa2014,yilmaz2014,sekizawa2015,ayik2015,sekizawa2016,vophuoc2016,bourgin2016,scamps2017a,scamps2017b,li2019,sekizawa2017,sekizawa2017a,regnier2018,ayik2018,yilmaz2018,ayik2019,ayik2019b,wu2019,sekizawa2020,godbey2020b,simenel2020,wu2020,ayik2021,roy2022,wu2022,arik2023,ayik2023a,ayik2023b,zhang2024a,li2024b}, quasi-fission \cite{wakhle2014,oberacker2014,ayik2015a,umar2015a,hammerton2015,sekizawa2016,umar2016,morjean2017,yu2017,guo2018c,zheng2018,godbey2019,li2019,simenel2021,li2022,stevenson2022,mcglynn2023,lee2024,scamps2024b,li2024c},
breakup \cite{ass09}, deep-inelastic collisions \cite{sim11,wang2016,umar2017,roy2018,williams2018,simenel2020,ayik2020b,jiang2025,gao2025}, reactions at Fermi energy \cite{besse2020}, as well as the dynamics of $\alpha$-clusters \cite{uma10a,iwata2015,schuetrumpf2017,ren2020a,ren2022b,stevenson2020b} and actinide collisions \cite{gol09,ked10,ayik2017,umar2018,ayik2020,zhang2024b} have been investigated. 

In this section we describe numerical aspects of solving time-dependent mean-field equations, with a focus on TDHF.
The inclusion of pairing correlations via solving the TDHFB equation is discussed in section~\ref{sec:TDHFBnum}.

\subsubsection{Numerical method}

Most of these applications were performed on a three dimensional cartesian grid using a time iterative method. 
We give here a summary of the main steps which are usually followed to treat the collision of two nuclei:
\begin{enumerate}
\item Static Hartree-Fock (HF) calculations are performed to determine the initial condition where the nuclei are usually assumed to be in their HF ground state. 
\item The nuclei are placed in a larger box, avoiding any overlap of the HF solutions. This latter condition allows to construct a single Slater determinant from the two initial HF states.
\item A Galilean boost\footnote{In case of a single nucleus, for instance to study its response to a specific excitation, the Galilean boost is replaced by the appropriate velocity boost generating the excitation (examples are given in section~\ref{chap:vib}). Alternatively, one can start with a constrained Hartree-Fock (CHF) solution obtained with an external constraint in the HF calculation. The response to the excitation is then studied by relaxing the constraint in the TDHF calculation~\cite{blo79}.}
 is applied at the initial time assuming that the nuclei followed a Rutherford trajectory prior to this time.
\item The TDHF equation is solved iteratively in time and expectation values of one-body observables are eventually computed at each time step to get their time evolution.
\end{enumerate}
Of course, variations of the main numerical steps described above are possible. 
For instance, one can question the validity of the assumption that the nuclei are in their HF ground state at initial time. In particular, heavy nuclei generate strong Coulomb fields which may induce long range excitations of the collision partners~\cite{sim04}. 


\subsubsection{Centre of mass corrections in heavy-ion collisions}

HF calculations are usually performed with centre of mass corrections to improve the description of the nucleus in its intrinsic frame. This is done by removing spurious centre of mass motion, i.e., by replacing the total kinetic energy $\oT$ by 
\oeqn
\oT-\frac{\ovP^2}{2Am}&=& \oT-\frac{\(\sum_{i=1}^{A}\ovp_i\)^2}{2Am}\nonumber\\
&=&\oT-\frac{1}{2Am} \[\sum_i\ovp_i^2+\sum_{i\ne j}\ovp_i\ovp_j\]\!.
\ceqn
This correction contains a one-body and a two-body contributions. 
Usually, the two-body term is neglected and only the one-body part of the correction is included in standard HF calculations~\cite{cha98}. 

The initial condition of a TDHF calculation uses static HF or constrained Hartree-Fock (CHF) solutions.
However, centre of mass corrections are difficult to incorporate in TDHF calculations of heavy-ion collisions~\cite{uma09c}. 
They are usually neglected to allow a consistent treatment of colliding partners. 
Indeed, these corrections are explicitly dependent on the number of nucleons $A_i$ of the collision partner $i$ and would induce a different treatment of the single-particle wave-functions depending on which nucleus they come from. 

To treat structure and dynamics on the same footing, one can also neglect the centre-of mass corrections in the initial HF calculations~\cite{kim97}. However, one should then use an EDF which has been fitted {\it without} these corrections.
This is the case of the SLy4$d$ \cite{kim97} and UNEDF1 \cite{kortelainen2012} parametrisations of the Skyrme EDF.

\subsubsection{Numerical approximations and algorithm}

Most of the numerical approximations and techniques used in modern TDHF codes are based on similar algorithms, but may contain differences such as in the calculation of spatial derivatives.  For instance, the latter are computed with finite-difference formulae in the \textsc{tdhf3d}~\cite{kim97} and \textsc{HIT3D}~\cite{shi2024} codes, while spline and fast-Fourier transform (FFT) techniques are used in Ref.~\cite{uma05} and in \textsc{Sky3D} \cite{mar05,maruhn2014}, respectively.  
Typical regular mesh spacing with $\Delta x\simeq0.6$~fm \cite{sim09}, 0.8~fm \cite{kim97} and 1.0~fm \cite{uma05,mar05} are used, although adaptive grids have been also considered~\cite{nak05}.
Codes based on the Gogny interaction often use harmonic oscillator basis \cite{has12}, although the TDHFB code in \cite{hashimoto2013} uses one-dimensional spatial grid points of Lagrange mesh in one direction and two-dimensional harmonic-oscillator basis functions in the other two directions.

Modern TDHF codes contain all the time-odd and even terms of standard Skyrme EDF (see sec.~\ref{sec:Skyrme}).
The inclusion of time-odd terms is indeed crucial for a proper description of translational motion and to avoid spurious excitations~\cite{mar06}.
As time-reversal symmetry is not assumed, the code contains no degeneracy of single-particle wave-functions. 
This means that up to $\sim500$ (for, e.g., two actinides without pairing) wave-functions are evolved in time. 

Due to the self-consistency of the  mean-field, the TDHF equation needs to be solved iteratively in time 
with a small time step increment $\Delta t$ with typical values ranging from $\sim5\times10^{-25}$~s \cite{nak05,sim09} to $\sim1.5\times10^{-25}$~s \cite{kim97,uma05}.
Over small time intervals $\[t,t+\Del t\]$, the HF Hamiltonian is assumed to be constant.
However, to conserve  energy, the numerical algorithm should be symmetric with respect to time-reversal operation.
This implies to consider the Hamiltonian  at time $t+\frac{\Del t}{2}$ for the evolution of single-particle wave-functions from $t$ to $t+\Del t$~\cite{bon76}
\oeq
|\nu(t+\Del t)\> \approx e^{-i \frac{\Del t}{\hb} \oh\(t+\frac{\Del t}{2}\)} \sdf |\nu(t)\>.
\ceq
A schematic illustration of a possible real time propagation is
\oeq
\begin{array}{ccc}
\{ |\nu_1^{(n)}\> \cdots |\nu_N^{(n)} \>\} & \Rightarrow &\ro^{(n)}\\
\Uparrow && \Downarrow\\
|\nu_i^{(n+1)}\> =  e^{-i\frac{\Delta t}{\hb}  \oh^{(n+\frac{1}{2})}}  |\nu_i^{(n)}\>
 && \oh^{(n)}\equiv\oh[\ro^{(n)}] \\
\Uparrow && \Downarrow\\
 \oh^{\(n+\frac{1}{2}\)}\equiv\oh\[\ro^{\(n+\frac{1}{2}\)}\] &&
   |\tilde{\nu}_i^{(n+1)}\> = e^{-i\frac{\Delta t}{\hb} \oh^{(n)}}  |\nu_i^{(n)}\> \\ 
\Uparrow & &   \Downarrow \\
\ro^{\(n+\frac{1}{2}\)}= \frac{\ro^{(n)} + \tro^{(n+1)}}{2} & \Leftarrow & \tro^{(n+1)} \\
\end{array} 
\label{eq:algo}
\ceq
where $|\nu^{(n)}\>$ corresponds to an approximation of $|\nu(t_n=n\Del t)\>$.
In this algorithm, starting from the density at time $t$, a first estimate of the density at time $t+\Del t$, denoted by  $\tro^{(n+1)}$ is obtained.
The Hamiltonian used in the propagator is computed using the average density obtained from  $\ro^{(n)}$ and $\tro^{(n+1)}$. 
Then, the new density at time $t+\Del t$ is obtained using this Hamiltonian. 
An approximate form of the exponential function is generally used which in some cases, breaks the unitarity, and orthonormalisation of the single particle states must be controlled.

\subsubsection{Boundary conditions}

TDHF calculations on a spatial grid require boundary conditions. 
Hard and periodic boundary conditions may lead to 
spurious effects when low-density tails reach the box boundaries, causing problems in the conservation of energy and angular momentum \cite{guo08}.

A multigrid approach \cite{degiovannini2012} may be used to extend the boundaries.
Alternatively, emitted particles may be removed via radiating or exact boundary conditions \cite{boucke1997,mangin-brinet1998,pardi2013,pardi2014}. 
The twist-averaged boundary conditions (TABC) are also useful to remove the spurious finite-volume effects for
periodic boundary conditions \cite{schuetrumpf2015a,schuetrumpf2016}. 
However, implementations of such methods are often numerically costly while approximate methods such as 
the absorbing boundary conditions (ABC) has proven efficient \cite{nak05,rei06}.

\subsubsection{Inclusion of pairing correlations\label{sec:TDHFBnum}}

The development of a realistic TDHFB spherical code with a full Skyrme EDF and a density-dependent pairing effective interaction was  initiated in Ref.~\cite{ave08} and applied to the study of pairing vibrations. 
More recently, 3-dimensional codes have been developed to study the effect of the pairing dynamics at the BCS level~\cite{eba10,sca12} and solving the  TDHFB equation with a Skyrme functional~\cite{ste11} and with the Gogny effective interaction~\cite{has12}. 

While solving TDHF+BCS equations can be done by extending existing TDHF codes to incorporate the  evolution of the occupation numbers 
through the coupled equations (\ref{eq:TDBCS1}) and (\ref{eq:TDBCS2}), TDHFB solvers are usually developed on their own. 
Thanks to the same form of the TDHF and TDHFB equations [compare Eqs.~(\ref{eq:tdhf}) and~(\ref{eq:tdhfbR})], similar algorithms can be used, such as Eq.~(\ref{eq:algo}) with a truncated evolution operator. 

In practice, however, TDHFB codes are  more computationally demanding due to the much larger dimension of the generalised density and Hamiltonian matrices in Eqs.~(\ref{eq:Rmatrix}) and~(\ref{eq:Hmatrix}), respectively. 
In the \textsc{lise} code \cite{jin2021}, for instance, the dimension of the quasiparticle Hamiltonian and the number of quasi-particle wave-functions are $4N_xN_yN_z$ (which could reach $\sim10^{-6}$), while a TDHF code typically requires a number of wave function of the order of the number of nucleons. 
Realistic applications of TDHFB are nevertheless feasible with modern computational power, e.g., by making use of graphic processing units (GPU). See, e.g., Ref.~\cite{jin2021} for further details.

\subsection{Beyond TDHF approaches\label{sec:persp}}

\begin{figure}
\includegraphics[width=8.8cm]{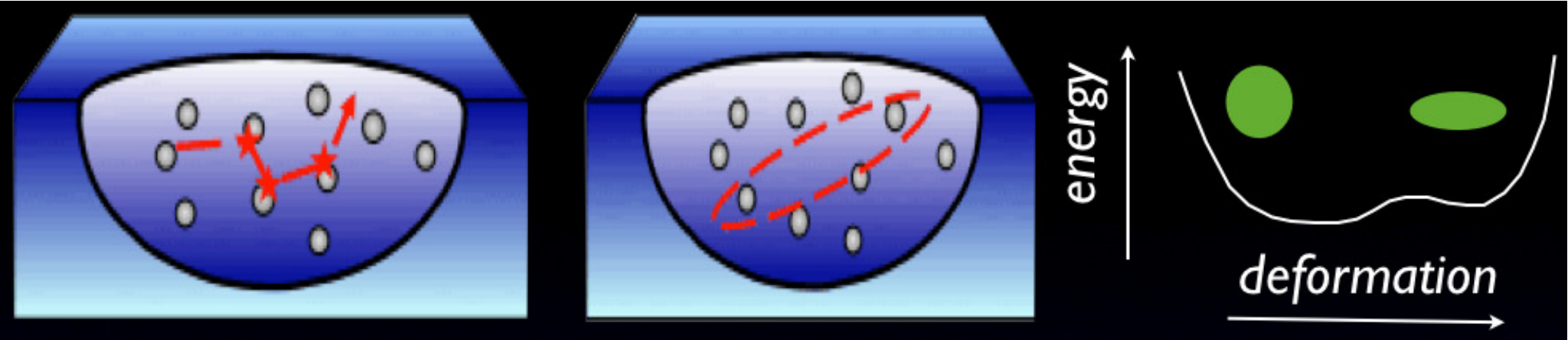}
\caption{(courtesy of D. Lacroix). Illustration of the different kinds of correlations. From left to right: in medium particle-particle collisions, pairing correlations, and large amplitude dynamical correlations.
\label{fig:correlations}}
\end{figure}

The TDHF theory is a microscopic quantum mean-field approach of independent particles. 
Although some approaches could in principle provide the exact dynamics, such as  the stochastic mean-field based on functional integrals \cite{car01,jui02}, they remain prohibitive for realistic applications. 
Extensions of the TDHF theory, e.g., the extended TDHF \cite{won78,won79,dan84,bot90,ayi80,lac99} and stochastic TDHF theories \cite{rei92,ayi01,lac01,lac04}, include correlations which are not present at the mean-field level. 
Table~\ref{tab:sum_micro} gives a summary of beyond TDHF approaches, together with the types of correlations they include.
These correlations are typically of three kinds illustrated in Fig.~\ref{fig:correlations}.

\subsubsection{In medium particle-particle collisions} 
In a first approximation, 
this collision term can be neglected thanks to the Pauli blocking. However, in violent collisions, where the Pauli principle is less efficient to block collisions between nucleons of the two colliding fragments, this term is expected to affect the dynamics.  It is also responsible for the thermalisation of the compound nucleus and for the spreading width of giant resonances. 
In medium collisions are included, e.g., in the time-dependent density-matrix (TDDM) theory \cite{cas90,bla92}.
See also Ref.~\cite{jiang2025} for a recent study of the role of nucleon-nucleon collisions on dissipation and thermalisation within the relaxation-time approximation. 

\subsubsection{Pairing correlations} 
They generate a superfluid phase in nuclei. They have a strong effect on the structure of mid-shell nuclei (e.g., odd-even mass staggering) and in transfer reactions where they favour the transfer of paired nucleons. 
As discussed in section~\ref{sec:pairingTh}, pairing dynamics can be described with generalised time-dependent mean-field approaches such as TDHFB and TDHF+BCS. 

Numerical aspects of solving the TDHFB equation were presented in \ref{sec:TDHFBnum}.
Applications of TDHFB and TDHF+BCS are discussed in the following sections. 
Note that pairing is also treated as part of the two-body correlations included in TDDM~\cite{ass09}.

\subsubsection{Large amplitude dynamical correlations} 
Unlike a single Slater determinant which is usually localised in a potential energy surface (PES), the correlated state may be described by a configuration mixing of localised states across the entire PES. A typical example is the zero point motion along a collective coordinate. These correlations allow also for the state to be in a classically forbidden region of the PES, and, then, are necessary to treat quantum tunnelling of the many-body wave function, e.g. in sub-barrier fusion.

The Balian-V\'en\'eroni prescription discussed before is an example where fluctuations and correlations of specific observables, not included at the TDHF level, are described dynamically. 
In the classification used above, the included correlations belong to the class of dynamical correlations, although, in this case, they are of small amplitude nature. 
In fact, the fluctuations obtained in Eq~(\ref{eq:CiiBV}) are those included in the time-dependent random phase approximation (TDRPA) which is obtained assuming small fluctuations of the density matrix around the average evolution~\cite{bal92,ayi08}. 

A possible extension of TDHF  including dynamical correlations of large amplitude involves the path integral technique with the stationary phase approximation (SPA) \cite{neg82}.
The SPA assumes that the path integral is dominated by the classical action (for a single particle), or, equivalently, by the TDHF action for a many-fermion system. 
The method provides an elegant way to include fluctuations around the mean-field trajectory. 
Unlike the BV prescription, however, these fluctuations are not limited to the small amplitude limit.  
This approach leads to a self-consistent eigenvalue problem in four space-time dimensions. 
It is in fact similar to a HF eigenvalue problem with time as an additional dimension. 
In case of a vibrational motion, the problem involves the (difficult) task to find periodic solutions of the TDHF equations \cite{wu1997,abada1992}. 

Another important possible application of this path integral approach is to treat quantum tunnelling of the many-body wave function through a barrier. 
In this case,  quantum many-particle closed trajectories in imaginary time need to be found.
Up to now, realistic applications in imaginary time have faced the difficulties brought by the limitations of computational power, and only simple cases, such as the spontaneous fission of the $^{8}$Be in two~$\alpha$ have been studied~\cite{neg82}.
However, the recent increase of computational power, should lead to a revival of these techniques.
A toy model application of imaginary time-dependent Hartree-Fock is presented in \cite{mcglynn2020}.

An alternative way to include large amplitude dynamical correlations is to derive collective Hamiltonians, such as the Bohr Hamiltonian, from microscopic calculations using the generator-coordinate method (GCM)~\cite{hil53,gri57} and its time-dependent extension (TDGCM)~\cite{rei83}. Recent applications of the GCM to low-lying collective excitations can be found in Refs.~\cite{ben08,del10,rod10}. The TDGCM has also been applied with the Gaussian overlap approximation (GOA) to investigate the fission process \cite{gou05,schunck2016,schunck2022}.
Solvers such as \textsc{felix} \cite{regnier2018b} provide numerical solutions to TDGCM+GOA equations for applications to fission.
Removing the GOA is an ongoing challenge in fission theory \cite{bender2020,verriere2020}. 
As a first step, it requires removing discontinuities in PES \cite{lau2022} to ensure continuous propagation of the collective wave-function. 
In these applications of TDGCM, the variational space is made of static mean-field states. 

\subsubsection{TDGCM with time-dependent basis states \label{sec:TDGCM}}

Following the seminal work of Reinhard {\it et al} \cite{rei83}, some groups have recently endeavoured to apply  TDGCM with time-dependent basis states and applied the approach to various problems, including vibrations \cite{li2023b,marevic2023,marevic2024}, nucleon transfer reactions \cite{regnier2019a}, and sub-barrier fusion \cite{hasegawa2020}.
In particular, interesting beyond TDHF effects were predicted, such as multi-phonon  excitations \cite{marevic2023}, spreading width of giant resonances \cite{li2023b}, and many-body tunnelling \cite{hasegawa2020}.
These  successes motivate a more detailed discussion of the formalism, also called multiconfiguration TDDFT \cite{regnier2019a}.

The correlated state is written as a coherent superposition of independent (quasi)particle states
$$|\Psi(t)\>=\sum_\alpha f_\alpha(t) |\phi_\alpha(t)\>,$$
where $\alpha$ represents one or a set of collective coordinates.
Requesting the stationarity of the action 
$$S=\int_{t_0}^{t_1}dt\,\<\Psi(t)|\oH-i\hb\partial_t|\Psi(t)\>$$
leads, in principle, to coupled equations for both $f_\alpha(t)$ and $|\phi_\alpha(t)\>$. 
As a result of this coupling, the $|\phi_\alpha(t)\>$ do not necessarily obey the TDHF(B) equation. 
In particular, this coupling can induce a subset of $|\phi_\alpha(t)\>$ to encounter tunnelling~\cite{hasegawa2020}.
For simplicity, however, it is often assumed that the $|\phi_\alpha(t)\>$ obey time-dependent mean-field equations. 
In this case, their evolutions can be determined with a standard TDHF(B) code, while the variational principle provides the time evolution of the mixing coefficients $f_\alpha(t)$. 

A complication comes from the fact that the $|\phi_\alpha(t)\>$ are usually not orthogonal. 
It is then necessary to compute the overlap kernel 
$$\mN_{\al\al'}(t)=\<\phi_\al(t)|\phi_{\alpha'}(t)\>$$ 
as well as the Hamiltonian kernel 
$$\mH_{\al\al'}(t)=\<\phi_\al(t)|\oH|\phi_{\alpha'}(t)\>.$$ 
The latter requires, in principle, a multi-reference EDF as both states lead to different density dependences. 
A common approximation is to use the average density prescription~\cite{schunck2019}.
The variational principle then leads to 
\oeq i\hb\dot{g}=\[\mN^{-\frac{1}{2}}\(\mH-\mD\)\mN^{-\frac{1}{2}}+i\hb\dot{\mN}^{\frac{1}{2}}\mN^{-\frac{1}{2}}\]g, \label{eq:TDGCM}\ceq
where 
$$\mD_{\al\al'}(t)=\<\phi_\al(t)|i\hb\partial_t|\phi_{\alpha'}(t)\>$$ 
is the derivative kernel and 
$$g_\al(t)=\sum_{\al'}\mN^{\frac{1}{2}}_{\al\al'}(t)f_{\al'}(t)$$
is the normalised collective wave-function whose time evolution is obtained by solving Eq.~\ref{eq:TDGCM}

The choice of collective coordinates $\alpha$ depends on the physical problem. 
These could be, e.g., boost velocities  \cite{marevic2023} and boost multipolarities \cite{li2023b} in vibration studies, or centre of mass energy in reaction studies \cite{hasegawa2020}. Chemical potential difference between the collision partners has also been proposed as a generator coordinate inducing various transfer channels \cite{simenel2014b}.

\begin{table*}[th]
    \begin{center}
{ 
\begin{tabular}{|l|l|l|l|} \hline
{\it Name} & {\it Approximation} & {\it Variational}  & {\it Associated}  \\
&&{\it space}&{\it observables}\\
\hline
TDHF & mean-field (m.-f.) & indep. part. & one-body \\ 
&&&\\
\hline
BV prescription & m.-f. + small ampl.  & indep. part & one-body \\ 
(TDRPA)&fluctuations&&+fluctuations\\
\hline
Path integrals & m.-f. + fluctuations  & indep. part. & one-body \\ 
&&&+fluctuations\\
\hline
(TD)GCM & m.-f. + fluctuations & correlated & one-body \\ 
&&states&+fluctuations\\
\hline
TDHF-Bogoliubov & m.-f. + pairing  & indep. & generalised  \\
&&quasipart.  &one-body\\
\hline
Extended-TDHF & m.-f. + collision  & correlated  & one-body  \\ 
&(dissipation)&states&\\
\hline
Stochastic-TDHF & m.-f. + collision  & correlated  & one-body   \\ 
&(dissipation+fluctuations)&states&\\
\hline
Time-dependent & m.-f. + two-body & correlated  & one- and \\ 
density matrix & correlations&states&two-body\\
\hline
Stochastic m.-f. & Exact & correlated  & all   \\ 
(Functional integrals) &(within statistical errors)&states&\\
\hline
\end{tabular}
}
\end{center}
\caption{The TDHF approach and several possible extensions.}
\label{tab:sum_micro}
\end{table*}

\section{Collective vibrations \label{chap:vib}}

\subsection{Introduction\label{sec:intro_vib}}

A particular interest in strongly interacting systems is their ability to present disorder or chaos, 
and, in the same excitation energy range, well-organised motion.
Atomic nuclei are known to show both behaviors~\cite{boh75}.
In particular, they exhibit a large variety of  vibrations,
from low-lying collective modes to giant resonances (GR) with excitation energy usually above the particle emission threshold~\cite{har01}.

Baldwin and Klaiber observed the isovector giant dipole resonance (GDR) 
in photofission of uranium nuclei~\cite{bal47}, 
interpreted as a vibration of neutrons against protons~\cite{gol48}. 
Other kinds of GR have been discovered, such as the isoscalar giant quadrupole resonance (GQR) 
associated with an oscillation of the shape between prolate and oblate deformations~\cite{fuk72},
and the isoscalar giant monopole resonance (GMR) corresponding to a breathing mode~\cite{mar76,har77,you77}. 

GR are usually associated with the first phonon of a small-amplitude harmonic motion.
In the harmonic picture, it corresponds to a coherent sum of one-particle one-hole ($1p1h$) states~\cite{boh75}.
GR decay by particle ($p$, $n$, $d$, $\alpha$, $\gamma$...) emission, leading to an escape width in their energy spectra. 
The escape width is then due to a coupling of the correlated $1p1h$ states to the continuum. 
Other contributions to GR width are the Landau damping and the spreading width. 
Landau damping occurs due to a one-body coupling to non-coherent $1p1h$ states~\cite{pin66}. 
The spreading width is due to the residual interaction coupling $1p1h$ states to $2p2h$ states.
The $2p2h$ states can also couple to $3p3h$ states and more complex $npnh$ configurations until an equilibrated system is reached. 
These contributions to the GR width are illustrated in Fig.~\ref{fig:damping}.

\begin{figure}
\includegraphics[width=8.8cm]{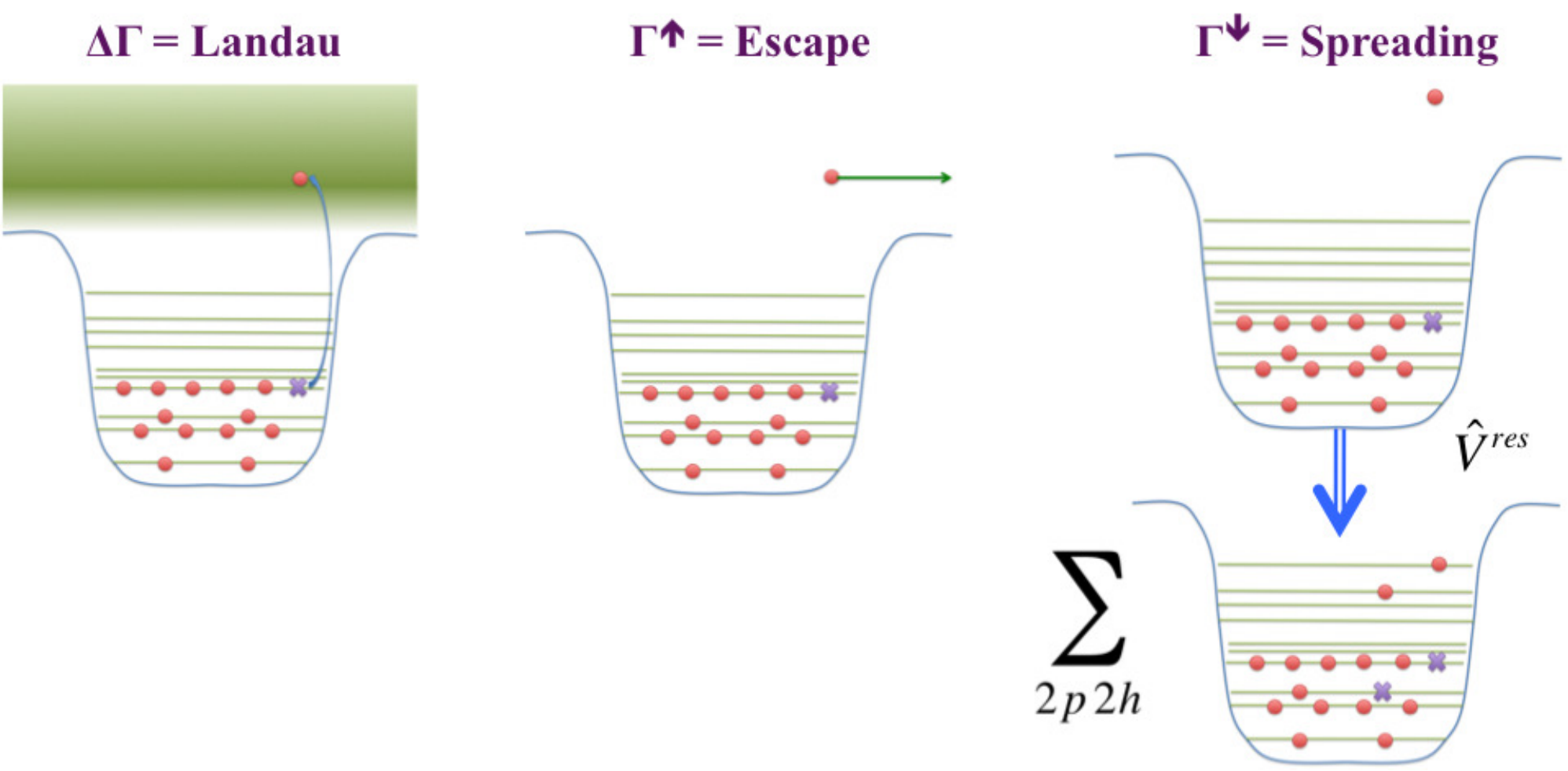}
\caption{Illustration of the three contributions to the width of the GR (see text). (left) Landau damping due to coupling to incoherent $1p1h$ states. (middle) Escape width due to direct decay. 
(right) Spreading width due to coupling to $2p2h$ states.
\label{fig:damping}}
\end{figure}

The proof of the vibrational nature of GR came with the observation 
of their two- and three-phonon states~\cite{cho95,aum98,sca04}.
Multiphonon studies also provided a good test to the harmonic picture. 
In particular, anharmonicity was found in an abnormally large excitation probability of these states~\cite{cho95}.
The coupling between different phonon states is predicted to be an important source of this anharmonicity~\cite{vol95,bor97,sim03,fal03,cho04,sim09,lan06}. 

Coherent motion of fermions such as collective vibrations in nuclei
can be modeled by time-dependent mean-field approaches like the 
TDHF theory. 
In fact, in its linearised version, TDHF is equivalent to the
Random Phase Approximation (RPA)~\cite{rin80} which is the basic tool to understand the
collective vibrations in terms of independent phonons.
In particular, the time evolution of one-body (collective) observables, which can be estimated using a TDHF code, can be used to evaluate vibrational energy spectra.
See \cite{burrello2019} for a detailed numerical comparison between RPA and TDHF calculations of vibrational spectra.

Direct decay contributing to the escape width can be studied within the continuum-(Q)RPA model\footnote{The quasiparticle-RPA (QRPA) is an extension of the RPA including pairing correlations and can be obtained from a linearisation of the TDHFB equation.}~\cite{kre74,liu76,kam98,mat01,hag01,kha02}. 
GR direct decay can also be investigated within the TDHF framework~\cite{cho87,pac88,avez2013,sca12}.
Indeed, the TDHF  theory allows for the evaporation of unbound components of the single particle wave-functions.
Due to its one-body nature, Landau damping is also included in RPA. 
However, they do not contain the residual interaction responsible for the spreading width.

An interesting feature of TDHF applications is that they are not limited to small amplitude vibrations, unlike RPA, allowing for investigations of non linear effects in collective motions. 
In particular, couplings between collective modes have been invoked as a possible source of anharmonicity~\cite{sim03,cho04,sim09}. 

In principle, TDHF and RPA codes can be used to study the vibrational spectra of any nucleus. 
However, most of the applications have focused on doubly magic nuclei. 
The main reason is that these nuclei are well described at the HF level, as there is usually no pairing correlation in their ground-state. 
Mid-shell nuclei, however, are better described with the HF+BCS or Hartree-Fock-Bogoliubov (HFB) approach. 
TDHF codes have been extended to study the role of pairing correlations on collective vibrations at the BCS level~\cite{eba10,sca12} and within the TDHFB theory~\cite{ave08,ste11}.

In this section, we  briefly introduce the linear response theory. 
For an illustrative purpose, applications of the latter to both low-lying vibrational states and GR  are given within the TDHF framework. 
These standard applications to the linear response theory are followed by an investigation of the direct decay by nucleon emission and its link to the GR microscopic structure.
Recent TDGCM calculations that allow for a description of the spreading width are then discussed. 
Non-linearities in collective vibrations are studied to investigate a possible source of anharmonicity in GR multiphonon spectra. 
As a last application, we return to the linear response theory and study pairing vibrations with a TDHFB code.
Finally we conclude this section and present some perspectives to the study of collective vibrations.

\subsection{Linear response theory}

The linear response theory has been used since the early days of TDHF
studies of collective vibrations in nuclei~\cite{blo79,str79,cho87,pac88,chinn1996}.
In this theory, one computes the time evolution of an observable $Q(t)$
after an excitation induced by a small boost on the ground state $|\Psi_0\>$,
\oeq
|\Psi(0)\> = e^{-i\epsilon \oQ /\hbar} |\Psi_0\>=|\Psi_0\>-\frac{i\epsilon}{\hb}\sum_\nu q_\nu|\Psi_\nu\>+O(\epsilon^2), \label{eq:boost}
\ceq
where  $q_\nu=  \<\Psi_\nu| \oQ |\Psi_0\>$ is the transition amplitude between 
the ground state and the eigenstate $|\Psi_\nu\>$ of the Hamiltonian with eigenenergy $E_\nu$.
The time-evolution of the state reads
\oeqn
|\Psi(t)\> &=& e^{-i\oH t/\hb}|\Psi(0)\> \nonumber \\
&=&e^{-iE_0t/\hb} \( |\Psi_0\> - \frac{i\epsilon}{\hb}\sum_\nu q_\nu e^{-i\omega_\nu t} |\Psi_\nu\>  \) + O(\epsilon^2),\nonumber \\
&&
\ceqn
with $\hb\omega_\nu = E_\nu-E_0$.

The response $Q(t)=\<\Psi(t)|\oQ|\Psi(t)\>-\<\Psi_0|\oQ|\Psi_0\>$ to this excitation can be written
\oeq
Q(t)=-\frac{2\epsilon}{\hb}\sum_\nu |q_\nu|^2\sin \omega_\nu t+O(\epsilon^2).
\label{eq:linresp}
\ceq
The latter can be decomposed
into various frequencies $\omega$, giving the strength function
\begin{eqnarray}
R_{Q}(\omega) &=&\lim_{\epsilon\rightarrow 0} \frac{-\hbar }{\pi \epsilon}\,
\int_{0}^{\infty}  dt\, {Q}(t) \, \sin (\omega t). \label{eq:strengthlin} \\
 &=& \sum_\nu \, |q_\nu |^2 
 \delta (\omega - \omega_\nu). \label{eq:strengthfinal}
\end{eqnarray}
We see in eqs.~(\ref{eq:strengthlin}) and (\ref{eq:strengthfinal})  that the strength function is obtained for small $\epsilon$.
In practice, it is sufficient to check that the amplitude $Q_{max}$ of $Q(t)$ evolves linearly with $\epsilon$ [see Eq.~(\ref{eq:linresp})]. 

It is interesting to note that the first phonon energy is obtained with an amplitude of the oscillation which is much smaller than the one associated with the first phonon. 
Indeed, in a coherent picture, and for a single mode with transition amplitude $q$, the number of excited phonons reads~\cite{sim03}
$$n=\(\frac{2Q_{max}}{q}\)^2.$$
In particular, the amplitude associated with the first phonon, $Q_{max}^{1ph}=q/2$ may be well beyond the linear regime. 

\subsection{Low-lying vibrations}

\begin{figure}
\begin{center}
\includegraphics[width=6.5cm]{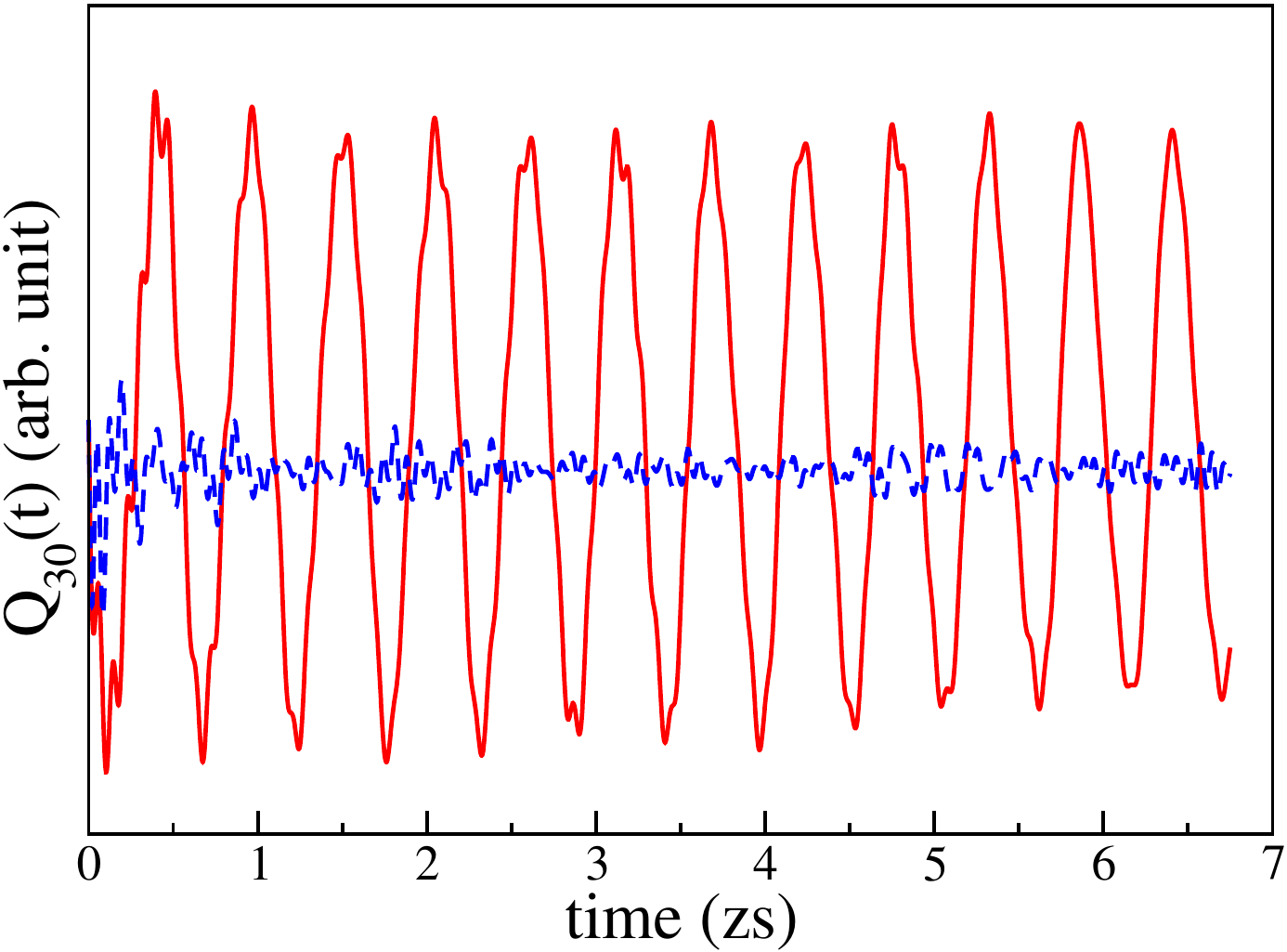}\\
\vspace{1.5cm}
\includegraphics[width=6.5cm]{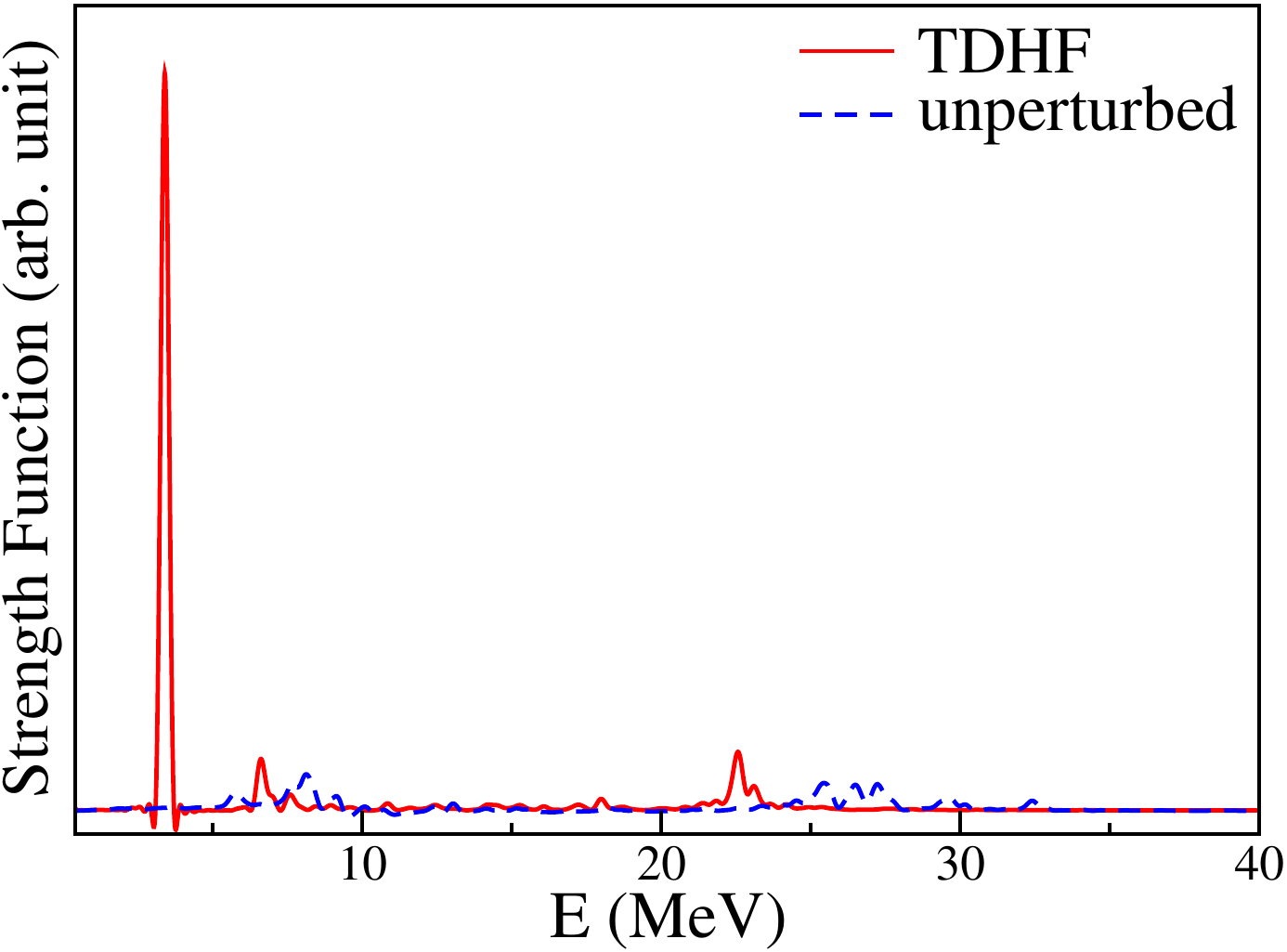}
\caption{(top) Time evolution of the octupole moment in $^{208}$Pb after an octupole boost obtained with the \textsc{tdhf3d} code for both TDHF (solid line) and unperturbed (dashed line) responses (see text). (bottom) Associated strength function.
\label{fig:PbQ3}}
\end{center}
\end{figure}

TDHF solvers were  used to study low-lying vibrations in Refs.~\cite{simenel2013a,simenel2013b,scamps2013b,simenel2016b,vophuoc2016,burrello2019}.
Figure~\ref{fig:PbQ3} gives an example for the octupole vibrational modes in $^{208}$Pb obtained with the \textsc{tdhf3d} code~\cite{kim97} (see also Ref.~\cite{nak05} for a study of this mode in $^{16}$O).
An octupole boost is applied at initial time on the HF ground state of $^{208}$Pb. 
The latter has spin-parity $J^\pi=0^+$ and the boost induces a transition with $\Delta L=3$, exciting vibrational states with $J^\pi=3^-$. 

We can see on the upper panel of Fig.~\ref{fig:PbQ3} the oscillation of the octupole moment induced by the boost (solid red line). 
The associated strength function, shown in the lower panel, exhibits a strong peak at an energy of $\sim3.4$~MeV. 
It corresponds to the main oscillation seen in the time evolution of $Q_{30}(t)$. 
Note that this state is clearly bound, as can be seen from the undamped nature of the oscillation. 
This peak is associated with the low-lying $3^-$ state in $^{208}$Pb. 
The energy of this state is overestimated with the SLy4 parametrisation of the Skyrme EDF, as the experimental value gives 2.6~MeV.
Note that the tensor interaction can also affect the energy of low-lying collective vibrational states \cite{guo2018}.

However, its collective nature is unambiguous. 
This can be seen from a comparison with the unperturbed response of the same boost.
The latter is obtained by freezing the mean-field to its initial HF value, i.e., neglecting the self-consistency of the mean-field in the dynamics. 
This procedure removes the residual interaction which is responsible for the collectivity of vibrations in TDHF (and RPA). 
We see that this peak disappears in the unperturbed spectrum, proving its collective nature. 

\begin{figure}
\begin{center}
\includegraphics[width=8cm]{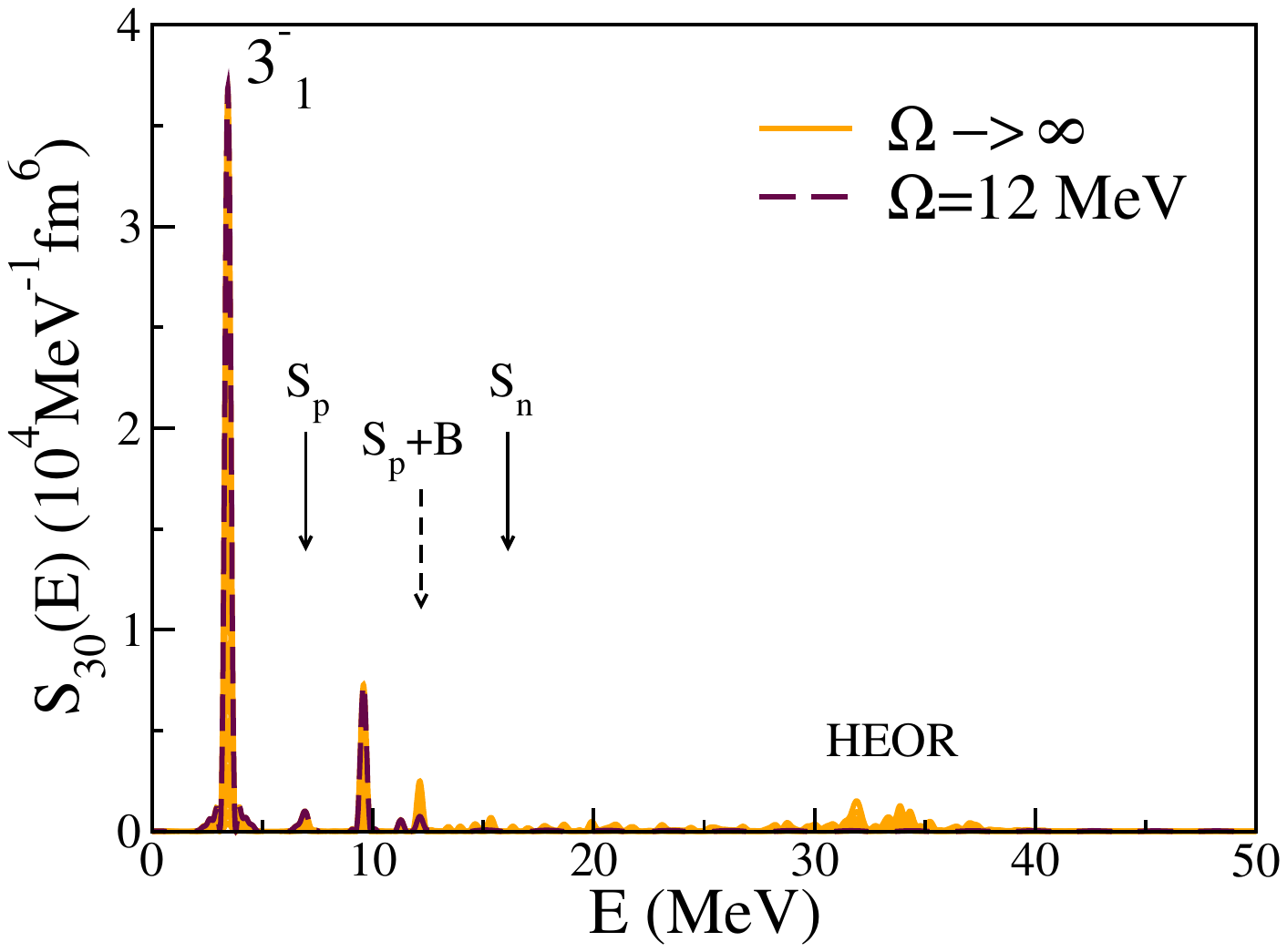}
\caption{Strength function of the octupole moment in $^{40}$Ca for different choices of $\Omega$ (see text). From Ref.~\cite{simenel2013b}.
\label{fig:3-_filter}}
\end{center}
\end{figure}

Unbound states may also be populated by the boost, as can be seen at $E\simeq23$~MeV in the spectrum of Fig.~\ref{fig:PbQ3}, corresponding to a high-energy octupole resonance (HEOR). 
These states decay by particle emission into the continuum.
To avoid spurious contributions from the reflexion of emitted particles (which could re-enter the nucleus), one can introduce a time-dependent function $f(t)$ in the boost operator to filter out unbound-states. 
Indeed, the boost excitation in Eq.~(\ref{eq:boost}) is equivalent to applying a perturbation with a time-dependence $f(t)\propto\delta(t)$. 
The Fourier transform of $\delta(t)$ being constant, the boost can populate states at any energy. 
Instead, replacing the boost by a time-dependent perturbation $\hat{V}(t)=\epsilon f(t) \hat{Q}$ with\footnote{The Fourier transform of $f(t)$ is a step function with $\tilde{f}(\omega)=1$ for $\Omega/\hb>\omega>0$ and 0 for $\omega>\Omega/\hb$.} $f(t)=\frac{\sin(\Omega t/\hb)}{\pi t}$ will only excite states with $E<\Omega$, as illustrated in Fig.~\ref{fig:3-_filter}~\cite{simenel2013b}.

\subsection{Giant resonances}

\begin{figure}
\begin{center}
\includegraphics[width=6cm]{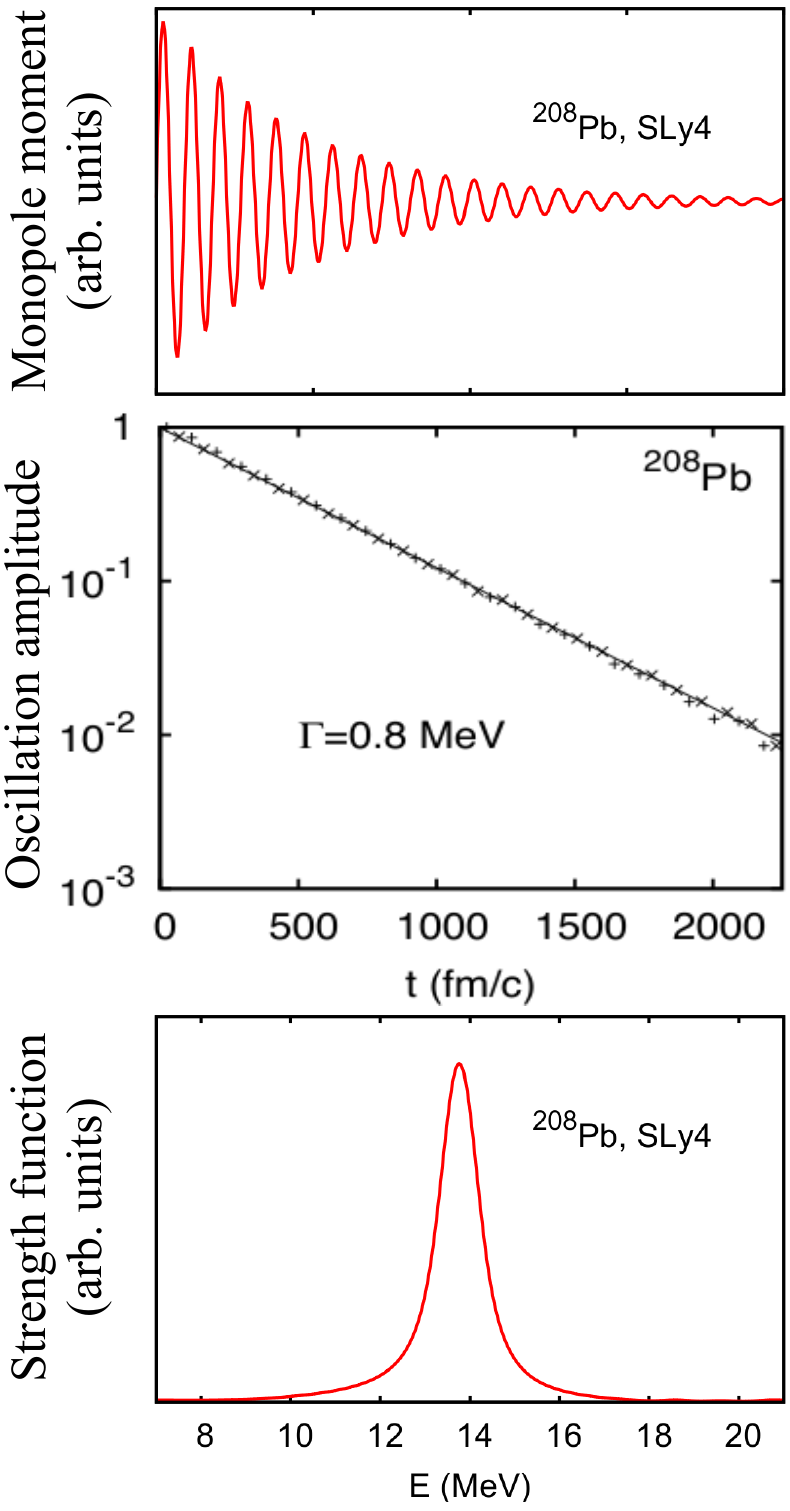}
\caption{(top) Time evolution of the mean square radius in $^{208}$Pb after a monopole boost obtained with the \textsc{tdhfbrad} code (without pairing). (middle) Time evolution of the oscillation amplitude (log scale). 
(bottom) Associated strength function. From Ref.~\cite{ave09}.
\label{fig:PbQ0}}
\end{center}
\end{figure}

Giant resonances are collective unbound states which have been thoroughly studied with TDHF solvers\footnote{See also Ref. \cite{geng2025} for a recent time-dependent relativistic HF study of the isoscalar giant monopole resonance with spherical symmetry.} 
\cite{sim03,cho04,ste04,nak05,uma05,mar05,rei07,bro08,sim09,ste11,has12,fracasso2012,avez2013,scamps2013b,pardi2013,goddard2013,scamps2014a,stetcu2015,burrello2019,shi2020,aitbenmennana2020,aitbenmennana2021,li2023b,aitbenmennana2023,marevic2023,shi2023}. An  example  is shown in Fig.~\ref{fig:PbQ0}.
Here, a monopole boost is applied on the $^{208}$Pb ground-state.
The evolution has been obtained with a TDHF code\footnote{This is the \textsc{tdhfbrad} code, developed by B. Avez as an extension of the \textsc{hfbrad} code~\cite{bennaceur2005}, which solves the TDHFB equation with spherical symmetry. However, the pairing interaction is set to zero in the calculation shown in Fig.~\ref{fig:PbQ0}.} in spherical symmetry (allowing for large boxes)~\cite{ave09}. 
In this case, almost all the strength (if not all) goes into the GMR which is unbound.
Such a resonance can then decay by particle emission.
In particular, the direct decay induces an escape width.
The latter is included in the TDHF framework thanks to evaporation of unbound components of the single particle wave functions~\cite{cho87,pac88,avez2013}.
Note that the emission of nucleons in the continuum is quadratic with the boost intensity $\epsilon$. 
As a result, it does not affect the evolution of $Q(t)$ in the linear regime. 
However, special care should be taken with possible spurious numerical effects due to the reflection of the nucleons on the box edges~\cite{rei07}.

The decay induces an exponential decrease of the oscillation amplitude (see middle panel of Fig.~\ref{fig:PbQ0}). 
The corresponding escape width is $\Gamma^\uparrow=0.8$~MeV.
This escape width is responsible for the main part of the width of the peak in the lower panel of Fig.~\ref{fig:PbQ0}.
Indeed, the total width of the peak is $\Gamma^{tot.}\simeq1.1$~MeV.
The difference between the TDHF predictions of $\Gamma^{tot.}$ and $\Gamma^\uparrow$ could be attributed to Landau damping. 

Experimentally, the energy of the GMR in $^{208}$Pb is~\cite{you04}
$$E_{GMR}^{exp.}\equiv \frac{m_1}{m_0}= 14.0\pm0.2 \mbox{ MeV,}$$
where the energy weighted moment $m_k$ is defined as~\cite{rin80}
\oeq
m_k = \sum_\nu \omega_\nu^k |q_\nu|^2.
\ceq
This value is in excellent agreement with the TDHF result determined from the $m_1/m_0$ ratio: $E_{GMR}^{TDHF}\simeq13.9$~MeV \cite{ave09}.
However, the total width obtained from TDHF, $\Gamma^{tot.}_{TDHF}\simeq1.1$~MeV,  underestimates the experimental value $\Gamma^{tot.}_{exp.}=2.9\pm0.2$~MeV \cite{you04}. 
This is due to the fact that TDHF does not take into account the spreading width (see sec.~\ref{sec:spreading}).

\begin{figure}
\begin{center}
\includegraphics[width=6cm]{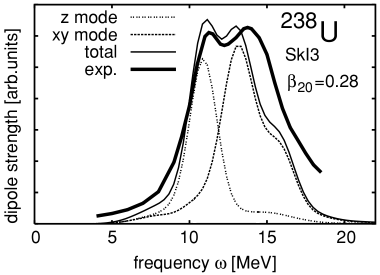}
\caption{Dipole strength in the prolately deformed $^{238}$U nucleus with the SkI3 parametrisation of the Skyrme functional. The  strength is shown along the symmetry axis ($z$ mode) and perpendicular to the symmetry axis ($xy$ mode). The total strength is shown with the thin-solid line and compared with experimental data from Ref.~\cite{die88} (thick-solid line). Adapted from Ref.~\cite{mar05}.
\label{fig:U-GDR}}
\end{center}
\end{figure}

TDHF calculations of giant-resonances are not limited to the study of monopole modes with spherical symmetry. Indeed, 3-dimensional TDHF codes have also been used to study other modes, such as the isovector GDR \cite{bro08,has12,fracasso2012,nak05,sim03,rei07,sim09,ste11,goddard2013,burrello2019,shi2020,aitbenmennana2020,aitbenmennana2021,aitbenmennana2023} and the isoscalar GQR~\cite{has12,uma05,sim03,rei07,sim09,scamps2013b,scamps2014a,li2023b,marevic2023,marevic2024}.
An example of such calculations is shown in Fig.~\ref{fig:U-GDR}, where the TDHF dipole strength function is plotted in $^{238}$U for different excitation modes~\cite{mar05} and compared with experimental data from Ref.~\cite{die88}. 
The $^{238}$U being prolately deformed, the dipole response is investigated along and perpendicular to the deformation axis, exhibiting the well known splitting of the GDR peak, i.e., a lower (higher) energy along (perpendicular to) the deformation axis~\cite{har01}. 
The qualitative agreement between the theoretical prediction of the peak positions and experimental data is good. However,  part of the TDHF width is due to the finite time window used in the Fourier transform. It is then difficult to draw any conclusion on the width of the GDR.

\begin{figure}
\begin{center}
\includegraphics[width=8cm]{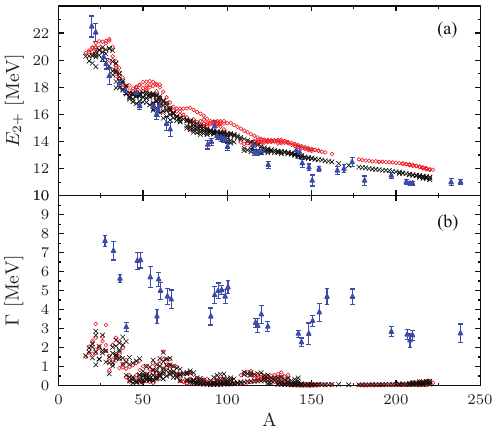}
\caption{Isoscalar GQR energy (a) and width (b) for spherical nuclei with the SkM* (black crosses) and Sly4 (red circles) functionals. Experimental data are shown with blue triangles \cite{bertrand1981}. From Ref. \cite{scamps2013b}.
\label{fig:GQR_spher}}
\end{center}
\end{figure}

Systematics studies of GQR in spherical nuclei with TDHF+BCS are reported in Fig.~\ref{fig:GQR_spher} \cite{scamps2013b}.
The overall decrease of the GQR energy with the mass of the nucleus is quantitatively well reproduced by theory.
The GQR width, however, is significantly underestimated as the mean-field approach does not account for correlations respondible for the spreading width. 

\begin{figure}
\begin{center}
\includegraphics[width=8cm]{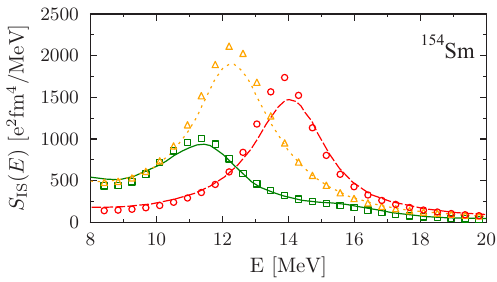}
\includegraphics[width=8cm]{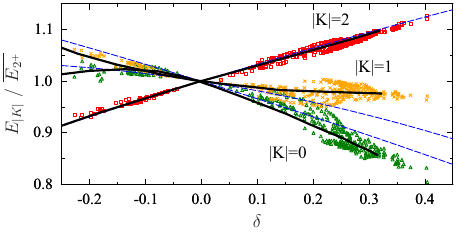}
\caption{(top) QRPA (lines) and TDHF+BCS (symbols) predictions of GQR spectra in $^{154}$Sm for different modes $|K|$. (bottom) Systematics of the energy splitting of the $|K|$ modes as a function of the quadrupole deformation parameter $\delta$. The color code is the same in both panels. From Ref. \cite{scamps2014a}.
\label{fig:GQR_def}}
\end{center}
\end{figure}

Calculations of vibrational modes in deformed nuclei are more complex as one needs to consider boosts in different directions. 
An example is shown in Fig.~\ref{fig:GQR_def}(top) for the isoscalar GQR in $^{154}$Sm. 
The various modes $|K|=0$, 1 and 2 are associated\footnote{The $K$ and $-K$ modes are degenerate for axially symmetric systems.} with the observables $\hat{Q}_{2K}$. 
Note that both QRPA and TDHF+BCS predictions agree well. 
Larger quadrupole deformations lead to larger splittings between the energies of these modes, as shown by the TDHF+BCS systematics in Fig.~\ref{fig:GQR_def}(bottom)\cite{scamps2014a}. 

\subsection{Direct decay\label{sec:decay}}

Giant resonances usually lie above the proton and neutron emission thresholds.
As mentioned above, their direct decay induces an escape width contributing to the total width of the GR. 
Such escape widths can be studied with the TDHF formalism. 
Indeed, TDHF codes have been used in the past to investigate the direct decay of GMR~\cite{cho87,pac88}.
In addition to its contribution to the escape width, the GR direct decay is particularly interesting as it brings informations on the microscopic structure of the GR~\cite{har01}. 
This is the main purpose of the present section.

\begin{figure}
\includegraphics[width=8.8cm]{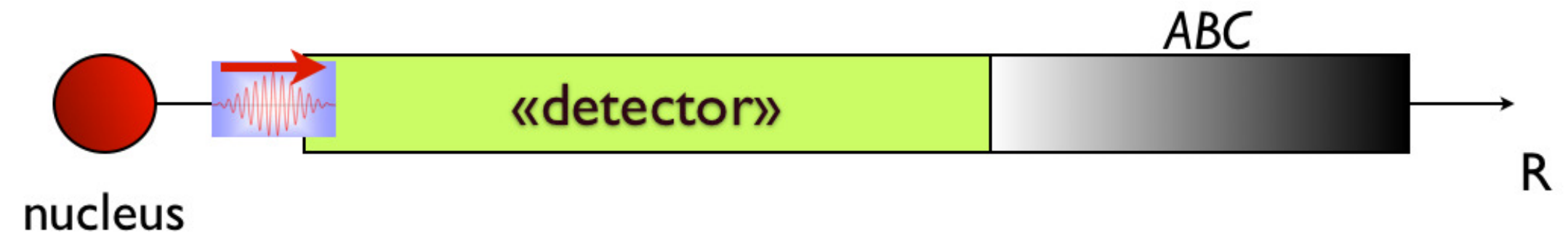}
\caption{Schematic description of the spatial repartition of numerical elements to compute spectra of emitted nucleons. The excited nucleus is in the centre of a (spherical) box. Unbound parts of single-particle wave-functions are emitted in the continuum. The ''detector''  shows the region of space where the energy of the emitted wave functions is computed. Absorbing boundary conditions (ABC) are used to absorb particles leaving the detector and to avoid spurious reflection on the box boundary.
\label{fig:detector}}
\end{figure}

TDHF calculations of GR direct decay are performed with large spatial grids to construct spectra of emitted nucleons with a  good precision. 
It is then easier to use spherical TDHF codes, although the applications are limited to monopole vibrations only.  
Let us introduce a numerical ''detector'' corresponding to the region of space where the energy of the emitted nucleons is computed from Fourier transform of their spatial wave-functions. 
This detector should be away from the centre of the box to avoid any nuclear interaction of the emitted nucleons with the nucleus. 
Fig.~\ref{fig:detector} shows a schematic representation of this numerical setup. 
Absorbing boundary conditions (ABC) with an imaginary potential may be used to avoid any spurious interaction with particles reflected on the box boundary~\cite{nak05,rei06}.

Fig.~\ref{fig:spectra_ivGMR} shows an example of a calculation of emitted nucleon spectra for the isovector GMR in $^{40}$Ca.
The calculations have been performed with the same spherical code and simplified Skyrme EDF as in Ref.~\cite{sim05}.
The upper panel of Fig.~\ref{fig:spectra_ivGMR} shows the proton spectra at different times. 
The first protons to reach the ''detector'' are obviously  the fastest, i.e., with the highest kinetic energy. 
They also leave quickly the detector while slower protons reach it. 
Put together, these proton spectra form an envelope\footnote{This envelope is  defined by the maxima of the spectra obtained at different times.}, shown in black thick solid line in the lower panel of Fig.~\ref{fig:spectra_ivGMR}.
The neutron spectrum is also shown (dashed line). 
The fact that neutrons are more bound than protons in $^{40}$Ca, together with the absence of Coulomb barrier for neutrons, explain that the neutron spectrum shows a more important contribution at low energy than the proton one.

\begin{figure}
\begin{center}
\includegraphics[width=7cm]{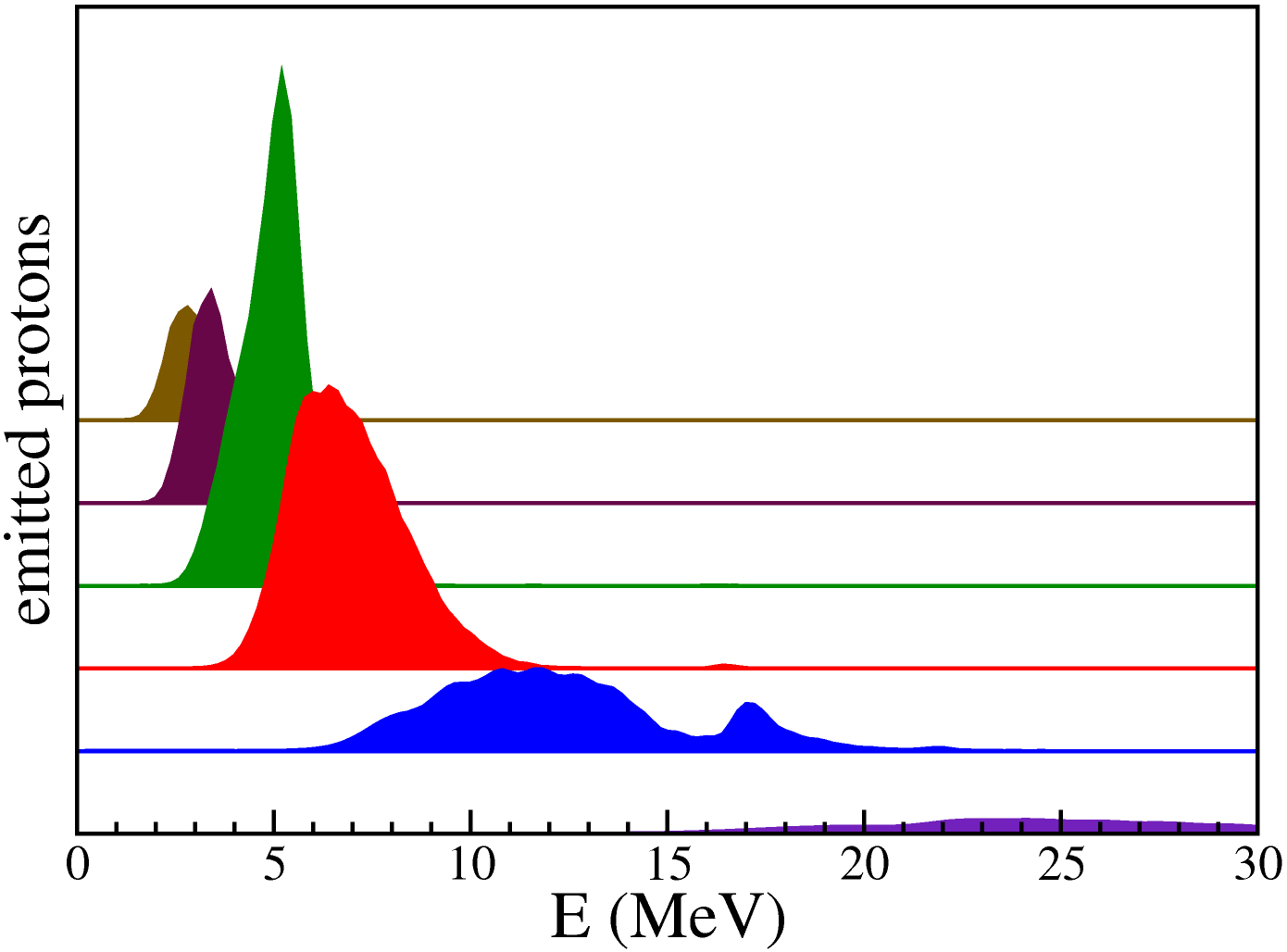}
\includegraphics[width=7cm]{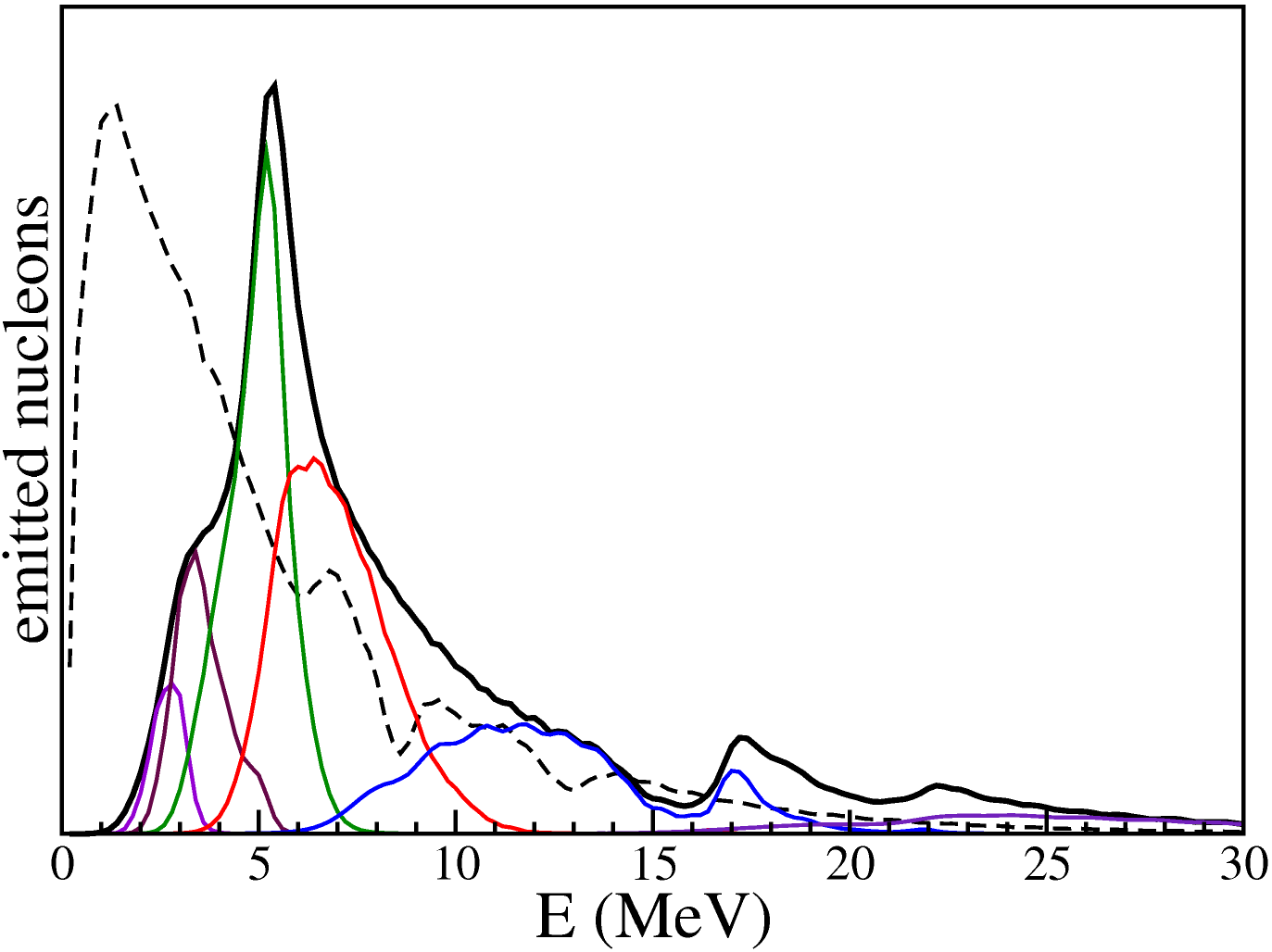}
\caption{Spectra of protons in the ''detector'' (see text and Fig.~\ref{fig:detector}) at different times following an isovector monopole boost in $^{40}$Ca. (top) Each spectrum is shifted vertically for clarity (time increases from bottom to top). The time delay between two consecutive spectra is $\Delta T=5$~zs. (bottom) The proton spectra at different times form an envelope (thick solid black line). The similar envelope obtained for neutron is also shown with a dashed line. 
\label{fig:spectra_ivGMR}}
\end{center}
\end{figure}

We can see in Fig.~\ref{fig:spectra_ivGMR} that both proton and neutron spectra exhibit some structures which cannot be explained by a simple hydrodynamical model.
Instead, one should seek for an explanation in terms of the microscopic structure of the GR. 
This motivated a more detailed investigation with a realistic TDHF spherical code and a full Skyrme EDF~\cite{avez2013}.
A brief summary of the results for the GMR in $^{16}$O is presented here (see Ref.~\cite{avez2013} for more details and for more results on, e.g., tin isotopes).
In this study, the \textsc{tdhfbrad} code~\cite{bennaceur2005,ave08} is used with the SLy4 parametrisation~\cite{cha98} of the Skyrme EDF without pairing. 
Fig.~\ref{fig:16Oa} shows the time evolution (top) of the monopole moment after an isoscalar monopole boost, and the associated spectrum (bottom) obtained within the linear response theory. 
The GMR spectrum exhibits structures which are associated with different single-particle orbitals. 
For instance, the high energy shoulder around 31~MeV is due to $s_{1/2}$ particle-hole excitations\footnote{The monopole excitations is associated with a $\Delta L=0\hb$ angular momentum transfer so that particle and hole have the same quantum numbers at the time of the excitation.}.

\begin{figure}
\begin{center}
\includegraphics[width=8.5cm]{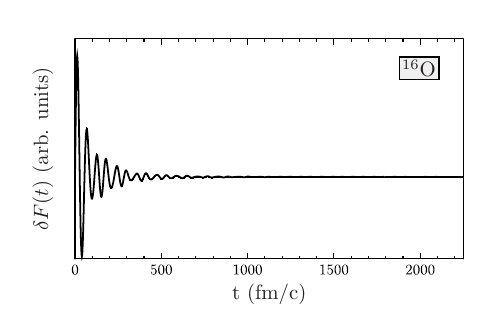}
\includegraphics[width=8.5cm]{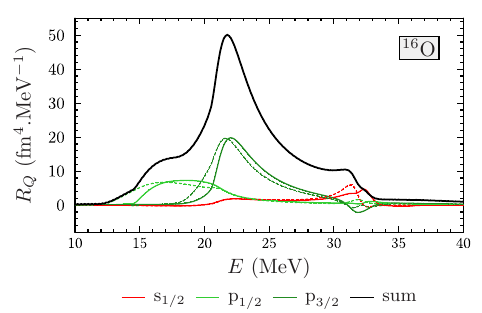}
\caption{(top) Time evolution of the monopole moment in $^{16}$O after a monopole boost obtained with the \textsc{tdhfbrad} code~\cite{ave08}. (bottom) Associated strength function and its decomposition onto single-particle quantum numbers $l$ and $j$ (spectroscopic notation).
Solid (dashed) lines show neutron (proton) contributions. 
\label{fig:16Oa}}
\end{center}
\end{figure}

The spectra of emitted protons and neutrons are shown in the upper panel of Fig.~\ref{fig:16Ob}.
The latter depend strongly on the associated single-particle quantum numbers. 
In particular, no $s_{1/2}$ nucleons are emitted. 
This is due to the fact that the  $1s_{1/2}$ hole state is deeply bound (-32.4~MeV for  protons and -36.2~MeV for neutrons according to the HF initial configuration~\cite{avez2013}).
In fact, the high energy shoulder of the GMR spectrum (see Fig.~\ref{fig:16Oa}-bottom) does not have enough energy to bring the initial $1s_{1/2}$ particle into the continuum.

\begin{figure}
\begin{center}
\includegraphics[width=8.5cm]{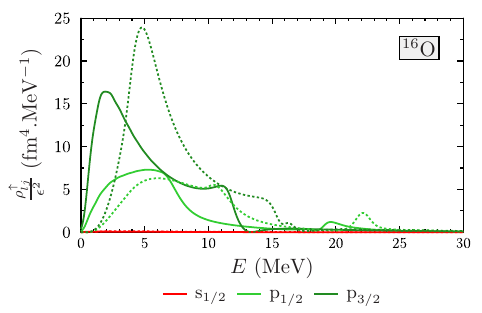}
\includegraphics[width=8.5cm]{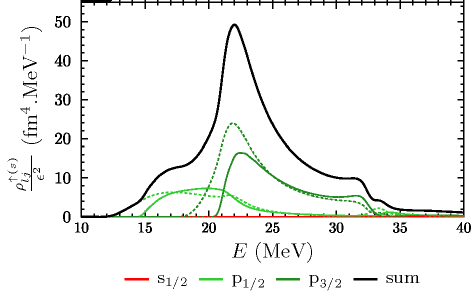}
\caption{(top) Neutron (solid lines) and proton (dashed lines) direct-decay spectra. (bottom) 
Same spectra ''shifted'' by the energy of 
the initially occupied single-particle state. Their 
sum is shown in black solid line. 
\label{fig:16Ob}}
\end{center}
\end{figure}

The lower panel of Fig.~\ref{fig:16Ob} shows the same quantity as the upper panel, with a shift in energy (different for each $l_j$ contribution) corresponding to the binding energy of the hole state. 
The sum of each shifted $l_j$ contribution gives a spectrum which is very close to the GMR spectrum (compare with the bottom panel of Fig.~\ref{fig:16Oa}). 
The agreement is excellent for both the shape and the magnitude of the spectra. 
The origin of the structures in the direct  emission spectra is entirely due to the shell structure of the nucleus. 

It is also interesting to note that, although the high energy shoulder in the $^{16}$O GMR spectrum is due to the excitation of a {\it bound} $1p1h$ state, it appears in the ''shifted'' spectrum (Fig.~\ref{fig:16Ob}-bottom) due to the {\it emission} of particles in $p$-states.
In fact, the TDHF (or RPA) residual interaction is responsible for the coupling  between the bound  $1p1h$ $s$-state and unbound $1p1h$ $p$-states. 
Note that similar couplings have been obtained in tin isotopes~\cite{avez2013}.
For instance, the GMR in $^{100}$Sn decays by protons only, while it is associated with a collective oscillation of both protons and neutrons. 

Coincidence experiments between particles emitted in the GR decay and the ejectile resulting from the GR excitation process have been performed in the past to investigate GR properties~\cite{har01}. 
The present theoretical analysis of GR direct decay allows, in principle, a direct comparison between theoretical and experimental spectra. 
However, for quantitative comparisons, one should use a more elaborated approach than the TDHF theory.
Indeed, the fact that TDHF does not include $2p2h$ residual interaction is a strong limitation, as the latter has been shown to be crucial to reproduce the width and the fragmentation of GR spectra~\cite{lac04}.
The present analysis of GR decay should then be repeated with, e.g., the extended-TDHF (ETDHF) or the time-dependent density-matrix (TDDM) approaches (see table~\ref{tab:sum_micro} and Ref.~\cite{sim10a}). 
Note also that calculations should be performed with 2D or 3D codes in order to study the decay of GR with higher multipolarity than the GMR. 

\subsection{Spreading width \label{sec:spreading}}

The description of the spreading width requires  residual interaction and is then beyond the scope of mean-field theories. 
Calculations at the semi-classical level~\cite{sur89} and with the TDDM approach~\cite{luo99,bla92} have indeed shown an increase of the damping thanks to the introduction of a collision term.

Recently, it was shown that the TDGCM with time-dependent basis states  (see section~\ref{sec:TDGCM}) provides a framework able to account for the spreading width of giant resonances \cite{li2023b}. 
This was achieved by considering a set of time-dependent relativistic mean-field (RMF) states with different multipole moment constraints in their initial conditions.
Specifically, the initial Slater determinants were determined from static constrained RMF calculations with monopole, quadrupole, octupole and hexadecapole deformations.
Each state was evolved following its own mean-field dynamics after removing the constraint.
The TDGCM equation~\ref{eq:TDGCM} was solved to determine the collective wave-functions describing coherent superpositions of the resulting time-dependent Slater determinants. 
The initial condition for the collective wave-function was determined according to the multipole vibration to study. 

\begin{figure}
\begin{center}
\includegraphics[width=7.5cm]{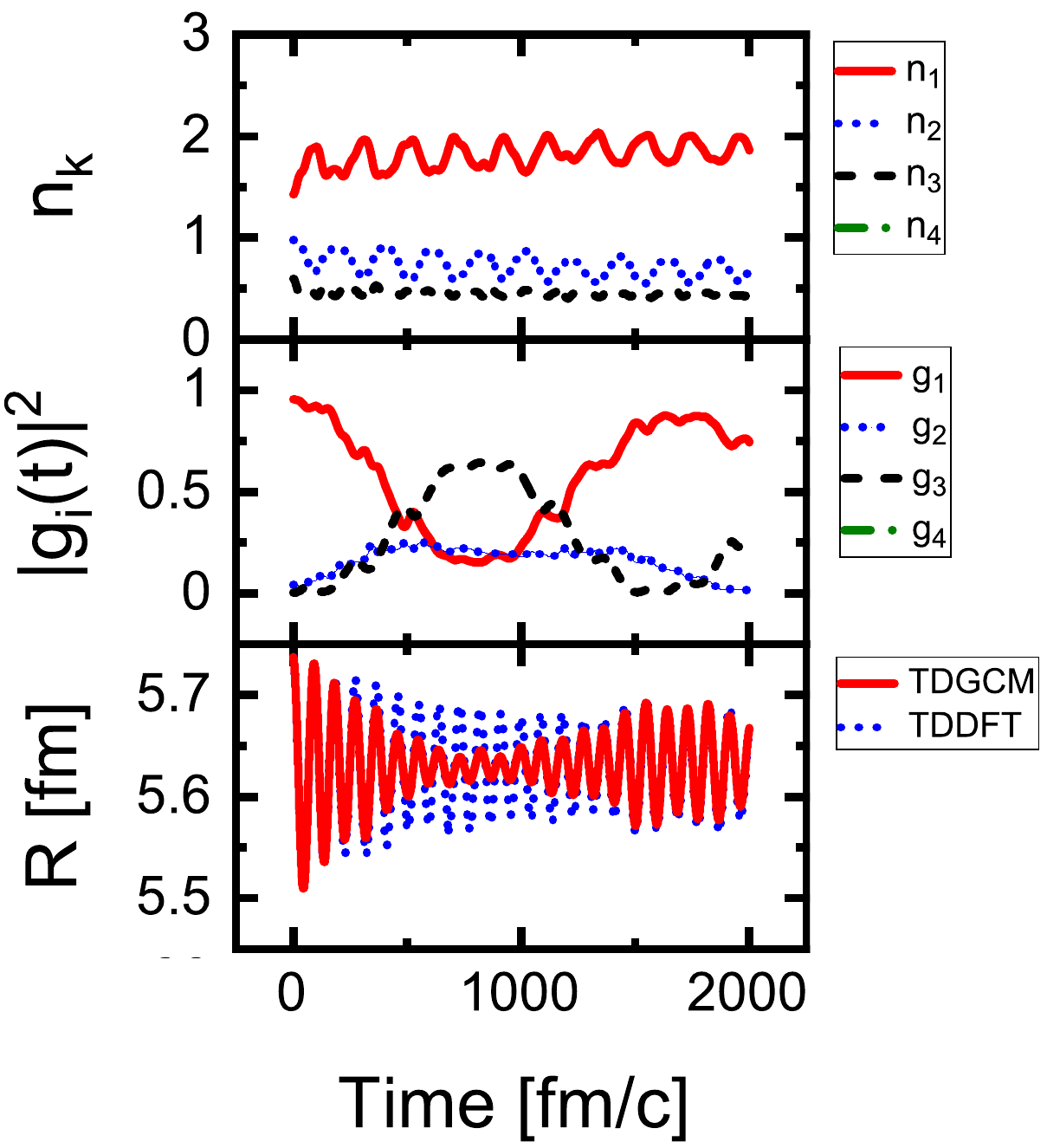}
\caption{Monopole vibration in $^{208}$Pb from TDGCM. (a) Square moduli of components of the collective wave-function $g_i(t)$ associated with initial monopole ($i=1$), quadrupole ($i=2$) and octupole ($i=3$) deformations. (b) Mean-field (TDDFT) and TDGCM predictions of the monopole moment $R(t)$. Adapted from Ref.  \cite{li2023b}. \label{fig:TDGCM-Li1}}
\end{center}
\end{figure}

For instance, the TDGCM solution for monopole vibrations shown in Fig.~\ref{fig:TDGCM-Li1} was obtained by setting $g_i(0)=1$ for the Slater determinant $|\phi_{i=1}(t)\>$ built with an initial monopole constrained, while  $g_{i\ne1}(0)=0$ was chosen for the other multipolarities. 
A significant mixing between the different mean-field solutions is present after few oscillations of the monopole moment, inducing a beating pattern in the TDGCM prediction of the time evolution of the monopole moment. 
The resulting strength distribution is in much better agreement with experiment than the mean-field prediction (see Fig.~\ref{fig:TDGCM-Li2}). 
It is concluded that the fluctuations included in the TDGCM with time-dependent mean-field basis states account for at least part of the spreading width of giant reasonances.

\begin{figure}
\begin{center}
\includegraphics[width=7.5cm]{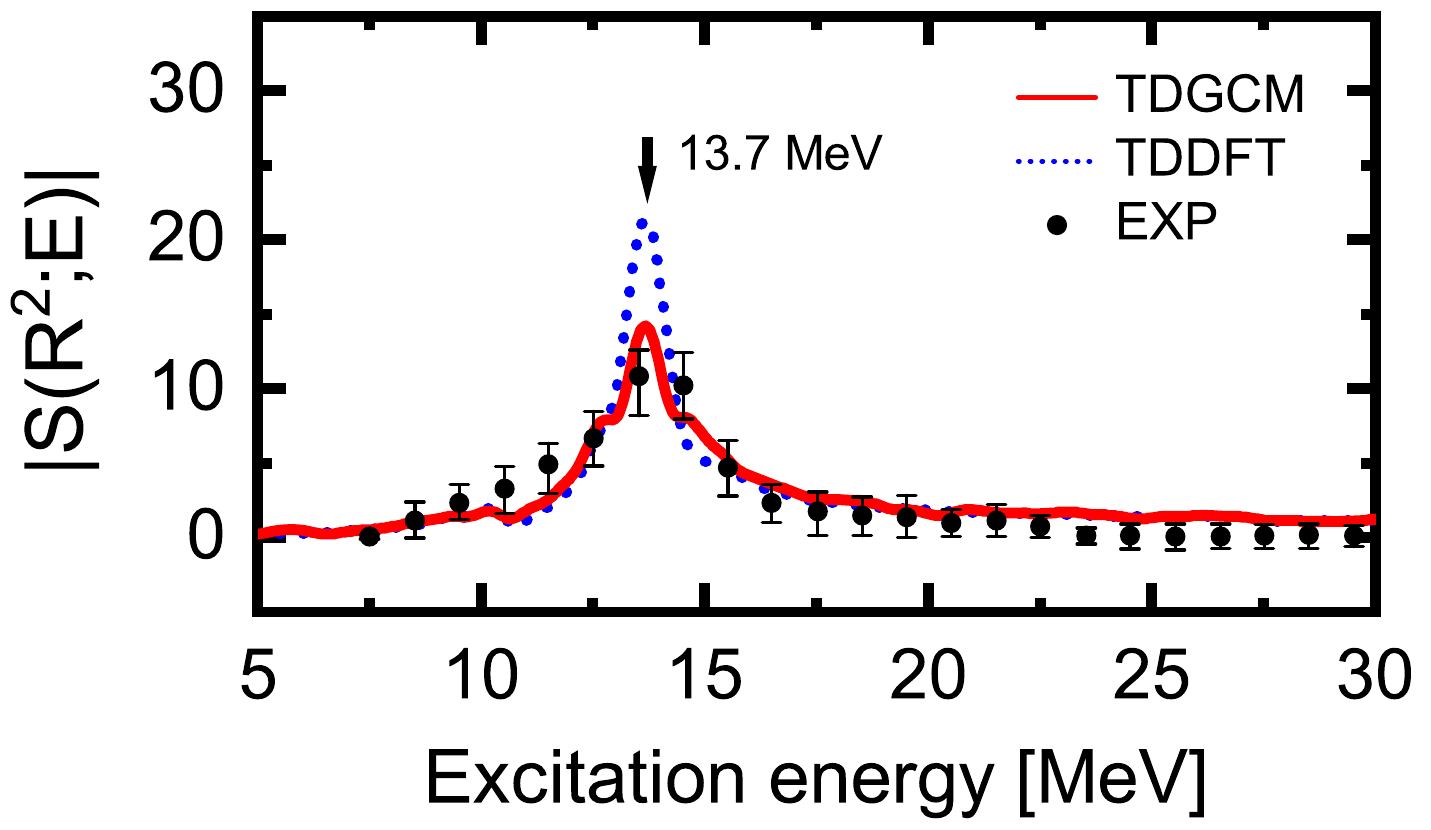}
\caption{The experimental isoscalar GMR stregth function in $^{208}$Pb \cite{patel2014} is compared to time-dependent mean-field (TDDFT) and TDGCM predictions. Adapted from Ref.  \cite{li2023b}. \label{fig:TDGCM-Li2}}
\end{center}
\end{figure}

\subsection{Anharmonicity\label{sec:anharm}}

In the harmonic picture, a GR is the first phonon eigenstate of an harmonic oscillator describing the collective motion, and corresponds to a coherent sum of  $1p1h$ states \cite{boh75}.
Experimental observations of 2 and 3-phonon states prove the vibrational nature of GR.
However, they also show limitations of the harmonic picture~\cite{cho95,aum98,sca04}.
In particular, the excitation probability of multi-phonon states is  larger than predicted by the harmonic picture.
This indicates that different phonon states may be coupled by the residual interaction~\cite{vol95,bor97,sim03,fal03,cho04,lan06,sim09}. 

The TDHF approach takes into account some effects of the residual interaction if the considered phenomenon can be observed in the time evolution of a one-body observable. 
In particular, the nonlinear response in TDHF contains the couplings between one- and two-phonon states coming from the $3p1h$ and $1p3h$ residual interaction~\cite{sim03}.
In that sense, it goes beyond the RPA, which is a harmonic picture and contains only $1p1h$ residual interaction. 

The couplings leading to the excitation of a GMR or a GQR (resp. a GMR) on top of a GDR (resp. a GQR) were investigated in Ref.~\cite{sim03} using the nonlinear response to an external field in the TDHF theory.  
As a continuation to this work, different techniques to compute the matrix elements of the residual interaction responsible for these couplings were introduced in Ref.~\cite{sim09}.

These matrix elements can be written $v_{\mu}=\<\nu|\oV|\nu\mu\>$ where the residual interaction $\oV=\oH-\oH_0$ is the difference between the full Hamiltonian $\oH$ and the HF+RPA Hamiltonian $\oH_0$. 
$|\nu\>$ and $|\nu\mu\>$ are 1 and 2-phonon eigenstates of $\oH_0$ with eigenenergies $E_\nu=E_0+\hb\omega_\nu$ and $E_{\nu\mu}=E_0+\hb\omega_\nu+\hb\omega_\mu$, where $\omega_{\nu,\mu}$ denote the collective frequencies and $E_0$ is the ground state energy. 
The state $|\nu\mu\>$ can be seen as one phonon of the GR $\mu$ (e.g., a GQR) excited on top of one phonon of the GR $\nu$ (e.g., a GDR).

In addition to the original technique based on the non-linearities of the time-dependent response~\cite{sim03},  two other methods were introduced in Ref.~\cite{sim09}.
A brief summary of these three methods is given below (see Ref.~\cite{sim09} for more details):
\begin{itemize}
\item {\it method 1:} A boost $e^{-ik_\nu\oQ_\nu}$ applied on the ground state induces, at lowest order in $k_\nu$, an oscillation of $Q_\nu(t)=\<\oQ_\nu\>(t)$  linear in $k_\nu$, and an oscillation of $Q_\mu(t)$ quadratic in $k_\nu$ and proportional to $v_\mu$. Computing the response $Q_\mu(t)$ to such a boost with TDHF gives then access to~$v_\mu$.
\item {\it method 2:} The same boost can be applied on a HF state obtained with a small constraint $\lambda\oQ_\mu$. The linear response $Q_\nu(t)$ oscillates then with a frequency $\omega_\nu(\lambda)$. The variation $\frac{\partial\omega_\nu}{\partial\lambda}$ is proportional to $v_\mu$. Computing  $\frac{\partial\omega_\nu}{\partial\lambda}(\lambda=0)$ with a TDHF or a deformed RPA code\footnote{This technique involves linear response only. The matrix element $v_\mu$ can then be computed with a RPA code  allowing initial deformations (generated by the constraint $\lambda\oQ$) of the ground-state.} gives also access to $v_\mu$.
\item {\it method 3:} A ''double'' boost  $e^{-ik_\nu\oQ_\nu}e^{-ik_\mu\oQ_\mu}$ is applied on the ground state $|0\>$. Define the ''coupling response function'' as\footnote{This function differs from the standard response function [see Eq.~(\ref{eq:strengthlin})] essentially by the cosine function instead of a sine function.} 
\oeq
R_\nu^c(\omega) = \frac{-1}{\pi k_\nu k_\mu}\int_0^\infty dt\, \cos(\omega t) Q_\nu(t).
\label{eq:R_nu}
\ceq
At lowest order in $k_{\mu,\nu}$, we can show that $R_\nu^c(\omega)$ is proportional to $v_\mu$ and  exhibits peaks at $\omega_\nu$ and $|\omega_\nu\pm\omega_\mu|$ with opposite signs.
The amplitude of these peaks provides  a third way to extract $v_\mu$.
\end{itemize}

\begin{figure}
\begin{center}
\includegraphics[width=6.5cm]{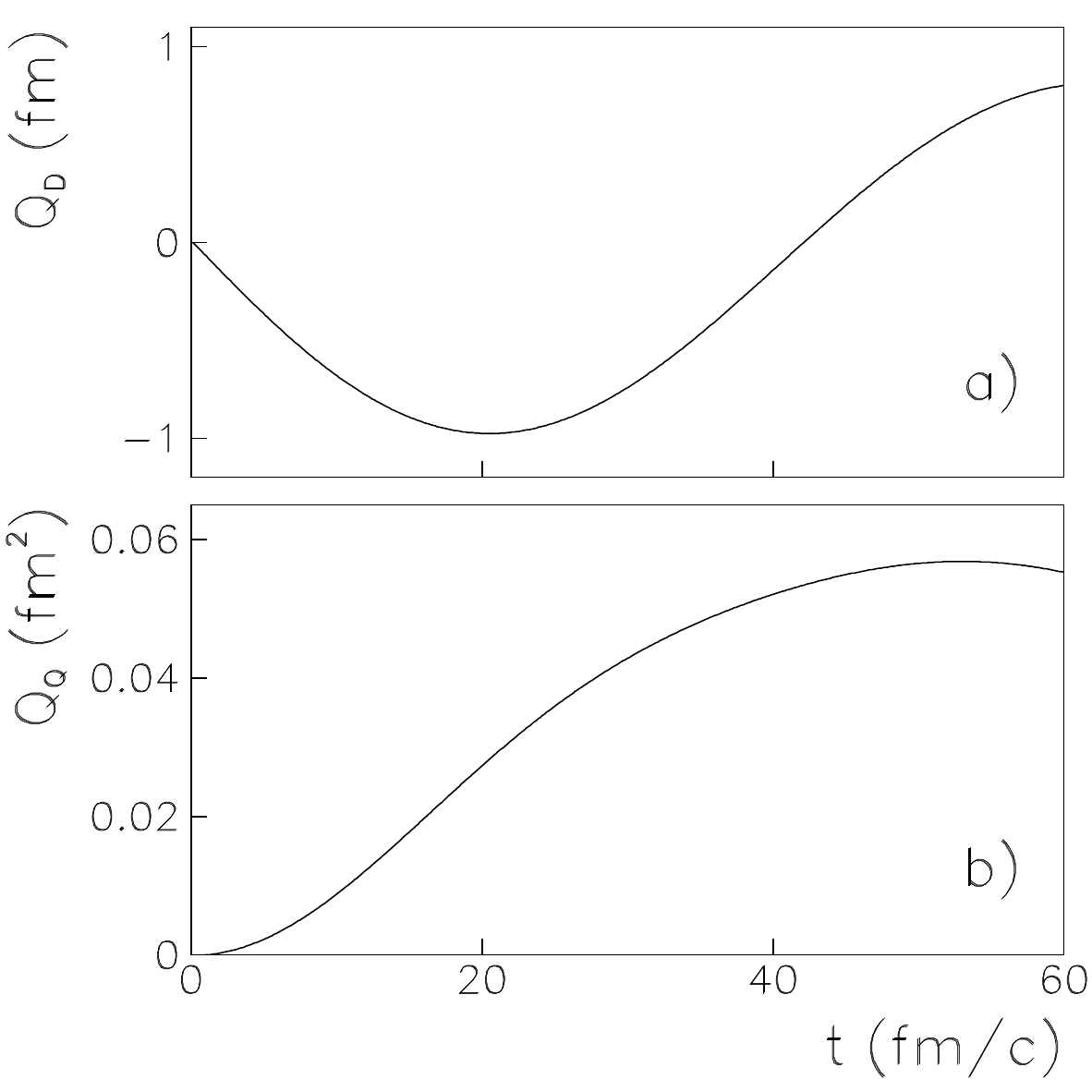}
\caption{Time evolution of the dipole (a) and quadrupole 
(b) moments in $^{132}$Sn after a dipole boost.
\label{fig:quadratic1}}
\end{center}
\end{figure}

\begin{figure}
\begin{center}
\includegraphics[width=6.5cm]{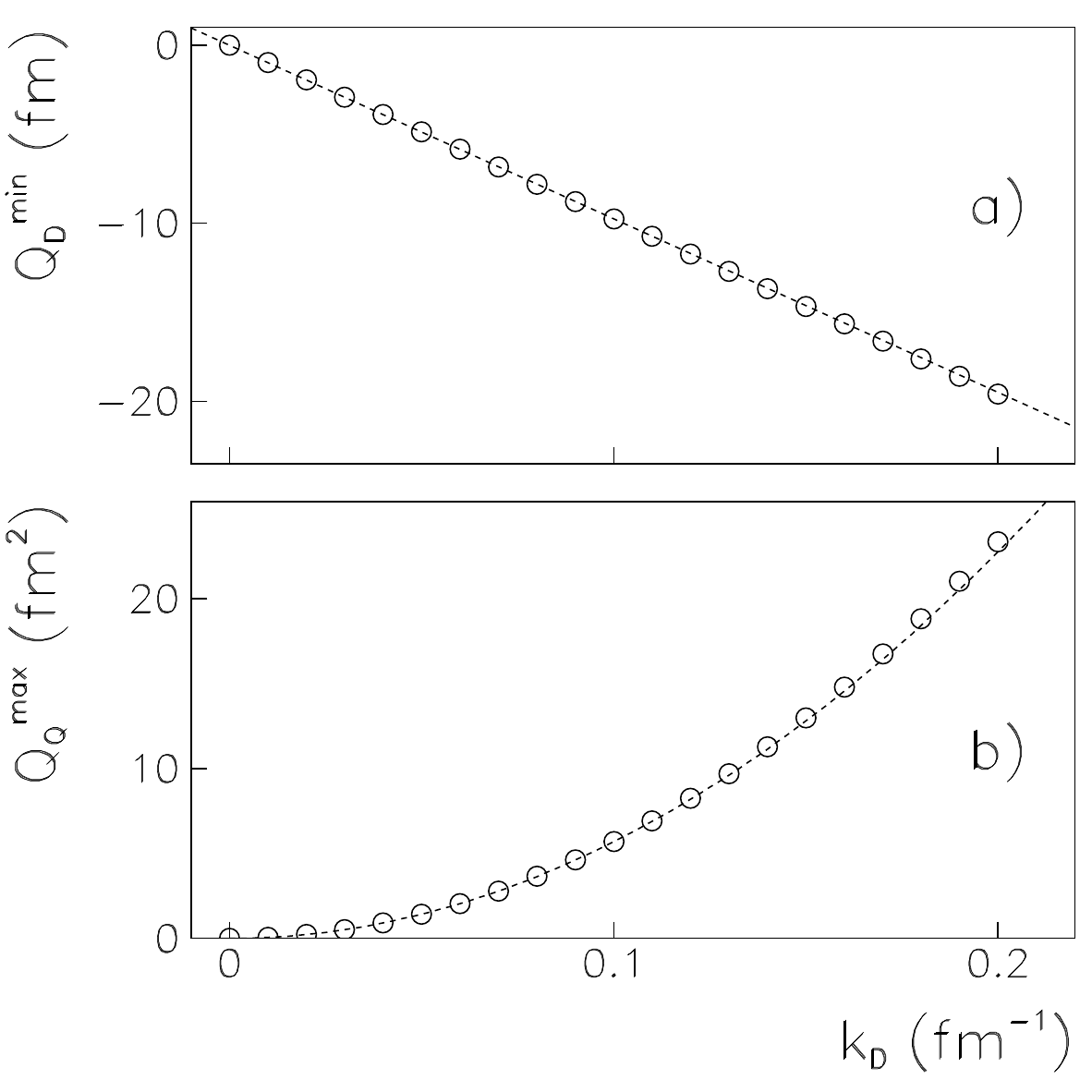}
\caption{Circles indicate the first minimum and maximum of the dipole (a) and quadrupole (b) moment evolution, respectively, following a dipole boost in $^{132}$Sn with a boost velocity~$k_D$. 
Dashed lines show linear and quadratic extrapolations at $k_D\rightarrow0$ of the dipole (a) and quadrupole (b) amplitudes, respectively.
\label{fig:quadratic2}}
\end{center}
\end{figure}

An illustration of the first method applied to the $^{132}$Sn nucleus with $\nu\equiv$GDR and $\mu\equiv$GQR is given in figs.~\ref{fig:quadratic1} and~\ref{fig:quadratic2}.
On Fig.~\ref{fig:quadratic1}, we observe an oscillation of both the dipole and quadrupole moments, although the boost contains only the dipole moment. 
The oscillation of the quadrupole moment is, in fact, induced by the residual interaction. 
The right panel shows that, as expected, the amplitude of the dipole (resp. quadrupole) oscillation is linear (quadratic) in the boost velocity~$k_D$. 
Numerical application gives a matrix element of the residual interaction $v_Q^{(1)}\simeq-0.61$~MeV. 
The two other methods give $v_Q^{(2)}\simeq-0.56$ and $v_Q^{(3)}\simeq-0.68$~MeV, respectively, showing a relatively good agreement between the three methods~\cite{sim09}.

Couplings have been computed in other tin isotopes~\cite{sim09} and in other nuclei ($^{40}$Ca, $^{90}$Zr, and $^{208}$Pb) with TDHF~\cite{sim03} and with a boson mapping method~\cite{fal03}.
Refs.~\cite{sim03,fal03} also discuss couplings involving the GMR built on top of the GQR or the GDR. 
The TDHF results provide a confirmation to the amplitude of the couplings computed with the  boson mapping method (see discussion in Ref.~\cite{sim03}).

Another conclusion of Ref.~\cite{sim09} is that there is no (or little) dependence of the coupling between dipole and quadrupole motion with isospin.
However,  an overall decrease of the coupling is obtained with increasing mass, indicating that the couplings are mediated by the surface~\cite{sim03,fal03,sim09}. 

Overall, the couplings are small but significant compared to the GR energies (e.g., $v/\omega\sim5\%$ in $^{132}$Sn). 
Their effect on the first phonon is negligible, but becomes sensible on the second and third phonons, with a typical shift in $\hb\omega$ of the order of $\sim0.5$~MeV as compared with the harmonic picture~\cite{fal03}.
How the anharmonicities induced by these couplings affect the excitation probability to the multiphonon states have been investigated by Lanza {\it et al.} within a semiclassical coupled-channels formalism~\cite{lan06}.
This model, based on the boson mapping method for the multi-phonon properties, allows for calculations of inelastic cross sections for the multiple excitation of giant resonances induced by heavy-ion probes.
Their calculations show that these anharmonicities induce an increase of the inelastic cross-section (as compared to the harmonic model) in the multi-phonon region, in good agreement with experimental data.

The role of pairing correlations, neglected so far, should also be considered. 
For instance, fully self-consistent quasi-particle-RPA (QRPA) codes allowing for static deformations (see, e.g., the code developped by S. P\'eru~\cite{per08}) could be used to obtain the couplings between the GQR and the GMR built on top of it. 
Couplings with exotic modes such as the pygmy dipole resonance~\cite{lan09} should also be investigated with the present methods. 

Anharmonicities can also be studied with beyond mean-field approaches such as the TDGCM with time-dependent basis states (see section~\ref{sec:TDGCM}).
Indeed, we see in Fig.~\ref{fig:TDGCM-Li1} that this approach predicts that an initial monopole excitation can induce a time-evolution of quadrupole and octupole moments  \cite{li2023b}. 
Such couplings induce a beating pattern in the evolution of the monopole moment and contribute to the  spreading width (see Fig.~\ref{fig:TDGCM-Li2}). 
As the TDHF response predicts a smaller width, it would be interesting to use TDGCM to evaluate the matrix elements for the couplings between vibrational modes and compare with TDHF predictions. 
Similar implementation of the TDGCM also predicts the excitation of second and third phonons of the same vibrational mode, opening an interesting perspective for studying the role of anharmonicities in the properties of multi-phonon states \cite{marevic2023}. 

\subsection{Pairing vibrations\label{sec:pairing}}

The TDHF calculations presented in the previous sections were applied to ''normal'' vibrations, i.e., vibrations of the one-body density (also called {\it normal} density) $\rho(t)$ with matrix elements 
\oeq
\rho_{\al\be}(t)=\<\Psi(t)|\oad_\be\oa_\al|\Psi(t)\>.
\ceq
These vibrations do not probe directly the pairing correlations between nucleons. 
Inclusion of pairing is possible in the small amplitude limit with the QRPA based on HFB vacua. 
The HFB +QRPA has been widely used in nuclear structure studies~\cite{eng99,kha02,fra05,per08,peru2014,nakatsukasa2016}.

Similarly to the fact that the TDHF approach is an extension to the HF+RPA, a natural extension to the HFB+QRPA is the TDHFB theory~\cite{bla86}.
In particular, the TDHFB theory provides a fully self-consistent\footnote{Here, the self-consistency refers to the fact that the HFB vacuum and the residual interaction inducing the collective dynamics are derived from the same EDF.} response to an external excitation including pairing dynamics and non-linearities. 
This section provides a brief presentation of an application of  TDHFB to pairing vibrations (see~\cite{ave08} for more details).

Pairing vibrations are a particular manifestation of the dynamics of pairing correlations~\cite{boh75,rin80,bes66}.
They are probed in two-nucleon transfer reactions~\cite{rip69,oer01,kha04,pll11,shi11,potel2013}. 
Pairing correlations are then expected to induce a collectivity which manifests itself as an increase of transition amplitude toward states associated with pairing vibrations. 
Starting with an even-even nucleus ground-state with $A$ nucleons and spin-parity $0^+$, and assuming a $\Delta L=0$ direct pair transfer reaction, pair vibration states with $J^\pi=0^+$ are populated in the $A+2$ (pair addition) and/or $A-2$ (pair removal) nuclei. 

Such a process can be simulated within the TDHFB formalism using an initial boost with a Hermitean pair-transfer operator~\cite{bes66}
\begin{eqnarray}
\oF = 
\int d\mathbf{r} \, f(r) \left( \oad_{\mathbf{r},\downarrow} 
\oad_{\mathbf{r},\uparrow} 
+ \oa_{\mathbf{r},\uparrow} \oa_{\mathbf{r},\downarrow} \right),
\label{eq:pairtrans}
\end{eqnarray}
where the arrows label the spin of the single-particles (we omit the isospin to simplify the notation). 
In the present application, $f(r)$ is a Fermi-Dirac spatial distribution containing the nucleus and cutting at 4~fm outside the nucleus. 
Its role is to remove unphysical high energy modes associated with  creation of pairs outside the nucleus.

In this approach, it is assumed that the spectroscopy of the $A-2$, $A$ and $A+2$ nuclei can be obtained from the same quasiparticle vacuum (the $A$ ground-state). 
Improvements have been proposed by Grasso~{\it et al.} where this limitation is overcome for ground-state to ground-state transitions by using different vacua for the parent and daughter nuclei~\cite{gra12}.

\begin{figure}
\begin{center}
\includegraphics[width=4.2cm,angle=-90]{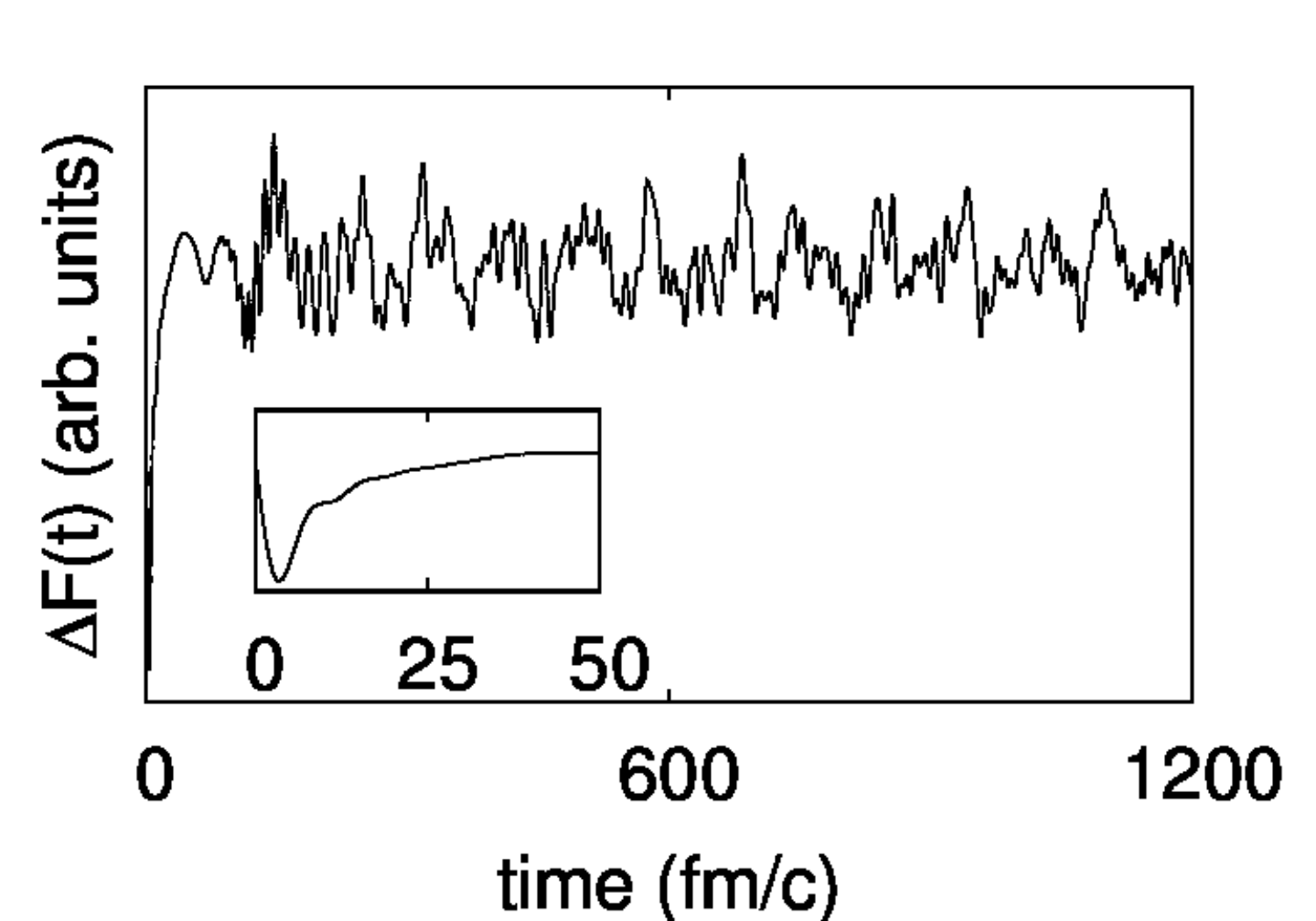}
\includegraphics[width=7cm,angle=-90]{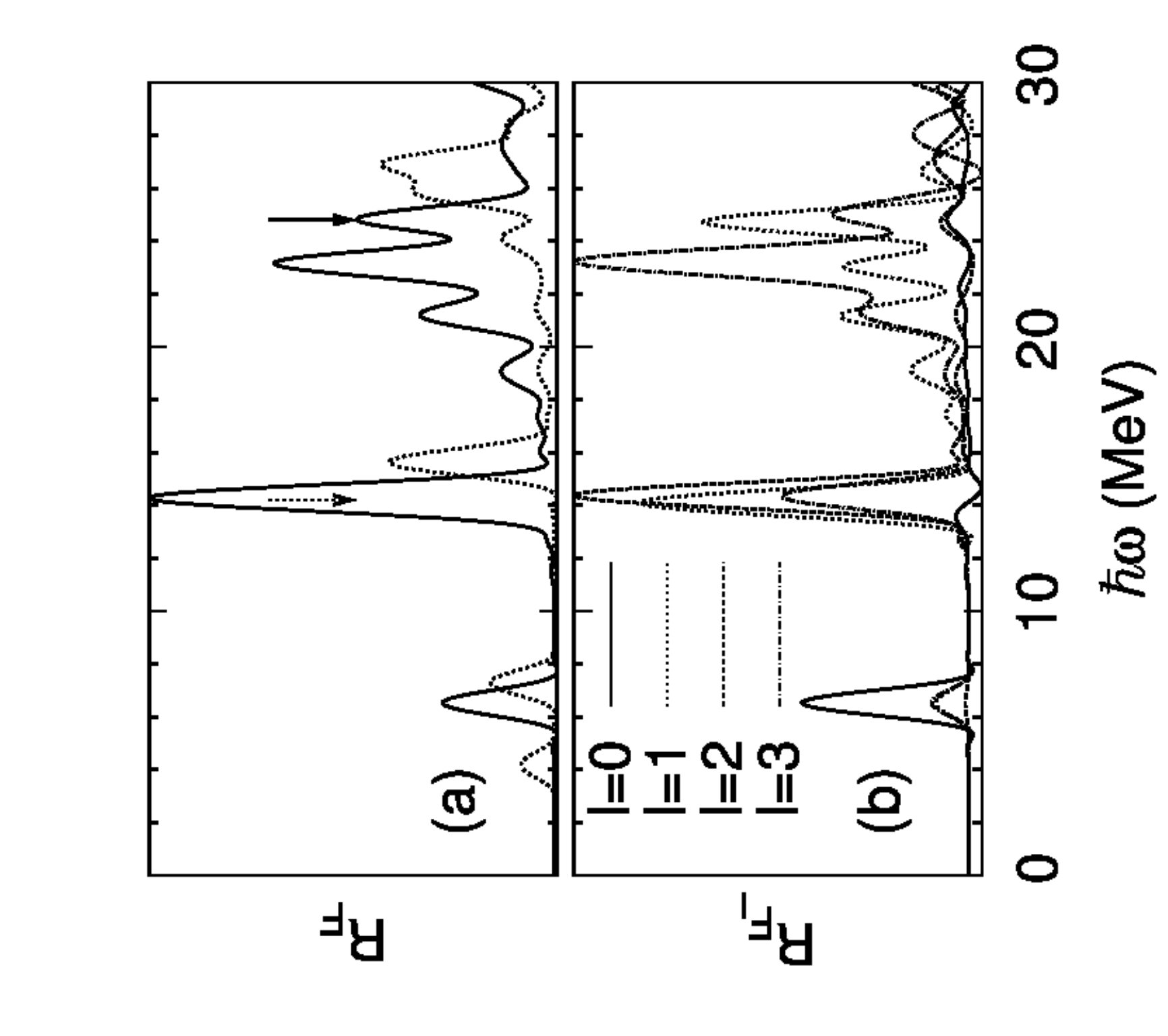}
\caption{(top) Evolution of $\<F\>(t)$ after a pair transfer type excitation on $^{18}$O. The inset shows the same quantity at early times. (middle) Associated TDHFB strength function (solid) compared with the unperturbed spectrum (dashed). 
The arrows indicate pair removal transitions from the $1p_{3/2}$ (solid) and $1p_{1/2}$ (dotted) deep hole states. 
(bottom) TDHFB strength function decomposed into single-particle orbital angular momentum $l$-components. From Ref. \cite{ave08}.
\label{fig:pairvib}}
\end{center}
\end{figure}

The \textsc{tdhfbrad} code has been developed to solve the TDHFB equation in spherical symmetry with a full Skyrme EDF and density-dependent pairing effective interaction. 
It is an extension to the static solver \textsc{hfbrad}~\cite{bennaceur2005}.
As a first application, the linear response of $\<\oF\>(t)$ has been computed in several oxygen and calcium isotopes~\cite{ave08}.
The time-evolution of $\<\oF\>(t)$ is shown in the upper panel of Fig.~\ref{fig:pairvib} for a $^{18}$O vacuum.
The apparent chaotic behaviour of $\<\oF\>(t)$ is due to the simultaneous excitation of several pair vibrations, as we can see from the strength function (solid line) in the middle panel. 
Both pair additional and pair removal (indicated by the arrows) modes are present. 
A comparison with the unperturbed strength function (dashed line) obtained by removing the selfconsistency of the generalised mean-field shows two features in the TDHFB spectrum:
\begin{itemize}
\item an increase of the strength,
\item and a lowering of the transition energies. 
\end{itemize}
Both are compatible with the attractive nature of the dynamical pairing residual interaction. 
In particular, the increase of the strength is a clear signature for collective effects. 
Note that similar conclusions were drawn from  continuum QRPA calculations by Khan {\it et al.}~\cite{kha04}.

The bottom panel of Fig.~\ref{fig:pairvib} shows a decomposition of the response in terms of single-particle orbital angular momentum $l$. 
Together with the structure of the initial HFB vacuum, this decomposition allows for an understanding of each peak in terms of their main particle and hole contributions. 
See Ref.~\cite{ave08} for a detailed microscopic analysis. 
See also Refs.~\cite{ave08,ave10} for an analysis of other nuclei (oxygen, calcium and tin isotopes).

This first realistic application of the TDHFB theory to nuclear systems has confirmed previous QRPA calculations of pairing vibrations~\cite{kha04}. 
The development of 3-dimensional codes~\cite{ste11,eba10,has12} opens also new perspectives for the study of pairing dynamics in deformed nuclei as well as for $L\ne0$ vibrations. 

\subsection{Conclusions and perspectives}

Real time mean-field calculations have been performed to investigate collective vibration properties. 
The response to an external excitation has been obtained with 3D and spherical TDHF codes and associated strength functions have been computed within the linear response theory. 

Direct decay of GR have been analysed from energy spectra of emitted nucleons. 
Within the TDHF approach, the latter contains enough information to reconstruct the strength function if the hole structure of the nucleus is known. 
A comparison between the microscopic decompositions of the strengths obtained from the time response of the excitation operator and from the emitted nucleon spectra shows that the residual interaction couples bound $1p1h$ states with unbound ones, allowing for particle emission in the continuum. 

Non linear vibrations were also studied within the TDHF framework. 
They are used to quantify the coupling between one GR phonon and two (different) GR phonon states.
The large values of the couplings which have been obtained in different nuclei confirm that these are a probable source of anharmonicities as observed in GR multi-phonon experiments. 

The inclusion of pairing correlations in the mean-field dynamics was performed with a realistic spherical TDHFB code.
Applications to pairing vibrations excited in pair-transfer reactions were discussed.
Comparisons with unperturbed calculations show that the dynamical pairing residual interaction included in TDHFB is attractive and enhance collectivity of pairing vibrations. 
Possible extensions to these works include non-linear vibrations and particle decay with pairing, and the study of $L\ne0$ vibrations with 3-dimensional TDHFB codes.

In the present studies, we focused on collective motion at zero temperature, in particular vibrations built on top of the ground state. 
The role of finite temperature on collective motion (e.g., the so-called ''hot GR'') has been widely discussed in the nuclear physics community~\cite{bor98}.
Questions such as the effect of temperature on pairing dynamics and on the couplings between GR multi-phonon states could be addressed with an extension of the present calculations to finite temperature systems. 
TDHF studies of giant resonances at finite temperature are indeed possible~\cite{vau87} starting from an initial hot HF solution~\cite{bon84}.
In section \ref{sec:charge-eq}, we also discuss the particular case of GDR excited in the fusion of two nuclei. 

The calculations presented in this section and the possible perspectives discussed above are based on a mean-field approach. 
For a more realistic comparison to experimental data, extensions to theories going beyond the one-body picture are often necessary, in particular to reproduce the GR fragmentations and widths~\cite{lac04}.
Possible approaches include extended-TDHF~\cite{lac98}, second RPA~\cite{dro90,lac00}, time-dependent density matrix~\cite{toh01,toh02a,toh02b,wan85},
or stochastic one-body transport~\cite{rei92,lac01,jui02} theories.
The TDGCM with time-dependent mean-field basis states has recently been applied to nuclear vibrations, showing promising properties such as spreading width \cite{li2023b} and multi-phonon excitations \cite{marevic2023}.

\section{Fission dynamics \label{chap:fission}}

Microscopic modelling of fission has gained momentum with the TDGCM method with static basis mean-field states \cite{gou05,schunck2016,schunck2022}.
In particular, this method provides a fully quantum mechanical treatment of the evolution through the saddle point.
Nevertheless, standard implementations of the TDGCM lack non-adiabatic effects to fully describe the fragment formation and their evolution post-scission. 
Such non-adiabatic effects are in principle included in TDHF and its extensions. 
As a results, several solvers have been used to study fission mechanisms  \cite{simenel2014a,scamps2015a,tanimura2015,goddard2015,goddard2016,bulgac2016,tanimura2017,scamps2018,scamps2019,bulgac2019,bulgac2019b,pancic2020,bulgac2021,qiang2021a,qiang2021b,bulgac2022,scamps2022,ren2022,iwata2022,ren2022b,scamps2024a,tong2022b,abdurrahman2024,huang2024a,huang2024b,qiang2025}.

The early work of Negele and collaborators \cite{negele1978} demonstrated  the importance of pairing dynamics in fission, enabling rearrangement of single-particle occuppation numbers and playing the role of a lubricant in the evolution from saddle to scission. 
However, realistic applications of time-dependent mean-field  approaches (and their extensions) to nuclear fission have only emerged in the past ten years.

\subsection{Time-dependent mean-field  (and beyond) methods in the context of fission}

\begin{figure}
\begin{center}
\includegraphics[width=8.5cm]{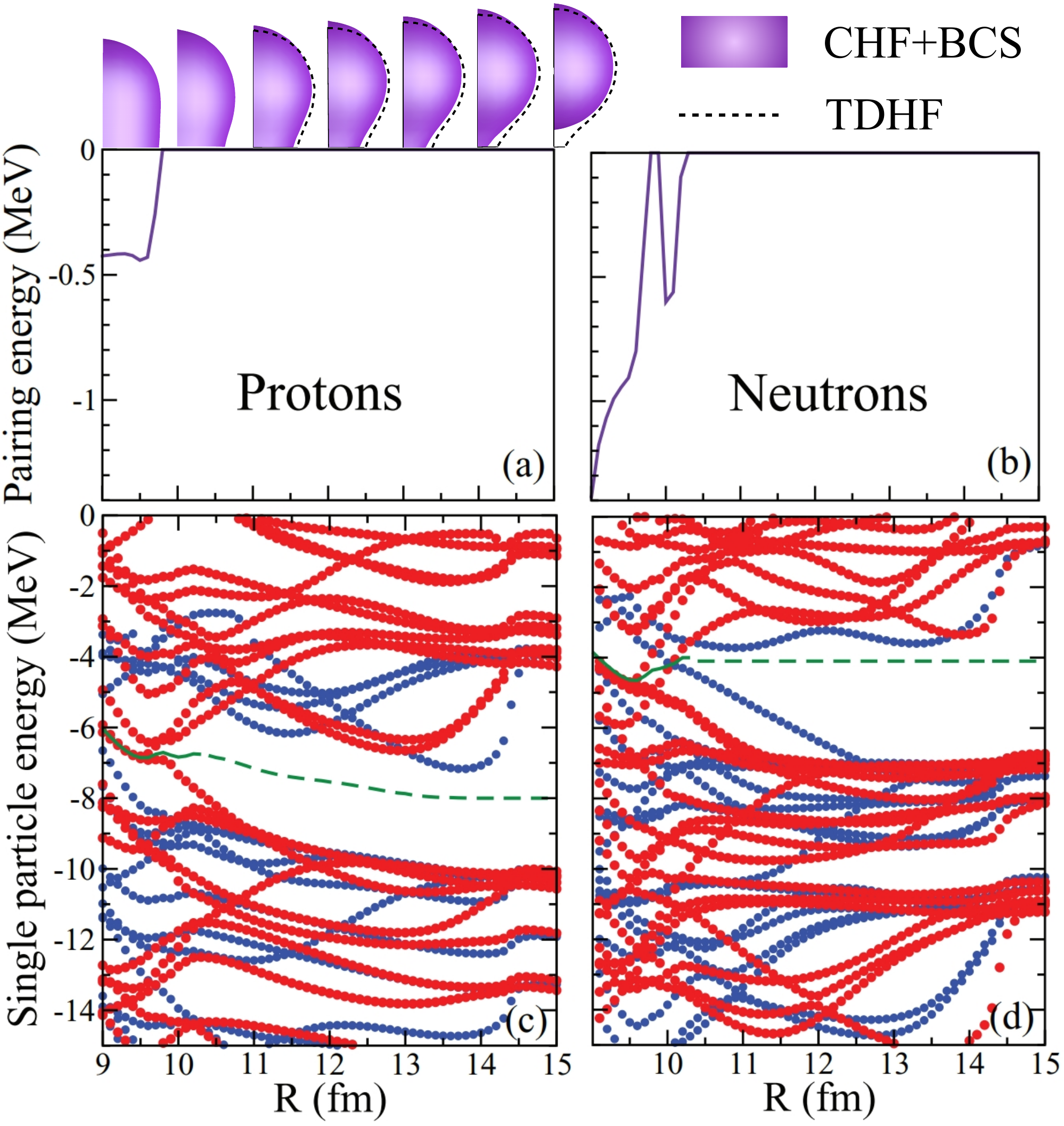}
\caption{Proton (a) and neutron (b) pairing energies as functions of  the distance $R$ between the centres of mass of the  fragments  in $^{264}$Fm symmetric fission. Proton (c) and neutron (d) single-particle energies for states with positive (blue) and negative (red) parity. The green solid lines indicate the Fermi levels. These are continued by green dashed lines, which represent the Fermi levels located arbitrarily in the magic gaps. 
Constrained Hartree-Fock+BCS (CHF+BCS) adiabatic isodensities at 0.08~fm$^{-3}$ are shown on the top at $R = 9, \cdots ,15$~fm.
 Half a fragment is represented, the fission axis being vertical. TDHF isodensities are represented by dashed lines. 
 Adapted from Ref. \cite{simenel2014a}.
\label{fig:264Fm}}
\end{center}
\end{figure}

Fission fragment formation dynamics was studied for the $^{264}$Fm symmetric fission \cite{simenel2014a}. 
This is a special situation where two doubly magic $^{132}$Sn fragments are formed. 
As the latter have no pairing correlations at the mean-field level, this study could be done with TDHF. 
The initial condition for the time-dependent mean-field evolution  was determined by the appearance of magic gaps in the fragments, correlated with the disappearance of pairing energy, as shown in Fig.~\ref{fig:264Fm}.
Comparison of adiabatic (CHF+BCS) and TDHF isodentisites demonstrates that non-adiabatic effects delay scission that occurs for longer elongations, even in a specific case as this one where the fragments are clearly preformed with well defined shell gaps. 
 
 \begin{figure}
\begin{center}
\includegraphics[width=8cm]{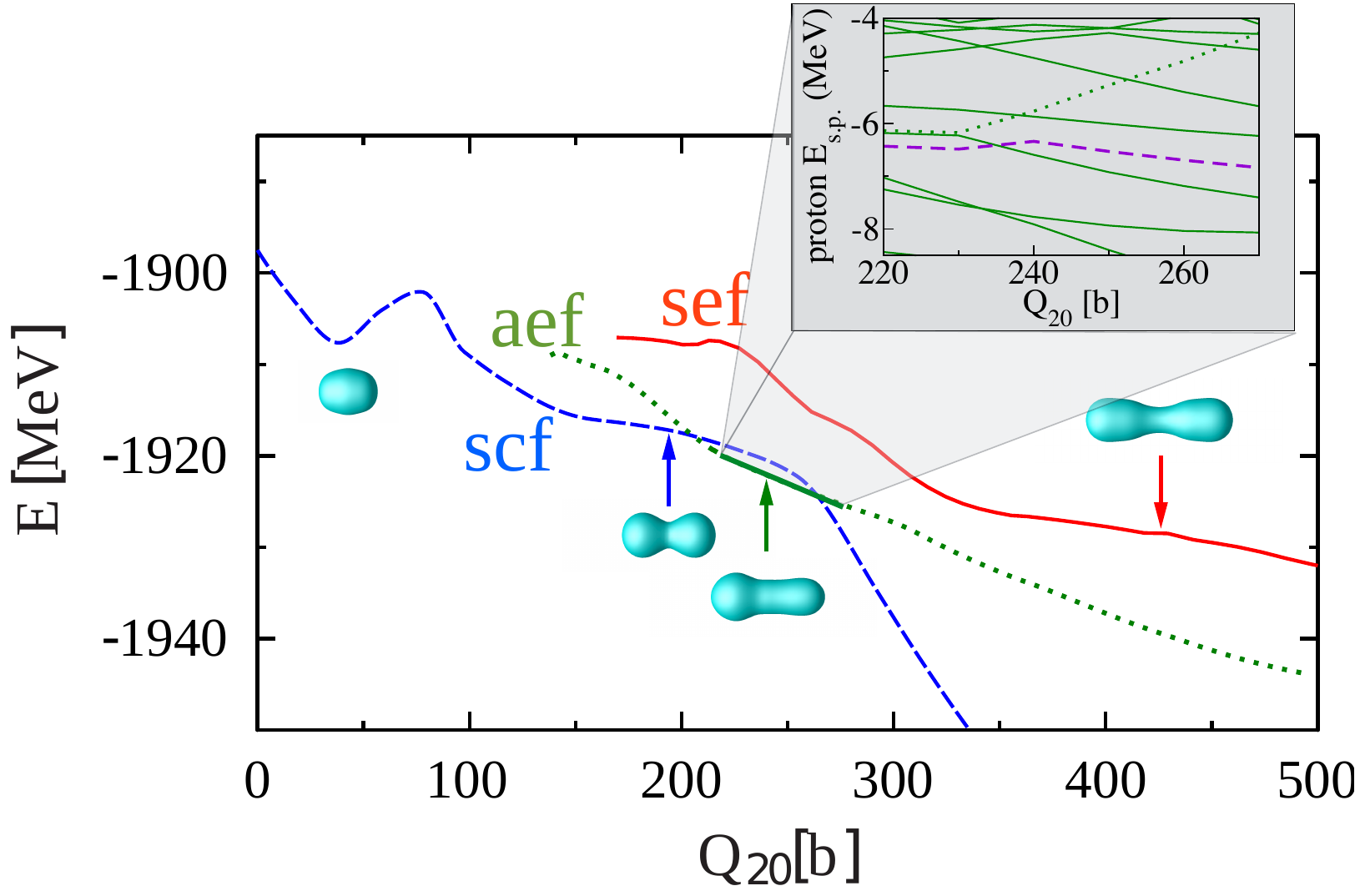}
\caption{One-dimensional HF+BCS adiabatic paths in $^{258}$Fm for the symmetric elongated (sef, solid line), asymmetric elongated (aef, dotted line) and symmetric compact (scf, dashed line) fission modes. The inset shows a subset of single-particle proton energies near the Fermi level (dashed line) in the aef mode. The dotted line represents a specific single particle level whose energy (occupation number) increases (decreases) with elongation. 
From Ref. \cite{scamps2015a}.
\label{fig:Fm258_pot}}
\end{center}
\end{figure}

Systems in which pairing correlations are present in the initial configuration have also been studied with BCS correlations using the frozen occupation approximation (FOA) (see section \ref{sec:pairingTh}) during the time evolution \cite{goddard2015,goddard2016,huang2024a,huang2024b}.
The lack of possibility to rearrange occupation numbers, however, sometime prevents the system to encounter fission. 
This is illustrated in the case of $^{258}$Fm asymmetric fission \cite{scamps2015a} in the inset of Fig.~\ref{fig:Fm258_pot}: the dotted line represents an HF+BCS single particle-level with an occupation number $n\simeq0.4$ at $Q_{20}=220$~fm$^2$, with an increasing energy with respect to the Fermi level (dashed line) and an occupation number going to zero. 
In a TDHF evolution, occupation numbers are fixed and the resulting total energy for this system increases with elongation, preventing its fission. 
 
 \begin{figure}
\begin{center}
\includegraphics[width=8cm]{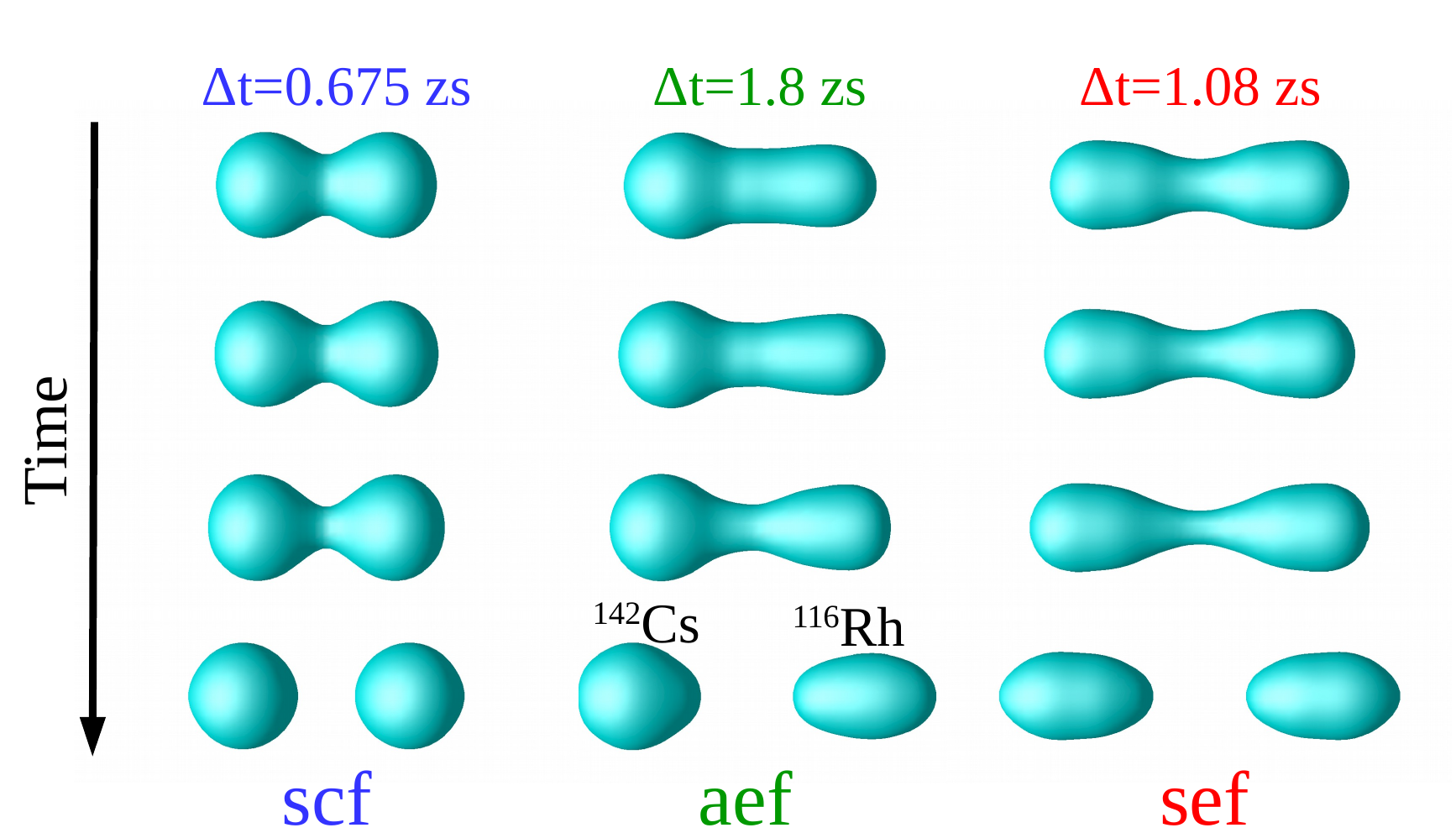}
\caption{Isodensities obtained from TDHF+BCS in the symmetric elongated (sef, solid line), asymmetric elongated (aef, dotted line) and symmetric compact (scf, dashed line) fission modes of $^{258}$Fm. Each isotensity is separated in time by $\Delta t$ whose values depend on the modes and are indicated in the top. From Ref. \cite{scamps2015a}.
\label{fig:Fm258_dens}}
\end{center}
\end{figure}

Realistic dynamical pathways to fission therefore often require a dynamical treatment of pairing correlations, either with TDHF+BCS or TDHFB (see section \ref{sec:pairingTh}) , in order to allow time-dependent evolutions of the occupation numbers. 
This was first achieved in Ref.~\cite{scamps2015a} with TDHF+BCS in the case of $^{258}$Fm fission modes as shown in Fig.~\ref{fig:Fm258_dens} representing isodensity time evolutions for initial states along the one-dimensional fission paths indicated in Fig.~\ref{fig:Fm258_pot}. 
The $^{240}$Pu system was later studied both with TDHFB \cite{bulgac2016,bulgac2019b} and TDHF+BCS \cite{scamps2018}. Although the dynamics is impacted by the pairing strength \cite{bulgac2019b}, the similarities between TDHF+BCS and TDHFB outcomes indicate that the main benefit of a dynamical account of pairing correlations is to enable variation of occupation numbers, thus allowing the system to find a path to fission, while the details of the treatment of these correlations are of secondary importance. 
In fact, thermal fluctuations can play a similar role as pairing in terms of enabling fission, as recently shown by Qiang {\it et al}  \cite{qiang2021a}.

 \begin{figure}
\begin{center}
\includegraphics[width=7.5cm]{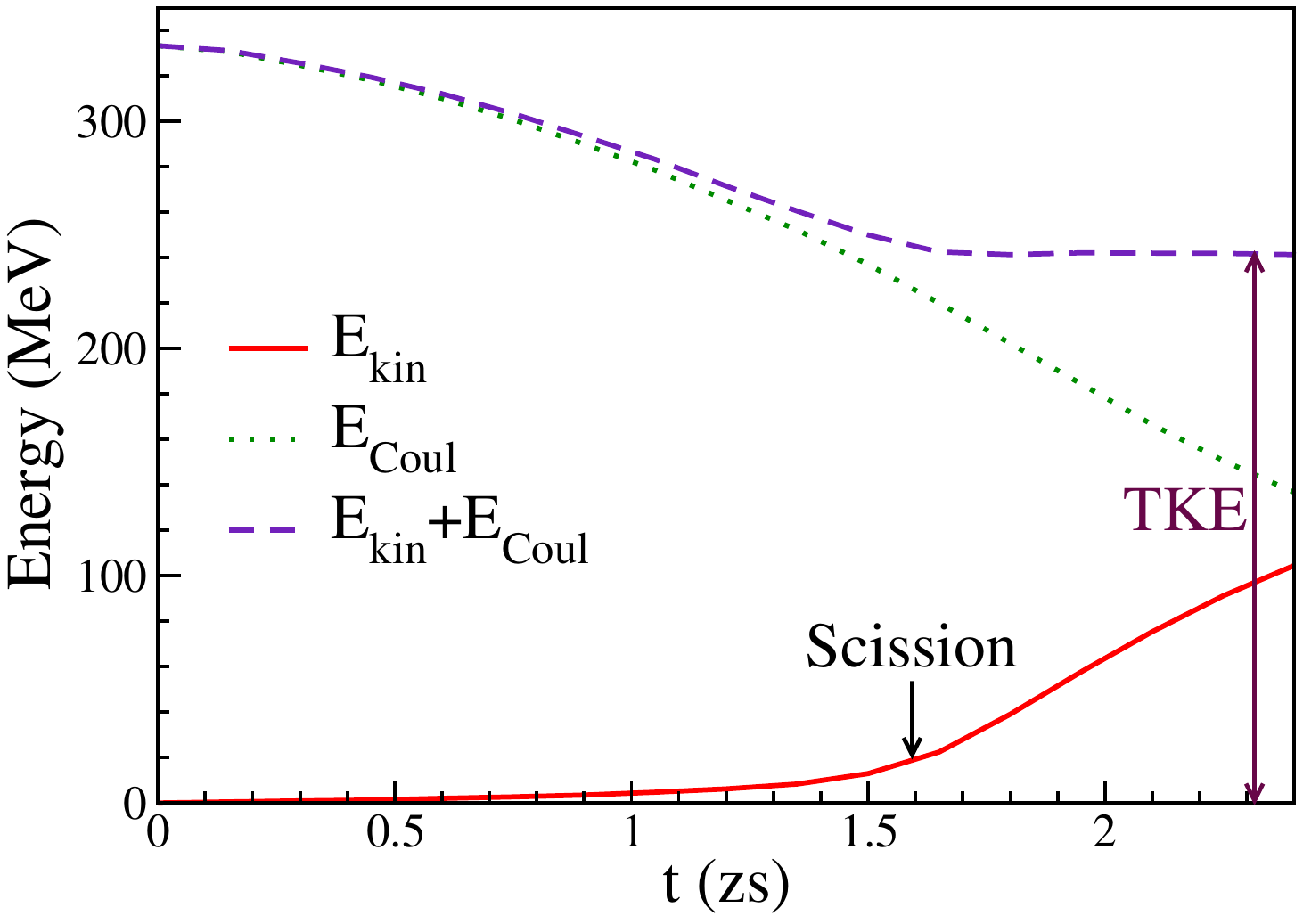}
\caption{Coulomb (dotted line) and  kinetic (solid line) energies and their sum (dashed line) in TDHF evolution of $^{264}$Fm symmetric fission once the $Z=50$ and $N=82$ shell gaps are formed in the pre-fragments. From Ref. \cite{simenel2014a}.
\label{fig:264FmE}}
\end{center}
\end{figure}
 
An advantage of such time-dependent methods is that they could be carried post-scission, allowing for a precise determination of the total kinetic energy (TKE) of the fragments as well as their deformations and vibrational properties~\cite{simenel2014a}. 
Figure~\ref{fig:264FmE} shows the time evolution of Coulomb and kinetic energies of the fragments in $^{264}$Fm symmetric fission. After scission, the sum of these two energies becomes constant as the short range nuclear attraction between the fragments vanishes, and can be identified with the TKE of the fragments. 

 \begin{figure}
\begin{center}
\includegraphics[width=8cm]{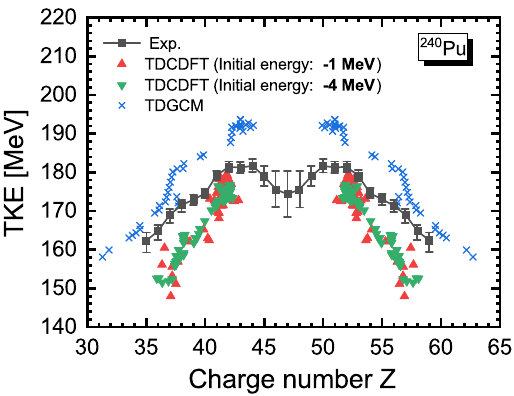}
\caption{$^{240}$Pu fission fragment TKE  from TDHF+BCS with covariant EDF (green and red triangles) compared with experimental data from \cite{caamano2015} (black squares) and TDGCM+GOA predictions (blue crosses). From Ref.  \cite{ren2022}. 
\label{fig:TKE240Pu}}
\end{center}
\end{figure}

Fission fragment TKE are then commonly extracted from various time-dependent microscopic methods, e.g., TDHF+BCS with FOA \cite{goddard2015}, TDHF+BCS \cite{scamps2015a}, TDHFB \cite{bulgac2019b}, and SMF \cite{tanimura2017}. 
Figure~\ref{fig:TKE240Pu} shows an example of application to $^{240}$Pu fission fragment TKE with covariant EDF \cite{ren2022}. The TDHF+BCS calculations reproduce the experimental data reasonably well at $Z\approx52$ but underestimate the TKE for more asymmetric repartitions. Similar observations were made from TDHF+BCS with Skyrme functional \cite{scamps2018} and with the FOA \cite{huang2024a}.
The TDGCM+GOA prediction, however, overestimates the TKE for all fragments. 

 \begin{figure}
\begin{center}
\includegraphics[width=8cm]{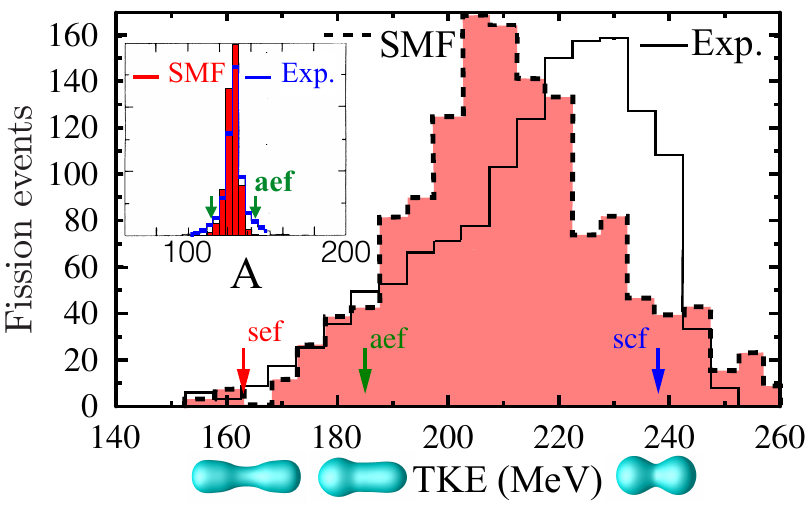}
\caption{$^{258}$Fm fission fragments TKE distribution from SMF \cite{tanimura2017,tanimura2017err} (dashed line) and experiment \cite{hulet1986} (solid line). The arrows indicate the TDHF+BCS predictions \cite{scamps2015a} of Fig.~\ref{fig:Fm258_dens}. The inset shows fragment mass distributions from SMF (red histogram) and experiment (blue solid line). 
Adapted from Refs. \cite{tanimura2017,tanimura2017err}.
\label{fig:Fm258_SMF}}
\end{center}
\end{figure}

One drawback of time-dependent mean-field calculations is that they lack quantum fluctuations.
Although for each fission mode they can be used to predict, e.g., the average number of protons and neutrons in the fragments or their average TKE, the lack of fluctuations prevents from making realistic descriptions of the full distributions of these quantities. 
This is illustrated in Fig.~\ref{fig:240PuFiss} where TDHFB fission trajectories in the $Q_{20}-Q_{30}$ plane are plotted on top of the potential energy surface (PES) for various initial conditions near the outer fission barrier. 
All trajectories mostly follow the fission valley, leading to the prediction of similar fission fragments. 
The resulting fission fragment mass and charge distributions are usually narrower than experimental ones. 

 \begin{figure}
\begin{center}
\includegraphics[width=8cm]{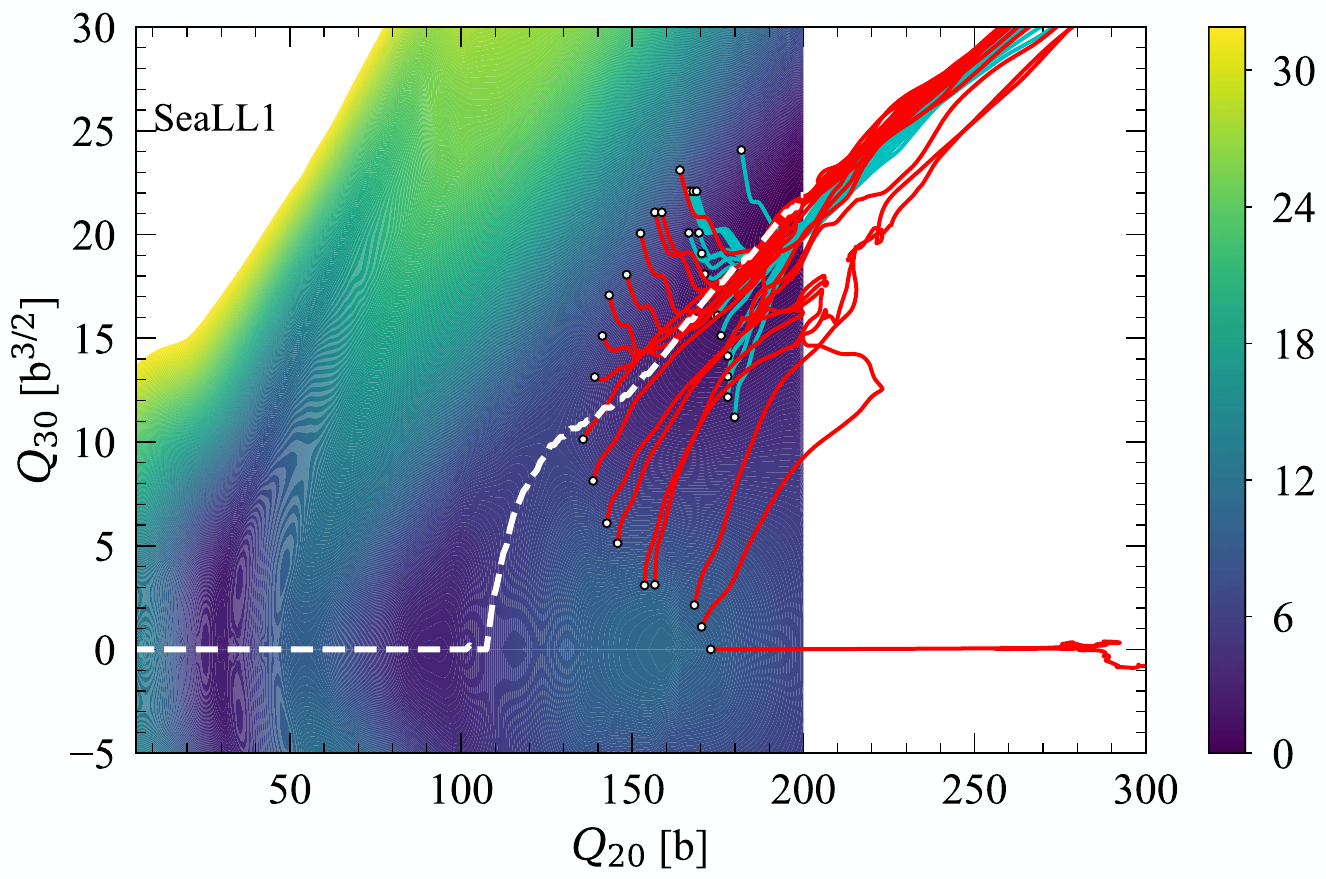}
\caption{$^{240}$Fm fission trajectories from TDHFB. Initial conditions are chosen around the outer fission barrier. The underlying PES is also shown (colour scale in MeV). 
Adapted from Ref. \cite{bulgac2019b}.
\label{fig:240PuFiss}}
\end{center}
\end{figure}

To overcome this limitation, beyond mean-field approaches are required. 
Although the TDRPA has been used to evaluate the widths of fragments mass and charge distributions in $^{264}$Fm symmetric fission \cite{scamps2015a}, the main recent breakthrough in incorporating beyond mean-field fluctuations has been achieved with the stochastic mean-field (SMF) approach by Tanimura {\it et al}
\cite{tanimura2017,tanimura2017err}. 
The width of both fragment mass and TKE distributions is indeed well reproduced by SMF, as shown in Fig.~\ref{fig:Fm258_SMF}.

Although TDHF+BCS, TDHFB and SMF are appropriate tools to describe fission dynamics starting from configurations beyond the fission saddle point, they lack correlations to simulate the whole evolution starting from the compound nucleus. 
A possible approach to overcome this limitation is to combine TDGCM+GOA for the first stage of the evolution, from the compound nucleus to configurations outside the saddle point, with TDHF+BCS describing further evolution towards the formation of two fragments. 
This was recently achieved with the relativistic energy density functional PC-PK1 and a monopole pairing interaction \cite{ren2022}.
However, it was shown that turning off the quantum fluctuations in the TDHF+BCS stage results in too narrow charge distributions. 
Applying a double projection technique \cite{scamps2013a} induces a finite width of the fragment mass and charge distributions for each mean-field evolution \cite{huang2024a}, as well as an odd-even mass effect enhanced with the inclusion of the tensor force \cite{huang2024b}, but the approach still lacks fluctuations in order to reproduce the experimental widths.  
A potential path forward would be to replace the TDHF+BCS calculations by TDGCM with time-dependent basis states (see Sec.~\ref{sec:TDGCM}) in order to maintain these fluctuations in the non-adiabatic evolution towards and at scission.

\subsection{Applications to fission mechanisms}

The origin of mass asymmetric fission of actinides has been debated since its first observation more than 80 years ago \cite{meitner1939}.
Shell effects are responsible for driving the system towards asymmetry as, without them, the dynamics could be described classically by a liquid drop fissioning into two symmetric fragments. 
These shell effects can occur in the compound system \cite{gustafsson1971,bernard2023} as well as in the fragments \cite{hulet1986,wilkins1976,zhang2016,sadhukhan2016,scamps2018}, stabilising their final number of protons and neutrons. 
Shell effects are also invoked to explain asymmetric fission in sub-lead nuclei experimental \cite{andreyev2010,prasad2020,swinton2020b,mahata2022,kaur2024} and theoretical \cite{moller2012,ichikawa2012,andreev2012,panebianco2012,warda2012,andreev2013,mcdonnell2014,scamps2019,bernard2024} studies.

TDHF+BCS calculations were preformed to investigate fission dynamics in several actinides \cite{scamps2018}. 
Several initial conditions were considered along the one dimensional asymmetric fission path obtained by assuming that the system follows a minimum energy path on its way to fission. 
The resulting distributions of proton number $Z$ in the fragments are in good agreement with experimental data as shown in Fig.~\ref{fig:distribution}. 

\begin{figure}
\begin{center}
\includegraphics[width=7.5cm]{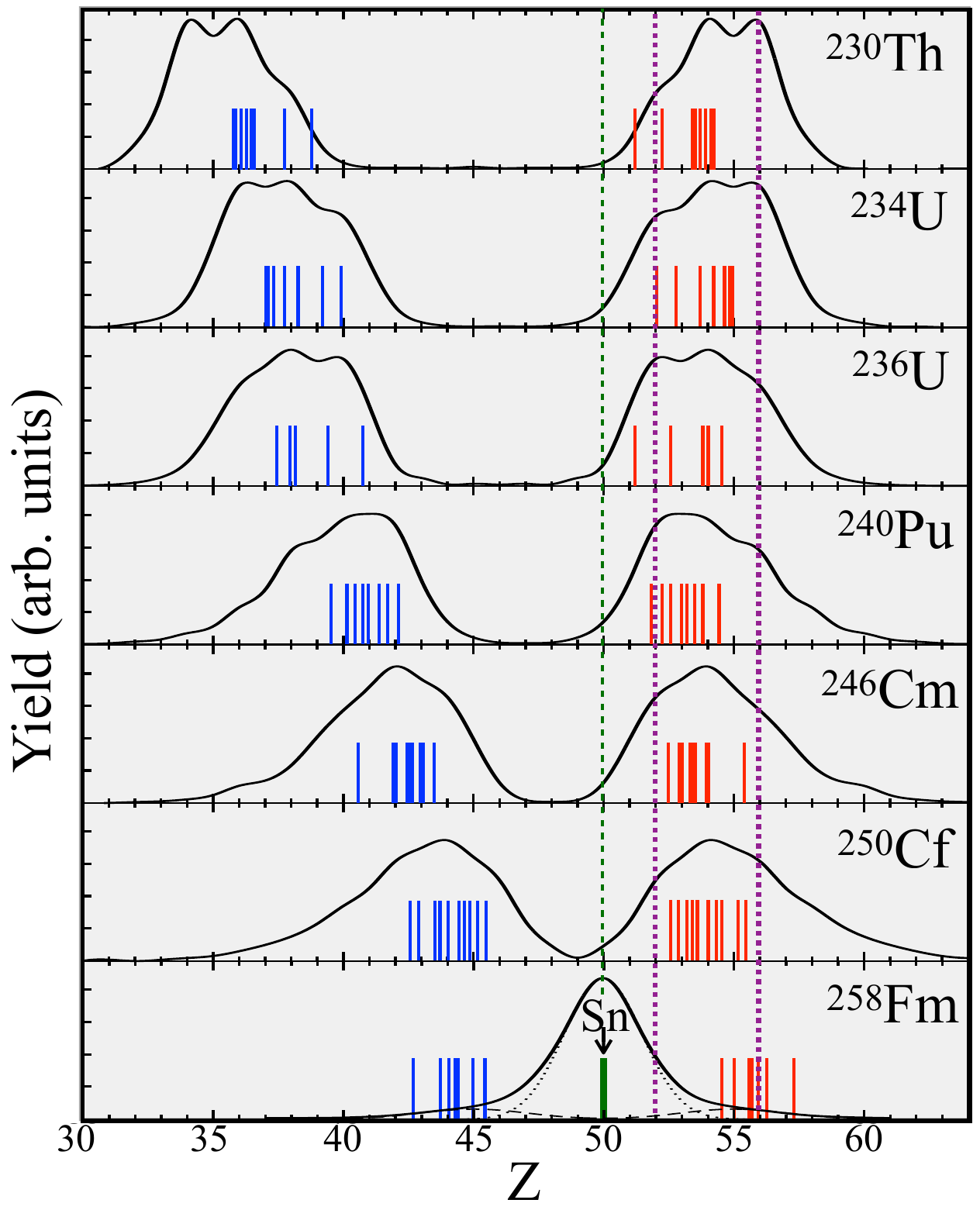}
\caption{Experimental charge distributions in actinide fission (solide black lines) \cite{unik1974,hulet1986}. Vertical blue and red lines show the expectation values of the number of protons $Z$ in the light and heavy fragment, respectively, as obtained from TDHF+BCS calculations with various initial conditions along the one-dimensional asymmetric fission path. The green vertical dashed line shows expected spherical shell effects at $Z=50$. The purple dotted lines show the expected octupole deformed shell effects at $Z=52,56$.  Adapted from Ref. \cite{scamps2018}.
\label{fig:distribution}}
\end{center}
\end{figure}

Depending on the chosen initial condition in the potential energy surface, the ``Standard 1'' (S1) or the ``Standard 2'' (S2) mode \cite{brosa1990} are populated in TDHF evolutions \cite{huang2024a}.
These Brosa modes are centred around $Z\approx52.5$ and $Z\approx55$ protons in the heavy fragment, respectively \cite{bockstiegel2008}. 
Although these modes are expected to be induced by different  shell effects, they both contribute to the same asymmetric fission valley \cite{mcglynn2024}. 
As argued in \cite{scamps2018}, these shell effects may arise from octupole deformed shell gaps at $Z=52$ and $Z=56$. Near scission, fragments take on pear shapes due to a competition between short-range neck attraction and long-range Coulomb repulsion (see before scission isodensities in Fig.~\ref{fig:Fm258_dens}). Octupole shell effects facilitate these deformations. They are still present after inclusion of the tensor force \cite{huang2024b}.

However, the case of the $Z=50$ spherical shell gap is different as it induces a rapid increase of energy with octupole deformation, acting against the formation of such fission fragments \cite{scamps2018}. Nevertheless, spherical shell effects induce a ``supershort'' symmetric Brosa mode in neutron-rich fermium isotopes \cite{hulet1986}, where both fragments approach the doubly magic $^{132}$Sn nucleus. This mode is also described by TDHF+BCS calculations (see left column in Fig.~\ref{fig:Fm258_dens}).
Note that, while S1 and S2 contribute to the same asymmetric fission valley \cite{mcglynn2024}, the symmetric valley associated with the supershort mode in neutron-rich fermium isotopes is separated from the asymmetric one by a potential barrier (or ridge) \cite{bernard2023}. As a result, the TDHF+BCS fission trajectories in $^{258}$Fm \cite{scamps2015a,scamps2018} produce distinct groups of fission fragments: one at $Z\simeq50$ (supershort) and one at $Z_{L/H}\simeq 44/56$ (S2), as shown in the bottom panel of Fig.~\ref{fig:distribution}.

 \begin{figure}
\begin{center}
\includegraphics[width=5cm]{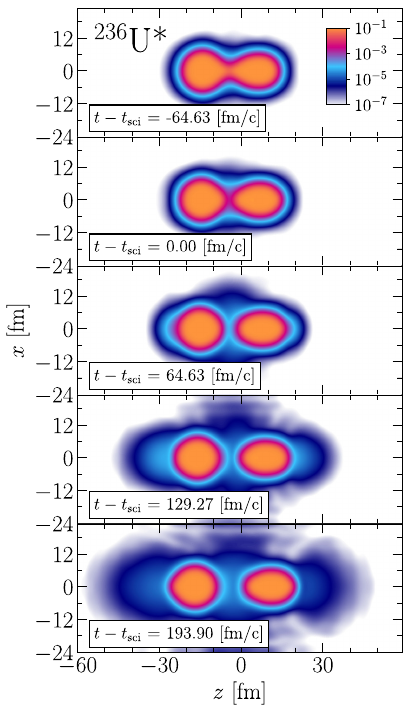}
\caption{Neutron density is represented at various time for a $^{236}$U fission evolution from TDHFB. From Ref. \cite{abdurrahman2024}.
\label{fig:236UFissEvap}}
\end{center}
\end{figure}

Non-adiabatic effects play a significant role near scission and thus time-dependent microscopic studies have brought new  insights into the emission of neutrons at scission \cite{abdurrahman2024} and alpha-cluster formation in the neck~\cite{ren2022b}.
Figure~\ref{fig:236UFissEvap} shows the density of neutrons emitted near scission in $^{236}$U as predicted by TDHFB calculations \cite{abdurrahman2024}.
Scission neutrons are predicted to be  emitted in roughly equal numbers in the equatorial plane and along the fission axis. 

\begin{figure}
\begin{center}
\includegraphics[width=8cm]{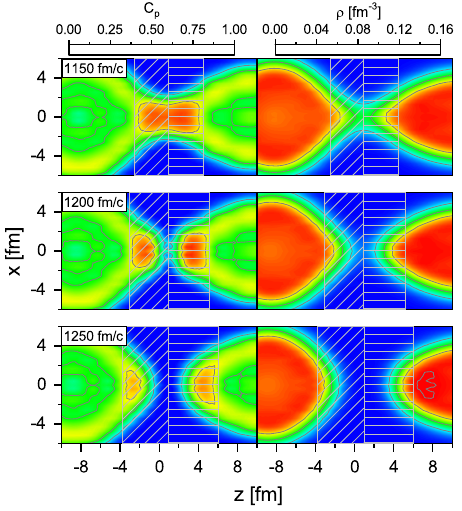}
\caption{TDHF+BCS calculations with covariant EDF showing proton localisation function $C_p$ (left) and total density $\rho$ (right) in $^{240}$Pu asymmetric fission at various times near scission. The color scales are such that $C_p=0$ and $\rho=0$ (blue), while  $C_p=1$ and $\rho=0.16$~fm$^{-3}$ (red).  From Ref. \cite{ren2022b}.
\label{fig:alpha-neck}}
\end{center}
\end{figure}

Figure~\ref{fig:alpha-neck} shows the evolution of the density in the neck region (right panels) during scission of $^{240}$Pu in two asymmetric fragments~\cite{ren2022b}. 
The left panels show the proton localisation function obtained from~\cite{becke1990,reinhard2011} (see Sec.~\ref{sec:DCFHF} for details)
\oeq
C_{q\sigma}(\vr)=\[1+\(\frac{\tau_{q\sigma} \rho_{q\sigma}-\frac{1}{4}|\vna\rho_{q\sigma}|^2-\vj^2_{q\sigma}}{\rho_{q\sigma}\tau_{q\sigma}^{TF}}\)^2\]^{-1} \label{eq:localisation}
\ceq
with isospin $q=p$ and summing over spins $\sigma$. 
See Eqs. (\ref{eq:def_rho}), (\ref{eq:def_tau}), and (\ref{eq:def_j}) for the definitions of the various densities. 
$\tau_{q\sigma}^{TF}=\frac{3}{5}(6\pi^2)^{2/3}\rho_{q\sigma}^{5/3}$ is the Thomas-Fermi kinetic energy density. 
$C_p(\vr)\approx1$ indicates that two protons of same spins are unlikely to be found in $\vr$ due to their spacial localisation (and similarly for neutrons). 
This is what is expected in an alpha cluster, and thus $C_p\approx1$ indicates that the presence of an alpha cluster is likely. 
These calculations then suggest the formation of two alpha clusters in the neck. 
The fact that each alpha ends up in a different fragment provides a possible picture for the neck breaking mechanism. 

After scission, the fragments repel and acquire angular momentum. 
The generation of angular momentum in fission fragments has recently been revisited experimentally, raising interesting questions on the correlations between the spins of the two fission fragments~\cite{wilson2021}. 
The mechanisms at play and their dynamical origin have been investigated in several time-dependent microscopic works \cite{bulgac2021,bulgac2022,scamps2022,scamps2023b,scamps2024a}.
In particular, angular momentum projection techniques have been implemented in this context in order to extract fragment spin distributions, both in static \cite{bertsch2019c,marevic2021} and time-dependent calculations \cite{bulgac2021,bulgac2022,scamps2023b}.

Finally, it is worth noting that fission fragments remain entangled after scission, providing a unique example of quantum entanglement in strongly interacting many-body systems.
The effect of entanglement on fragment properties has been recently studied by combining random fluctuations and particle number projection technique in TDHF+BCS calculations of the fission of $^{240}$Pu \cite{qiang2025}.
In particular, it is shown that dynamical quantum entanglement plays a crucial role in the appearance of sawtooth distributions of neutron multiplicities.

\section{Heavy-ion collisions \label{chap:HIC}}

We now investigate the collision of two atomic nuclei. 
Dynamical mean-field theories like TDHF are well suited to the study of low-energy reaction  mechanisms, such as fusion, multi-nucleon transfer and quasi-fission. 
Indeed, at low energies, the Pauli principle blocks collisions between nucleons, increasing their mean-free path to the order of the size of the nuclear system. 
In addition, heavy-ion collisions often induce a transfer of relative motion into internal excitation via one-body mechanisms well treated by the TDHF approach.
Completed with beyond mean-field approaches such as TDRPA and SMF to account for quantum fluctuations, TDHF solvers can also be used to investigate deep-inelastic collisions (DIC) well above barrier. 

The outcome of a heavy-ion collision depends essentially on few properties of the entrance channel: energy, masses, angular momentum, deformation and orientation, $N/Z$ asymmetry, and internal structure (e.g., magicity). 
Thus, we discuss different reaction mechanisms and their dependence on entrance channel properties.

\subsection{Direct evaluation of fusion thresholds and cross-sections from TDHF(B) \label{sec:TDHFfusion}} 
 
Early TDHF codes have been successfully applied to describe above-barrier fusion reactions in light systems~\cite{bon78}.
However, these calculations also predicted a lower limit to the angular momentum for fusion. 
For smaller angular momenta, a ``transparency'' was observed in the calculations (two fragments are emitted along the collision axis). 
This prediction was never confirmed experimentally.
In fact, it was shown by Umar and collaborators that this so-called ``fusion-window'' problem was solved with the inclusion of the spin-orbit interaction~\cite{umar1986a,umar1989,reinhard1988}.
Indeed, the latter was shown to be an important source of dissipation in heavy-ion collisions. 
Modern TDHF calculations are now performed with a full Skyrme EDF including spin-orbit terms~\cite{kim97,mar05,nak05,uma05}. 
The tensor force also impacts transparency thresholds \cite{stevenson2016}. 
Naturally, two-body dissipation is also expected to play a role on fusion. 
This can be included in beyond mean-field approaches such as TDDM, leading to an increase of friction coefficients, including at low incident energy  \cite{wen2018}. 

TDHF codes and their extensions can be used to evaluate the capture threshold between two nuclei as well as  above barrier fusion cross-sections.
These studies have shown a rich variety of phenomena induced by the structure of the reaction partners, such as their deformation and orientation, their mass and charge, their neutron-to-proton ratio, and 
superfluid phases. 
Numerous studies of fusion thresholds and direct evaluation of above barrier fusion cross-sections (i.e., without the intermediate evaluation of a nucleus-nucleus potential)  have been performed with TDHF solvers~\cite{kim97,sim01,uma06a,uma06c,sim08,was08,leb12,simenel2013a,vophuoc2016,hashimoto2016,magierski2017,guo2018,ren2020b,magierski2022,yao2024}.

\subsubsection{Fusion of spherical nuclei \label{sec:fus-spher}}

Because the TDHF theory does not allow for quantum tunnelling of the many-body wave function, the TDHF fusion barrier can be identified as the capture threshold for central collisions, above which a compound system is formed and below which the two fragments re-separate. 
Due to the finite time of the TDHF evolutions, one has to define a maximum computational time\footnote{This time may depend on the system. For medium mass systems such as $^{16}$O+$^{208}$Pb, a typical time of $10^3$~fm/$c$ (1~zs=300~fm/$c$) is used.} above which the final configuration (i.e., one compound system or two fragments) is assumed to be reached. 

\begin{figure}
\includegraphics[width=8.8cm]{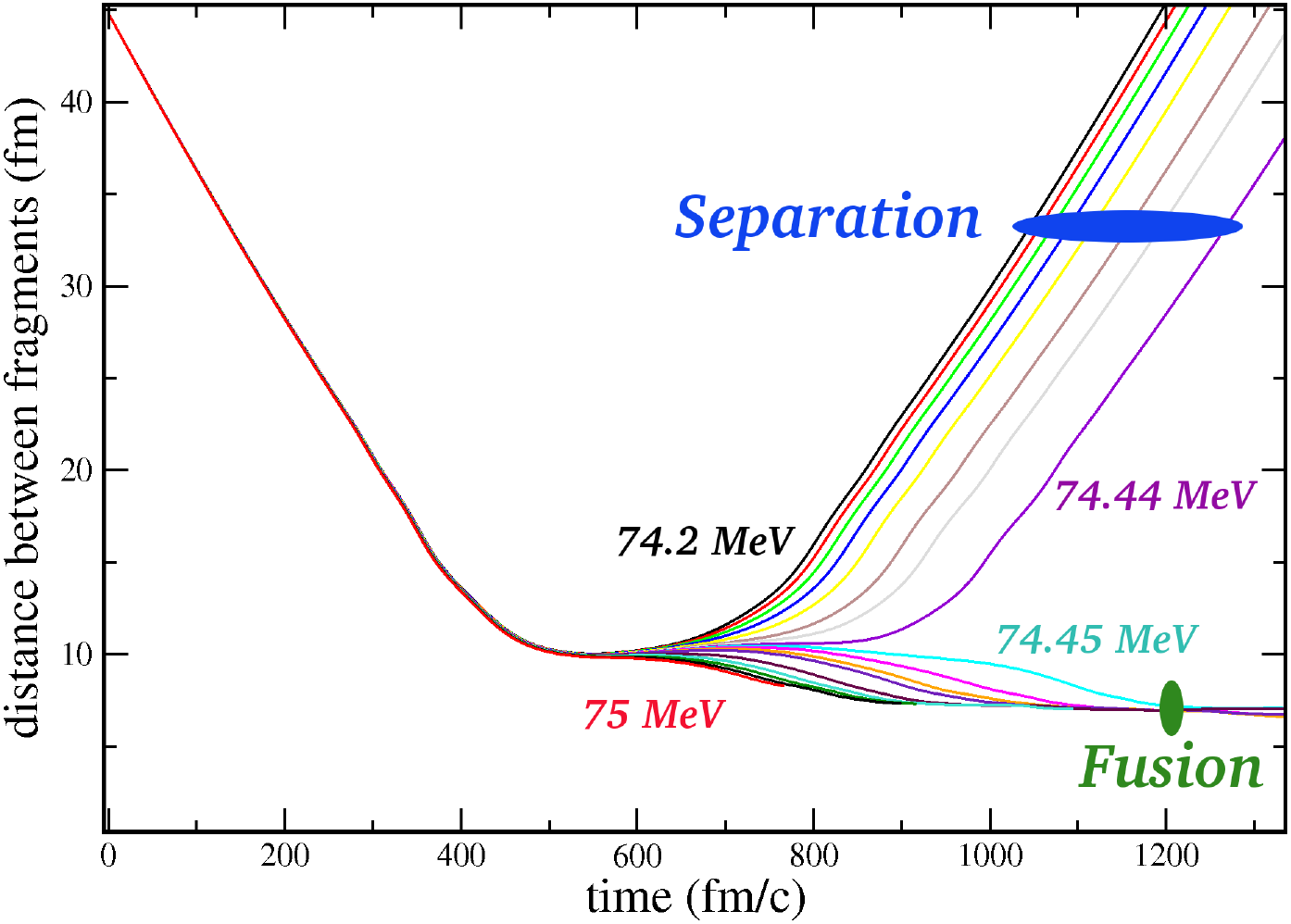}
\caption{Relative distance between the fragments as a function of time for head-on 
$^{16}$O+$^{208}$Pb reactions computed with the \textsc{tdhf3d} code~\cite{kim97}.\label{fig:distances}}
\end{figure}

Figure~\ref{fig:distances} shows the evolution of the relative distance between fragment centres of mass in central $^{16}$O+$^{208}$Pb collisions at different energies around the capture threshold. 
We clearly identify two sets of trajectories associated with capture (fusion) and with re-separation of the fragments. 
As shown in Fig.~\ref{fig:dist_barr}, these calculations predict a fusion threshold of $V_B^{TDHF}=74.445\pm0.005$~MeV that is 
$\sim1.5$~MeV lower than the frozen HF prediction (see Sec.~\ref{sec:FHF}) due to dynamical effects included in TDHF.
We also observe in Fig.~\ref{fig:dist_barr} a good agreement between the TDHF prediction and the centroid of the experimental barrier distribution. 
Other methods based on a macroscopic reduction of the mean-field dynamics, namely the dissipative dynamics TDHF \cite{was08} and the density constrained TDHF \cite{uma09b,umar2014a} (see  Sec.~\ref{sec:DCTDHF}),
also find similar results.

\begin{figure}
\begin{center}
\includegraphics[width=7.5cm]{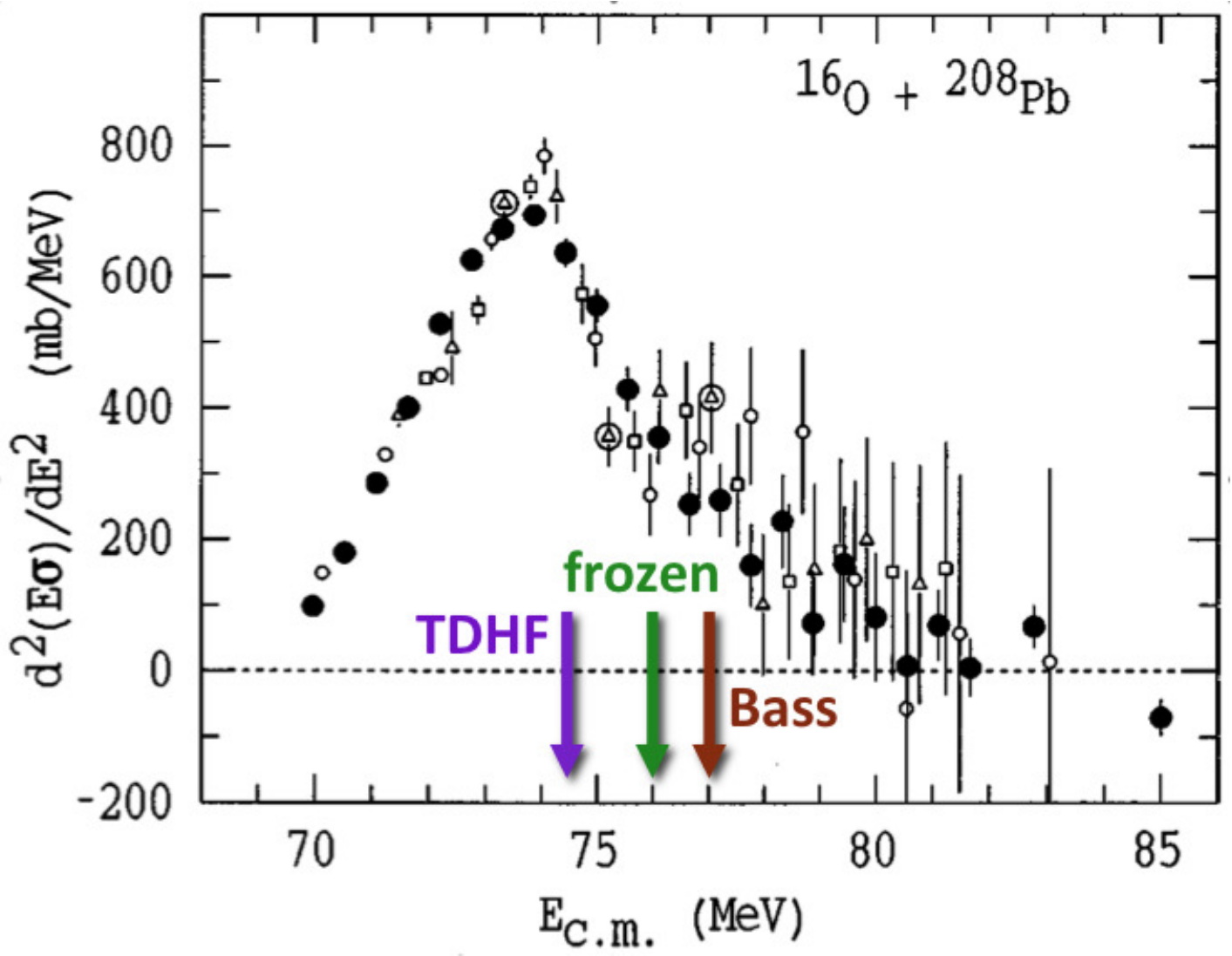}
\caption{Experimental fusion barrier distribution of the $^{16}$O+$^{208}$Pb system from Ref.~\cite{mor99}. The frozen HF (see Sec.~\ref{sec:FHF}), Bass \cite{bas77} and TDHF  (see Fig.~\ref{fig:distances}) barriers are shown with arrows. 
\label{fig:dist_barr}}
\end{center}
\end{figure}

\begin{figure}
\includegraphics[width=8.8cm]{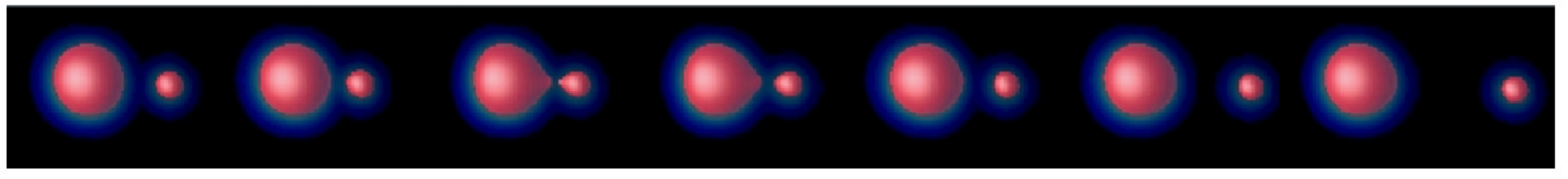}
\includegraphics[width=8.8cm]{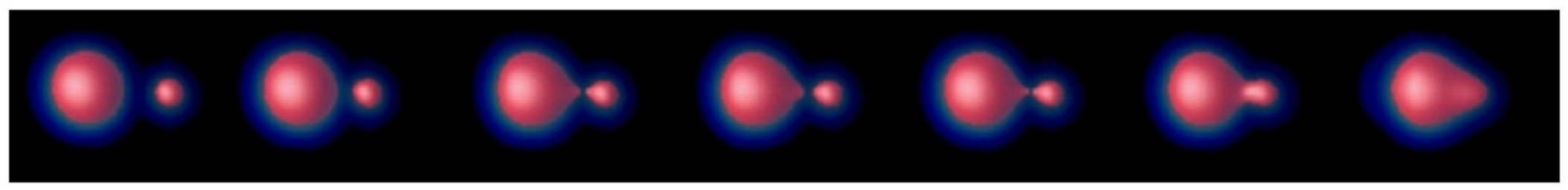}
\caption{(top) Density evolution for the reaction $^{16}$O+$^{208}$Pb corresponding to a head-on collision at a centre of mass energy $E_{c.m.}=74.44$~MeV (just below the fusion barrier). The red surfaces correspond to an iso-density at half the saturation density ($\rho_0/2=0.08$~fm$^{-3}$). Each figure is separated by a time step of 135~fm/c. Time runs from left to right. (bottom) Same at $E_{c.m.}=74.45$~MeV, i.e., just above the fusion threshold.\label{fig:dens}}
\end{figure}

To get a deeper insight into these dynamical effects,  the density evolutions at $E_{c.m.}=74.44$ and 74.45~MeV are plotted in Fig.~\ref{fig:dens}.
In the first case, a ``di-nuclear'' system is formed during a relatively long time ($\sim500$~fm/c) before re-separation. 
In the second case, the system overcomes the fusion barrier. 
More generally, the two figures illustrate the richness of physical phenomena contained in TDHF calculations, e.g., surface diffuseness, neck formation, quadrupole and octupole shapes of the compound system.

The observed lowering of the fusion barrier due to dynamical effects could be partly explained by a coupling of the relative motion to a transfer mechanism~\cite{sim08}.
In fact, the outgoing channel of $^{16}$O+$^{208}$Pb at $E_{c.m.}=74.44$~MeV (see top of Fig.~\ref{fig:dens}) is, in average, $^{14}$C+$^{210}$Po. 
This two-proton transfer channel effectively lowers the barrier by decreasing $Z_1Z_2$ and, then, the Coulomb repulsion. 
Transfer reactions  are discussed in more details in section~\ref{sec:transfer}. 
Note  that low-lying collective vibrations, such as the first $3^-$ state in $^{208}$Pb (see Fig.~\ref{fig:PbQ3}) also affect the fusion barrier distribution~\cite{mor99} (see section \ref{sec:CC}).
In addition, the energy of these states can be modified by the tensor interaction, and thus affect the fusion threshold accordingly \cite{guo2018}.

\begin{figure}
\begin{center}
\includegraphics[width=8cm]{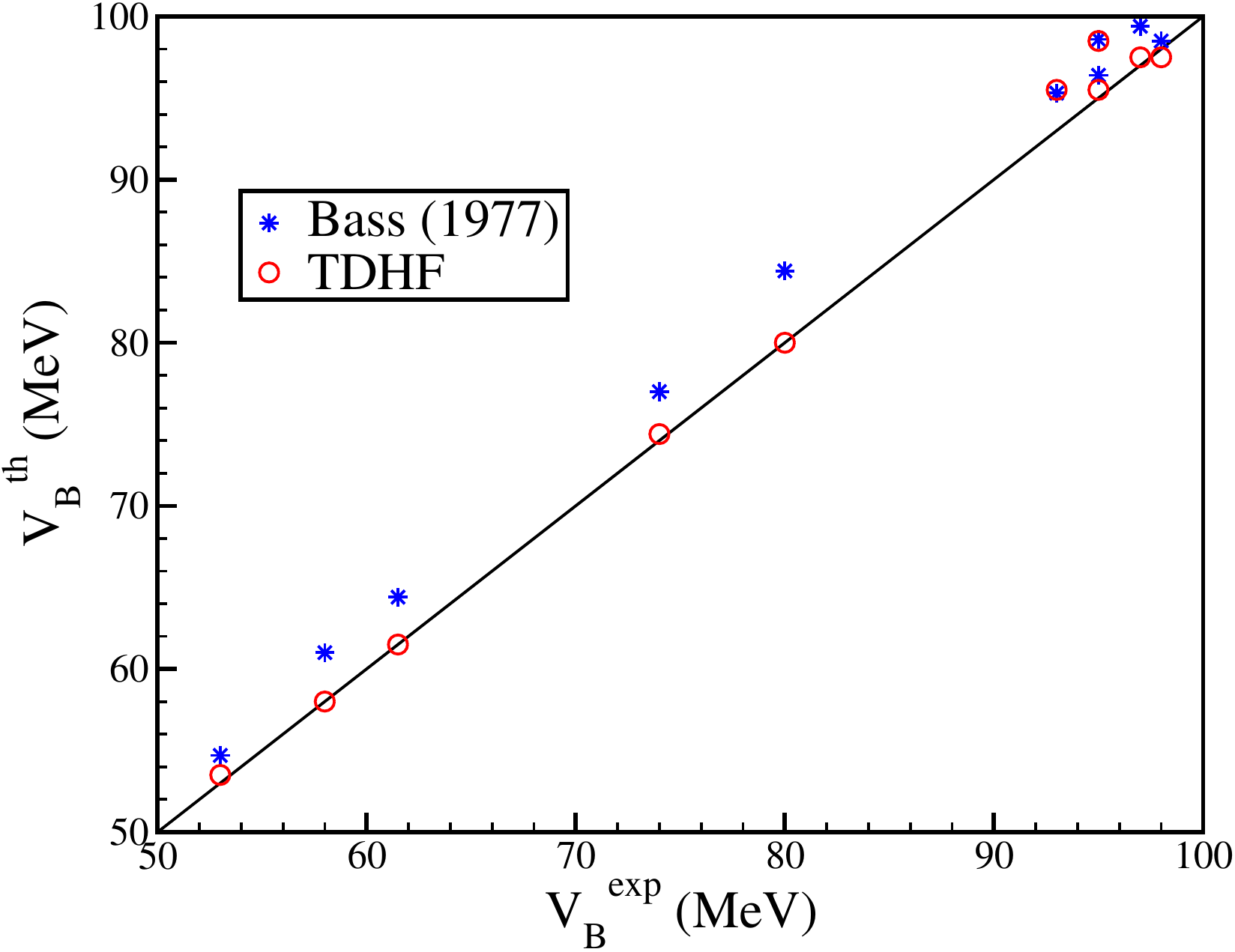} 
\caption{Bass barriers~\cite{bas77} (stars) and barriers 
extracted from TDHF calculations (circles) as a function of experimental barriers (centroids of fusion barrier distributions).  
}
\label{fig:th_exp}
\end{center}
\end{figure}

Systematic calculations of fusion barriers have been performed for medium mass systems involving spherical nuclei~\cite{sim08}. 
A summary of the results is shown in Fig.~\ref{fig:th_exp}.
A good reproduction of the barrier distribution centroids has been obtained (better than the Bass parametrisation \cite{bas77}) for all the studied systems. 
Other calculations with 3-dimensional TDHF codes confirmed the predictive power of the TDHF approach for the determination of fusion barriers~\cite{was08,guo2012,yao2024}. 

Above barrier fusion cross-sections can be  computed directly from TDHF calculations. 
The fusion cross-section is given by
\begin{equation}
\sigma_{fus}(E) = \frac{\pi\hbar^2}{2\mu E} \sum_{l=0}^\infty (2l+1) P_{fus}(l,E)\;,
\label{eq:cs}
\end{equation}
where $\mu$ is the reduced mass of the system, and $P_{fus}(l,E)$ is the fusion probability for the partial
wave with orbital angular momentum $l$ at the centre-of-mass energy $E$.
The fact that fusion probabilities are either 0 or 1 implies that cross sections are obtained using the ``quantum sharp cutoff formula''~\cite{bla54}
\begin{eqnarray}
\si_{fus} (E) &=& \frac{\pi\hbar^2}{2\mu E} \sum_{l=0}^{l_{max}(E)} (2l+1) \nonumber \\
&=& \frac{\pi \hbar^2}{2\mu E} \sdf [l_{max}(E)+1]^2,
 \end{eqnarray}
where the fusion probability is $0$ for $l>l_{max}(E)$ and 1 for $l\le l_{max}(E)$.
To avoid discontinuities due to the integer values of  $l_{max}(E)$,
$[l_{max}(E)+1]\hbar$ is generally approximated by its semi-classical equivalent $\mL_c=\sqrt{2\mu E}\, b_c$.
The latter corresponds to the classical angular momentum threshold for fusion and $b_c$ denotes the maximum impact parameter below which fusion takes place \cite{bas80}.
We finally obtain the standard classical expression for fusion cross sections
 $\si_{fus}(E) \simeq \pi \mL_c^2/2\mu E = \pi b_c^2$.

\begin{figure}
\begin{center}
\includegraphics[width=8cm]{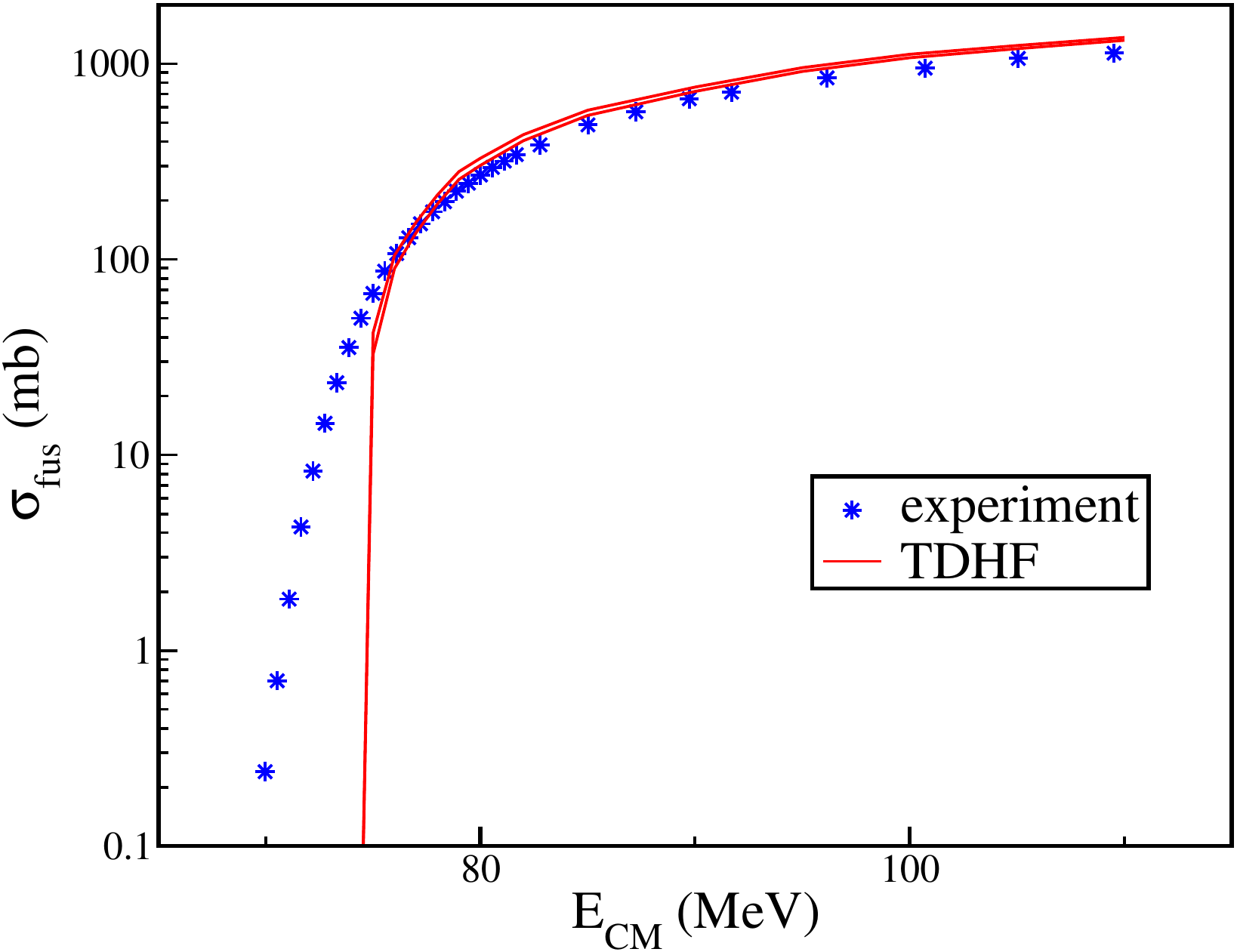} 
\caption{Experimental fusion cross sections from Ref.~\cite{mor99}
(stars) compared to cross sections deduced from TDHF calculations (lines) of $^{16}$O+$^{208}$Pb collisions.
The two lines correspond, respectively, to lower and upper limits of theoretical 
cross sections.}
\label{fig:fus}
\end{center}
\end{figure}

An example of application is shown in Fig.~\ref{fig:fus} for the $^{16}$O+$^{208}$Pb system~\cite{sim08}.
We see that TDHF predictions overestimate experimental data by about $16\%$ above the barrier. 
Although this discrepancy is small for a theory which has no parameter adjusted on reaction mechanisms, its origin is unclear.

\begin{figure}
\begin{center}
\includegraphics[width=7.5cm]{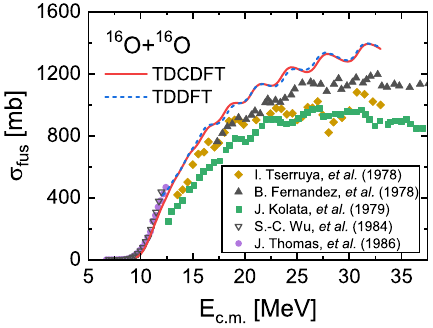} 
\caption{$^{16}$O$+^{16}$O fusion cross-sections from TDHF with Skyrme EDF (blue dashed line) \cite{simenel2013a} and covariant EDF PC-PK1 (red solid line) \cite{ren2022b}. Experimental data are from \cite{tserruya1978,fernandez1978,kolata1979,wu1984,thomas1986}. From Ref.  \cite{ren2022b}.
}
\label{fig:O+O}
\end{center}
\end{figure}

For symmetric systems with $0^+$ ground-states, fusion can only occur for even values of the angular momentum.
The cross-section with the sharp cut-off formula then reads
\begin{equation}
\sigma_{fus}(E_{\mathrm{c.m.}}) = \frac{\pi\hbar^2}{2\mu E_{\mathrm{c.m.}}} [L_{max}(L_{max}+3)+2]\;.
\label{eq:cseven}
\end{equation}
Tunnelling can be accounted in an approximate way by evaluating the barrier penetration probability according to the Hill-Wheeler
formula~\cite{hil53} with a Fermi function:
\begin{equation}
P_{fus}(l,E)\simeq \frac{e^{x_l}}{1+e^{x_l}}\;,
\label{eq:PfusHW}
\end{equation}
with $x_l=[E-V_B(l)]/\varepsilon$ and a decay constant $\varepsilon=0.4$~\cite{esbensen2012}
Figure~\ref{fig:O+O} shows an example for the reaction $^{16}$O$+^{16}$O. 
The oscillations observed at above barrier energies are induced by the individual angular momentum barriers $V_B(l)$. 
The predictions with the Skyrme EDF \cite{simenel2013a} and with the covariant EDF PC-PK1 are very close to each other. 

As TDHF is able to describe the angular momentum dependence of the fusion barrier, it provides a useful mean-field reference for the search for beyond-mean-field signatures in heavy-ion fusion reactions. 
In particular, a good benchmark for such comparison is provided by the DCTDHF technique (see Sec.~\ref{sec:DCTDHF}) which improves the comparison with experimental data with respect to the standard TDHF plus Hill-Wheeler approach, as recently shown with  $^{16,17,18}$O$+^{12}$C systems \cite{desouza2024}.

Naturally,  TDHF is unable to directly predict sub-barrier fusion cross-sections due to its semi-classical nature. 
The inclusion of quantum tunnelling of the many-body wave function is clearly one of the biggest challenges in the microscopic treatment of low-energy nuclear reactions\footnote{The problem of describing tunnelling of interacting particles is not limited to nuclear physics. For instance, tunnelling probability of an electron through an electrostatic barrier increases thanks to radiative corrections \cite{flambaum1999}. A description of particle tunnelling including couplings to other fields should be based on relativistic quantum field theory \cite{zielinski2024b}. A major difficulty, however, is the non-perturbative aspects of tunnelling \cite{deleo2009}, limiting such approach to scalar fields so far \cite{zielinski2024a}.}.

\subsubsection{Fusion with deformed nuclei \label{sec:fus-def}}

\begin{figure}
\includegraphics[width=8.8cm]{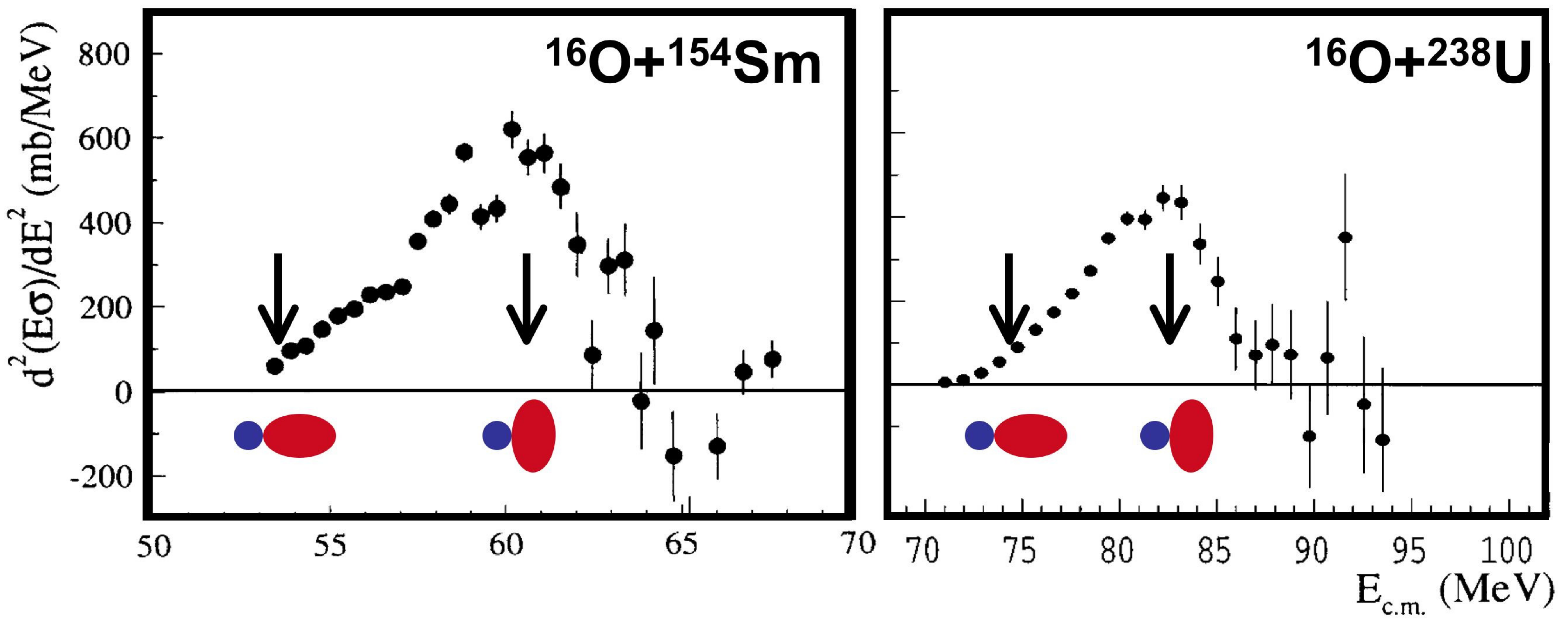} 
\caption{(left) Experimental barrier distributions for $^{16}$O+$^{154}$Sm~\cite{lei95}.
(right) Same for the $^{16}$O+$^{238}$U system~\cite{hin96}.
The arrows indicate the barriers obtained from TDHF calculations for 
central collisions with the tip (lower barriers) and with the side (higher barriers) of the deformed nucleus. 
From Ref.~\cite{sim08}.}
\label{fig:def}
\end{figure}

We now consider collisions of a spherical nucleus on a deformed one. 
In such a case, the barrier depends on the orientation of the deformed
nucleus at the touching point, leading to a wider barrier distribution 
than the single barrier case~\cite{das98,row91}.

Figure~\ref{fig:def} shows two examples of experimental barrier distributions involving a prolatly deformed heavy target \cite{lei95,hin96}. 
The TDHF fusion thresholds for the two different orientations are represented with arrows in Fig.~\ref{fig:def}.
These are compatible with the standard interpretation  that the low (resp. high) energy part of the barrier distribution corresponds to collisions with the tip (side) of the deformed nucleus.
Indeed, collision with the tip (side) are associated with a larger (smaller) distance between the fragments, and thus with a smaller (larger) Coulomb repulsion. 
We conclude that, in addition to a good reproduction of the centroids,
modern TDHF calculations also reproduce the widths of the barrier distributions
generated by static deformations of heavy targets without any adjustment of parameters.

The case of a light deformed projectile on a heavy spherical target has been investigated in Ref.~\cite{sim04} both within the TDHF approach and with a coupled channel framework~\cite{hag99}.
For such systems, the barrier distribution is affected by the reorientation of the deformed nucleus in the Coulomb field of the target. 
This induces an anisotropy of the orientation axis distribution and results into a fusion hindrance at low energies~\cite{sim04,uma06c,uma07}.
Possible experimental evidences of this effect have been reported~\cite{nay07}.
Note that the reorientation is proportional to $A_{s}/(A_{s}+A_d)$, where $A_s$ ($A_d$) is the number of nucleons in the spherical (deformed) nucleus,  and could be neglected in the systems studied in Fig.~\ref{fig:def}.

Reactions with two deformed nuclei can also be studied with TDHF.
In this case, however, it is common to investigate only specific orientations. Indeed, summing over all possible orientations is computationally prohibitive. 
As an example, the $^{14,15}$C$+^{232}$Th reactions were studied to investigate the effect of a deformed halo on fusion and transfer dynamics \cite{sun2023}.

Finally, it is worth mentioning that, although TDHF calculations help to understand the structure of fusion barrier distributions, they would not be able to reproduce their detailed structure due to a lack of quantum effects associated with the collective degrees of freedom. 
In particular, the state of the deformed nucleus should be a coherent superposition of different orientations in the laboratory frame.  
This quantum coherence is lost at the mean-field level.
A possible improvement would be to restore this coherence using a time-dependent generator-coordinate method (TDGCM) \cite{rei83}, using the orientation of the nucleus as a collective coordinate. 

\subsubsection{Fusion of $N/Z$ asymmetric nuclei \label{sec:charge-eq}}
Charge equilibration occurs when two nuclei with different $N/Z$ collide, a process that has been widely investigated with TDHF solvers
\cite{bon81,sim01,sim07,uma07,iwa09,iwa10a,obe12,godbey2017,simenel2020,gumbel2023}.
In fusion, the $N/Z$ difference induces a net dipole moment at contact which can oscillate. 
This isovector dipole oscillation is also called preequilibrium GDR, as it is a collective motion occuring in the preequilibrium stage of the compound system, i.e., before a complete equilibration of its degrees of freedom is reached.  
Experimental signatures and properties of preequilibrium GDR are obtained from high energy gamma rays \cite{parascandolo2016,parascandolo2022}.

\begin{figure}
\begin{center}
\includegraphics[width=5.5cm]{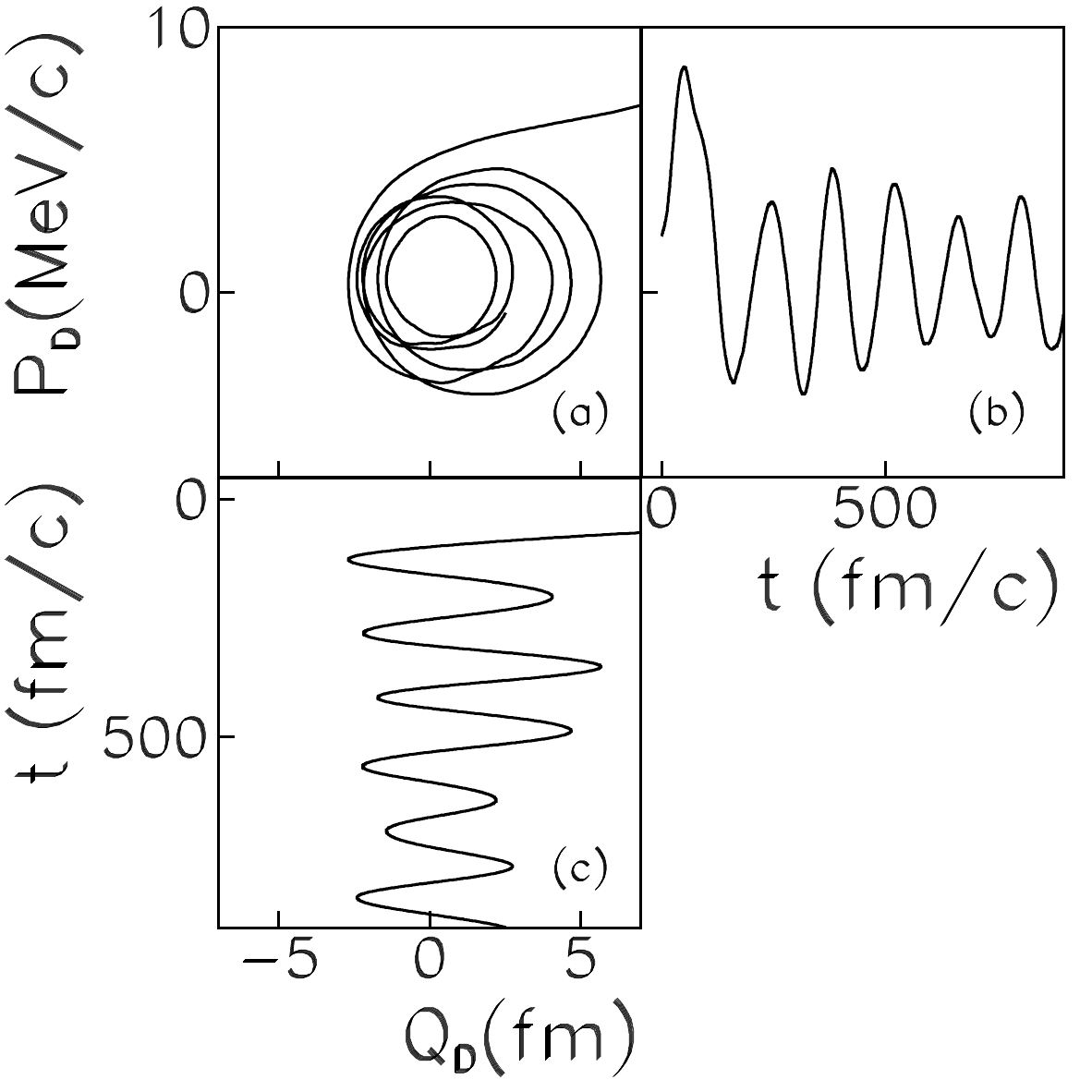} 
\caption{Evolution of the expectation value of the dipole moment $Q_D$ and its conjugated moment $P_D$ in the case of the $N/Z$ asymmetric reaction $^{40}$Ca+$^{100}$Mo at a centre-of-mass energy of 0.83 MeV/nucleon.}
\label{fig:fli}
\end{center}
\end{figure}

An example of such preequilibrium dipole motion is shown in Fig.~\ref{fig:fli}, where the time evolution of the dipole moment $Q_D$ (proportional to the distance between the proton and neutron centres of mass) and its conjugated moment $P_D$ (proportional to their relative velocity) are shown in the case of the $N/Z$ asymmetric reaction $^{40}$Ca +$^{100}$Mo \cite{sim07}. 
$P_D$ and $Q_D$ oscillate in phase quadrature. They exhibit a spiral in the plot of $P_D$ as a function of $Q_D$ due to the damping of the dipole vibration. 

\begin{figure}
\begin{center}
\includegraphics[width=4.5cm]{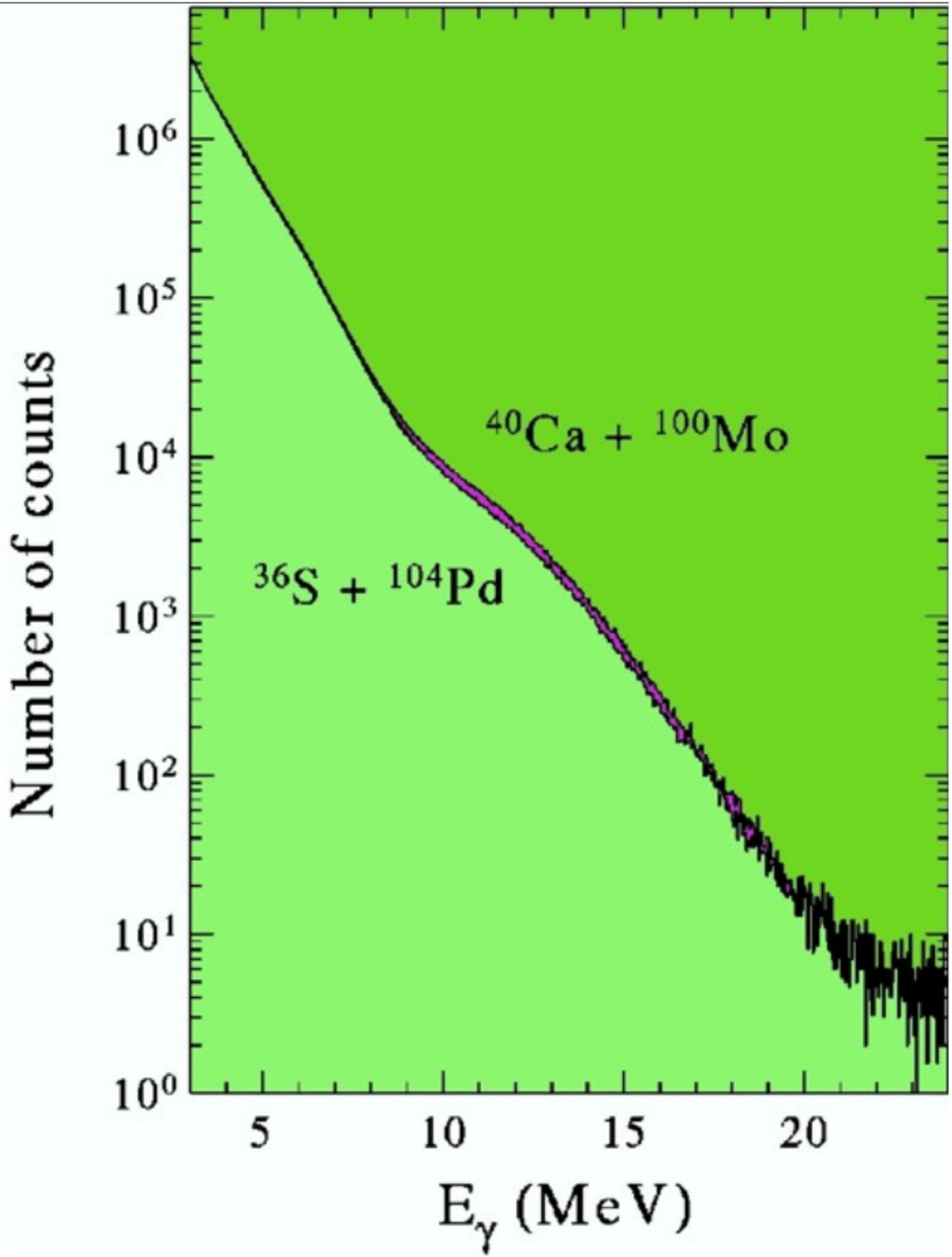} 
\includegraphics[width=4cm]{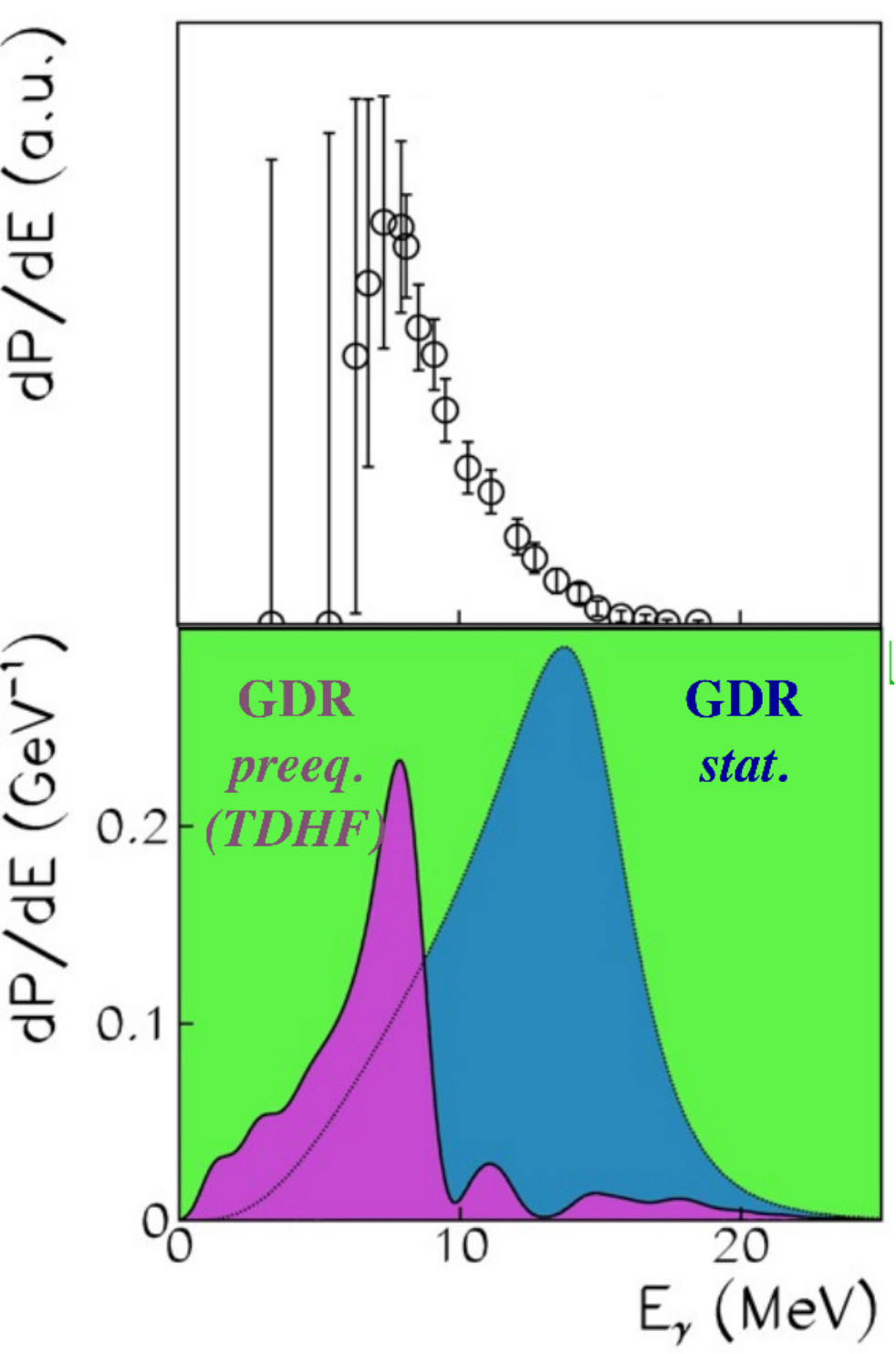} 
\caption{(left) $\gamma$-spectra measured in the $^{40}$Ca+$^{100}$Mo  and $^{36}$S+$^{104}$Pd reactions at a centre-of-mass energy of 0.83 MeV/nucleon~\cite{fli96}. (top-right) Preequilibrium GDR $\gamma$-decay spectrum obtained from the difference between the two $\gamma$-spectra in the top. (bottom-right) Theoretical $\gamma$ spectrum computed from the preequilibrium dipole moment evolution in Fig.~\ref{fig:fli} (solid line, purple area). The dotted line represents the first chance statistical $\gamma$-ray decay spectrum (blue area). Adapted from Ref. \cite{sim07}.}
\label{fig:fli_gamma}
\end{center}
\end{figure}

It is possible to compute the spectrum of~$\gamma$ emitted by the preequilibrium GDR using laws of classical electrodynamics. 
The preequilibrium GDR $\gamma$-ray spectrum is computed from the Fourier transform of the acceleration of the charges \cite{jac62,bar01}
\begin{equation}
\frac{dP}{dE_\gamma}(E_\gamma)=\frac{2\alpha}{3\pi}\frac{|I(E_\gamma)|^2}{E_\gamma}
\label{fourier}
\end{equation}
where $\alpha$ is the fine structure constant and
$$I(E_\gamma)=\frac{1}{c}\int_0^\infty \!\!\! dt \,\, \frac{d^2Q_D}{dt^2}\exp\left(i\frac{E_\gamma t}{\hbar}\right).$$
Such a $\gamma$-spectrum is shown in the bottom-right of Fig.~\ref{fig:fli_gamma} (purple area). 
A comparison with the first chance statistical GDR $\gamma$-ray decay spectrum\footnote{See Ref.~\cite{sim07} for details on the calculations of the latter.} is also shown (blue area). 
We observe that the preequilibrium GDR $\gamma$ are emitted at a lower energy than the statistical component. 
This is interpreted in terms of a large deformation of the compound nucleus in its preequilibrium phase~\cite{sim07}. Indeed, the preequilibrium dipole motion occurs along the prolate deformation axis of the compound system associated with a lower energy (see also Fig.~\ref{fig:U-GDR}).

Experimental $\gamma$-spectra are shown in the left panel of Fig.~\ref{fig:fli_gamma} for the $^{40}$Ca+$^{100}$Mo $N/Z$ asymmetric reaction and for the  $^{36}$S+$^{104}$Pd  reaction which is quasi-symmetric in $N/Z$. 
Only the first reaction is expected to exhibit a preequilibrium dipole motion. 
Indeed, more $\gamma$ are observed in this reaction. 
The difference (purple area in Fig.~\ref{fig:fli_gamma}-left) is interpreted in terms of $\gamma$-decay from the preequilibrium GDR~\cite{fli96}.
Subtracting the two $\gamma$-spectra, one obtains the preequilibrium GDR $\gamma$-spectrum which is shown in the top-right panel of Fig.~\ref{fig:fli_gamma}.
The energy of the peak is in good agreement with the spectrum computed from the TDHF response.
As mentioned above, this lowering of the preequilibrium GDR energy, by comparison to the hot GDR, is interpreted  as a signature of a strong deformation of the preequilibrium nucleus.
In particular, this means that the  equilibration of the shape is much slower than the charge equilibration (see also  Sec.~\ref{sec:timescales}).

To conclude, we see that the preequilibrium GDR contains informations on the structure of the preequilibrium compound system and, then, on the path to fusion. 
Here, the example of deformation has been discussed.
In Refs. \cite{sim01,sim07}, the preequlibrium GDR is also used to investigate other properties of the path to fusion, such as the role of rotation, of coupling with collective shape vibrations, of mass asymmetry...
It is also proposed that the decay of a preequilibrium GDR could serve as a cooling mechanism increasing the survival probability of the heaviest compound nuclei \cite{bar01,sim07}. 
This effect needs further theoretical and experimental investigations.

\subsubsection{Fusion with superfluid nuclei \label{sec:fus-superfluid}}

 In the collision process of two superfluid nuclei, their phase (i.e., $U(1)$ gauge angle) difference  could impact the reaction dynamics. 
TDHFB calculations with the Gogny interaction have demonstrated the dependence of the trajectories, pairing energies, and number of transferred neutrons on the relative gauge angle at the initial time \cite{hashimoto2016}. In symmetric systems, a differences of gauge angle $\Delta\az=\pi$ increases the fusion threshold in comparison to the $\Delta\az=0$ case. 
For the reaction $^{20}$O$+^{20}$O, the difference of the potential energy is about 0.4~MeV for the height of the barrier. 

\begin{figure}
\begin{center}
\includegraphics[width=8.5cm]{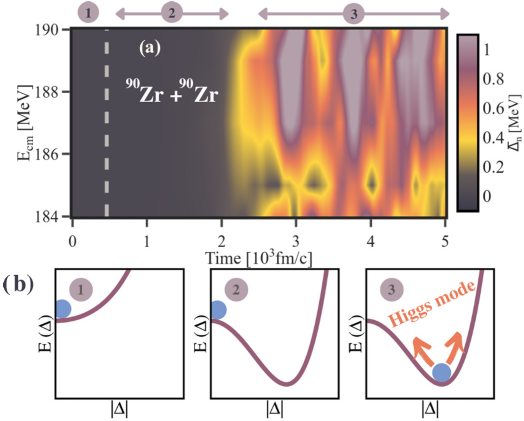} \caption{(a) Evolution of the average neutron pairing gap $\overline{\Delta}_n$ in $^{90}$Zr$+^{90}$Zr with $\Delta\az=0$. The collision occurs at $\sim0.5$~zs, indicated by the vertical dashed line. (b) Schematic  representation of the pairing Higgs mode excitation. From Ref.~\cite{magierski2022}.}
\label{fig:Higgs}
\end{center}
\end{figure}

Much larger differences of the order of  $30$~MeV were obtained in $^{90}$Zr+$^{90}$Zr with the Fayans energy density functional FaNDF0 without spin-orbit term \cite{magierski2017}. 
However, the effect was shown to be much smaller after inclusion of spin-orbit \cite{magierski2022}. 
The suppression of fusion was interpreted as an effect of the dynamics of the pairing field inducing solitonic excitations. 

Interestingly, the magnitude of pairing correlations can be dynamically enhanced after collision, a phenomenon that has been associated with a dynamical $U(1)$ symmetry breaking \cite{magierski2022}.
Figure~\ref{fig:Higgs} illustrates this mechanism in $^{90}$Zr+$^{90}$Zr with $\Delta\az=0$, leading to an absence of average neutron pairing field $\overline{\Delta}_n$  at initial time, but which exhibits large-amplitude oscillations after contact. This mechanism has been interpreted as an analog to the excitation of a pairing Higgs mode.

The role of pairing on fusion has also been studied via the effect of gauge angle difference on nucleus-nucleus potential barriers. 
For this purpose, the DCTDHFB method has been developed as an extension of DCTDHF including dynamical pairing effects \cite{scamps2019b} (see Sec.~\ref{sec:DCTDHF}). 
In particular, the DCTDHFB barriers were shown to be coherent with the TDHFB fusion thresholds. 
See also \cite{tong2022a}.

\subsubsection{Fusion hindrance in heavy systems \label{sec:fus-hindrance}}

Systems with charge product $Z_1Z_2>1600$ usually exhibit a fusion hindrance\footnote{This threshold is empirical. Based on his extra-push model, Swiatecki proposed an effective fissility, depending on both charges and masses of the nuclei, above which extra-push energy is needed to fuse~\cite{swi82}. Note that this is not the threshold for quasi-fission which may occur in lighter systems.
Indeed, quasi-fission has been observed in, e.g., $^{16}$O,$^{32}$S+$^{238}$U \cite{hin96,itk11,nas07}, and even in lighter systems such as $^{32}$S+$^{208}$Pb~\cite{nas07}.}.
 We now discuss TDHF studies that have investigated the reaction mechanism responsible for this fusion hindrance.
Fusion thresholds in such heavy systems were computed in Refs.~\cite{ave09,sim09b}.  
As an example, we consider the $^{90}$Zr+$^{124}$Sn system which has a charge product $Z_1Z_2=2000$. 
The proximity model~\cite{blo77} predicts a barrier for this system $V^{prox.}\simeq215$~MeV.
Figure~\ref{fig:ZrSn} shows the TDHF evolution of the relative distance between the fragments as a function of time for central collisions at different energies~\cite{ave09}. 
We see that the system encounters a fast re-separation at the energy of the barrier predicted by the proximity model. 
Long contact times possibly leading to fusion are observed at $E_{c.m.}\ge240$~MeV.

\begin{figure}
\begin{center}
\includegraphics[width=7.5cm]{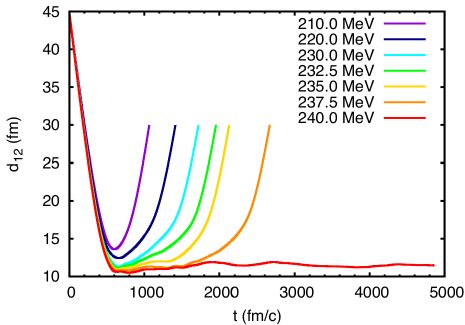} 
\caption{Distance between the centres of mass of the fragments as a function of time in head-on 
$^{90}$Zr+$^{124}$Sn collisions for different centre of mass energies. From Ref.~\cite{ave09}.}
\label{fig:ZrSn}
\end{center}
\end{figure}

Density profiles for this system at $E_{c.m.}=235$~MeV are shown in Fig.~\ref{fig:ZrSn_dens}.
A rapid neck formation is observed.
However the system keeps the shape of two fragments in contact during $\sim1400$~fm/$c$ before re-separation in two fission-like fragments. 
This is an example of quasi-fission reaction (see Sec.~\ref{sec:QF}), i.e.,  mechanism that differs from fusion followed by statistical fission as a compound system is not formed in the present case. 
In particular, the system is expected to keep the memory of its entrance channel. 

\begin{figure}
\begin{center}
\includegraphics[width=8.8cm]{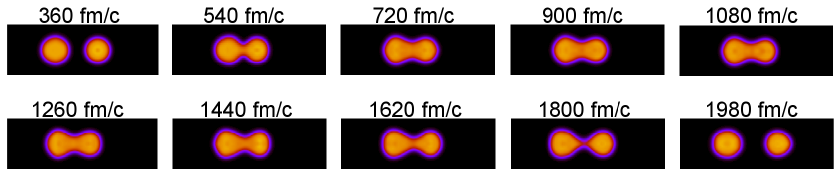} 
\caption{Density profile in the $^{90}$Zr+$^{124}$Sn  head-on collision at $E_{c.m.}=235$~MeV. From Ref.~\cite{ave09}.}
\label{fig:ZrSn_dens}
\end{center}
\end{figure}

Similar calculations have been performed for other systems with $Z_1Z_2>1600$~\cite{sim09b}. 
The TDHF fusion thresholds are shown in Fig.~\ref{fig:xpush}.
A comparison with the interaction barriers predicted by the proximity model~\cite{blo77} shows that dynamical effects included in TDHF induce a strong increase of the fusion threshold, in particular for the more heavy and symmetric systems. 
The order of magnitude of the additional energy needed to fuse is similar to the one predicted with the extra-push phenomenological model~\cite{swi82}. 

\begin{figure}
\begin{center}
\includegraphics[width=7.5cm]{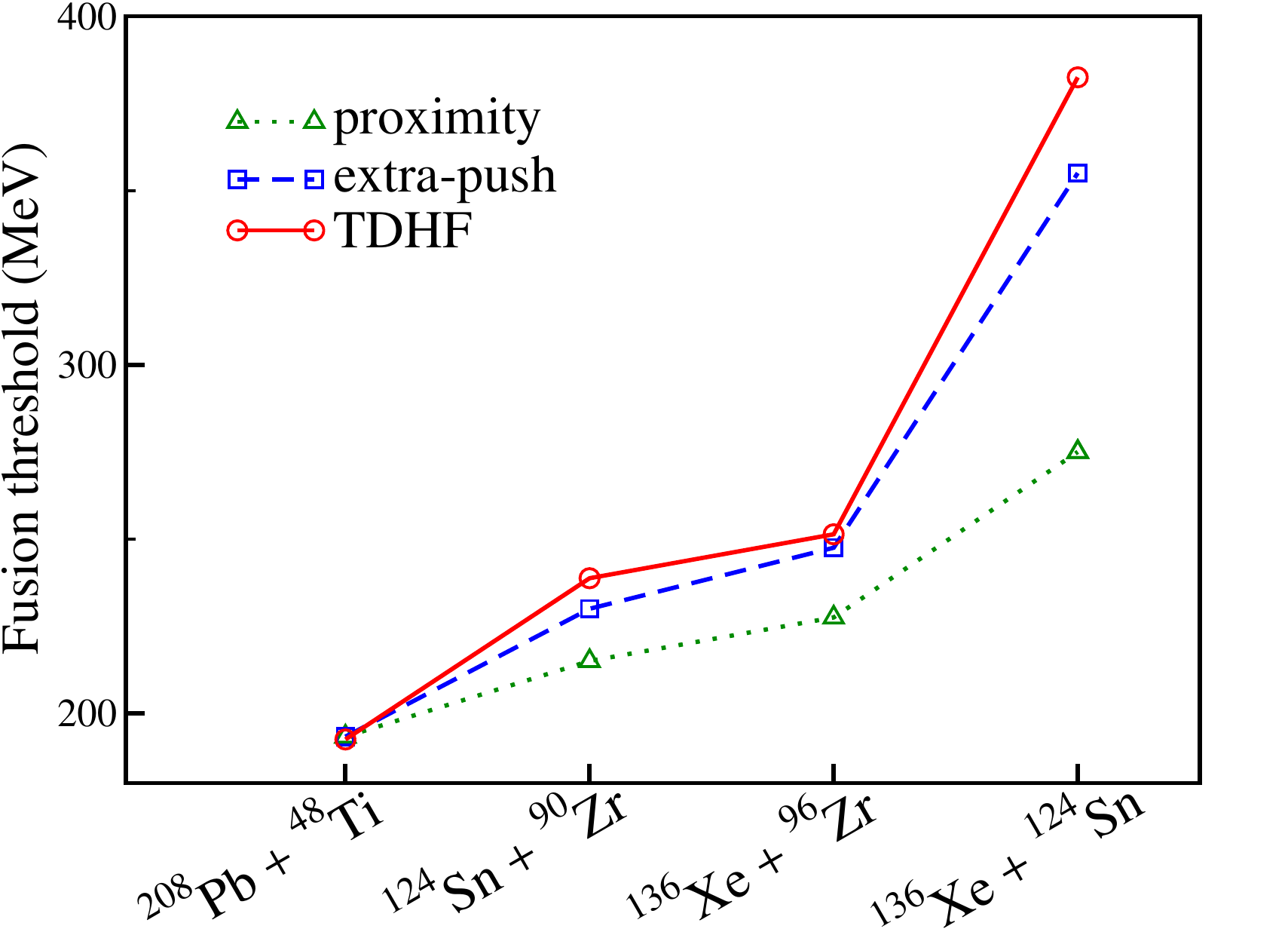} 
\caption{TDHF fusion thresholds for several heavy systems are compared with the proximity barrier~\cite{blo77} and with results from the extra-push model~\cite{swi82}.
}
\label{fig:xpush}
\end{center}
\end{figure}

Hybrid approaches that couple TDHF with dynamical diffusion models have been recently developed to investigate the competition between compound nucleus formation and quasi-fission \cite{sekizawa2019b,sun2022b,sun2023b}.
TDHF calculations are used to evaluate the distance of closest approach that determines the initial condition for the simulation of the capture process by the diffusion model \cite{sekizawa2019b}. 
TDHF has also been coupled to the coupled-channel approach (see \ref{sec:CC}) to evaluate the barrier transmission prior to the dynamical diffusion calculation to estimate the compound nucleus formation probability. This strategy was applied to cold fusion \cite{sun2022b} and hot fusion \cite{sun2023b} reactions. 
Coupling to statistical decay codes also allowed to evaluate the final reaction products \cite{sekizawa2017,umar2017,guo2018c}.

\subsection{Microscopic methods for nucleus-nucleus potentials \label{sec:NN-pot}}

In this section we briefly review microscopic methods to compute nucleus-nucleus potentials from EDF. 
``Frozen'' potentials neglects dynamical rearrangements during the collision. 
The frozen-Hartree-Fock (FHF) method gives a reference potential that does not account for the Pauli exclusion principle between nucleons of different collision partners. 
The density constrained FHF (DCFHF) includes this effect, leading to a Pauli repulsion.
The density constrained TDHF (DCTDHF) accounts for both Pauli repulsion and dynamical rearrangements. 
Here, we focus on the methods, while applications are discussed in later sections. 

\subsubsection{Frozen Hartree-Fock method \label{sec:FHF}}

A reference nucleus-nucleus potential could be obtained in the frozen approximation with HF (or HFB) densities~\cite{den02}, where the collision partners are assumed to keep their ground-state density during the approach \cite{brueckner1968}. 
The frozen potential can be computed with the same Skyrme EDF as in the TDHF calculations by translating the nuclei in their HF state~\cite{simenel2013b,stevenson2020b}.
Comparisons between TDHF and frozen fusion barriers allow to identify the combined role of dynamical effects and Pauli exclusion principle (between nucleons of different collision partners), which are included in TDHF but absent from the FHF approach. 

Writing the HF energy $E[\rho]$ as an integral of an energy density $\mH[\rho(\vr)]$, i.e., 
\oeq
E[\rho]=\int \d\vr \sdf \mH[\rho(\vr)],
\ceq
we get the expression for the frozen potential 
\oeq
V(\vR)=\int \d\vr \sdf \mH[\rho_1(\vr)+\rho_2(\vr-\vR)] - E[\rho_1] -E[\rho_2],
\label{eq:frozen}
\ceq
where $\vR$ is the distance between the centres of mass of the nuclei, and $\rho_{1,2}$ are the densities of their 
HF ground-state. 

Figure~\ref{fig:dist_barr} shows that, in the example of  $^{16}$O$+^{208}$Pb, the FHF barrier overestimates the centroid of the experimental fusion barrier distribution by $\sim1.5$~MeV, while the latter is well reproduced by the direct calculation of the TDHF fusion threshold.

\subsubsection{Density-constrained frozen Hartree-Fock method \label{sec:DCFHF}}

The FHF method does not account for the Pauli exclusion principle between nucleons belonging to different nuclei. 
The importance of this effect in heavy-ion collisions, also called ``Pauli orthogonalization'' was already recognised in early works \cite{fliessbach1971,brink1975,zint1975,fliessbach1975,beck1978,sinha1979}.
In order to take into account the Pauli principle between the nucleons of one nucleus and the nucleons of the other one in Eq.~(\ref{eq:frozen}), a proper treatment of the kinetic energy has to be considered, for instance using the Thomas-Fermi approach~\cite{den10a,den10b}.

The density-constrained method~\cite{cusson1985,umar1985} has been used to develop a new density-constrained FHF (DCFHF) approach \cite{simenel2017} for the construction of bare frozen-density nucleus-nucleus potentials incorporating
the Pauli exclusion principle by allowing the single-particle states to rearrange themselves. 
This reorganisation ensures they reach their minimum energy configuration and are properly antisymmetrised, as the many-body state is a Slater determinant of all occupied single-particle wave functions. 
The Hartree-Fock (HF) minimisation of the combined system is performed while ensuring that local proton ($p$) and neutron ($n$) densities remain unchanged:
\begin{equation} \delta \< \ H - \sum_{q=p,n}\int d\vr \ \lambda_q(\vr) \ [\rho_{1_q}(\vr)+\rho_{2_q}(\vr-\vR)] \ \> = 0\,, \label{eq:var_dens} \end{equation}
where the $\lambda_{n,p}(\vr)$ are Lagrange parameters at each point in space that constrain the neutron and proton densities. This equation determines the state vector (Slater determinant) $|\Phi(\vR)\>$.
The DCFHF potential, assumed to be central, is defined as:
\begin{equation} V_{\mathrm{DCFHF}}(R)=\<\Phi(\vR) | H | \Phi(\vR) \>-E[\rho_1]-E[\rho_2]\,. \label{eq:vr} \end{equation}

The Pauli exclusion principle is most significant when the nuclei overlap as some nucleons have no choice but to occupy a higher energy single-particle state. 
This results in a Pauli repulsion potential \cite{simenel2017} that can be obtained from the difference between the DCFHF and FHF potentials, noting that they  
use the same density.

An alternative way to evaluate the Pauli repulsion potential is via the Pauli kinetic energy (PKE) \cite{umar2021}.
The latter is obtained from the nucleon localisation function (NLF) \cite{reinhard2011} initially introduced in electron systems \cite{becke1990}. 
The probability of finding two nucleons with the same spin at  $\mathbf{r}$ and $\mathbf{r'}$ for
isospin $q$ is proportional to
\begin{equation}
	P_{qs}(\mathbf{r},\mathbf{r}') =
	\rho_q(\mathbf{r}s, \mathbf{r}s)\rho_q(\mathbf{r}'s, \mathbf{r}'s)
	-
	|\rho_q(\mathbf{r}s,\mathbf{r}'s)|^2\,,
\end{equation}
The conditional probability for finding a nucleon at
$\mathbf{r'}$ when we know with certainty that another nucleon with the same
spin and isospin is at $\mathbf{r}$ is proportional to
\begin{equation}
	\label{conditionalProb}
	R_{qs}(\mathbf{r},\mathbf{r}')
	=
	\frac{P_{qs}(\mathbf{r},\mathbf{r}')}{\rho_q(\mathbf{r}s,\mathbf{r}s)}\,.
\end{equation}
The leading term in the expansion of $R_{qs}$ with respect to $\mathbf{r}-\mathbf{r}'$ yields the localization measure
\begin{equation}\label{eq:prob_D}
    D_{qs_{\mu}} = \tau_{qs_{\mu}}-\frac{1}{4}\frac{\left|\boldsymbol{\nabla}\rho_{qs_{\mu}}\right|^2}{\rho_{qs_{\mu}}}
    -\frac{\left|\mathbf{j}_{qs_{\mu}}\right|^2}{\rho_{qs_{\mu}}}\,,
\end{equation}
where $\mu$ denotes the spin-quantisation axis.
In fact, $D_{qs}$ is a measure of the excess
of kinetic density due to the Pauli exclusion principle \cite{umar2021}. 

One can now define the net Pauli kinetic energy (PKE) for two nuclei
separated by a distance $R$ as
\begin{equation}\label{pke-1}
	\Delta E_{q\mu}^{\mathrm{P}}(R)=\frac{\hbar^2}{2m}\sum_{s_\mu}\int \!\! d^3r \left[ D_{qs_{\mu}}^\mathrm{DCFHF}(\mathbf{r},R)- D_{qs_{\mu}}^\mathrm{FHF}(\mathbf{r},R)\right],
\end{equation}
where the localisation measures are computed from Eq. (\ref{eq:prob_D}) with DCFHF and FHF density matrices, respectively. 
Summing over the isospin $q$ gives an estimate of the total Pauli repulsion 
\oeq \sum_q \Delta E^{\mathrm{P(F)}}_{q\mu}(R)\simeq V_{\mathrm{DCFHF}}(R)-V_{\mathrm{FHF}}(R). \ceq

\begin{figure}[!tb]
\centering
\includegraphics[width=8cm]{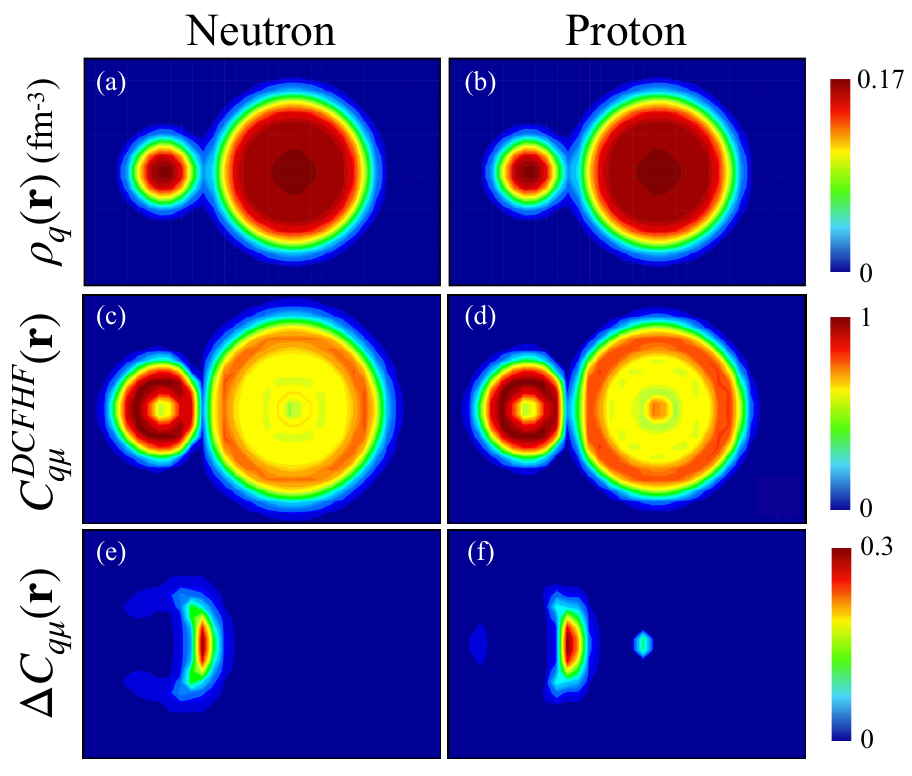}
\caption{\protect Frozen neutron (a) and proton (b) HF densities in $^{16}$O$+^{208}$Pb at the barrier radius  $R_B\simeq 12$~fm.	
Corresponding normalised NLF from Eq.~(\ref{eq:NLF})   are plotted in (c) and (d). The net PKE distribution from Eq.~(\ref{eq:PKEdist}) are shown  in (e) and (f). 
From Ref. \cite{umar2021}.}
\label{fig:PKElocal}   
\end{figure}

The localisation measure in Eq.~(\ref{eq:prob_D}) is also used to  visualise the NLF.
The localisation measure can be normalised with respect to the Thomas-Fermi kinetic density $\tau_{qs_\mu}^{\mathrm{TF}}(\mathbf{r})=\frac{3}{5}\left(6\pi^2\right)^{2/3}\rho_{qs_\mu}^{5/3}(\mathbf{r})$ as \cite{liu2019b,li2020}
\begin{equation}
	\mathcal{D}_{qs_\mu}(\mathbf{r})=\frac{D_{qs_\mu}(\mathbf{r})}{\tau_{qs_\mu}^{\mathrm{TF}}(\mathbf{r})}\,.
\end{equation}
The NLF can then be represented  by 
 \begin{equation}
	{C}_{qs_\mu}(\mathbf{r})=\left[1+\mathcal{D}_{qs_\mu}^{2}\right]^{-1}\,,
	\label{eq:NLF}
\end{equation}
giving Eq.~(\ref{eq:localisation}). Finally, a proxy for the spatial distribution of net PKE is obtained from the difference 
\oeq
\Delta C_{qs_\mu}=C_{qs_\mu}^{\mathrm{FHF}}-C_{qs_\mu}^{\mathrm{DCFHF}} \label{eq:PKEdist}
\ceq 
between FHF and DCFHF normalised NLF.
Figure \ref{fig:PKElocal} shows that the net PKE, and therefore the Pauli repulsion, is essentially localised in the neck formed by two nuclei in contact.

\subsubsection{Density-constrained TDHF(B) method \label{sec:DCTDHF}}

Unlike the FHF and DCFHF potentials, the DCTDHF technique and its generalisation with dynamical pairing (DCTDHF-Bogoliubov) allow for the computation of an energy-dependent nucleus-nucleus potential from TDHF trajectories that incorporate dynamical rearrangements of the density \cite{cus85,umar1985,uma06b}, including effect of the tensor force \cite{guo2018b,godbey2019c,sun2022}.
Starting from a TDHF simulation of a nuclear reaction, one-body density matrices $\rho^{DCTDHF}(R)$ are extracted, associated with  a distance $R(t)$ between nuclei. 
A static Hartree-Fock minimisation is then performed while holding the neutron and proton densities constrained to be the corresponding instantaneous TDHF densities, giving the ``density constrained energy''
\oeqn
E_{DC}(t)&=&\underset{\rho}{\min}\{E[\rho_n,\rho_p]^{~}_{~} \nonumber\\
&&+\sum_{q=p,n}\int {\mbox d}\mathbf{r}\; \lambda_q(\mathbf{r})[\rho_q(\mathbf{r})-\rho_q^{TDHF}(\mathbf{r},t)]\}.\nonumber
\ceqn
The nucleus-nucleus potential is then given by~\cite{uma06b}
\begin{equation}
V_{\mathrm{DCTDHF}}(R)=E_{DC}(R)-E[\rho_1]-E[\rho_2]\,.
\label{eq:vrdctdhf}
\end{equation}
Dynamical features such as neck formation, particle exchange, 
internal excitations, and deformation effects are naturally included in this  potential. 

The evaluation of transmission probabilities through nucleus-nucleus potential barriers via a Schr\"odinger equation also requires the computation of coordinate dependent masses $M(R)$. 
In the case of DCFHF, the mass is a constant and equal to the reduced mass, ${M(R)}=\mu$. 
In  DCTDHF, the  mass can
be calculated directly from TDHF dynamics~\cite{uma09b}.
Alternatively, the potential can be transformed by a scale factor~\cite{uma09b,goeke1983}
\begin{equation}
d\bar{R}=\left(\frac{{M(R)}}{\mu}\right)^{\frac{1}{2}}dR,
\end{equation}
while using $M(R)=\mu$ in the Schr\"odinger equation.

\begin{figure}[!tb]
\centering
\includegraphics[width=8cm]{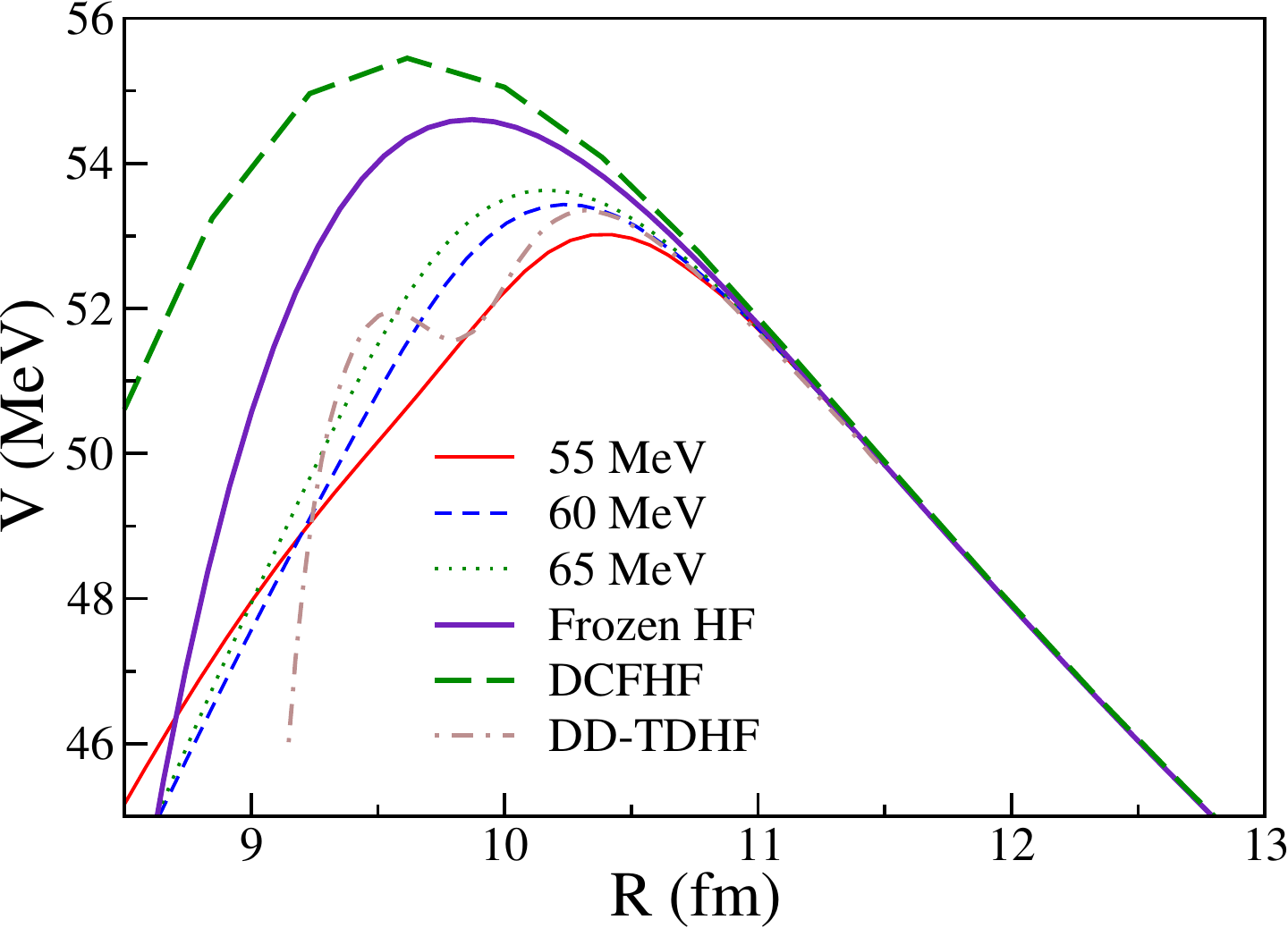}
\caption{\protect DCTDHF potentials in $^{40}$Ca$+^{40}$Ca obtained from TDHF central collisions at $E_{c.m.}=55$, 60 and 65~MeV, together with the FHF and DCFHF potentials. The dissipative dynamics TDHF (DD-TDHF) potential at $E_{c.m.}=55$~MeV from~\cite{washiyama2008} is also shown. From Ref.  \cite{simenel2018}.}
\label{fig:Pot}   
\end{figure}

Examples of DCTDHF potentials are shown in Fig. \ref{fig:Pot} for $^{40}$Ca$+^{40}$Ca at various TDHF energies~\cite{umar2014a}. 
The DCTDHF barriers are lower than the bare potential barriers obtained with the static FHF and DCFHF approaches. 
This is interpreted as a manifestation of the couplings to low-lying vibrational states near the barrier (see Sec.~\ref{sec:CC}).

We also see in Fig.~\ref{fig:Pot} that the DCTDHF potential depends on the energy of the collision simulated with TDHF \cite{umar2014a}. 
A similar energy dependence is observed with the dissipative dynamics TDHF (DD-TDHF) method~\cite{washiyama2008}.
A possible interpretation of this energy dependence  is that at high energy the nuclei have less time to polarise each other.
In fact, at  energies well above the barrier, the potential is expected to converge towards the FHF potential and not the DCFHF one 
as the momenta of the nucleons depend on which fragment they belong to, thus reducing the effect of the Pauli exclusion principle between them~\cite{simenel2017}. 

It is often useful to separate contributions from proton and neutron densities to the potential. 
Equivalently, this can be done with isocalar  ($\rho_{{I}=0} = \rho_n + \rho_p$) and isovector ($\rho_{{I}=1} = \rho_n - \rho_p$) densities.
One can  write the Hamiltonian density as 
\cite{dobaczewski1995}
\begin{equation}
\label{eq:edensity}
{\cal H}(\mathbf{r}) = \frac{\hbar^2}{2m}\tau_0
+  \sum_{{I}=0,1} {\cal H}_{I}(\mathbf{r})
+ {\cal H}_{Coulomb}(\mathbf{r})\,, 
\end{equation}
where ${\cal H}_I(\mathbf{r})$ depends on $\rho_I$.
The DCTDHF potential in Eq.~(\ref{eq:vrdctdhf}) can then be written as
\begin{equation}
V(R) = \sum_{{I}=0,1} v_{I}(R) + V_C(R)\, ,\label{eq:vIso}
\end{equation}
where $v_{I}(R)$ denotes the potential computed by using the isoscalar and isovector parts of
the Skyrme EDF~\cite{godbey2017}. 
The isovector potential $v_1$ vanishes in the FHF approximation and is then mostly induced by dynamical effects. 

\subsection{Applications with microscopic potentials \label{sec:NN-appli}}

The techniques described in Sec.~\ref{sec:NN-pot} to compute microscopic potentials from EDF have been used in numerous fusion studies \cite{uma06b,umar2012a,oberacker2013,desouza2013,simenel2013a,jiang2014,umar2014a,steinbach2014,umar2015a,reinhard2016a,godbey2017,guo2018b,scamps2019b,godbey2019b,godbey2019c,umar2021,tong2022a,sun2022,godbey2022,gumbel2023,tong2023,desouza2024}. 
Here, we review some of these applications. 

\subsubsection{Pauli repulsion \label{sec:Pauli}}

The introduction of the DCFHF method was motivated by the study of the role of the Pauli exclusion principle in nucleus-nucleus potentials \cite{simenel2017}.
Figure~\ref{fig:Pot} shows that, in $^{40}$Ca$+^{40}$Ca, the FHF and DCFHF potentials match outside the barrier, but differ at and inside the barrier. 
More precisely, the DCFHF potential exhibits a higher and wider barrier than the FHF one. This is a manifestation of the Pauli repulsion which increases with the overlap of the nuclei. 

\begin{figure}[!tb]
\centering
\includegraphics[width=8cm]{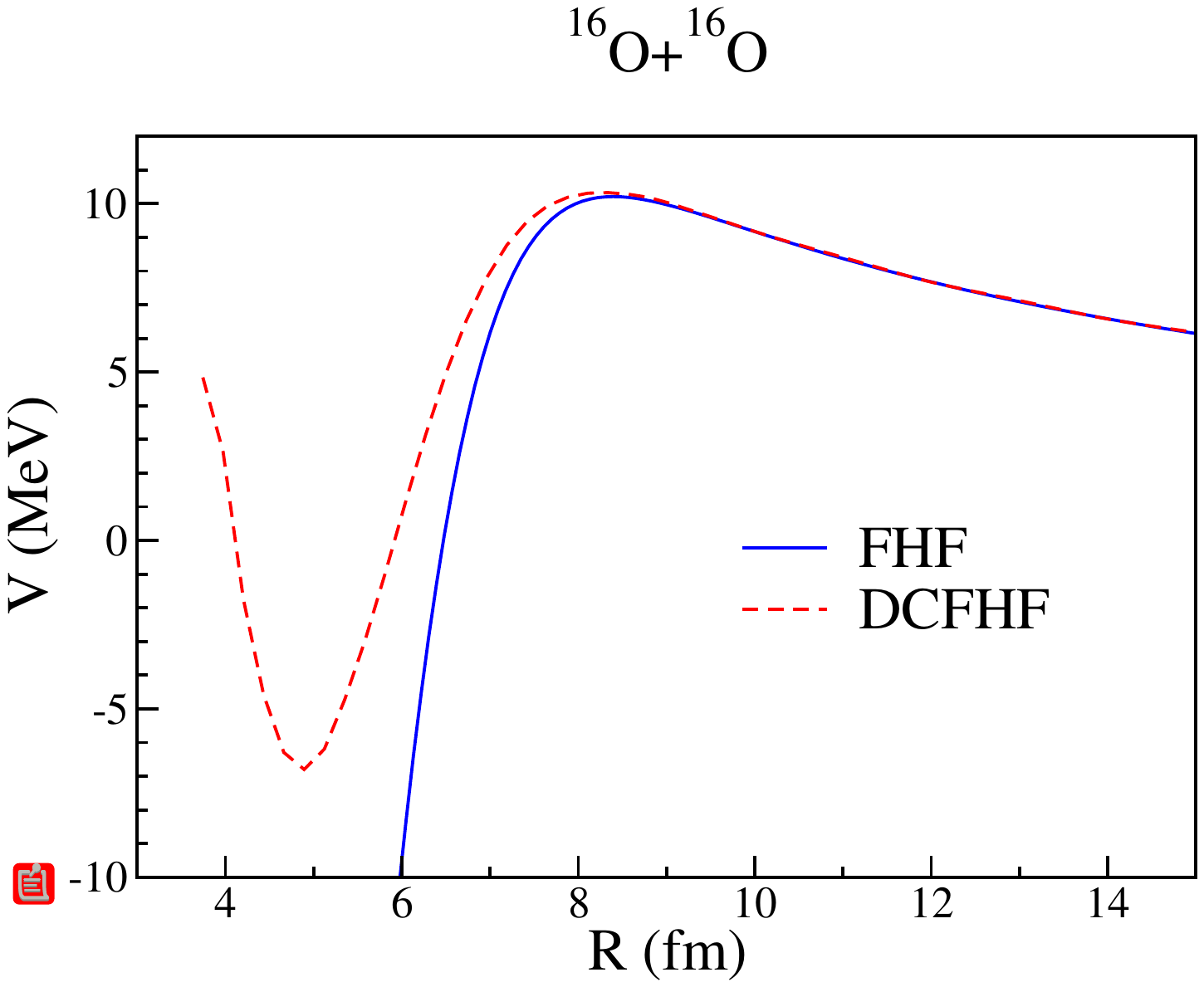}
\includegraphics[width=8cm]{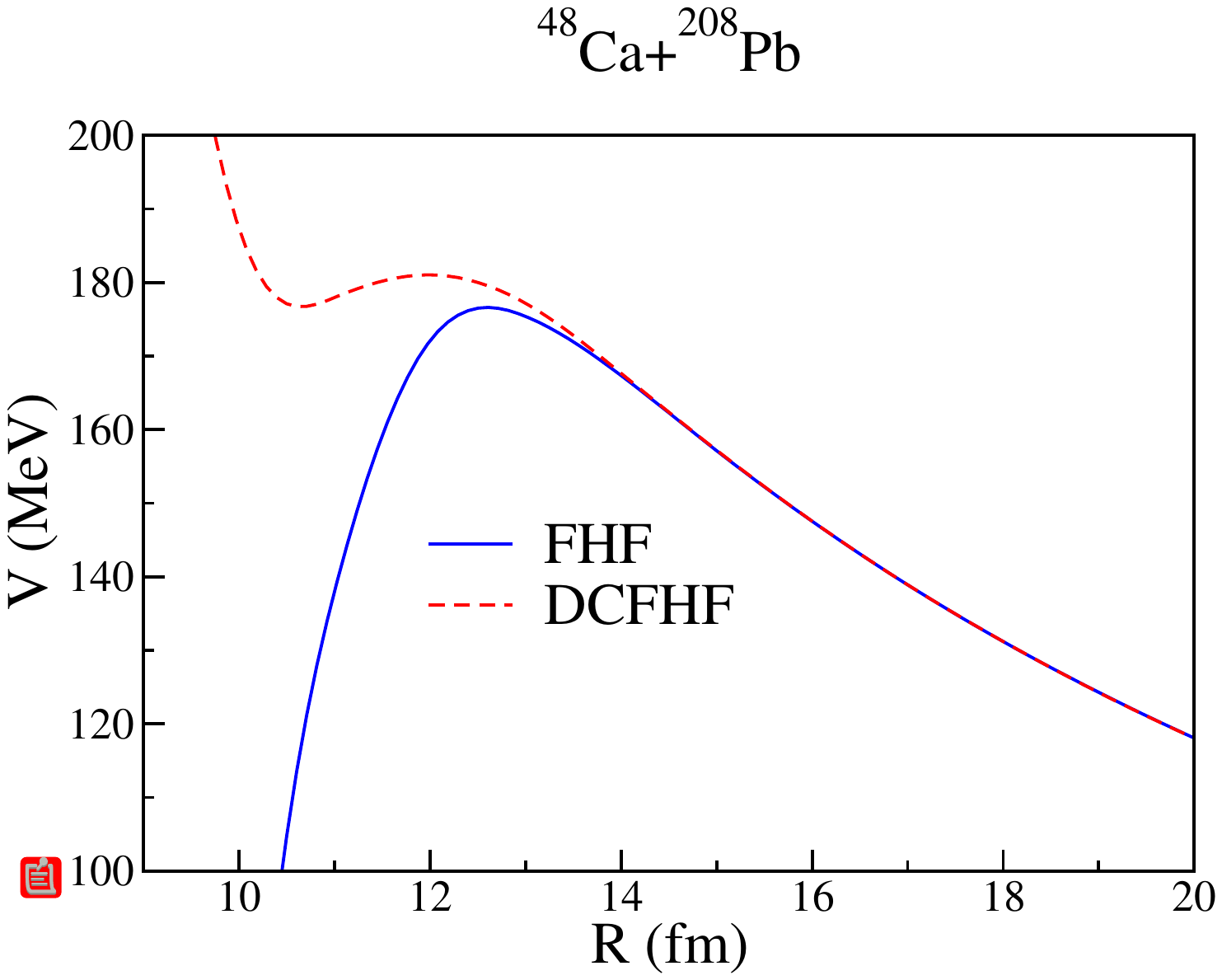}
\caption{\protect Comparison of FHF and DCFHF potentials $^{16}$O$+^{16}$O and $^{48}$Ca$+^{208}$Pb. From Ref. \cite{simenel2017}.}
\label{fig:FHF-DCFHF}   
\end{figure}

Naturally, reactions with larger charge products $Z_1Z_2$ require more overlap between the collision partners to reach the Coulomb barrier. 
This is required for the short range nuclear attraction to compensate for the stronger Coulomb repulsion. 
This larger overlap between reactants also results in a larger Pauli repulsion. 
As a result, while Pauli repulsion is not significant at near barrier energies in light systems like $^{16}$O$+^{16}$O, it has a much stronger impact in heavier systems such as $^{48}$Ca$+^{208}$Pb where the inner potential pocket almost disappears \cite{simenel2017}, as shown in Fig.~\ref{fig:FHF-DCFHF}.

At below barrier energy, the widening of the potential barrier due to Pauli repulsion reduces tunnelling probabilities and thus hinders sub-barrier fusion. 
Pauli repulsion then provides a natural (though only partial) explanation \cite{simenel2017} for the experimentally observed deep sub-barrier fusion hindrance \cite{jiang2002,dasgupta2007,stefanini2010} (see Ref. \cite{back2014} for a review).
As shown in Fig.~\ref{fig:Pauli}(a), the DCTDHF barrier for $^{16}$O$+^{208}$Pb is significantly lower than with the frozen approximations. 
However, its width is equivalent to the DCFHF one, indicating that the Pauli repulsion still plays a role once dynamical effects are included \cite{umar2021}.

\begin{figure}[!tb]
\centering
\includegraphics[width=5cm]{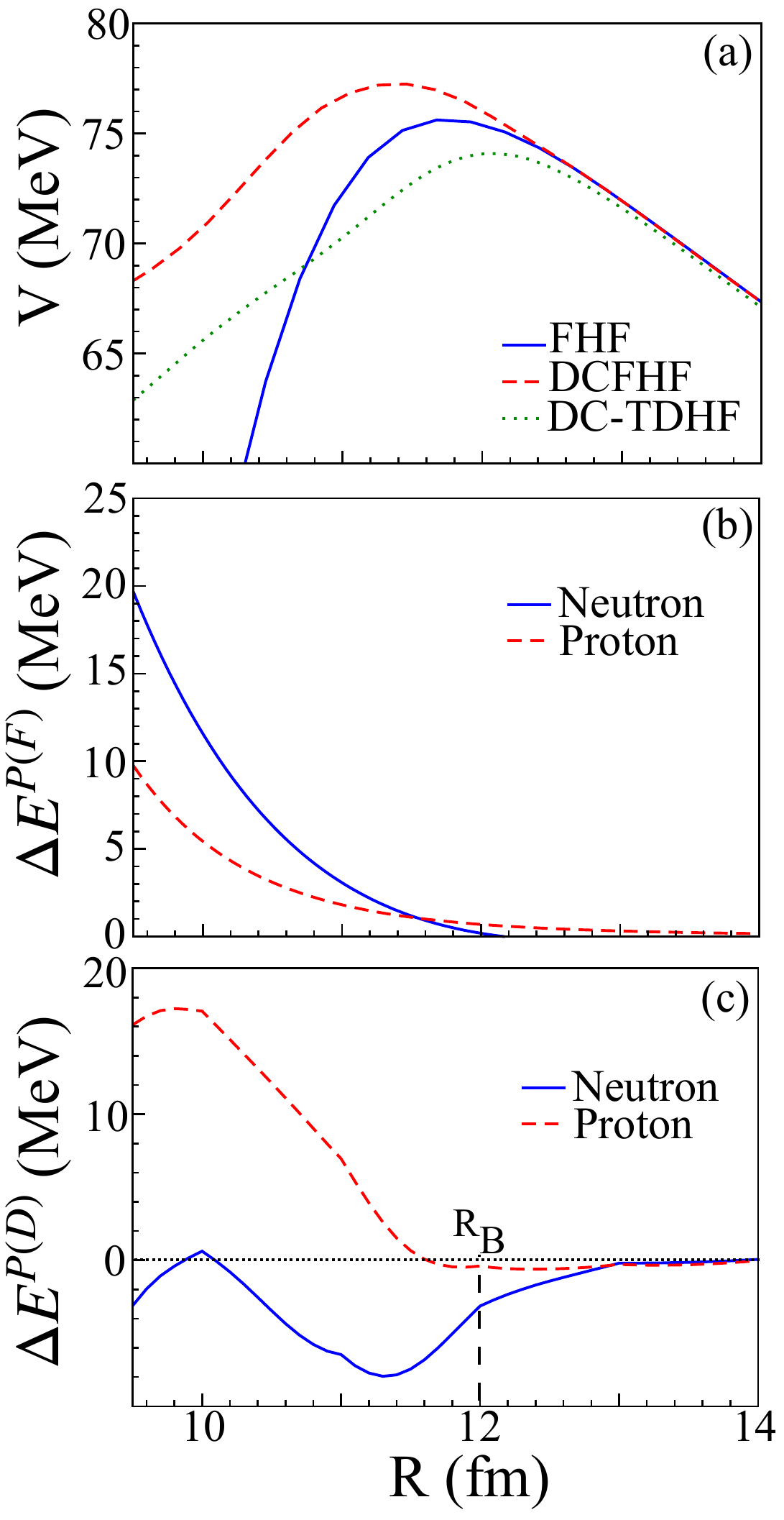}
\caption{\protect (a) $^{16}$O$+^{208}$Pb potentials from FHF, DCFHF, and DCTDHF methods. (b) Neutron and proton contributions to the Pauli repulsion from Eq. (\ref{pke-1}) in the frozen approximation. (c) Additional dynamical contributions to the Pauli repulsion computed from Eq. (\ref{pke-2}). $R_B$ is the DCTDHF barrier peak radius.  Adapted from Ref. \cite{umar2021}.}
\label{fig:Pauli}   
\end{figure}

The decomposition of the Pauli repulsion into contributions from protons and neutrons has been achieved with the use of the nucleon localisation function and Pauli kinetic energy (see Sec.~\ref{sec:DCFHF})
\cite{umar2021}.
 At the frozen level, Fig.~\ref{fig:Pauli}(b) shows that the neutron contributions in $^{16}$O $\!+^{208}$Pb, evaluated with Eq.~(\ref{pke-1}) is about twice that of protons and rapidly increases inside the barrier. 
It is also possible to evaluate the dynamical contribution to the net PKE, in analogy with Eq.~(\ref{pke-1}), using 
\begin{eqnarray}
	\Delta E_{q\mu}^{\mathrm{P(D)}}(R)&=\frac{\hbar^2}{2m}\sum_{s_\mu}\int \!\! d^3r &\left[ D_{qs_{\mu}}^\mathrm{DCTDHF}(\mathbf{r},R)\right.\nonumber\\
	&&\left.- D_{qs_{\mu}}^\mathrm{DCFHF}(\mathbf{r},R)\right].\,\,\,\,\label{pke-2}
\end{eqnarray}
This is represented in Fig.~\ref{fig:Pauli}(c). We see that the rearrangement of neutrons significantly lowers the Pauli repulsion, while that of protons increases the Pauli repulsion inside the barrier. 
The different dynamical behaviour between protons and neutrons could be a signature of $N/Z$ equilibration that induces proton and neutron transfer in opposite directions \cite{umar2021}.

\subsubsection{Coupled-channel effects \label{sec:CC}}

\begin{figure}[!tb]
\centering
\includegraphics[width=4cm]{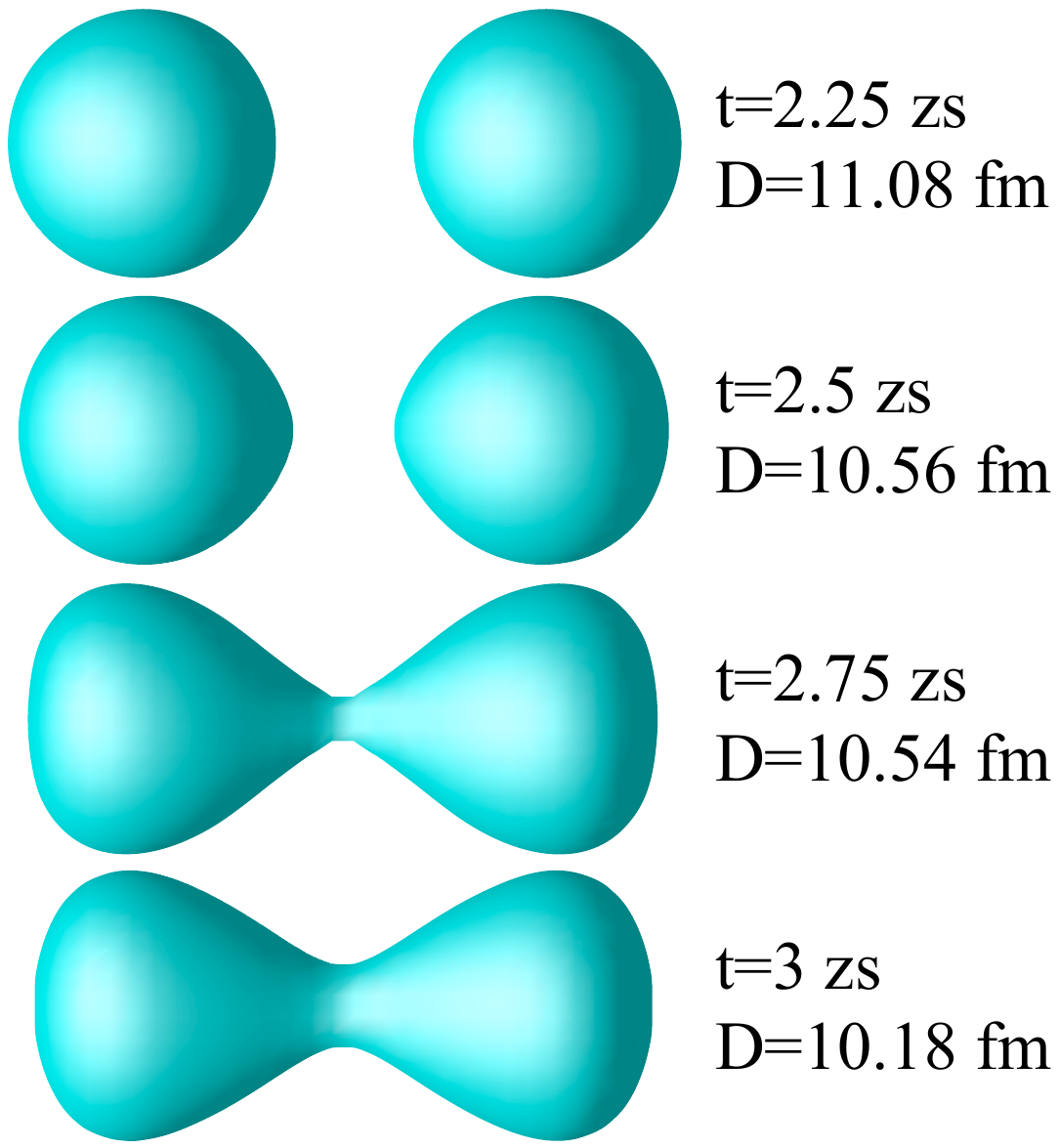}
\caption{\protect  Isodensity surfaces with $\rho_0/2 = 0.08$ fm$^{-3}$ for a $^{40}$Ca$+^{40}$Ca central collision at $E_{c.m.} = 53.3$ MeV. From Ref. \cite{simenel2013a}.}
\label{fig:dens_CaCa}   
\end{figure}

Dynamical effects influencing near-barrier fusion are often studied with the coupled-channel formalism that describes the coupling between the relative motion of the nuclei and their internal degrees of freedom
(see  \cite{hagino2022} for a review). 
These couplings are also included at the mean-field level in TDHF. An example is shown in Fig.~\ref{fig:dens_CaCa}
for a collision of two $^{40}$Ca nuclei at the barrier, where the nuclei acquire a strong octupole (pear) shape helping neck formation. 
This dynamical deformation is interpreted as an effect of the coupling to the low-lying collective octupole vibrational  $3^-_1$ state in $^{40}$Ca (see Fig.~\ref{fig:3-_filter}).

It is well known that couplings to low-lying collective states induce a distribution of fusion barriers \cite{row91}. 
In TDHF, however, only one fusion threshold is obtained (unless, e.g., the nuclei are deformed and several orientations are considered) due to the mean-field approximation. 
Coupled-channel calculations require inputs such as the nucleus-nucleus potential, and the energy and  transition strengths  of  excited states. 
While these quantities can be determined from experiment, they are also, in principle, accessible from microscopic calculations. 
Indeed, we saw in Sec.~\ref{chap:vib} that collective vibrations can be described in the linear response theory, while bare potentials can be computed with the frozen approximation (see Sec.~\ref{sec:NN-pot}). 

A method combining TDHF and coupled-channels approaches was then proposed to study the effect of collective vibrations on fusion \cite{simenel2013a}. 
It is based on the following steps:
\begin{itemize} 
\item The bare potential is computed with the FHF approach.
\item Energies and transition strengths of low-lying collective vibrational modes are computed with TDHF in the linear response theory.
\item The potential and vibrational state properties are used as input of a coupled-channel code such as CCFULL \cite{hag99} to determine fusion cross-sections over a range of energies.
\end{itemize}
The centroids of predicted fusion barrier distributions can also be compared with fusion thresholds computed directly from the TDHF method.
In particular, this can indicate if other couplings are necessary, such as couplings to transfer channels.
An example of application is shown in Fig.~\ref{fig:sigma_CaCa_log} for $^{40}$Ca$+^{40}$Ca where we see that couplings to the $3^-_1$ and GQR states in $^{40}$Ca are necessary to reproduce the cross-sections. Note that this agreement with experimental data is obtained with no adjustable parameters or input from experiment as the potential and properties of the vibrational states are entirely determined by microscopic calculations. 

\begin{figure}[!tb]
\centering
\includegraphics[width=8cm]{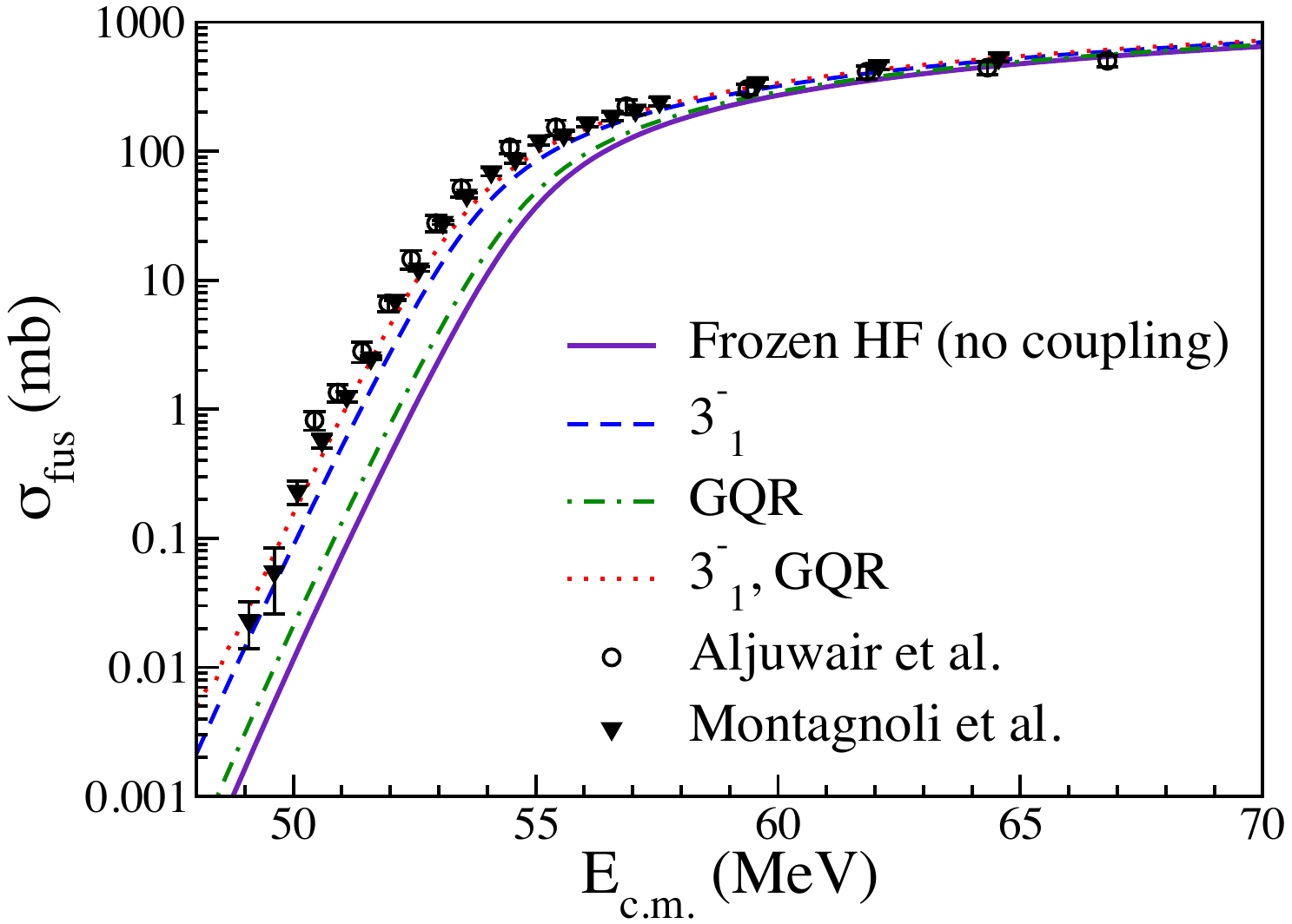}
\caption{\protect $^{40}$Ca$+^{40}$Ca fusion cross-sections. Experimental data are from \cite{aljuwair1984,montagnoli2012}. The coupled-channels 
calculations were performed with FHF potentials and couplings to $3^-_1$ and GQR states in $^{40}$Ca.  Adapted from Ref. \cite{simenel2013a}.}
\label{fig:sigma_CaCa_log}   
\end{figure}

\begin{figure}[!tb]
\centering
\includegraphics[width=6cm]{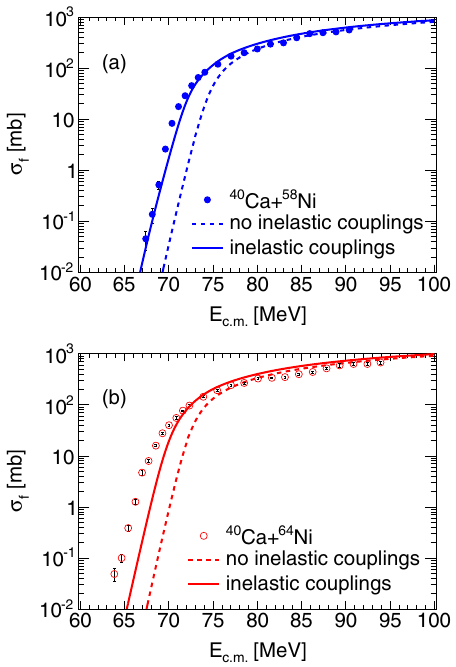}
\caption{\protect $^{40}$Ca$+^{58,64}$Ni fusion cross-sections. Experimental data are from \cite{bourgin2014}. The coupled-channels 
calculations were performed with FHF potentials and couplings to $3^-_1$ state in $^{40}$Ca  and to $2^+_1$ state in $^{58,64}$Ni.  From Ref. \cite{bourgin2016}.}
\label{fig:CaNi}   
\end{figure}

Another example of application is shown in Fig.~\ref{fig:CaNi} for $^{40}$Ca$+^{58,64}$Ni fusion cross-sections.
Here, only the FHF potential is determined microscopically, while the energy and deformation parameters of the $3^-_1$ state of $^{40}$Ca and of the $2^+_1$ states in $^{58,64}$Ni were determined from experiment. 
The good agreement in $^{40}$Ca+$^{58}$Ni illustrates the relevance of using FHF potentials in coupled-channel calculations. 
The difference between experiment and theory in $^{40}$Ca+$^{64}$Ni is attributed to transfer channels with positive $Q-$values not included in the calculations.

\subsubsection{Fusion reactions of astrophysical interest \label{sec:astro}}

The microscopic nature of TDHF, combined with its sole reliance on EDF parameters, makes it an attractive tool for predicting the outcomes of reactions in astrophysical environments that are challenging to replicate experimentally.
For instance, the fusion of highly neutron-rich nuclei, such as $^{24}$O$+^{24}$O, could impact the composition and heating of the crust in accreting neutron stars \cite{horowitz2008}.
The DCTDHF method has then been used to predict fusion cross-sections at deep sub-barrier energies for $^{12}$C,$^{16,24}$O$+^{24}$O exotic systems \cite{uma12b}.
Asymmetric reactions, in particular, were shown to have enhanced cross-sections compared to standard potential penetration models. 
This observation was interpreted as an effect of neutron transfer between the reactants. 

\begin{figure}[!tb]
\centering
\includegraphics[width=8cm]{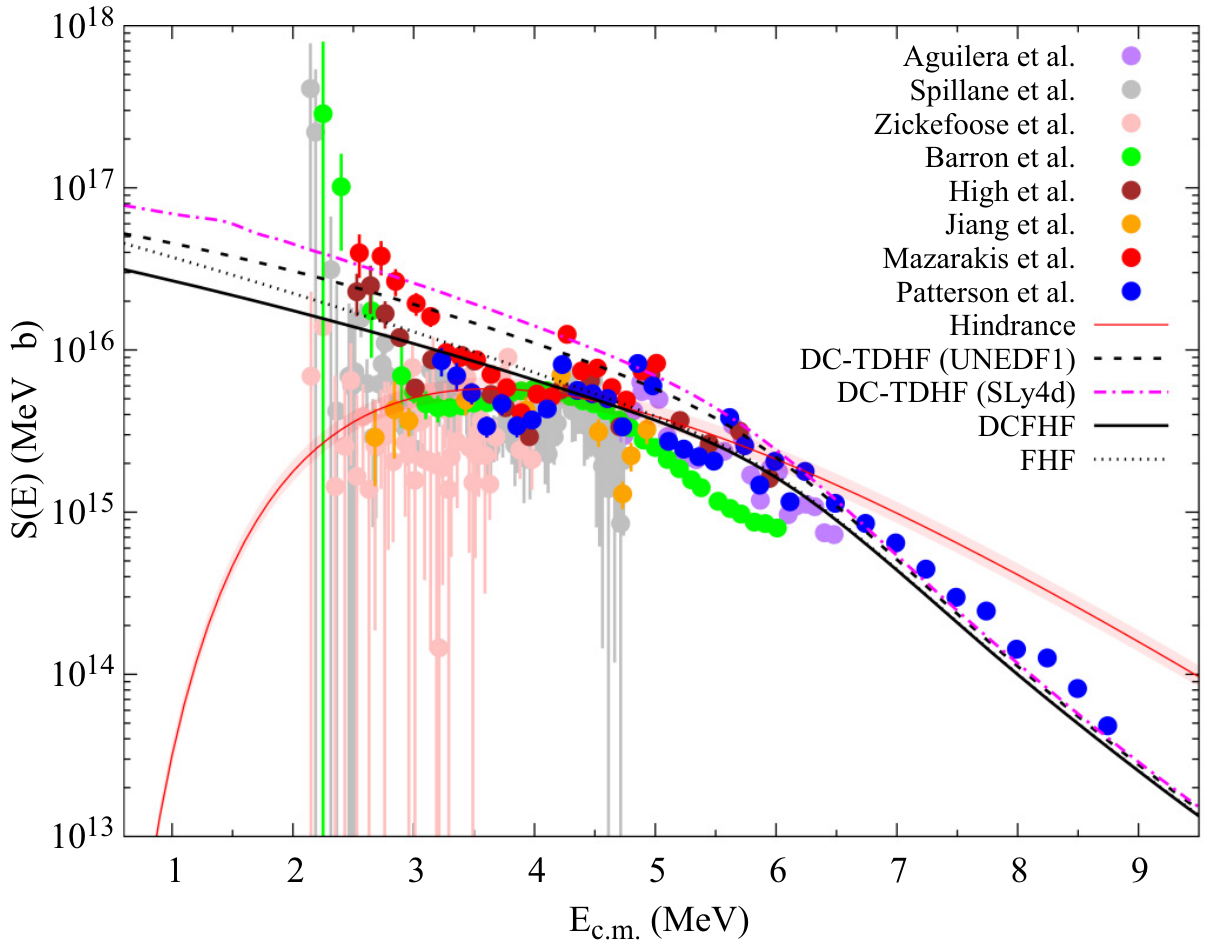}
\caption{\protect $^{12}$C$+^{12}$C fusion $S$ factor from FHF, DCFHF and DCTDHF. The hindrance model is from \cite{jiang2007}. See \cite{godbey2019b} for a full list of the experimental data. From Ref. \cite{godbey2019b}.}
\label{fig:12C12C}   
\end{figure}

The $^{12}$C $+^{12}$C reaction is another example of crucial system for astrophysical phenomena such as  neutron star superbursts \cite{cumming2001,strohmayer2002}. 
This reaction is also an important step towards the nucleosynthesis of heavier nuclei in the final stages of the helium burning process \cite{hoyle1954,monpribat2022,dumont2024}. 
As direct measurements of fusion cross-sections are particularly challenging at astrophysical energies, reliable theoretical predictions are desirable to assist astrophysical models. 
The FHF, DCFHF and DCTDHF methods were then applied to this system in Ref.~\cite{godbey2019b}. 
Note also the recent use of  a combination of time-dependent wave-packet method and DCTDHF to study this reaction \cite{close2025}.

The astrophysical $S$ factor, defined as 
$$S(E_{c.m.})=\sigma(E_{c.m.})E_{c.m.}e^{2\pi\eta},$$
where $\eta=Z_1Z_2e^2/\hb v$ is the Sommerfeld parameter, is shown in Fig.~\ref{fig:12C12C}. 
A comparison between the FHF and DCFHF approaches indicates that Pauli repulsion has a minor influence on this system and is insufficient to create a peak in the astrophysical S-factor.
Above barrier ($V_B\approx6$~MeV), all methods reproduce reasonably well the data, except for the phenomenological hindrance model \cite{jiang2007} that overestimates them. 
At deep sub-barrier energies, the DCTDHF predictions are in good agreement with the maximum cross-sections on-resonances, but overpredict them elsewhere.
This is expected as the barrier potential transmission coefficients are calculated  with incoming wave boundary conditions which cannot account for the sub-barrier resonances observed experimentally \cite{jiang2013}.
Further improvement of the method are then required to describe these resonances, as well as to account for the role of $\alpha-$clustering
in $^{12}$C \cite{umar2023}.

\subsubsection{Role of transfer and charge equilibration \label{sec:transNZ}}

(Multi)nucleon transfer occurs in the early stage of a heavy-ion collision. If triggered by an initial $N/Z$ asymmetry between the collision partners, protons and neutrons may even flow in opposite directions through the neck. These transfer  mechanisms will be studied in more details in Sec.~\ref{sec:transfer}. Here, we focus on the impact of  transfer on fusion and nucleus-nucleus potentials. 

\begin{figure}[!tb]
\centering
\includegraphics[width=6cm]{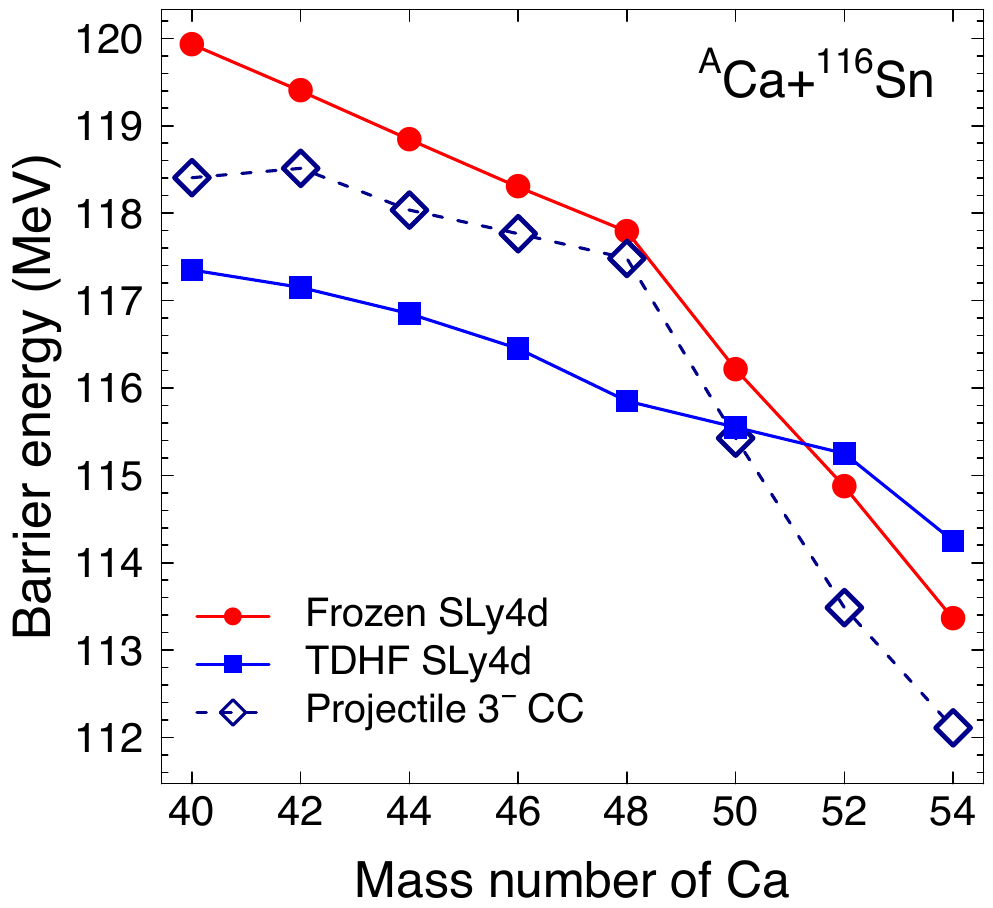}
\caption{\protect FHF (circles) Coulomb barriers and TDHF fusion thresholds (squares) in $^A$Ca$+^{116}$Sn. Coupled-channels average barriers with couplings to the $3^-_1$ state in calcium isotopes are also shown (diamonds).  From Ref. \cite{vophuoc2016}.}
\label{fig:CaSn}   
\end{figure}

Figure \ref{fig:CaSn} shows the FHF barriers and TDHF fusion thresholds in $^{40-54}$Ca$+^{116}$Sn \cite{vophuoc2016}. 
The change of slope induced by the neutron skin in heavy calcium isotopes is not present in TDHF calculations. 
Couplings to vibrational states do not reproduce this difference, which is attributed to transfer reactions. 

\begin{figure}[!tb]
\centering
\includegraphics[width=7cm]{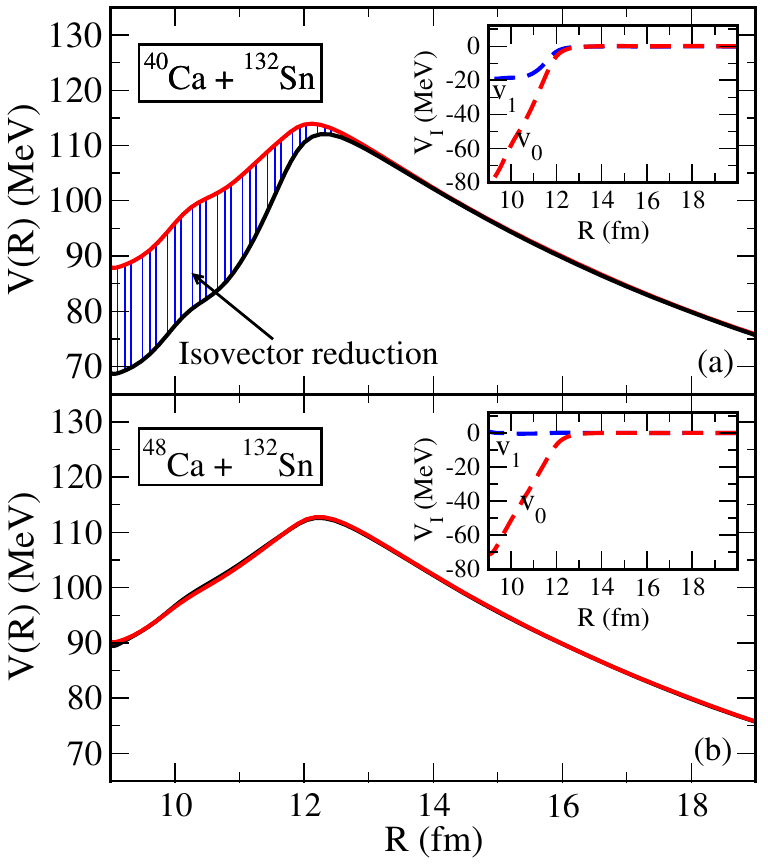}
\caption{\protect  DCTDHF potentials in (a) $^{40}$Ca$+^{132}$Sn and (b) $^{48}$Ca$+^{132}$Sn with TDHF collision energy $E_{c.m.}=120$~MeV. The total potential is shown in black, while red lines show its isoscalar contribution only. The blue-shaded region corresponds to the reduction originating from the isovector contribution. The inset show the isoscalar and isovector contributions to the potential (excluding Coulomb) as given in Eq.~(\ref{eq:vIso}). From Ref. \cite{godbey2017}.}
\label{fig:V_CaSn}   
\end{figure}

The DCTDHF technique has been used to investigate the role of transfer on the potential  \cite{uma07,obe12,godbey2017}.
As an example, Fig.~\ref{fig:V_CaSn} shows the isovector and isoscalar contributions to the potentials in $^{40,48}$Ca$+^{132}$Sn \cite{godbey2017}. 
The observed isovector reduction of the potential inside the barrier in $^{40}$Ca$+^{132}$Sn is interpreted as an effect of positive $Q-$value transfer channels.
In $^{48}$Ca$+^{132}$Sn, however, all transfer channels have negative $Q-$values which is compatible with the absence of isovector reduction to the potential. 
See also Ref.~\cite{tong2023} for a recent study of $^{40,48}$Ca$+^{124}$Sn and $^{58}$Ni$+^{132}$Sn systems. 

\subsubsection{Theoretical uncertainty quantification \label{sec:UQ}}

Although recent advances have been made to include rigorous uncertainty quantification for microscopic models of nuclear structure, very few work have evaluated uncertainties on predictions of observables from heavy-ion collisions. 
To address this gap, TDHF has been employed to assess model uncertainty in a series of low-energy, heavy-ion fusion reactions \cite{godbey2022}.

\begin{figure}[!tb]
\centering
\includegraphics[width=7cm]{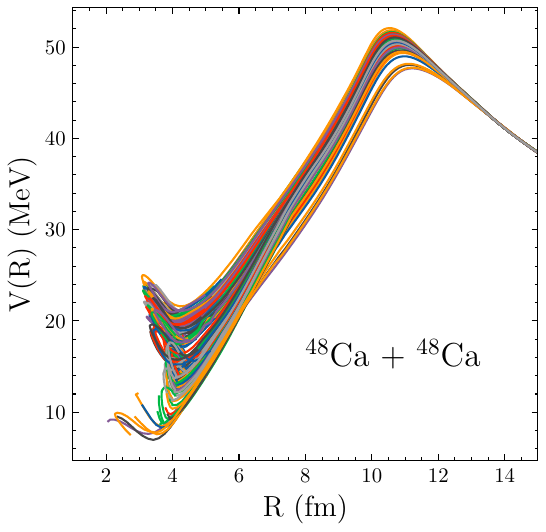}
\caption{\protect  DCTDHF potentials for $^{48}$Ca$+^{48}$Ca obtained with a TDHF simulation at $E_{c.m.} = 56$~MeV for 100 EDFs sampled from the posterior distributions for UNEDF1.
 From Ref. \cite{godbey2022}.}
\label{fig:48Ca48Ca_Pot}   
\end{figure}

As the uncertainties in predicting fusion cross-sections essentially stem from the EDF,  an ensemble of EDFs sampled from Bayesian posterior distributions was generated and used to compute DCTDHF potentials and associated fusion cross-sections \cite{godbey2022}. 
An example of resulting DCTDHF  $^{48}$Ca+$^{48}$Ca potentials is shown in Fig.~\ref{fig:48Ca48Ca_Pot}.
The barrier heights vary over a range of about $\sim5$~MeV and are anti-correlated with the barrier radius that varies by $\sim0.5$~fm.

\begin{figure}[!tb]
\centering
\includegraphics[width=7cm]{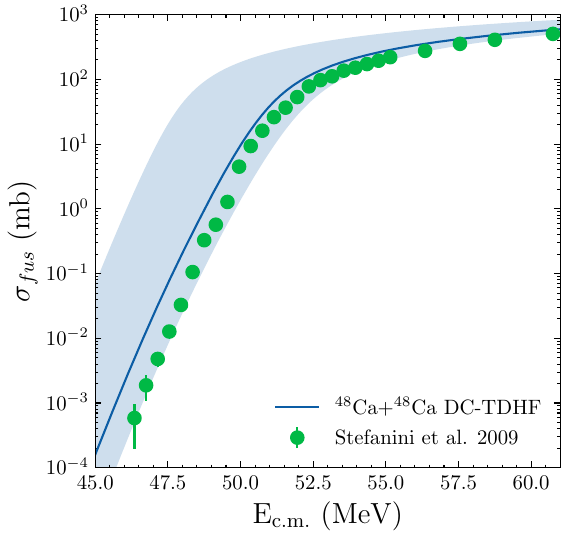}
\caption{\protect Fusion cross sections for $^{48}$Ca$+^{48}$Ca. The solid blue line represents the median results from the DCTDHF method, while the light blue shaded area indicates the uncertainty due to the distribution of the EDF parameters.  Experimental data are from \cite{stefanini2009}. From Ref. \cite{godbey2022}.}
\label{fig:48Ca48Ca_UQ}   
\end{figure}

Figure~\ref{fig:48Ca48Ca_UQ} shows the  cross-sections computed with these potentials. 
The variation of the potentials translate into large uncertainties in the cross-section, in particular at sub-barrier energies. 
This is not surprising as the transmission amplitudes vary exponentially with the potential characteristics in the tunnelling regime, and thus a small variation of the potential translates into large changes in sub-barrier fusion cross-sections. 
These large uncertainties in $^{48}$Ca$+^{48}$Ca  are interpreted in terms of uncertainties in the neutron radius of $^{48}$Ca that is particularly ill-constrained in the fitting protocol of Skyrme EDF. 
For example, the same method applied to $^{16}$O$+^{208}$Pb gives much smaller uncertainties, as shown in Fig.~\ref{fig:208Pb16O_UQ}, despite the influence of net nucleon  transfer (see Fig.~\ref{fig:dens}) that is absent in symmetric systems. 

\begin{figure}[!tb]
\centering
\includegraphics[width=7cm]{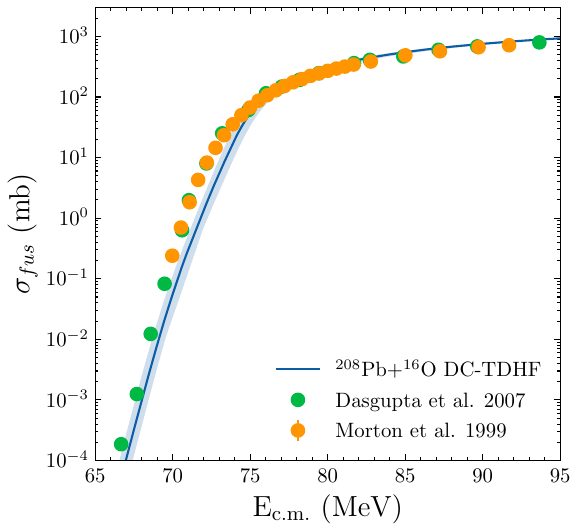}
\caption{\protect Same as Fig.~\ref{fig:48Ca48Ca_UQ} for $^{16}$O$+^{208}$Pb.  Experimental data are from \cite{morton1999,dasgupta2007}. From Ref. \cite{godbey2022}.}
\label{fig:208Pb16O_UQ}   
\end{figure}

\subsection{Multi-nucleon transfer reactions \label{sec:transfer}}

TDHF solvers have been used to investigate multi-nucleon transfer reactions in heavy-ion collisions  \cite{sim08,uma08a,was09b,sim10b,eve11,yil11,sim12,scamps2013a,sekizawa2013,sekizawa2014,yilmaz2014,sekizawa2015,ayik2015,sekizawa2016,vophuoc2016,bourgin2016,scamps2017a,scamps2017b,sekizawa2017,sekizawa2017a,regnier2018,ayik2018,yilmaz2018,ayik2019,ayik2019b,wu2019,sekizawa2020,godbey2020b,simenel2020,ayik2021,roy2022,wu2022,arik2023,ayik2023a,ayik2023b,zhang2024a,li2024b,gao2025}.
We discussed in section~\ref{sec:fus-spher} the interplay between fusion and transfer reactions in the $^{16}$O+$^{208}$Pb system around the barrier. 
In particular, we observed just below the barrier (see top of Fig.~\ref{fig:dens}) an average transfer of two protons. 
We now discuss such transfer reactions in more details.

\subsubsection{Average number of transferred nucleons \label{sec:transfer-average}}

\begin{figure}
\centering
\includegraphics[width=6cm]{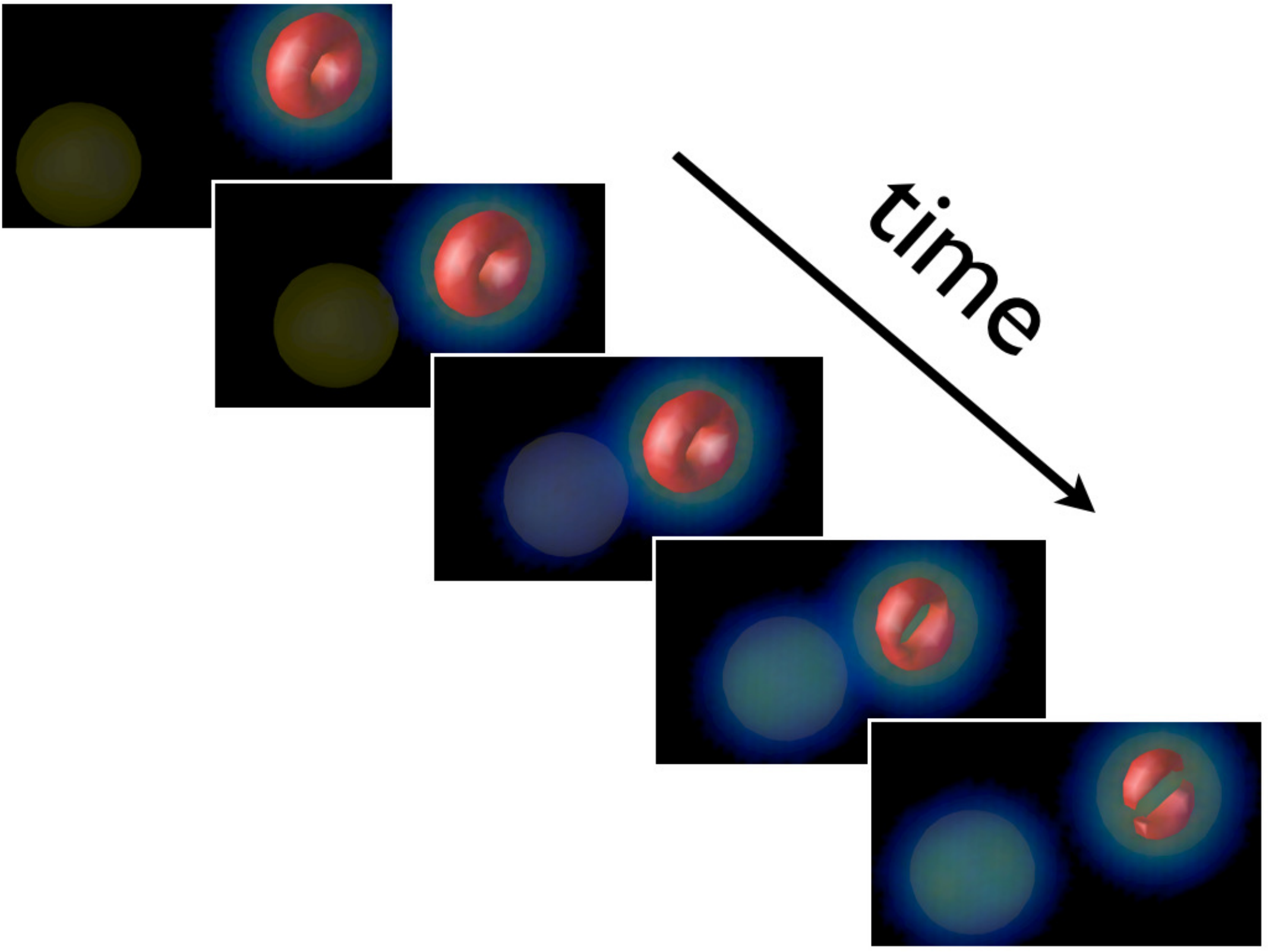} 
\caption{TDHF calculations of central sub-barrier 
collision of two $^{16}$O nuclei. The nuclei approach each 
other, and re-separate back-to-back due to the 
Coulomb repulsion. The evolution of a $p_{3/2}$ single-particle wave-function belonging initially to the nucleus in the right is 
shown. After the collision, part of this wave function 
has been transferred to its collision partner. Since this 
reaction is symmetric, a similar transfer occurs from 
the left to the right, and both fragments have the same 
particle number distributions. 
}
\label{fig:transfer_sp}
\end{figure}


The TDHF equation describes the evolution of single-particle wave-functions.
The latter, initially localised within one collision partner, may be partially transferred to the other fragment during the collision, as illustrated in Fig. \ref{fig:transfer_sp}.
Naturally, in a case of a symmetric collision, single-particle wave functions are transferred equally in opposite directions, leading to outgoing fragments with the same $Z$ and $N$ as in the entrance channel. 

\begin{figure}
\centering
\includegraphics[width=8cm]{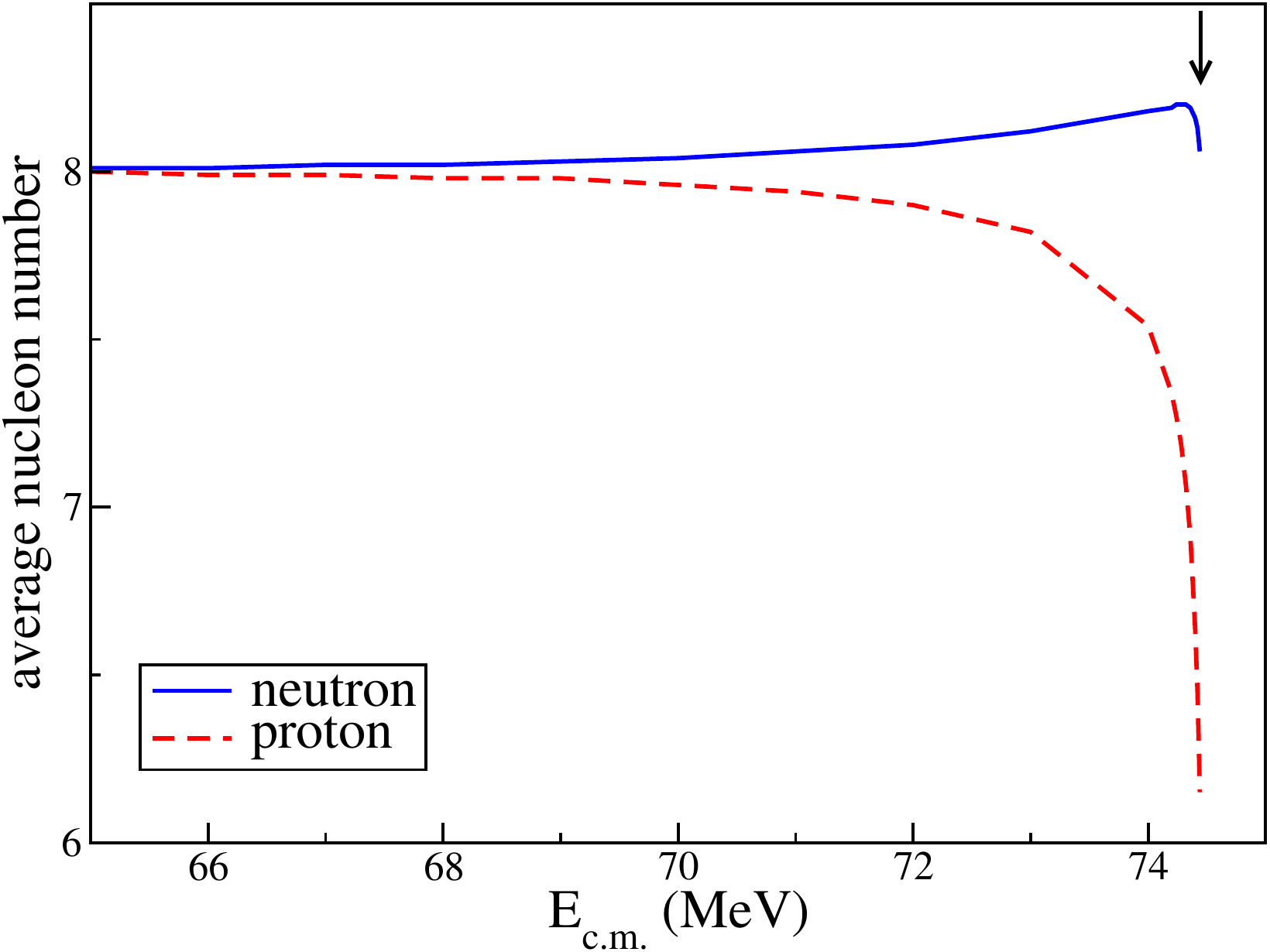} 
\caption{Transfer in $^{16}$O+$^{208}$Pb sub-barrier central collisions. 
The average number of protons and neutrons of the small fragment in the exit channel
are plotted as a function of $E_{c.m.}$. The arrow indicates the TDHF fusion barrier.}
\label{fig:average}
\end{figure}

In the case of an asymmetric collision, a change of the average particle number in the fragments in the exit channel is a clear signature that a transfer mechanism occurred in the reaction.
Figure~\ref{fig:average} gives  the expectation value of $\oZ$ and $\oN$ of the small fragment in the exit channel of $^{16}$O+$^{208}$Pb sub-barrier central collisions. 
At the barrier, $\sim2$~protons and no neutron, in average, are transferred (the corresponding evolution of the density is shown in the top of Fig.~\ref{fig:dens}), while at $\sim10\%$ below the barrier, $Z\simeq N\simeq8$ is obtained in average, indicating a dominance of (in)elastic scattering.
We see that the probability for proton stripping (transfer from the light to the heavy nucleus) is higher than for proton pickup (transfer from the heavy to the light nucleus), while neutron pickup is more probable 
than neutron stripping.
This qualitative observation is in agreement with experimental data~\cite{vid77,eve11,rafferty2016}.

\begin{figure}
\begin{center}
\includegraphics[width=7.5cm]{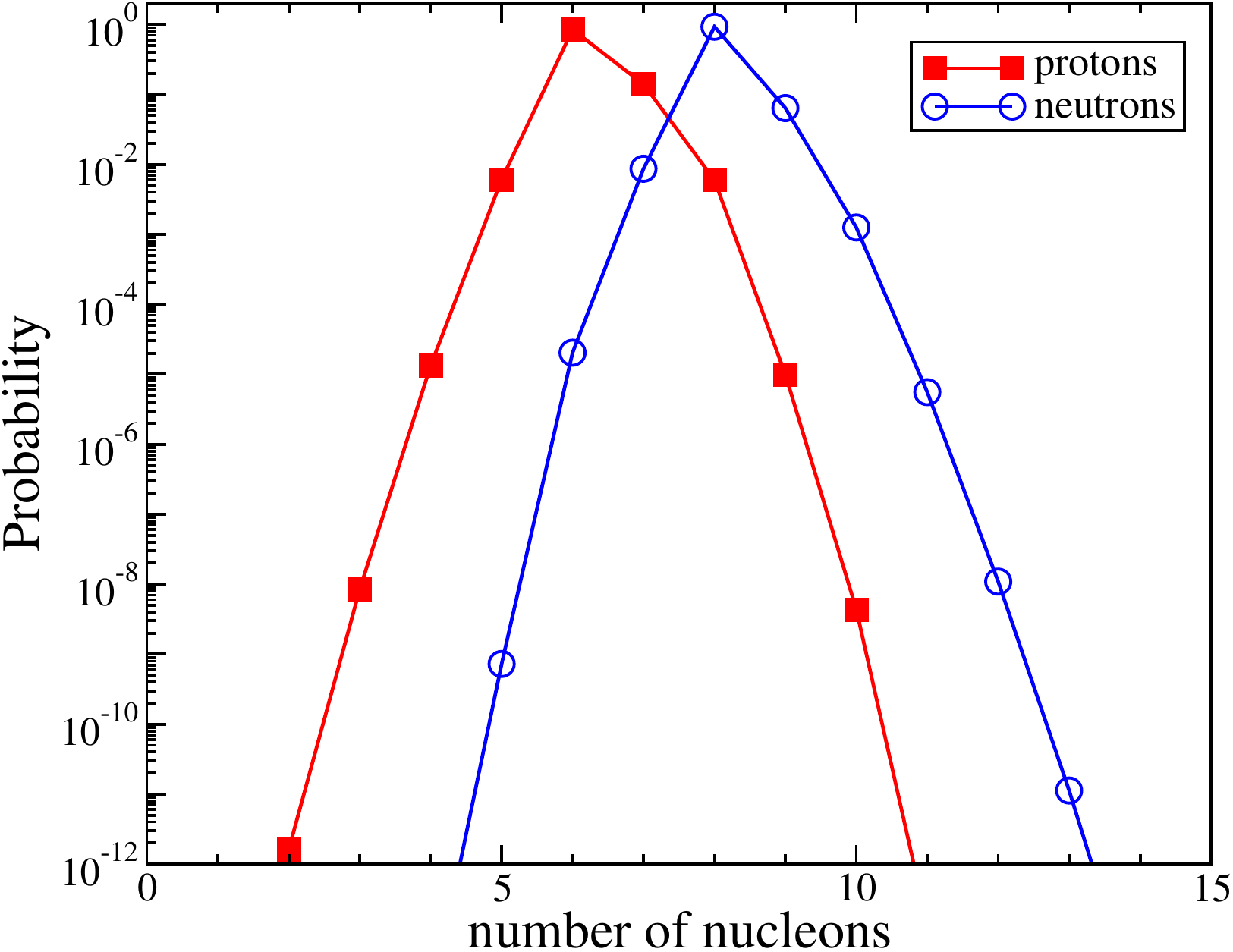} 
\includegraphics[width=7.5cm]{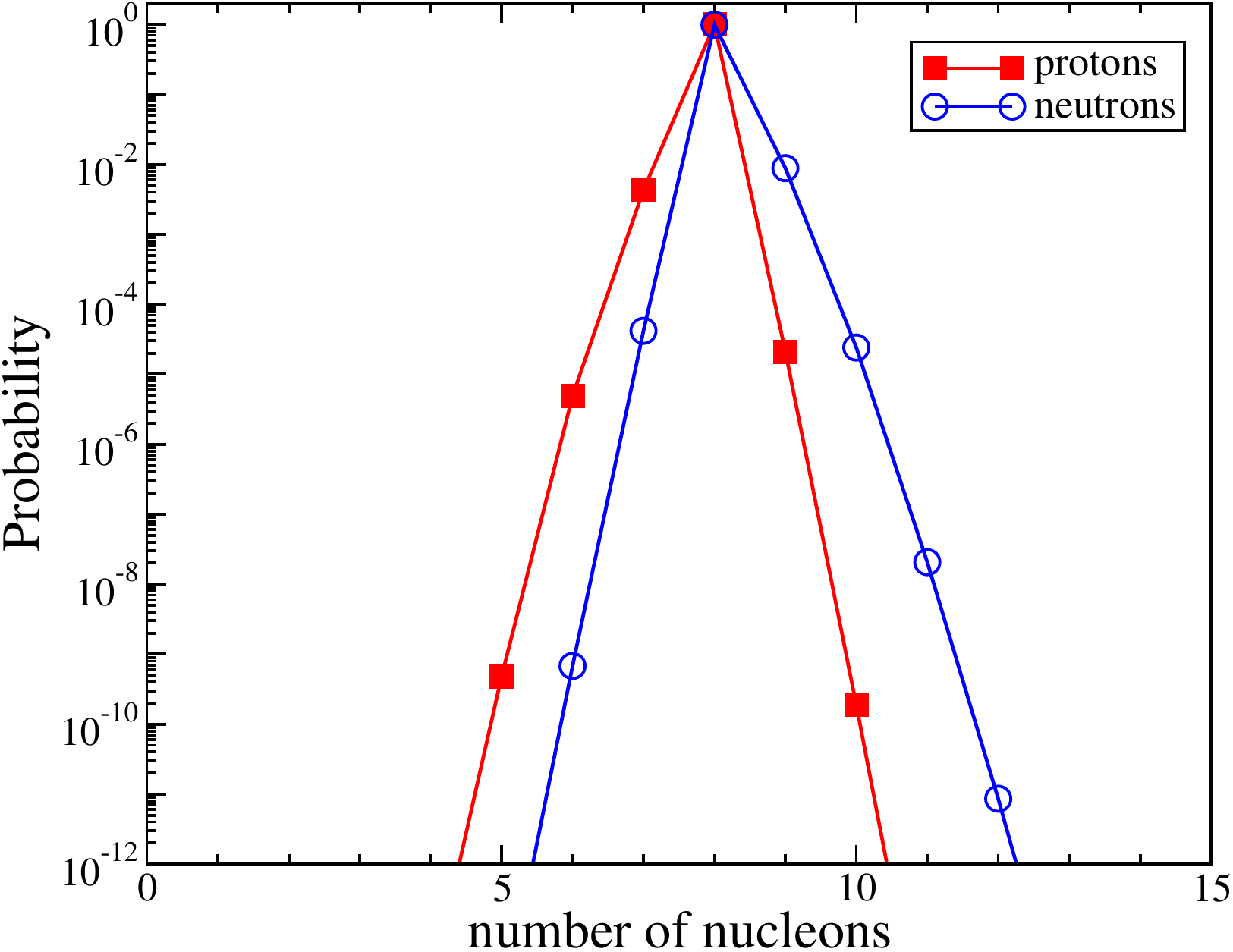} 
\caption{Neutron (circles) and proton (squares) number probability distributions 
of the lightest fragment in exit channel of a head-on 
$^{16}$O+$^{208}$Pb collision at $E_{c.m.}=74.44$~MeV (top) and $65$~MeV (bottom). Adapted from Ref.~\cite{sim10b}.}
\label{fig:proba}
\end{center}
\end{figure}

\subsubsection{Projection methods \label{sec:proj}}

To get a deeper insight into this transfer mechanism, the transfer probabilities are extracted at the TDHF level \cite{koo77,sim10b} thanks to a projection onto a good particle number technique\footnote{This technique is standard in beyond-mean-field models for nuclear structure when the number of particles is only given in average~\cite{rin80,lacroix2019}. See also~\cite{bulgac2019c}.} applied on the outgoing fragments\footnote{One could question this approach as the TDHF wave functions are not used for the calculation of expectation values of one-body operators. In particular, the width of the distributions should be underestimated~\cite{das79}. However, as we will see in section~\ref{sec:DIC}, the TDHF and BV widths are similar for non violent collisions such as sub-barrier transfer, justifying {\it a posteriori} this approach for quasi-elastic transfer.}.
It is possible to extract the component of the wave function associated with a specific transfer channel 
using a particle number projector\footnote{See also \cite{koo77} for an alternative combinatory approach.} onto $N$ protons or neutrons in the $x>0$ region where one fragment is located at the final time, the other one being in the $x<0$ region.
Such a projector is written~\cite{bender03}
\oeq
\oP_R(N)=\frac{1}{2\pi}\int_0^{2\pi} \stb \d \tet \stf e^{i\tet(\oN_R-N)},
\label{eq:projector}
\ceq
where
\oeq
\oN_R = \sum_{s} \sdf \int \stb \d \vr \stf \oad(\vr s) \sdf \oa(\vr s) 
\sdf \Theta(x)
\label{eq:NG}
\ceq
counts the number of particles in the $x>0$ region ($\Theta(x)=1$ if $x>0$ and 0 elsewhere).
Isospin is omitted to simplify the notation. 

The projector defined in Eq.~(\ref{eq:projector}) can be used to compute the probability to find $N$ nucleons in $x>0$ in the final state $\kfi$,
\oeq
\left|\oP_R(N)\kfi\right|^2=\frac{1}{2\pi}\int_0^{2\pi} \stb \d \tet \stf e^{-i\tet{N}}\bfi\phi_R(\tet)\>,
\label{eq:proba}
\ceq
where $|\phi_R(\tet)\>=e^{{i\tet\oN_R}}\kfi$ 
 represents a rotation of $\kfi$ by a gauge angle $\tet$ 
in the gauge space associated with the particle number degree of freedom.
Note that $|\phi_R(\tet)\>$ is an independent particle state. 
The last term in Eq.~(\ref{eq:proba}) is then the determinant of the matrix of the occupied single particle state overlaps:
\oeq
\bfi\phi_R(\tet)\>=\det (F)
\ceq
with
\oeq
F_{ij}= \sum_{s} \int \stb\d \vr \sdf{\az_i^s}^*(\vr) {\az_j^{s}}(\vr) e^{i\tet\Theta(x)}.
\ceq
The integral in Eq.~(\ref{eq:proba}) is discretised using $\tet_n=2\pi{n}/M$ with the integer $n=1\cdots{M}$.
Choosing $M=300$ ensures numerical convergence for the $^{16}$O+$^{208}$Pb system. 
Fig~\ref{fig:proba} shows the resulting transfer probabilities at (top) and well below the barrier at $E_{c.m.}=65$~MeV (bottom).
As expected from the average values (see Fig.~\ref{fig:average}), the most probable channels are $Z=6$ and $N=8$ at the barrier, and $Z=N=8$ with a small probability of neutron pickup or proton stripping (of the order of $10^{-2}$) well below the barrier.

A standard representation of experimental sub-barrier energy transfer data is to plot transfer probabilities as a function of the distance of closest approach $R_{min}$ between the collision partners~\cite{cor09}. 
$R_{min}$ is computed assuming a Rutherford trajectory~\cite{bro91}:
\oeq
R_{min}={Z_1Z_2e^2}[1+\mbox{cosec}(\theta_{c.m.}/2)]/{2E_{c.m.}}
\label{eq:R_min}
\ceq
where $\theta_{c.m.}$ is the centre of mass scattering angle.

\begin{figure}
\includegraphics[width=8.8cm]{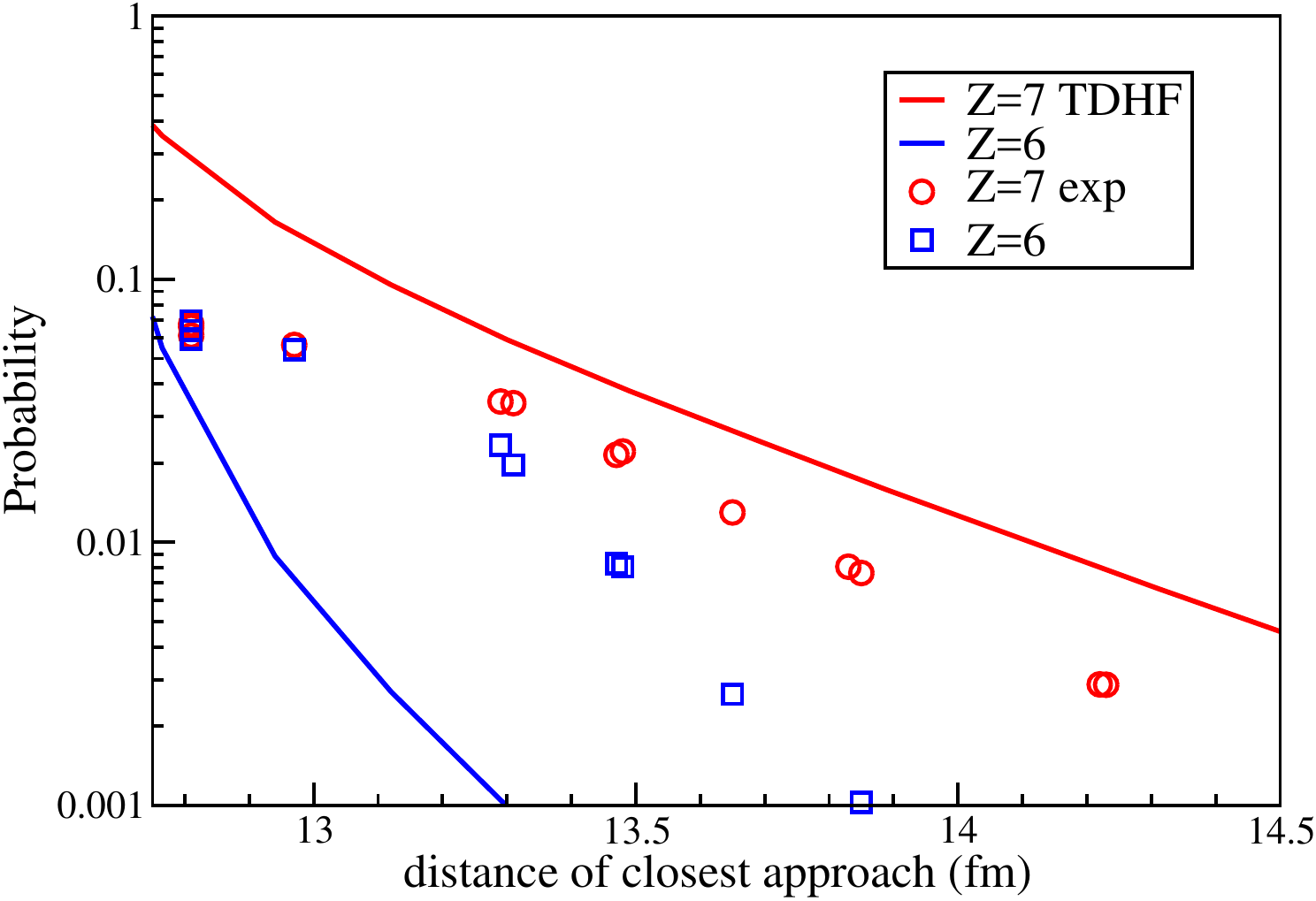} 
\caption{Proton number probability  
as function of the distance of closest approach  in the small outgoing fragment of the $^{16}$O+$^{208}$Pb reaction. TDHF results are shown with lines. Experimental data (open symbols) are taken from Ref.~\cite{eve11}.}
\label{fig:proba_TDHF+exp}
\end{figure}

A comparison of TDHF calculations with  data from Ref.~\cite{eve11} is shown in Fig.~\ref{fig:proba_TDHF+exp} for sub-barrier one and two-proton transfer channels in the $^{16}$O+$^{208}$Pb reaction.
We see that TDHF overestimates the one-proton transfer probabilities and underestimates the 2-proton transfer channel.
This discrepancy is interpreted as an effect of pairing interactions~\cite{sim10b,eve11}.
Indeed, paired nucleons are expected to contribute to two-nucleon transfer channels.
As a result, the two-nucleon (resp. single-nucleon) transfer probability increases (decreases). 


\begin{figure}
\centering
\includegraphics[width=8cm]{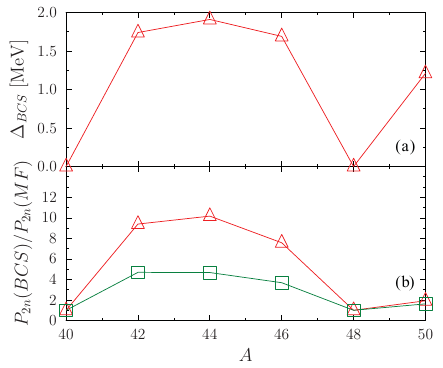} 
\caption{(a) Mean neutron pairing gap from BCS for calcium isotopes. (b) Ratio of the two-neutron transfer probability in $^{40}$Ca$+^A$Ca central collisions obtained with and without pairing at 4 MeV (squares) and 6 MeV (triangles) below the Coulomb barrier. From Ref.~\cite{scamps2013a}.}
\label{fig:2proj}
\end{figure}

In order to describe multi-nucleon transfer in superfluid systems with TDBCS or TDHFB, one needs to generalise the projection method to account for the fact that the total system does not have a good number of nucleons. 
This is done with a double projection technique \cite{scamps2013a}.
A first projection is performed on the total system to 
obtain a normalised state with good number of particles $N_0$:
\oeq
|N_0\>=\frac{\oP(N_0)|\Psi\>}{\sqrt{\<\Psi|\oP(N_0)|\Psi\>}},
\ceq
where $\oP$ is a particle number projector on the entire system. 
Then, a second projection is performed to compute the particle number  distribution in the fragments using
\oeq
\mP(N)=\<N_0|\oP_R(N)|N_0\>.
\ceq
Figure \ref{fig:2proj} gives an example of application to transfer reactions in $^{40}$Ca$+^{40,42,44,46,48,50}$Ca sub-barrier central collisions \cite{scamps2013a}. 
As expected, there is an enhancement of pair transfer induced by pairing correlations. 

\begin{figure}
\centering
\includegraphics[width=8.8cm]{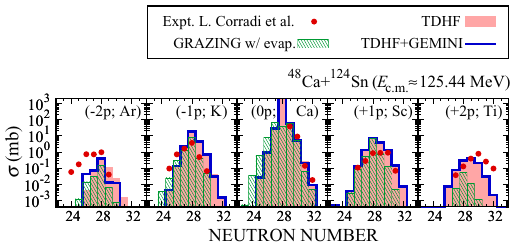} 
\caption{ Cross sections for fragment production in $^{48}$Ca$+^{124}$Sn at $E_{c.m.}\simeq125.44$~MeV with TDHF (primary products), TDHF+GEMINI, and  GRAZING. Experimental data are from \cite{corradi1997}. From Ref. \cite{sekizawa2017}.}
\label{fig:TDHFGEMINI}
\end{figure}

The particle number projection method has been also used to evaluate properties of the fragments such as their angular momentum and their energy \cite{sekizawa2014}.
This is particularly useful in order to couple TDHF with statistical models. Indeed, the angular momentum and excitation energy of the fragments are used as inputs to statistical decay codes.
Figure \ref{fig:TDHFGEMINI} shows an example of such calculations for the $^{48}$Ca$+^{124}$Sn reactions \cite{sekizawa2017}.
The results are in reasonable agreement with experiment.

\subsubsection{Deep-inelastic collisions \label{sec:DIC}}

Deep-Inelastic Collisions (DIC) occur essentially well above the barrier. Several studies with TDHF solvers have been dedicated to DIC  \cite{sim11,wang2016,umar2017,roy2018,williams2018,simenel2020,godbey2020b,ayik2020b,gao2025}.
The main characteristics of DIC exit channels are:
\begin{itemize}
\item A strong damping of the initial kinetic energy,
\item Large fluctuations of the fragment proton and neutron numbers around their initial value, 
\item  An angular distribution of the fragments following a $1/\sin\theta_{c.m.}$ behaviour.
\end{itemize}
The last point is due to a large orbiting of the fragments at contact. All $\theta_{c.m.}$ are then equiprobable in DIC. 
However the emission is not isotropic as it occurs essentially in the collision plane.
As a result, the differential cross-sections for DIC events obey 
$$\frac{d\sigma}{d\theta_{c.m.}}=2\pi\sin\theta_{c.m.}\frac{d\sigma}{d\Omega}\simeq\mbox{ constant}.$$
This leads to the $1/\sin\theta_{c.m.}$ behaviour, as 
$$\frac{d\sigma}{d\Omega}\propto \frac{1}{\sin \theta_{c.m.}}.$$

\begin{figure}
\centering
\includegraphics[width=7cm]{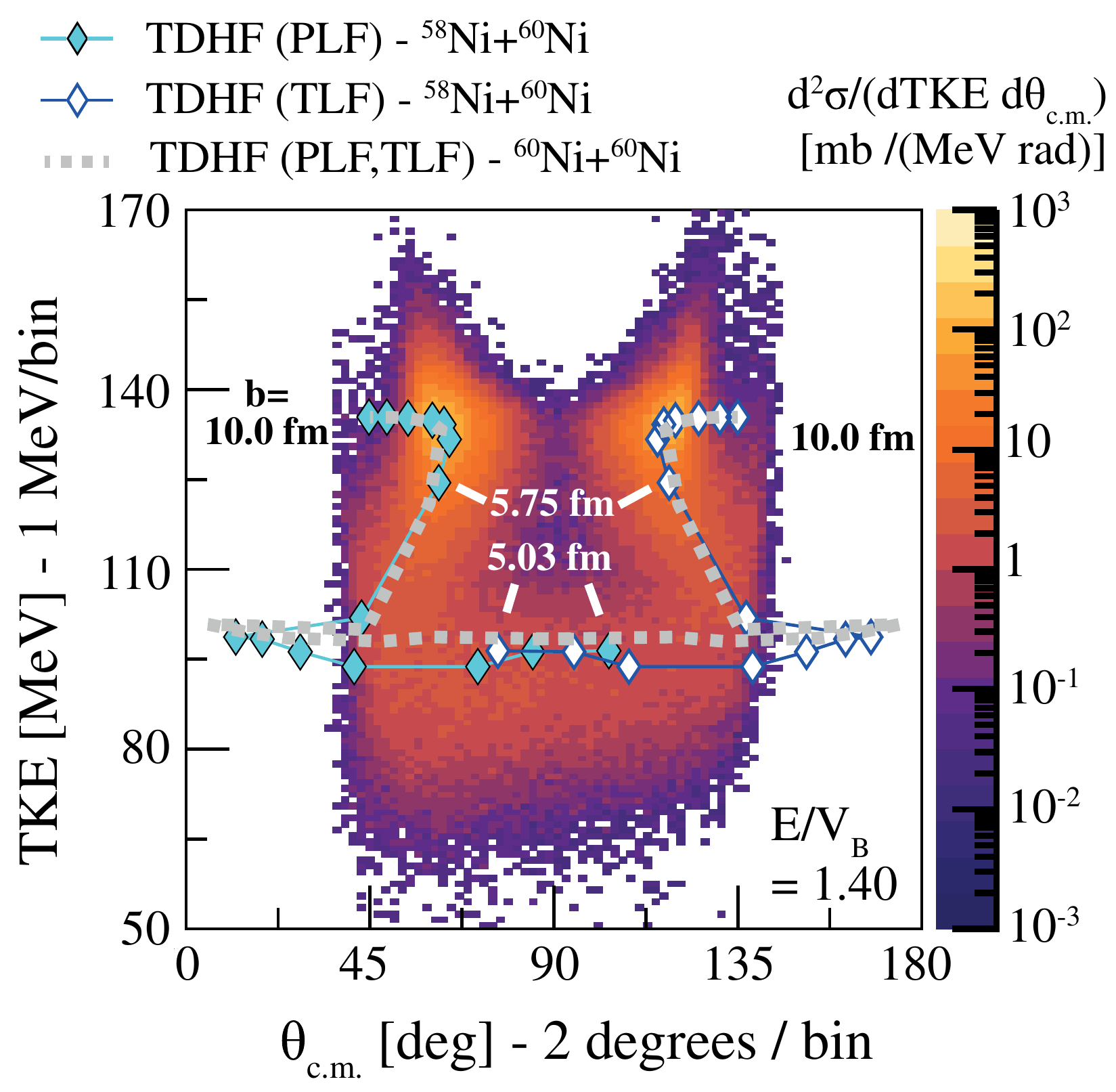} 
\caption{
Scattering angle  vs TKE differential cross section distributions   for $^{58}$Ni$+^{60}$Ni at $E_{c.m.}\simeq1.4V_B$ overlaid with   TDHF results. The impact parameter for selected TDHF calculations are shown.
From Ref.~\cite{williams2018}.}
\label{fig:NiNi}
\end{figure}

The condition of strong damping is illustrated in Fig. \ref{fig:NiNi} for $^{58}$Ni$+^{60}$Ni \cite{williams2018}.
The evolution of TKE with scattering angle observed experimentally is well reproduced by TDHF calculations.
A comparison between experiment and theory then allows to identify the range of impact parameters for which energy damping occurs. 

\begin{figure}
\centering
\includegraphics[width=7cm]{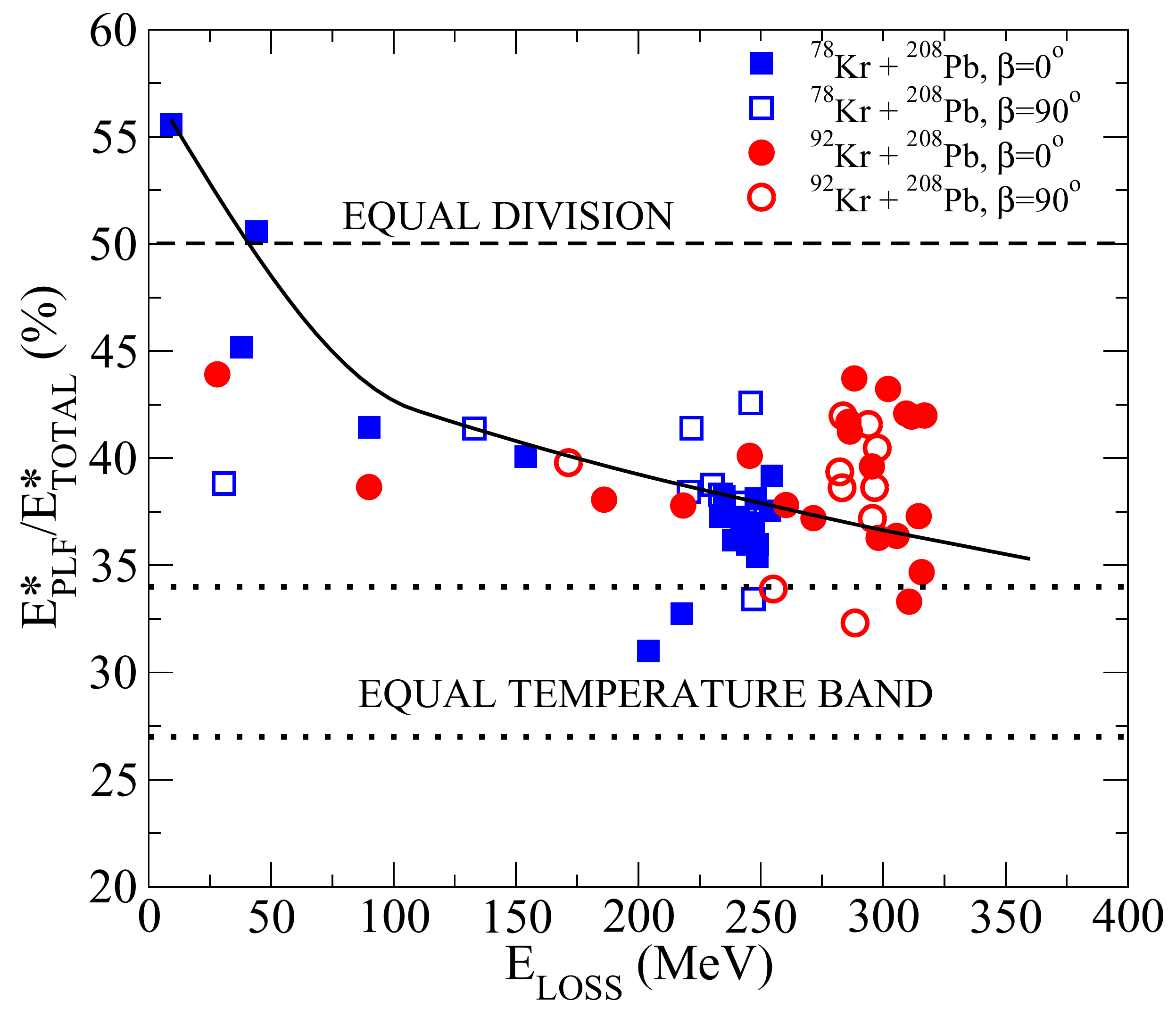} 
\caption{Percent of total excitation energy carried by the projectile like fragment (PLF) as a function of kinetic energy loss for various initial orientation of the deformed Kr nuclei. The dashed horizontal line marks equal sharing of excitation energy. The region between the dotted lines indicates the full thermalisation limit. 
From Ref.~\cite{umar2017}.}
\label{fig:KrPb}
\end{figure}

This damping of kinetic energy leads to an excitation of the fragments. 
The DCTDHF method is able to evaluate the amount of excitation energy stored in each fragment \cite{uma09a}. 
Figure~\ref{fig:KrPb} shows an application to the $^{78,92}$Kr$+^{208}$Pb reactions. 
We see that the final energy of the fragments is found in between the limits associated with equal repartition of excitation energy, and equal temperature, although the latter provides a better picture for the most damped reactions. 

Early TDHF calculations were able to reproduce fragment kinetic energies, mean masses, and scattering angles, but they failed to reproduce the observed large fluctuations of the fragment $Z$ and $N$ distributions~\cite{koo77,dav78}.
One way to overcome this is to include a collision term in the dynamics. 
However, Balian and V\'en\'eroni showed that the TDHF theory was in fact not optimised for the prediction of such fluctuations~\cite{bal81}. Instead, one should use the BV prescription given in Eq.~(\ref{eq:CiiBV}) which is derived from their variational principle and optimised for one-body fluctuations~\cite{bal84}.

The BV prescription (TDRPA) was used to compute particle number fluctuations in DIC in Refs. \cite{bon85,mar85,bro09,sim11,sim11b,williams2018,godbey2020b,gao2025}, and within a semi-classical approximation in Ref. \cite{zie88}.
Comparing with standard TDHF calculations, these works showed, indeed, an increase of the fluctuations. 
The increase of computational power and the inclusion of spin orbit interaction enabled direct comparisons with experimental data\footnote{The authors of Ref.~\cite{mar85} claim that their calculations are in good agreement with  experimental data. However, they only compute the fluctuations of $A=N+Z$ and compare with measured fluctuations of $Z$. In addition, their calculations do not include the spin-orbit interaction. They are performed at angular momenta leading to fusion when spin-orbit terms are included (see discussion in Ref.~\cite{bro09}).} \cite{sim11,williams2018,gao2025}. 

\begin{figure}
\begin{center}
\includegraphics[width=7cm]{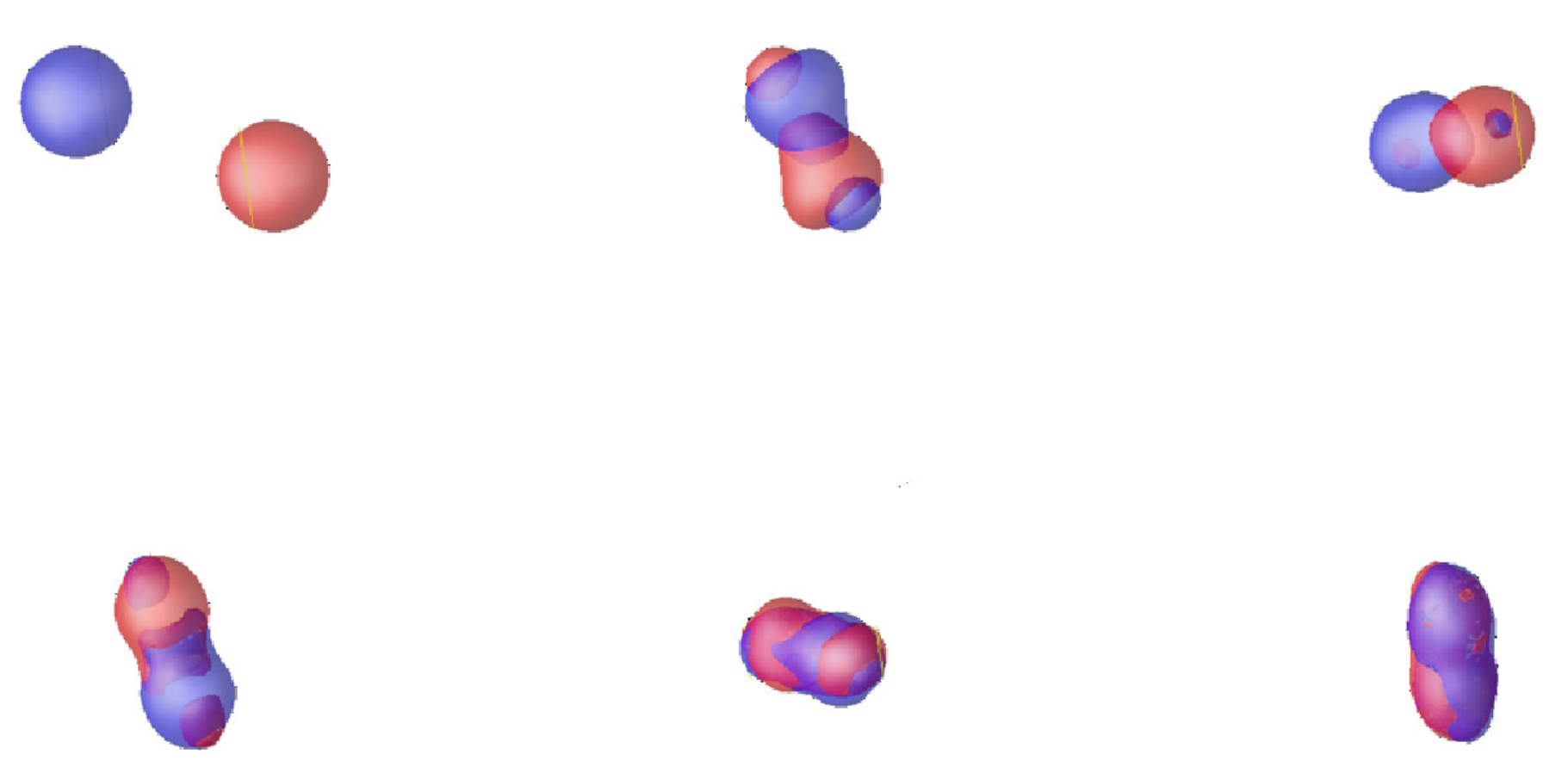} 
\caption{Density evolution for a $^{40}$Ca+$^{40}$Ca collision at $E_{c.m.}=128$~MeV and $L=60\hb$. Each snapshot is separated by 1.5~zs.}
\label{fig:40Ca+40Ca_L60}
\end{center}
\end{figure}

\begin{figure}
\begin{center}
\includegraphics[width=7cm]{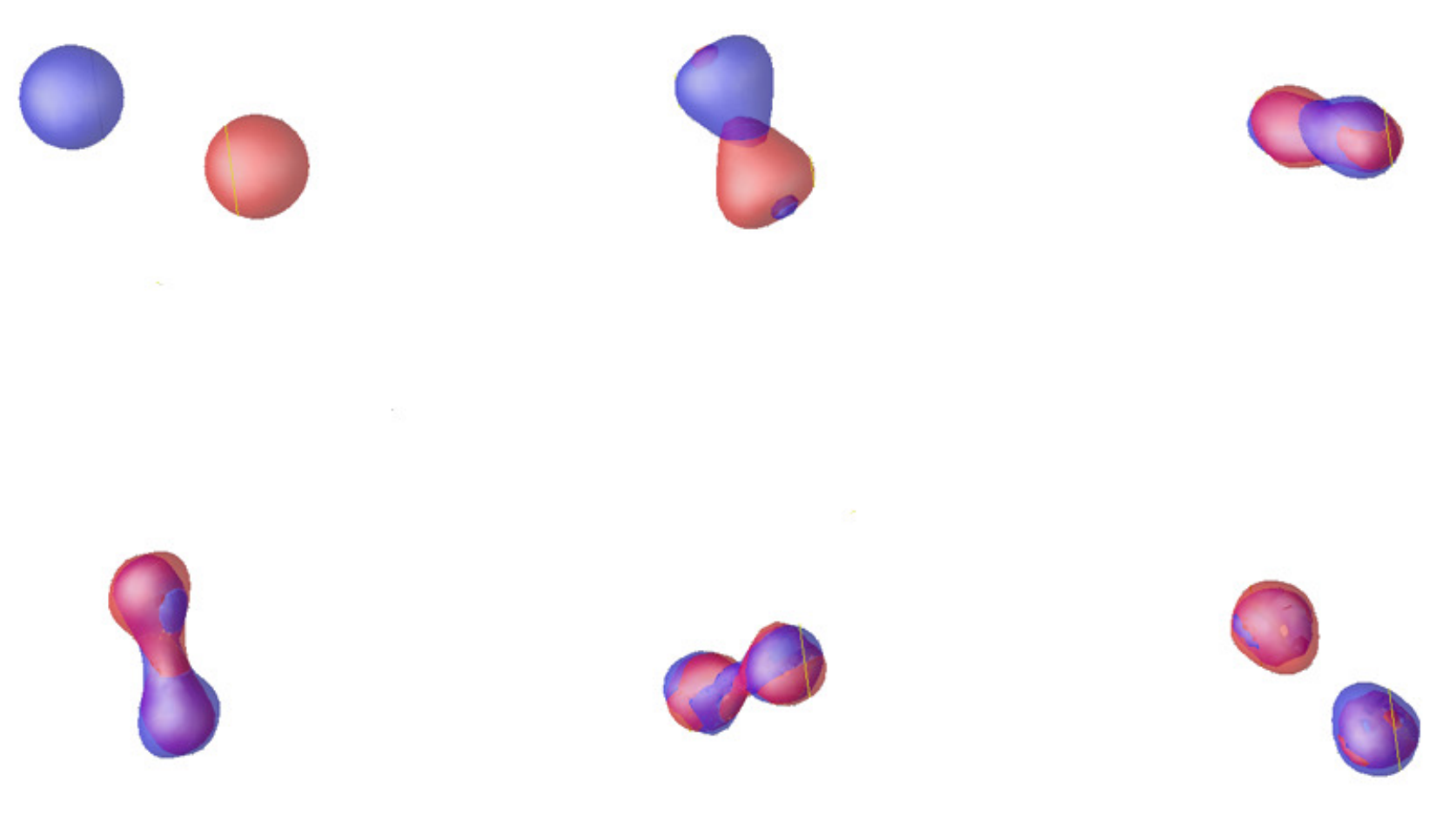} 
\caption{Same as Fig.~\ref{fig:40Ca+40Ca_L60} with $L=70\hb$.}
\label{fig:40Ca+40Ca_L70}
\end{center}
\end{figure}

\begin{figure}
\begin{center}
\includegraphics[width=7cm]{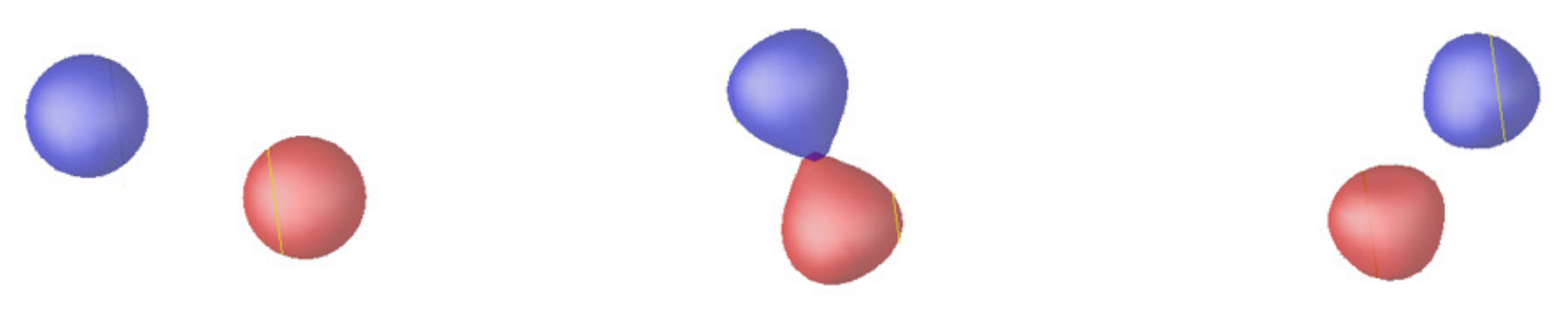} 
\caption{Same as Fig.~\ref{fig:40Ca+40Ca_L60} with $L=80\hb$.}
\label{fig:40Ca+40Ca_L80}
\end{center}
\end{figure}

\begin{figure}
\includegraphics[width=2.6cm]{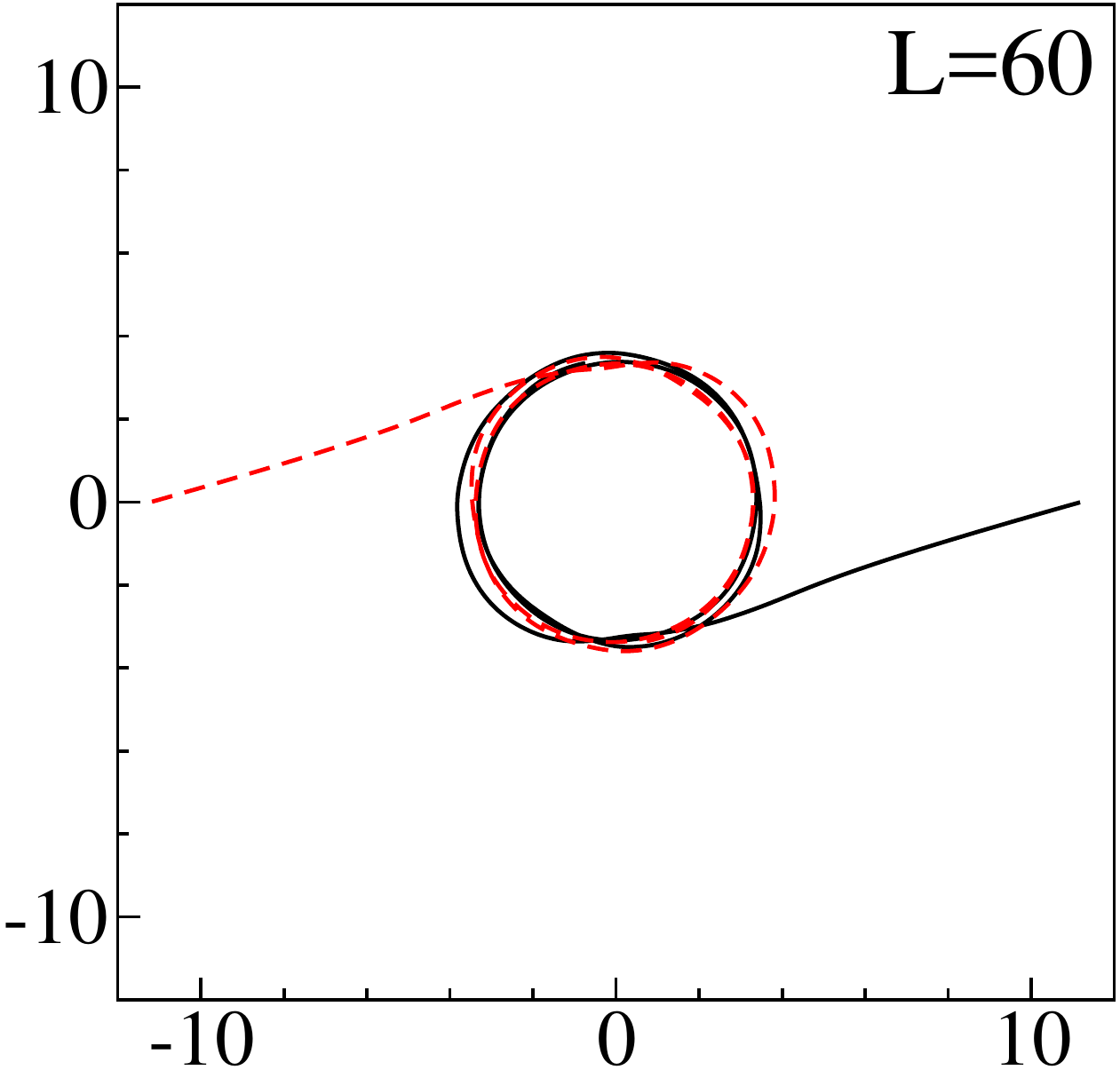}
\hspace{0.3cm} 
\includegraphics[width=2.6cm]{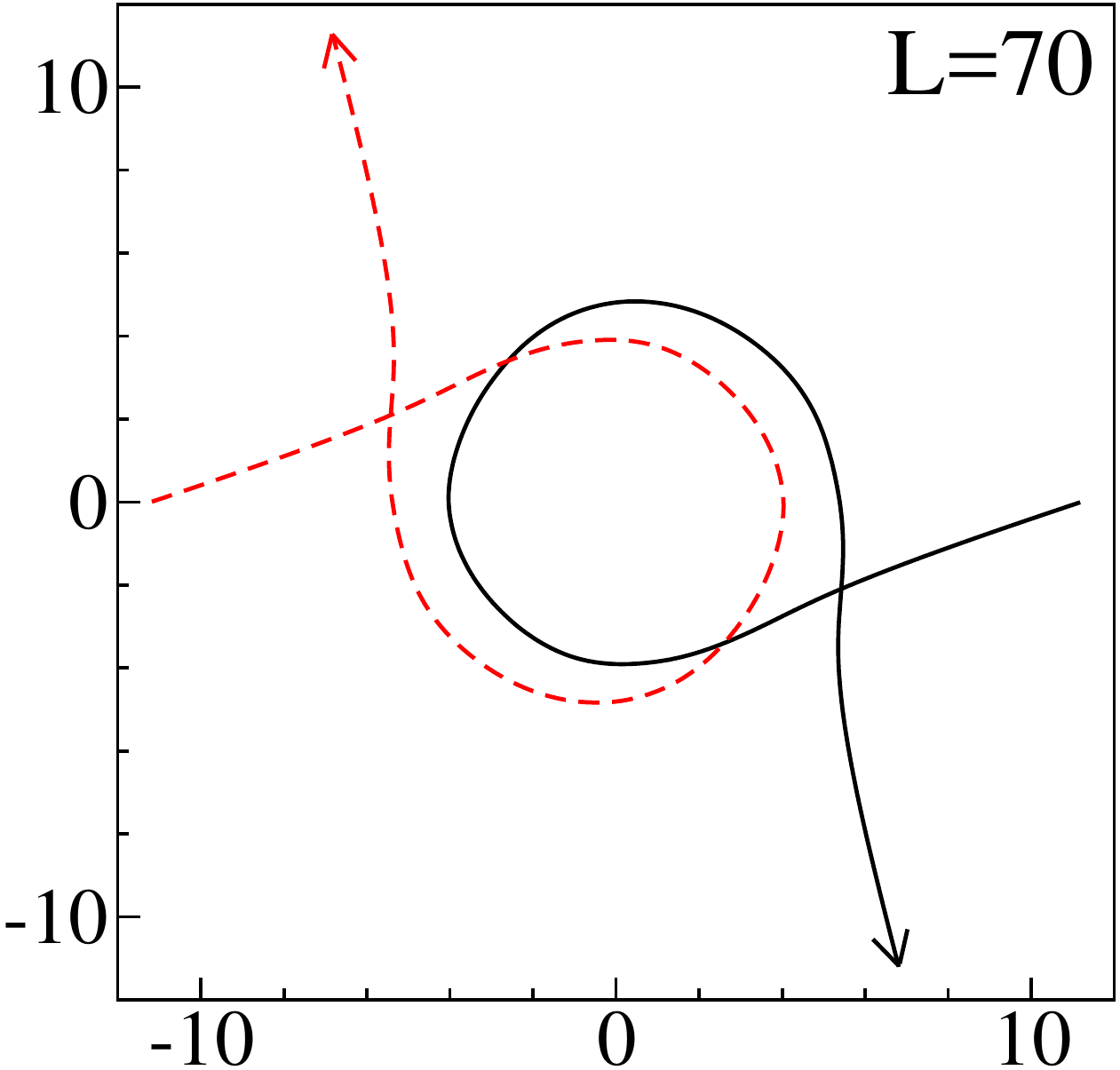} 
\hspace{0.3cm} 
\includegraphics[width=2.6cm]{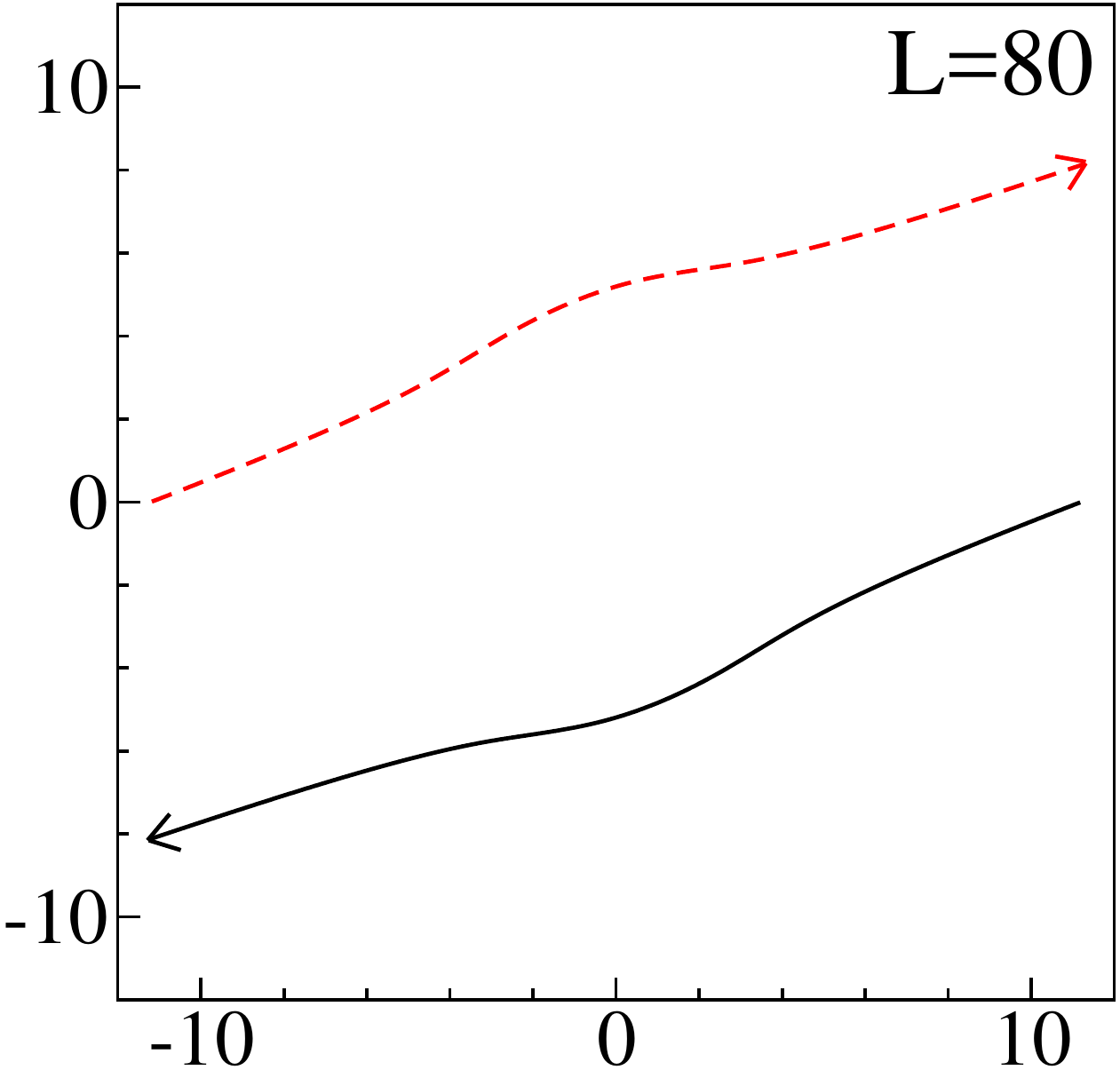} 
\caption{Trajectories of the centres of mass of the fragments in $^{40}$Ca+$^{40}$Ca collisions at $E_{c.m.}=128$~MeV.}
\label{fig:40Ca+40Ca_traj}
\end{figure}

We now discuss the main results of \cite{sim11} where $^{40}$Ca+ $^{40}$Ca collisions at $E_{c.m.}=128$~MeV ($\sim2.4$ times the barrier height) have been studied.  
Figures~\ref{fig:40Ca+40Ca_L60}, \ref{fig:40Ca+40Ca_L70}, and \ref{fig:40Ca+40Ca_L80} show density plots obtained with angular momenta $L=60\hb$, $70\hb$, and $80\hb$, respectively.
Three reaction mechanisms are observed: capture (at $L=60\hb$), DIC (at $L=70\hb$), and a partially damped collision (at $L=80\hb$).
In particular, the $L=70\hb$ case leads to an orbiting trajectory (see middle panel of Fig.~\ref{fig:40Ca+40Ca_traj}) which is characteristic of a DIC. 
This trajectory also corresponds to a strongly damped collision, as can be seen from the upper panel of Fig.~\ref{fig:sigma} where the total kinetic energy loss (TKEL) is plotted as a function of the initial angular momentum. 
Indeed, around $L\simeq70\hb$, the TKEL is $\sim60-70$~MeV.
These values are slightly below the Viola systematics~\cite{vio85} which predicts $\mathrm{TKEL_{Viola}}\simeq76$~MeV, indicating that these collisions are almost fully damped. 

\begin{figure}
\begin{center}
\includegraphics[width=6.5cm]{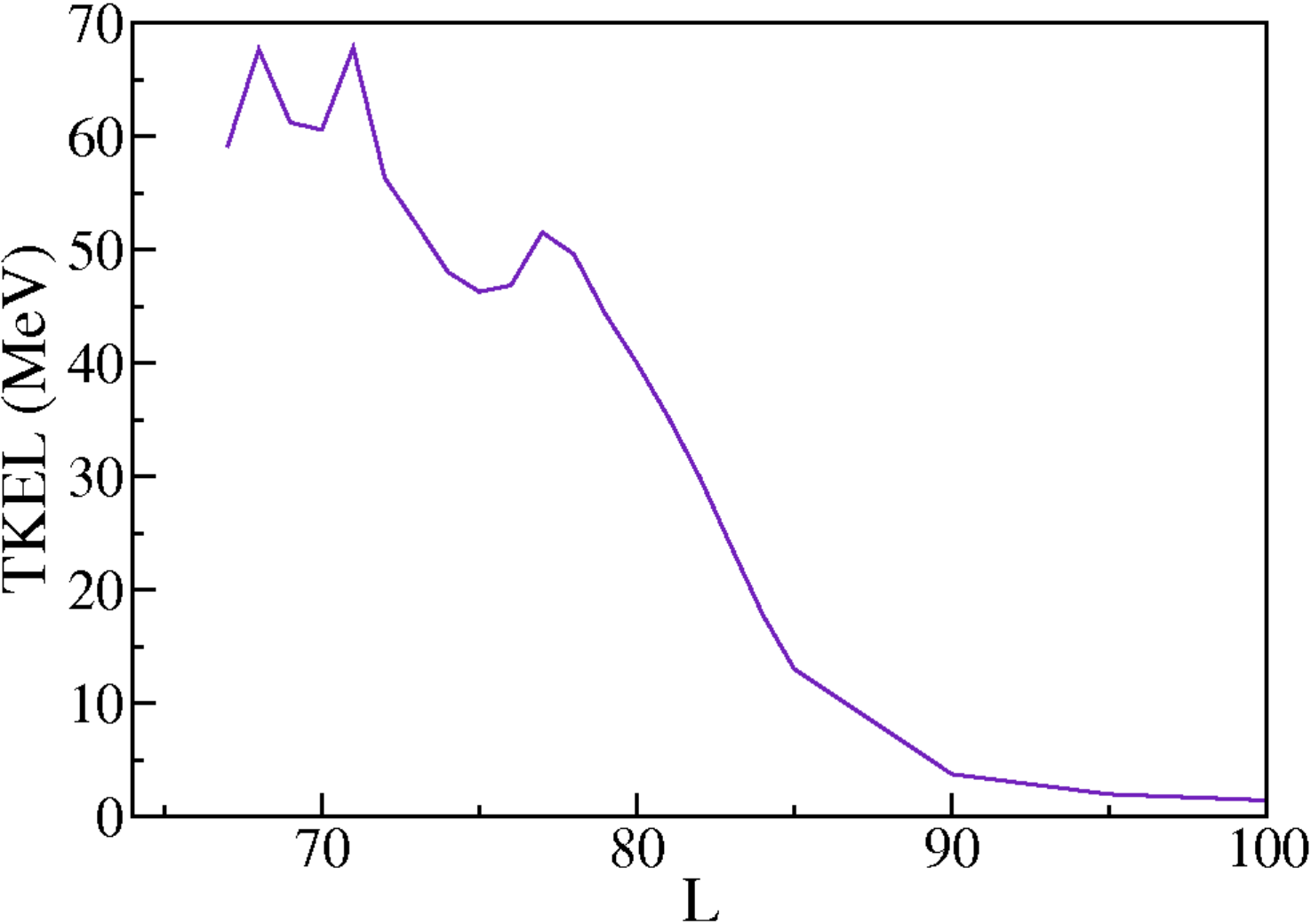} 
\includegraphics[width=6.5cm]{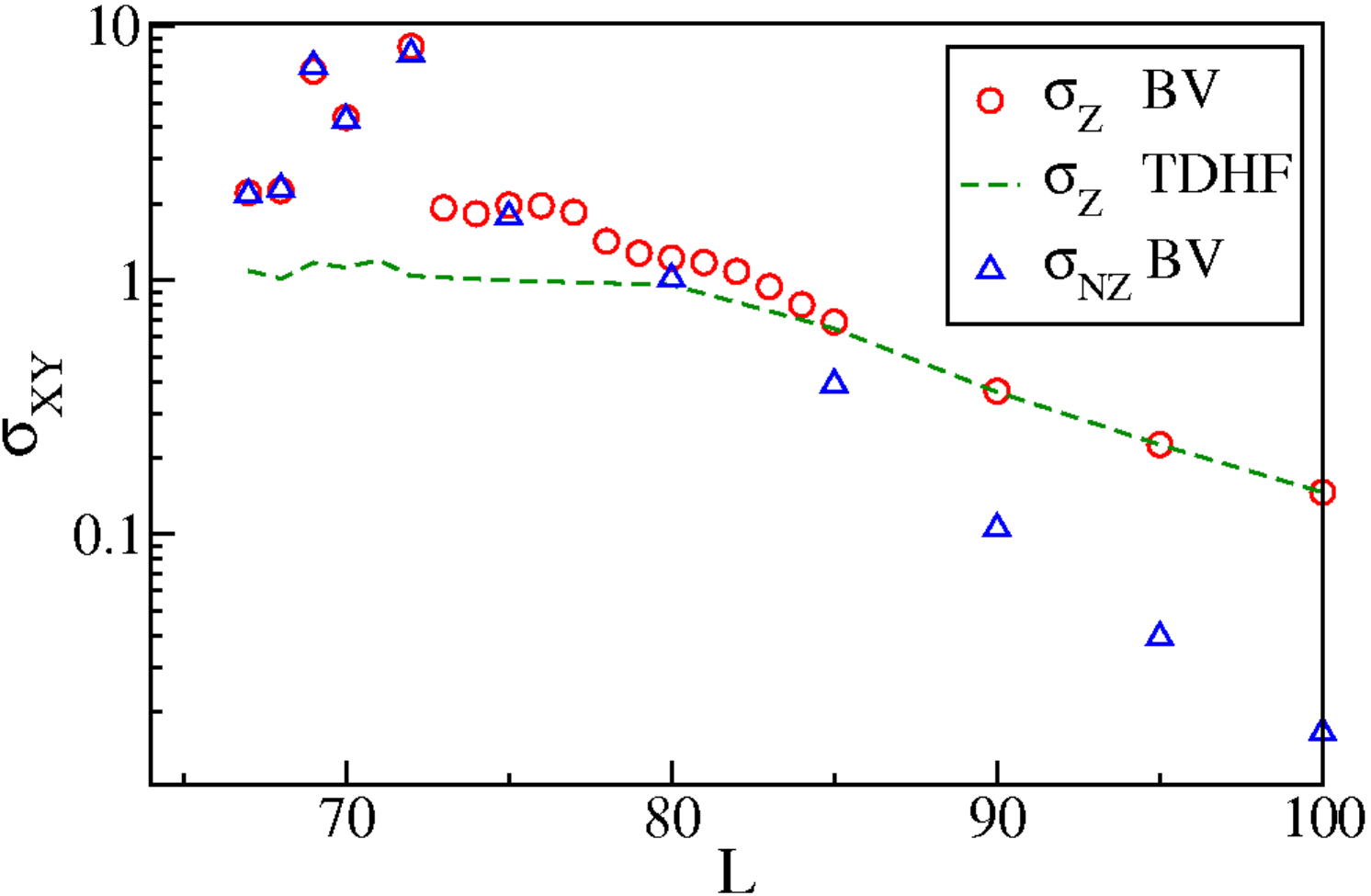} 
\caption{Properties of the exit channel of the $^{40}$Ca+$^{40}$Ca collisions at $E_{c.m.}=128$~MeV as a function of the initial angular momentum $L$. (top) Total kinetic energy loss from TDHF.
(bottom) TDHF (dashed line) and BV (circles) fluctuations of $Z$, 
and BV correlations between $N$ and $Z$ (triangles) of the outgoing fragments.
}
\label{fig:sigma}
\end{center}
\end{figure}

The lower panel of Fig.~\ref{fig:sigma} shows the evolution of the fluctuations $\sigma_Z$ of the number of protons\footnote{Neutron fluctuations are very close to the proton ones for this system. This is due to the fact that the collision partners are $N=Z$ nuclei.} in the outgoing fragments as a function of $L$.  
As expected\footnote{See early applications of the BV prescription for fluctuations where greater fluctuations than their TDHF counterparts were obtained~\cite{mar85,bon85,bro09}.}, the BV prescription [Eq.~(\ref{eq:CiiBV})] leads to larger fluctuations than standard TDHF fluctuations obtained from Eq.~(\ref{eq:CiiTDHF}).
This is particularly true for DIC.
However, TDHF and BV prescriptions converge at large $L$, e.g., for $L>90$ where the TKEL is less than 5~MeV, indicating a dominance of quasi-elastic scattering.
We conclude that TDHF calculations of particle number distributions give reasonable estimates for such non-violent collisions.
In particular, this justifies the calculation of transfer probabilities for sub-barrier collisions with TDHF (see section~\ref{sec:transfer}). 

Figure~\ref{fig:sigma} also shows an application of the BV prescription for the correlations between $N$ and $Z$ distributions. 
These correlations are determined from Eq.~(\ref{eq:CijBV}).
We observe increasing correlations with decreasing $L$. 
In particular, in DIC, the BV prescription predicts $\sigma_{N}\simeq\sigma_{Z}\simeq\sigma_{NZ}$, while, for quasi-elastic scattering, correlations are negligible. 

Let us recall the signification of such correlations. 
On one hand, uncorrelated distributions mean that the probability to transfer $z$ protons is independent of the probability to transfer $n$ neutrons, i.e., $P(z,n)\equiv P(z)P(n)$.
On the other hand, strongly correlated distributions mean that if we measure $n$ (resp. $z$), then we know what is $z$ ($n$).
This is the case, for instance, if all fragments have $N=Z$.
In this case, we would have $P(z,n)\equiv P(z)\delta_{n,z}\equiv P(n)\delta_{n,z}$.
The reality is usually in between and, assuming Gaussian distributions, we have
\oeqn
\stb\stb\stb P(z,n)&=&\(2\pi\sigma_N\sigma_Z\sqrt{1-\rho^2}\)^{-1} \nonumber\\
&&\exp\[ -\frac{1}{1-\rho^2} \( \frac{n^2}{\sigma_N^2} + \frac{z^2}{\sigma_Z^2} -\frac{2\rho nz}{\sigma_N\sigma_Z} \) \]\!,
\label{eq:Gauss}
\ceqn
where $\rho=\mathrm{sign}(\si_{NZ})\frac{\sigma_{NZ}^2}{\sigma_N\sigma_Z}$.
The case $\rho=0$ means no correlations between $N$ and $Z$ distributions, while the limit $\rho\rightarrow1$ ($\rho\rightarrow-1$) corresponds to maximum (anti) correlations.  

In the calculations of $^{40}$Ca+$^{40}$Ca at $E_{c.m.}=128$~MeV shown in Fig.~\ref{fig:sigma}, we have $\sigma_{NZ}\simeq\sigma_{N}\simeq{\sigma_Z}$ for DIC ($L<80\hb$).
This means that $N$ and $Z$ distributions of the fragments are strongly correlated in DIC.
However, quasi-elastic reactions ($L>90\hb$) have $\sigma_N\simeq\sigma_Z\gg\sigma_{NZ}$, meaning that the $N$ and $Z$ distributions are almost independent in this case. 

\begin{figure}
\begin{center}
\includegraphics[width=6.5cm]{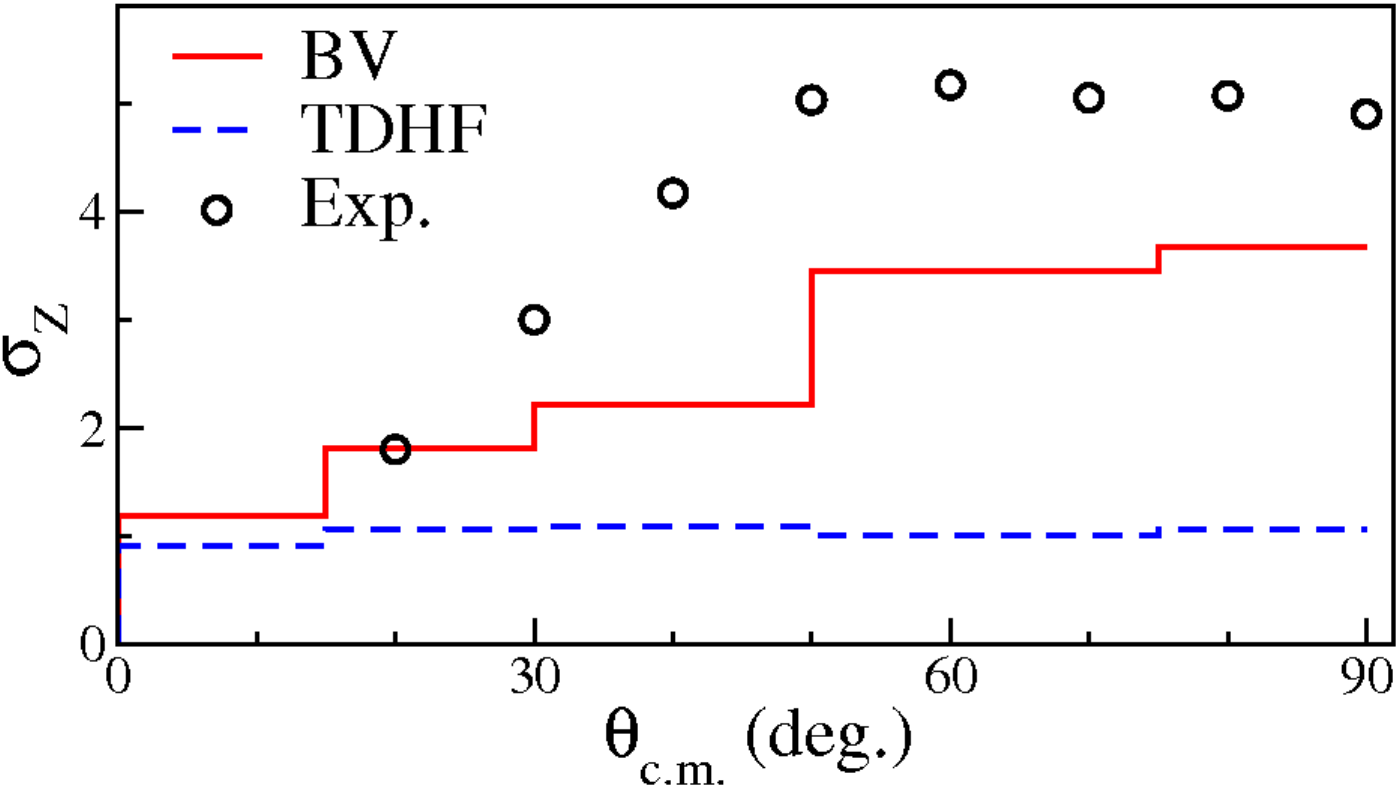} 
\caption{Comparison between BV (solid line) and TDHF (dashed line) predictions of $\sigma_{Z}$ for damped events (see text) as a function of $\theta_{c.m.}$ with data (circles) from~\cite{roy77}.
}
\label{fig:sig_tet}
\end{center}
\end{figure}

We now compare these results with the experimental data of Roynette {\it et al.}~\cite{roy77}.
The angle $\theta_{c.m.}$ between the fragments in the outgoing channel and the collision axis have been determined for each $L$. 
Figure~\ref{fig:sig_tet} shows theoretical and experimental evolutions of the charge fluctuations for damped events (defined, as in Ref.~\cite{roy77}, by a $\mathrm{TKEL}>30$~MeV) as a function of $\theta_{c.m.}$. 
Due to  orbiting, only DIC are expected to emit fragments at large angles, and the experimental plateau at $\theta_{c.m.}\ge50$~deg is then attributed to DIC. 
We see that TDHF fluctuations underestimate experimental results at all angles, except at very forward angles where quasi-elastic reactions dominate.
The results of the BV prescription are in better agreement, although they still underestimate the experimental data. 
This is probably due to fusion-fission events (not included in the calculations) leading to large fluctuations and, to a less extent, to the cooling down of the fragments by nucleon emission~\cite{sim11}. 

\begin{figure}
\begin{center}
\includegraphics[width=8cm]{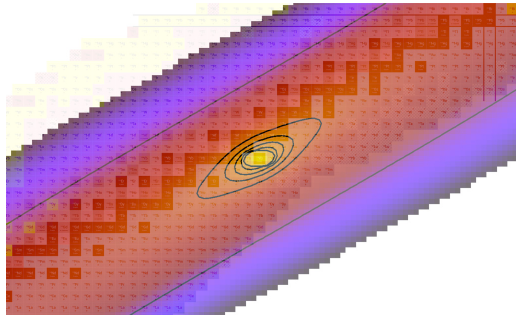} 
\caption{ Primary fragments production cross sections for $^{176}$Yb$+^{176}$Yb collisions at $E_{c.m.} = 660$~MeV in the side-side orientation overlaid onto the chart of nuclides. The innermost
contour corresponds to a cross section of 1 mb, with subsequent contours drawn every 0.2 mb. From Ref. \cite{godbey2020b}.
}
\label{fig:YbYb}
\end{center}
\end{figure}

Finally, Eq.~\ref{eq:Gauss} can be used to  represent distributions of primary DIC fragments on regions of the nuclear chart. 
An example is shown in Fig.~\ref{fig:YbYb} for the $^{176}$Yb$+^{176}$Yb reaction studied with TDRPA in \cite{godbey2020b}.
Such applications of TDRPA are, however, limited to symmetric collisions. Indeed, 
if one applies TDRPA to asymmetric reactions, fluctuations and correlations can be unphysically large.
The reason is that TDRPA fluctuations are computed in the small amplitude limit assuming that the distribution is symmetric around the average value, a fact that is not guaranteed in asymmetric collisions (see discussion in supplemental material of Ref.~\cite{williams2018}).

\subsection{Quasi-fission reactions \label{sec:QF}}

Quasi-fission occurs in heavy-ion collisions, leading to the formation of fission-like fragments without the intermediate formation of an equilibrated compound nucleus  \cite{boc82,tok85,she87}.
See \cite{hinde2021} for a recent review of experimental studies of quasi-fission. 
Due to the importance of understanding quasi-fission in order to optimise the entrance channel for superheavy element (SHE) synthesis, a large number of studies of quasi-fission have been dedicated over the past ten years with TDHF solvers \cite{wakhle2014,oberacker2014,ayik2015a,umar2015a,hammerton2015,sekizawa2016,umar2016,morjean2017,yu2017,guo2018c,zheng2018,godbey2019,simenel2021,li2022,stevenson2022,mcglynn2023,lee2024,scamps2024b,li2024c}.

\subsubsection{Quasi-fission characteristics}

The quasi-fission mechanism dominates near-barrier reactions in heavy systems with $Z_1Z_2>1600$, although it is also present in lighter systems (see Sec.~\ref{sec:fus-hindrance}). 
Quasi-fission reactions are characterised by the following properties:
\begin{itemize}
\item the reactions are fully damped,
\item a significant mass transfer occurs, usually towards mass equilibration,
\item typical contact times are of the order of few zeptoseconds ($1$~zs$=10^{-21}$~s) or few 10~zs. 
\end{itemize}

\begin{figure}
\begin{center}
\includegraphics[width=7.5cm]{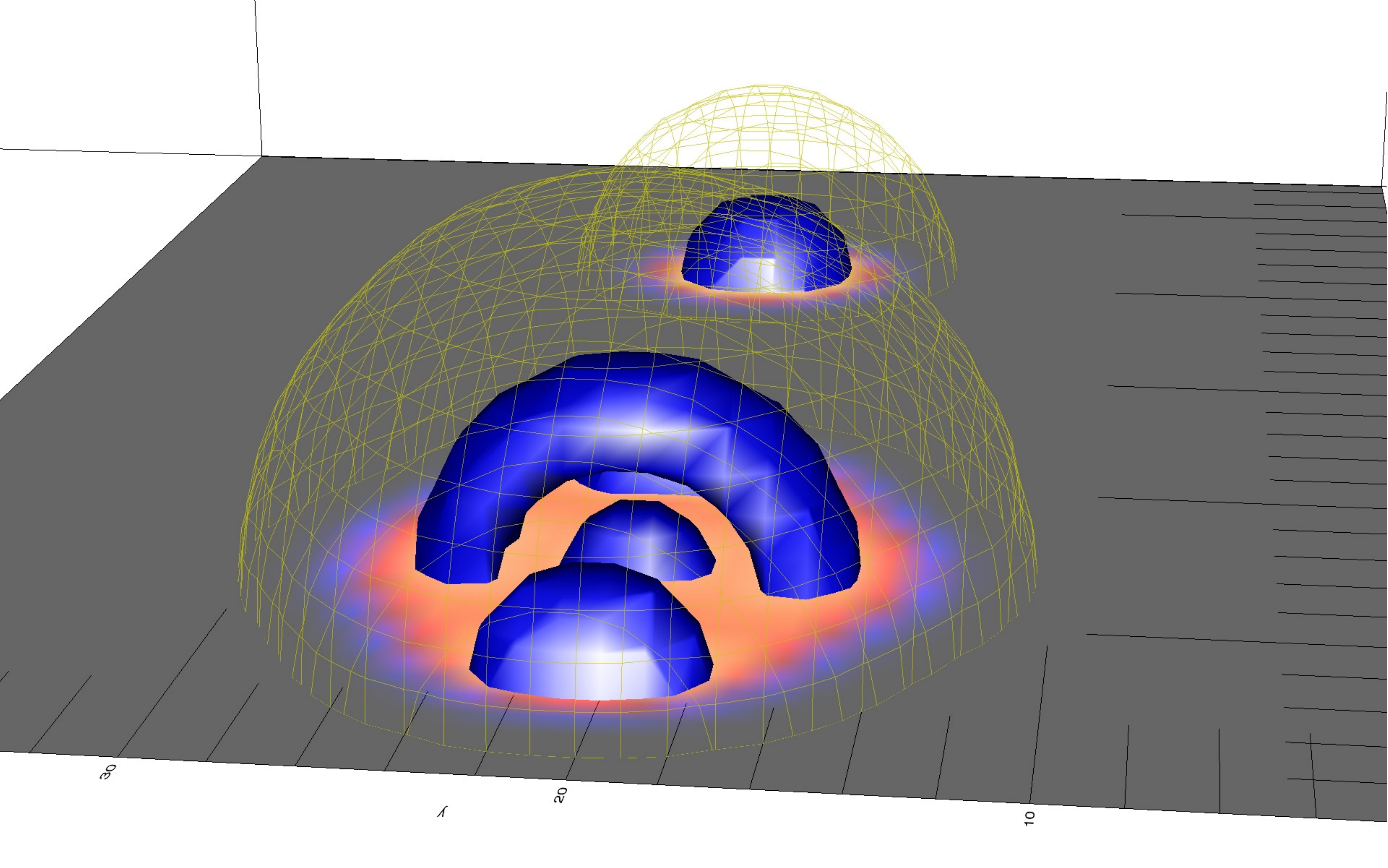}
\caption{Initial condition of a $^{40}$Ca+$^{238}$U collision visualised with the SDVision code~\cite{sdvision}. 
 Two isodensities are shown (yellow grid and blue area). A projection of the density is also shown on the $z=0$ plane.
\label{fig:sdvision}}
\end{center}
\end{figure}

\begin{figure}
\begin{center}
\includegraphics[width=8.8cm]{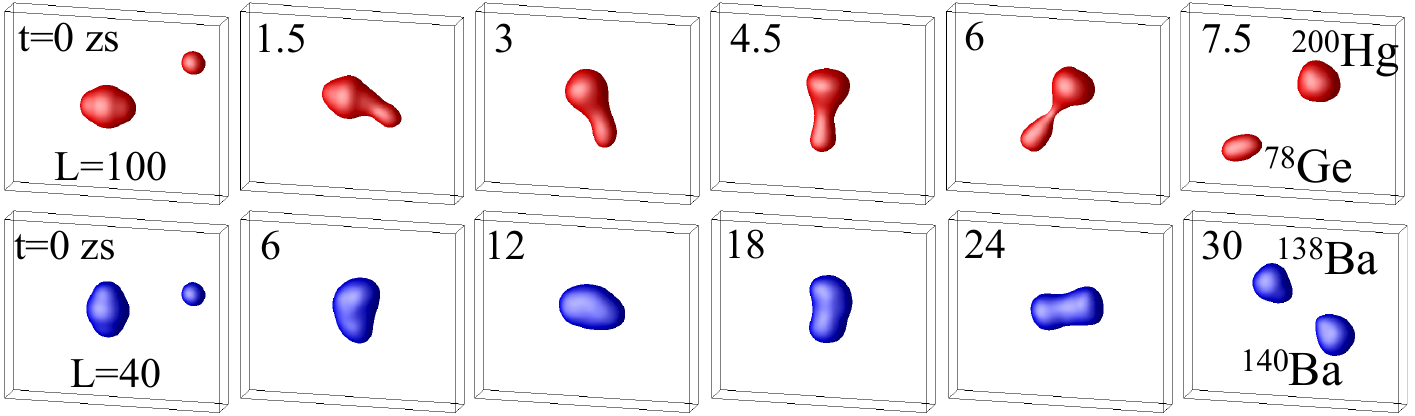}
\caption{Snapshots of the TDHF isodensity at half the saturation density in the $^{40}$Ca+$^{238}$U system for different initial conditions at a centre-of-mass energy $E=225.4$ MeV. From Ref. \cite{wakhle2014}.
}
\label{fig:densQF}
\end{center}
\end{figure}

An example of initial condition of the $^{40}$Ca+$^{238}$U collision is shown in Fig.~\ref{fig:sdvision} where we can clearly see internal structures in the $^{238}$U nucleus. 
The outcome of the collision depends on this initial condition, in particular on the orientation, the energy and angular momentum.
Examples of TDHF evolutions showing quasi-fission characteristics for this system following different initial conditions are shown in Fig.~\ref{fig:densQF} \cite{wakhle2014}.

\begin{figure}
\begin{center}
\includegraphics[width=8.8cm]{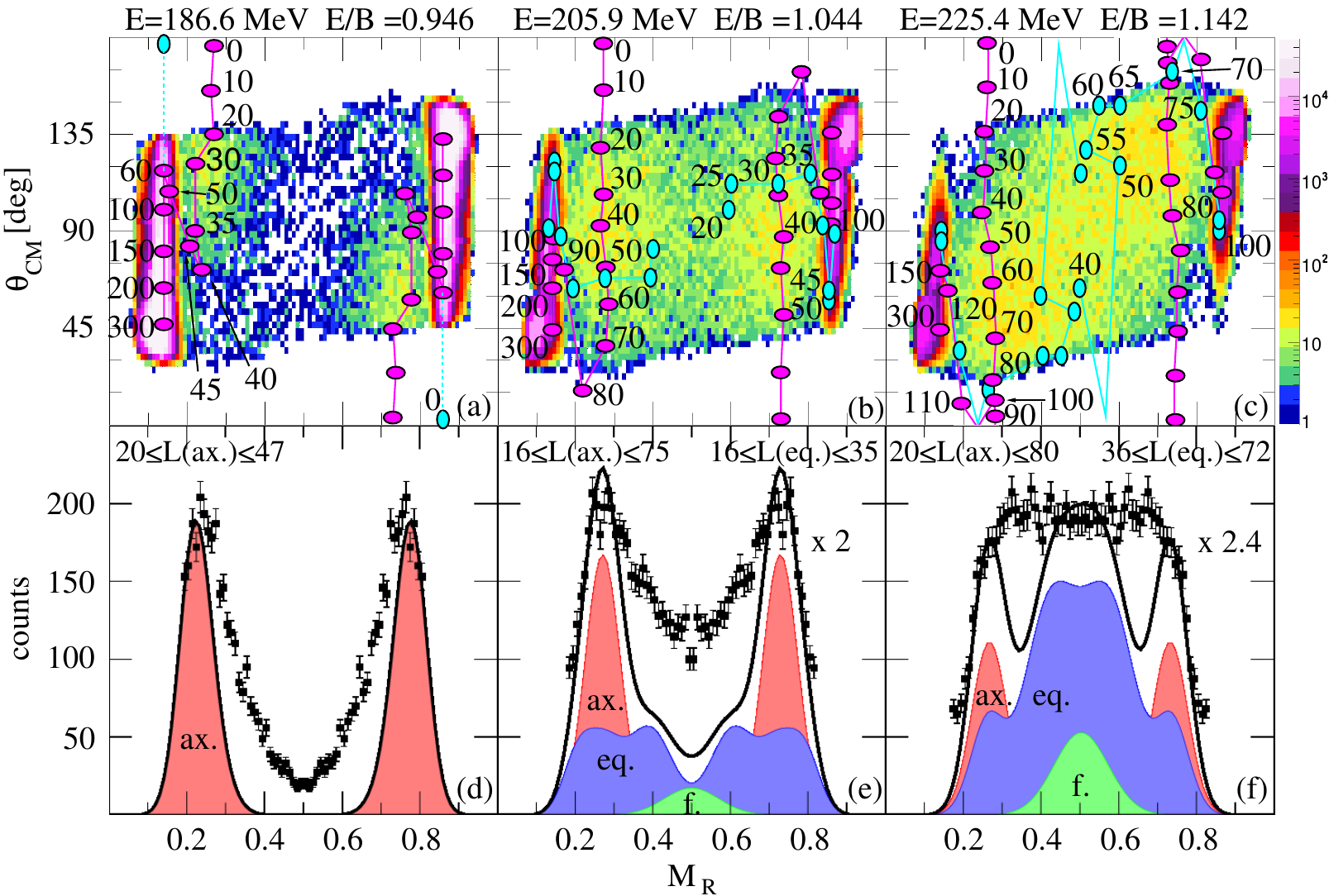}
\caption{ (a-c) Comparison between TDHF predictions and  experimental mass-angle distributions near the capture barrier
$B$.
The horizontal and vertical ellipsoids show the TDHF results for the tip and side orientations at contact, respectively. 
(d-f) Mass-ratio $M_R=M_{frag.}/M_{tot.}$ in the range $0.2<M_R<0.8$. 
The mass ratio distributions estimated from the TDHF results are shown with shaded areas for quasi-fission in the tip (ax.) and side (eq.) configurations, and for fusion-fission (f.) events. 
The solid line gives the  sum of these distributions. 
From Ref. \cite{wakhle2014}.
}
\label{fig:MAD}
\end{center}
\end{figure}

Figure \ref{fig:MAD} shows a comparison between TDHF predictions and experimental data for $^{40}$Ca+$^{238}$U quasi-fission \cite{wakhle2014}. 
Two orientations leading to collisions with the side and tip of $^{238}$U at contact were considered. 
The width of the mass distributions and the relative weights between orientations were determined phenomenologically.
Nevertheless, the good qualitative agreement between experiment and theory shows that TDHF accounts for the main quasi-fission  characteristics. 

\begin{figure}
\begin{center}
\includegraphics[width=7cm]{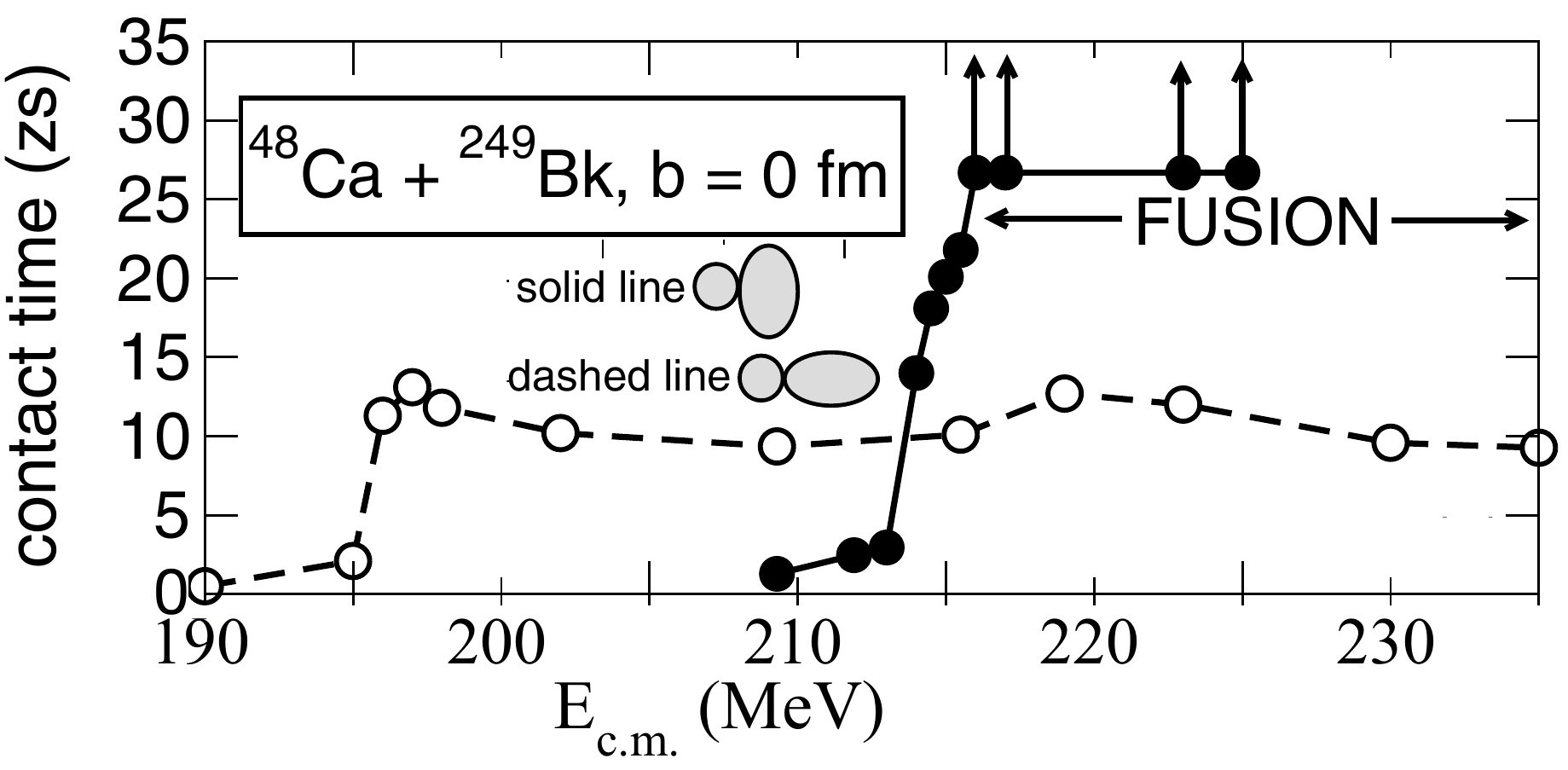}
\caption{  Contact times in $^{48}$Ca$+^{249}$Bk for central collisions as function of centre of mass energy for tip and side orientations. 
From Ref. \cite{umar2016}.
}
\label{fig:CaBktime}
\end{center}
\end{figure}

An interesting observation in Fig.~\ref{fig:MAD} is that quasi-fission is found down to central collisions for tip orientations, while side orientations lead to fusion at small angular momentum (provided the system has enough energy to reach contact). 
This feature is also found in other systems with actinide targets, such as  $^{48}$Ca$+^{249}$Bk \cite{umar2016} whose contact times are shown in Fig.~\ref{fig:CaBktime} for central collisions. 
In addition, while contact times in tip collisions saturate around 10~zs, significantly larger contact  times are found in side collisions before fusion occurs.

\begin{figure}
\begin{center}
\includegraphics[width=7cm]{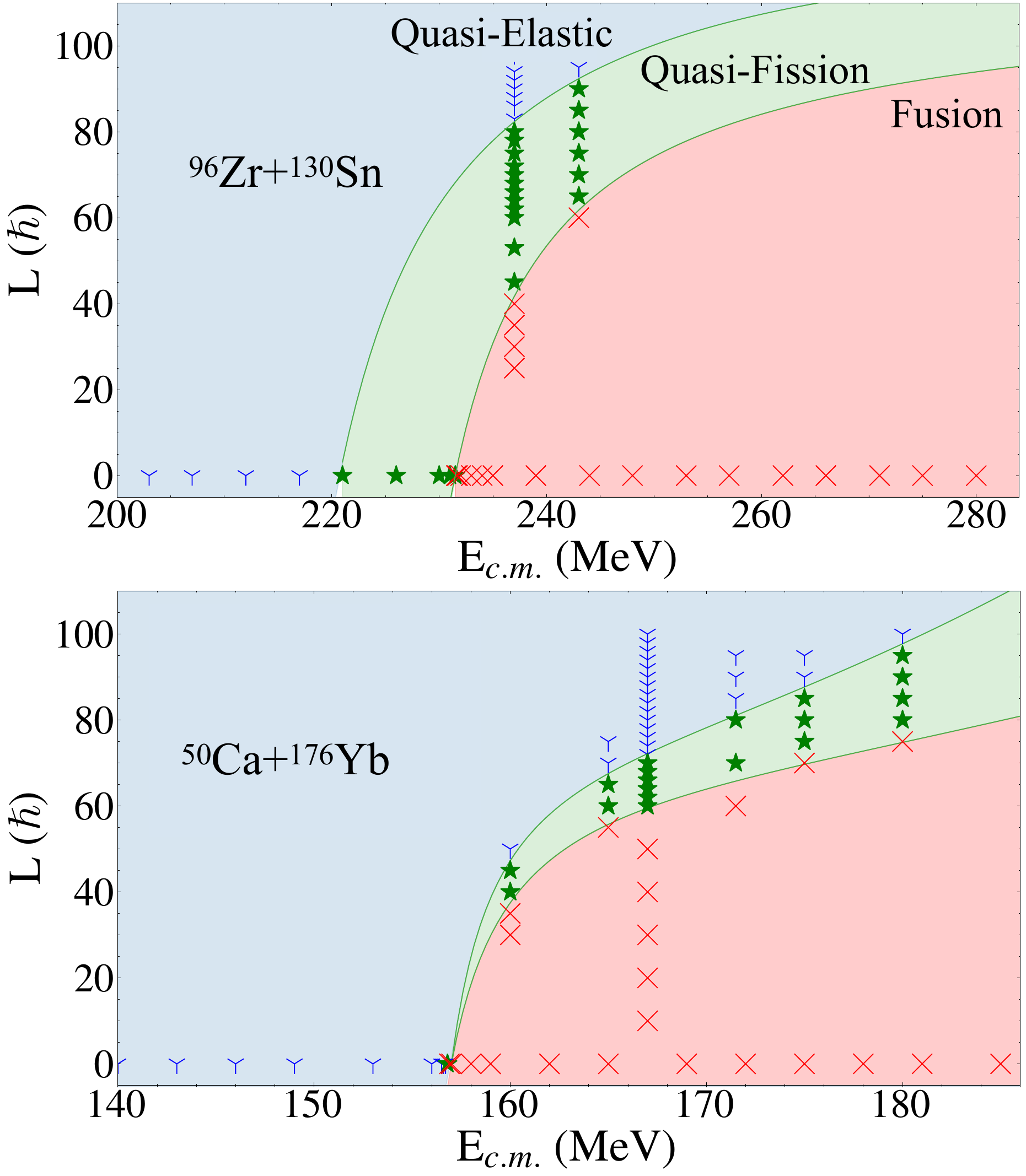}
\caption{ Outcomes of TDHF simulations for $^{96}$Zr$+^{130}$Sn (top) and $^{50}$Ca$+^{176}$Yb (bottom). ``Quasi-elastic'', ``quasi-fission'' and ``fusion'' refers to contact times $\tau_{QE}<2$~zs, $2<\tau_{QF}<30$~zs, and $\tau_F>30$~zs, respectively.
Adapted from Ref. \cite{lee2024}.
}
\label{fig:226ThQF}
\end{center}
\end{figure}

As shown in Fig.~\ref{fig:226ThQF}, quasi-fission is also expected in lighter systems, like $^{96}$Zr$+^{130}$Sn and $^{50}$Ca$+^{176}$Yb, both forming $^{226}$Th compound nucleus in case of fusion \cite{lee2024}.
However, a noticeable difference between both systems is that quasi-fission does not occur for central collisions in the more asymmetric system.
This could be interpreted as an effect of the centrifugal potential acting against fusion and thus favouring quasi-fission. 

\begin{figure}
\begin{center}
\includegraphics[width=7cm]{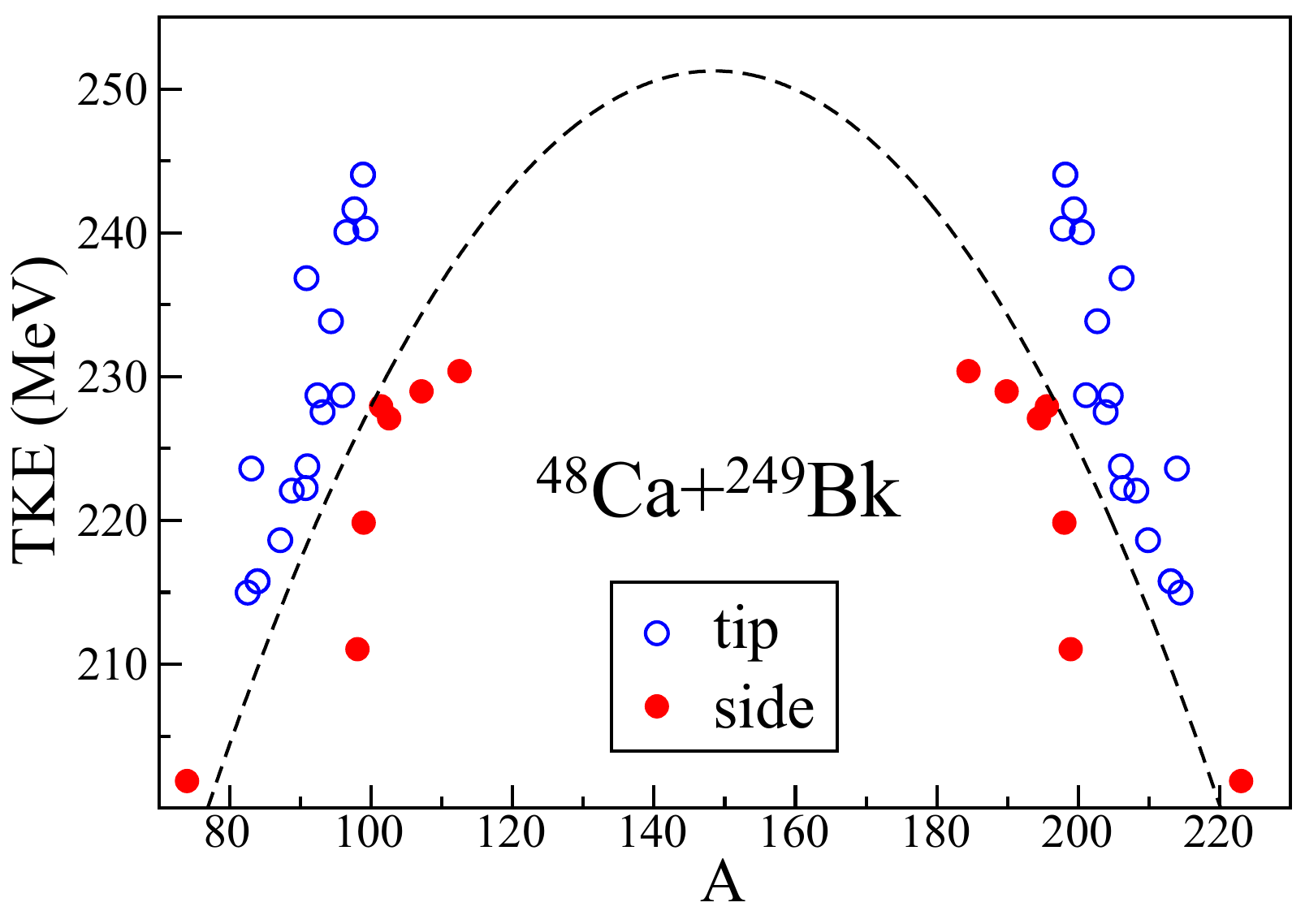}
\caption{ Total kinetic energy of quasi-fission fragments in $^{48}$Ca$+^{249}$Bk  central collisions  for tip and side orientations. The dashed line shows Viola systematics.
From Ref. \cite{umar2016}.
}
\label{fig:CaBkTKE}
\end{center}
\end{figure}

Quasi-fission reactions being fully damped, one would expect their TKE to match the Viola systematics for fission fragments \cite{vio85}. 
Figure~\ref{fig:CaBkTKE} shows that, although this is essentially the case in $^{48}$Ca$+^{249}$Bk, there is a non negligible orientation effect \cite{umar2016}. 
Indeed, the tip (side) collisions have slightly larger (smaller) TKE than expected from Viola systematics. 
This is reflecting the difference of elongation at contact with each orientation. 

\begin{figure}
\begin{center}
\includegraphics[width=7cm]{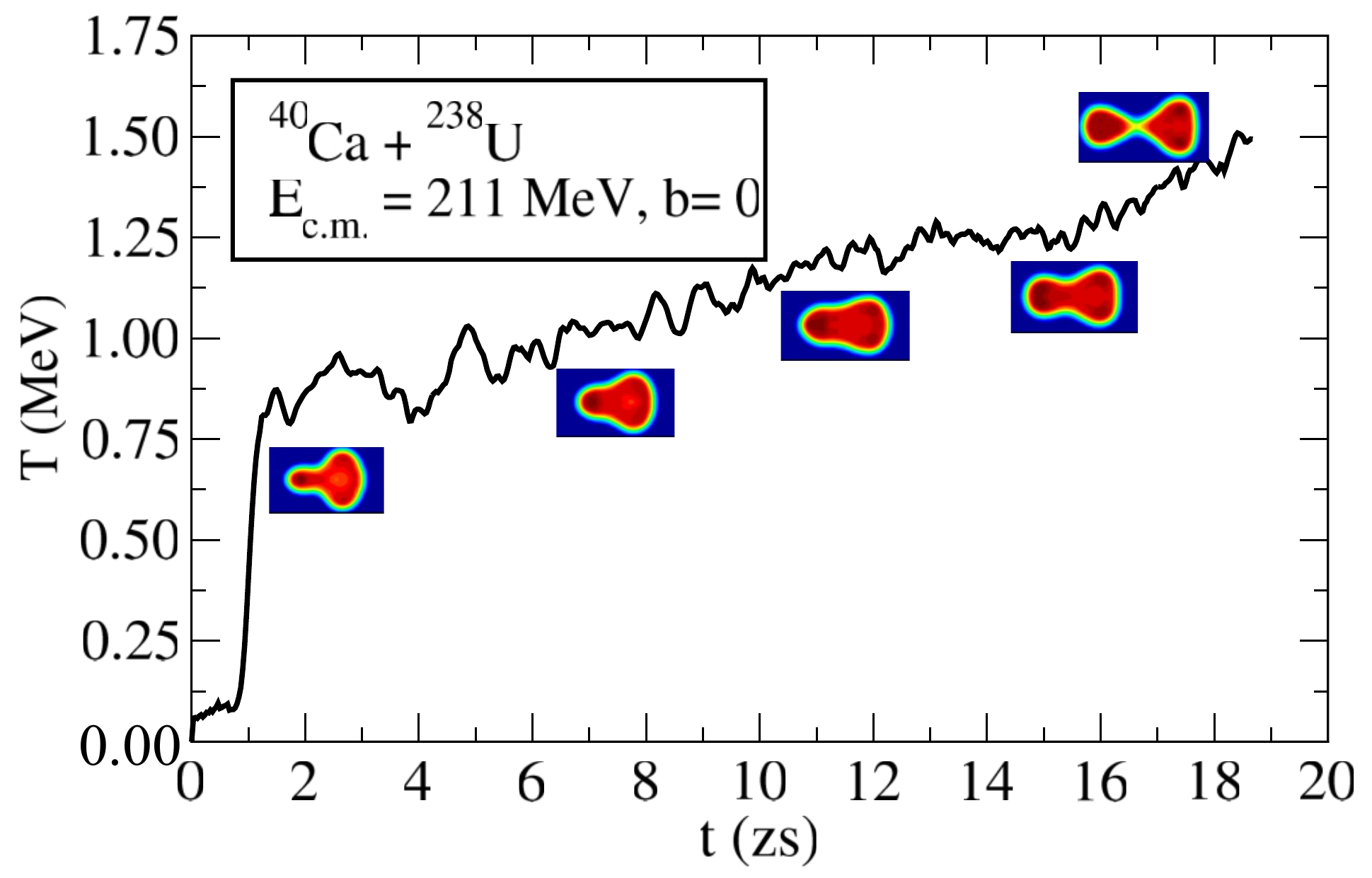}
\caption{ Time-dependence of the temperature for a $^{40}$Ca$+^{238}$U reaction from DCTDHF. From Ref. \cite{umar2015a}.
}
\label{fig:CaUtemp}
\end{center}
\end{figure}

Unlike reactions forming a compound nucleus, quasi-fission does not fully erase the memory of the entrance channel. 
Nevertheless, Fig.~\ref{fig:CaUtemp} shows that a steady rise of the temperature occurs while the nuclei are in contact. 
Here the temperature is evaluated from the excitation energy following $T(t)=\sqrt{E^*(t)/(A/8.5)}$ in MeV, where $E^*(t)$ is computed from DCTDHF. 

\subsubsection{Role of shell effects on quasi-fission}

The extra stability of $^{208}$Pb induced by spherical shell effect has   been invoked to explain experimental quasi-fission fragment mass distributions peaking near $A\simeq208$ \cite{itkis2004,nishio2008,wakhle2014,morjean2017,hinde2018}.
Possible influence of deformed shell effects around $^{100}$Zr and at $N=56$ (octupole deformed shell effect) have also been invoked to explain results of TDHF simulations of $^{48}$Ca$+^{249}$Bk \cite{godbey2019}. 
Interestingly, the inclusion of tensor terms seems to affect the competition between these shell effects \cite{li2022,li2024c}.

It has been suggested that the sequential fission of the target-like fragment could explain the asymmetric peaks observed in reactions on actinide targets, as events with three fragments in the exit channel are excluded \cite{jeung2022}. However, a recent experiment provided {\it ``clear evidence of quantum shell effects in slow quasi-fission processes,''} highlighting the role of shell-closed nuclei in the mass region $A\approx96$ \cite{pal2024} which could be attributed to the octupole deformed shell gap at $N=56$ \cite{godbey2019}.

\begin{figure}
\begin{center}
\includegraphics[width=7cm]{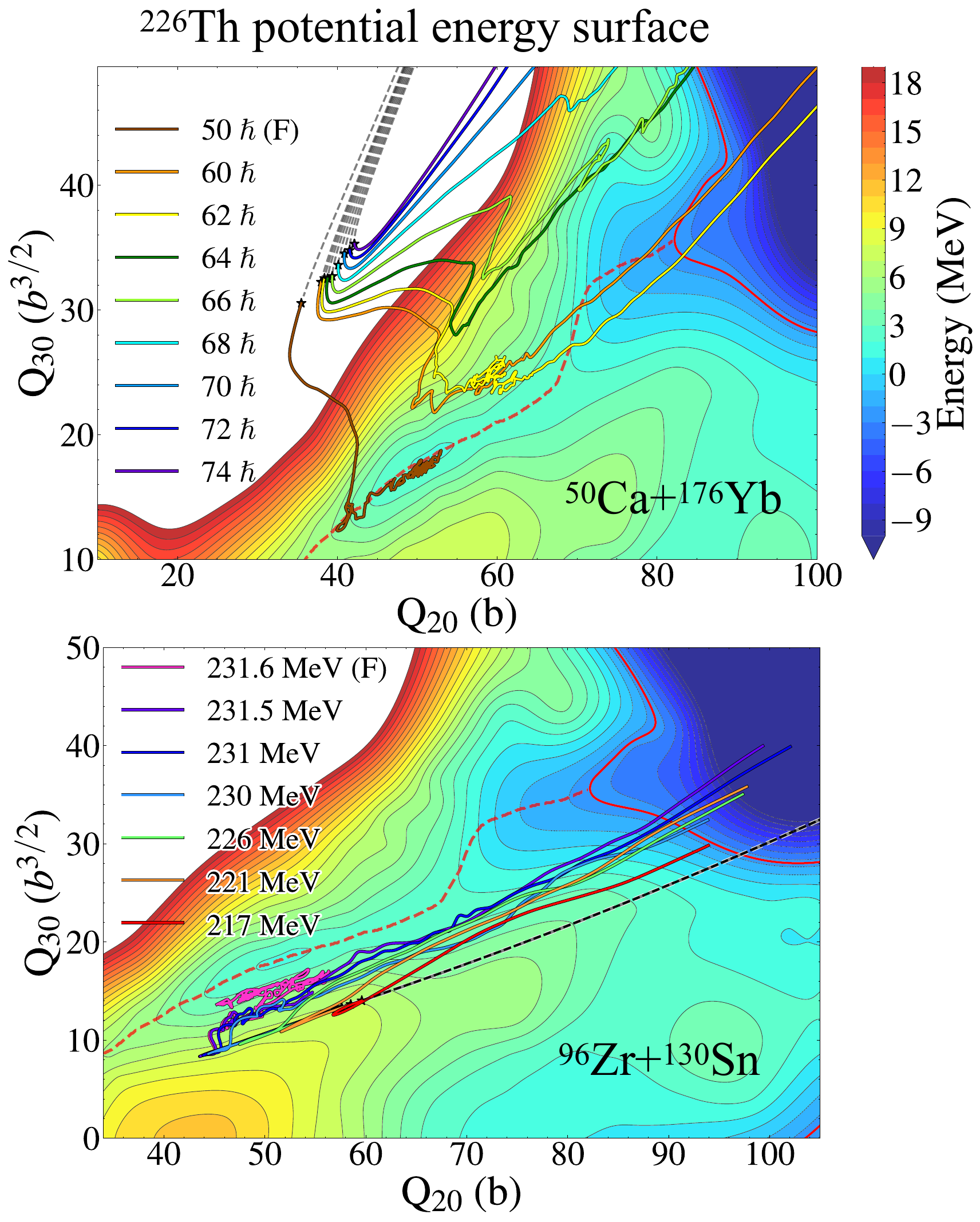}
\caption{Potential energy surface of $^{226}$Th near the asymmetric fission valley. The red dashed line shows the fission path.  TDHF trajectories for (top) $^{50}$Ca$+^{176}$Yb reactions at $E_{c.m.}=167$~MeV  with the side orientation of $^{176}$Yb, and for (bottom) $^{96}$Zr$+^{130}$Sn central collisions are   drawn in the $Q_{20}-Q_{30}$ plane on top of the PES. The entrance channel trajectories are shown by the black dashed lines and the entry (contact) points are represented by stars.  From Ref.  \cite{lee2024}. 
}
\label{fig:226ThPES}
\end{center}
\end{figure}

Reactions with sub-lead targets can circumvent sequential fission. Experimentally, the role of shell effects in such reactions remains unclear as studies report that {\it ``shell effects are clearly seen''} \cite{chizhov2003}, whereas others find only {\it ``weak evidence''} \cite{hinde2022}. 
Such reactions have been studied in TDHF \cite{simenel2021,lee2024,scamps2024b}, where it is found that shell effects stop mass equilibrations, forming fragments similar to those produced in asymmetric fission of the compound nucleus. 

An example is shown in the top of Fig.~\ref{fig:226ThPES} where $^{50}$Ca+$^{176}$Yb TDHF trajectories in the $Q_{20}-Q_{30}$ plane \cite{lee2024} are plotted on top of the $^{226}$Th PES obtained with the SkyAx code \cite{reinhard2021}.
The evolution of the fragment charge as function of contact time shown in the top of Fig.~\ref{fig:QF-Mtau} for this reaction shows that the mass drift towards symmetry stops when the system reaches $Z\simeq54$ as in asymmetric fission. 
This is an indication that, as in fission of actinides \cite{scamps2018},  fragment formation in quasi-fission is influenced by the same octupole deformed shell effects at $Z=52$ and 56. 

\begin{figure}
\begin{center}
\includegraphics[width=7cm]{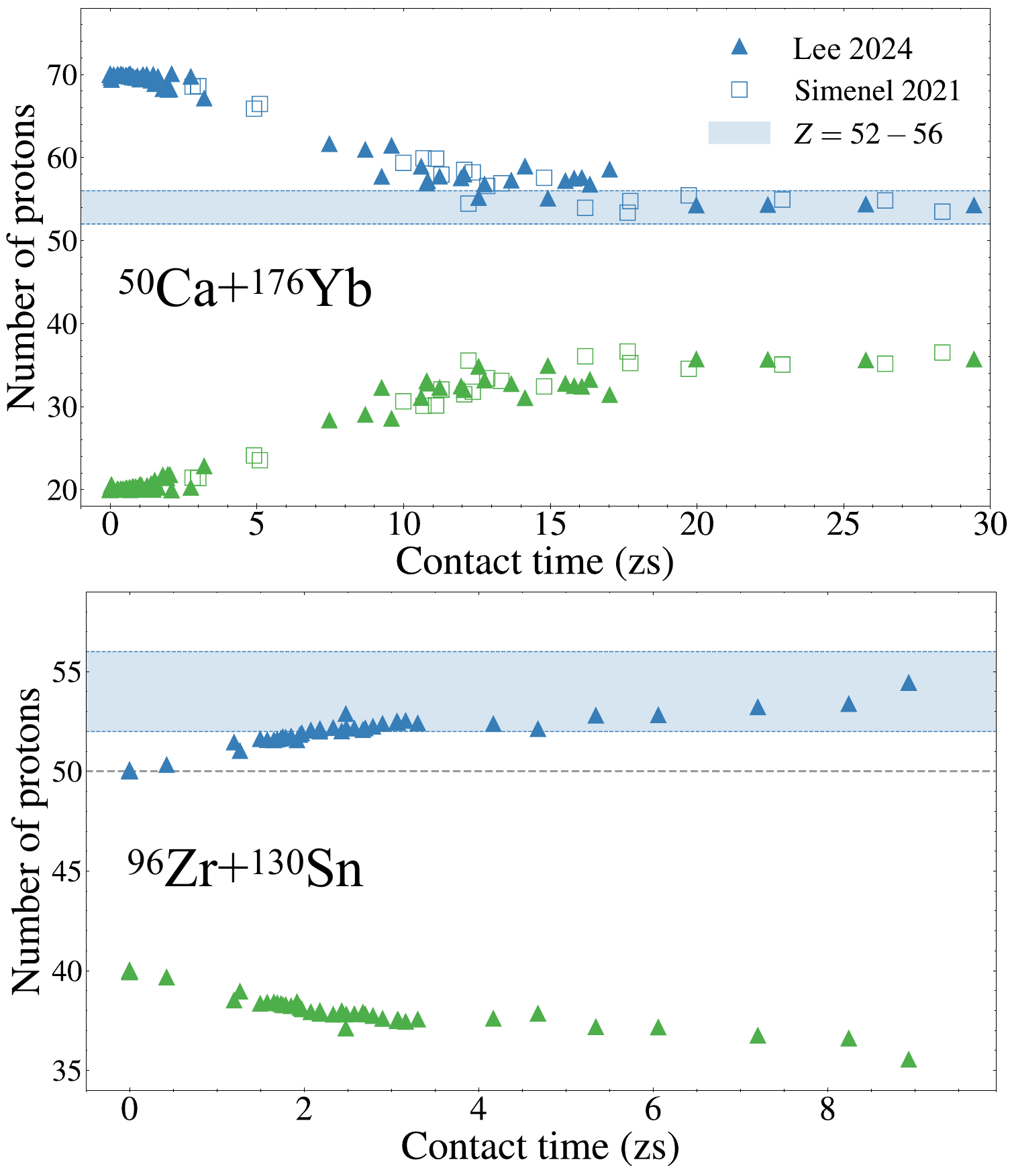}
\caption{ Proton  numbers in the heavy (blue symbols) and light (green symbols) fragments produced (top) in $^{50}$Ca$+^{176}$Yb  and (bottom) in $^{96}$Zr$+^{130}$Sn  as a function of contact time. The shaded regions represent expected octupole deformed shell effects.  Adapted from Ref.  \cite{lee2024}. 
}
\label{fig:QF-Mtau}
\end{center}
\end{figure}

As shown in the bottom panel of Fig.~\ref{fig:226ThPES}, the more symmetric reaction $^{96}$Zr$+^{130}$Sn enters the PES on the other side of the asymmetric fission valley.
Nevertheless, the outgoing fragments in this reaction are found to be more asymmetric than in the entrance channel (see bottom of Fig.~\ref{fig:QF-Mtau}). 
This ``inverse quasi-fission'' mechanism has also been discussed in other systems \cite{zag06,ked10} (see also Sec. \ref{sec:actinides}). 
Here, it is interpreted as an influence of the shell effects responsible for asymmetric fission, driving the system towards the asymmetric fission valley. 

Naturally, the comparison between TDHF trajectories and PES should be taken with care as the PES is computed at zero excitation energy, zero angular momentum, and assuming axial symmetry. 
Usually, these conditions do not hold in heavy ion collisions. Nevertheless, these comparisons have proven a powerful tool to get a deeper insight into the role of shell effects on quasi-fission \cite{lee2024,mcglynn2024}.

\subsection{Timescales of Equilibration, Dissipation and Fluctuation \label{sec:timescales}}

Properties of colliding nuclei tend to equilibrate over time when the nuclei are in contact. 
A key challenge is understanding this equilibration process, during which collective energy dissipates into internal degrees of freedom. 
In particular, mass and neutron-proton equilibration in colliding systems occurs through nucleon flow and is strongly influenced by nuclear viscosity which drives the dissipation of initial kinetic energy and angular momentum of the fragments \cite{randrup1982}. 
According to Einstein's fluctuation-dissipation theorem, multi-nucleon transfer during fragment contact generates quantum fluctuations, leading to broad particle number distributions in the final fragments \cite{randrup1978}. Equilibration, dissipation, and fluctuation are thus interconnected phenomena.

Systematic TDHF and TDRPA results of heavy-ion collision studies were compiled in Ref.~\cite{simenel2020} in order to get a deeper understanding of the relationship between equilibration, dissipation and fluctuation. 
To facilitate comparisons between systems with varying initial conditions, a ``normalised'' observable 
\oeq
\delta X(\tau) = \frac{X(\tau) - X_\infty}{X_0 - X_\infty}
\ceq
was defined, where $X(\tau)$ characterises equilibration, dissipation, or fluctuations as a function of the contact time $\tau$ between fragments before separation. 
The initial value is $X_0$, and for long contact times, $X(\tau)$ approaches its equilibrium value $X_\infty$, resulting in $\delta X \rightarrow 0$.  

\begin{figure*}
\begin{center}
\includegraphics[width=13cm]{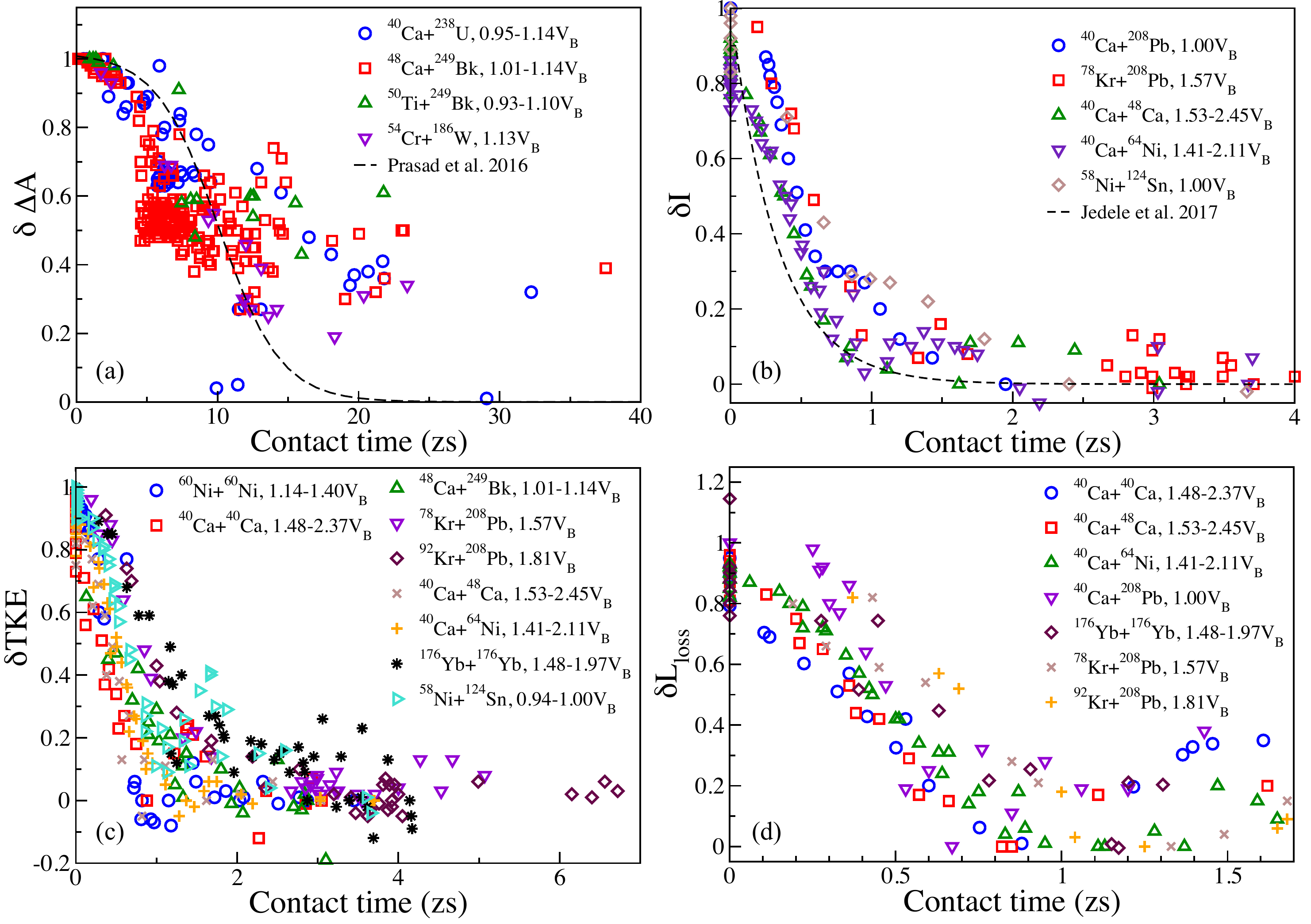}
\caption{
(a) Fragment mass asymmetry $\delta\Delta A$, (b) asymmetry between proton and neutron numbers $\delta I$, (c) ``normalised'' total kinetic energy $\delta$TKE, and (d) ``normalised'' angular momentum loss $\delta L_{loss}$ as a function of contact times from TDHF calculations \cite{oberacker2014,umar2016,godbey2019,hammerton2015,sim12,umar2017,williams2018,sim11,godbey2020b,wu2019,simenel2020}.
Energy ranges are given as function of the barrier height $V_B$~\cite{swiatecki2005}. 
The dashed lines  show the expected equilibration assuming (a) Fermi-type mass drift determined experimentally by Prasad \textit{et al.}~\cite{prasad2016} and (b) the rate constant of 3~zs$^{-1}$ determined experimentally by Jedele \textit{et al.}~\cite{jedele2017}. Adapted from Ref. \cite{simenel2020}.}
\label{fig:Equil}
\end{center}
\end{figure*}

Equilibration of mass asymmetry $\delta\,\Delta A(\tau)$ with $$\Delta A= A_1-A_2$$ and $A_{1,2}$ the number of nucleons in the outgoing fragments is shown in Fig.~\ref{fig:Equil}(a). 
Despite large fluctuations, all systems exhibit an equilibration of mass asymmetry.
A full symmetry ($\Delta A_\infty=0$) is expected to be reached at about 20~zs contact time in average. 
Mass equilibration, as occurring in quasi-fission, is then a relatively slow process.

\begin{figure}
\begin{center}
\includegraphics[width=7cm]{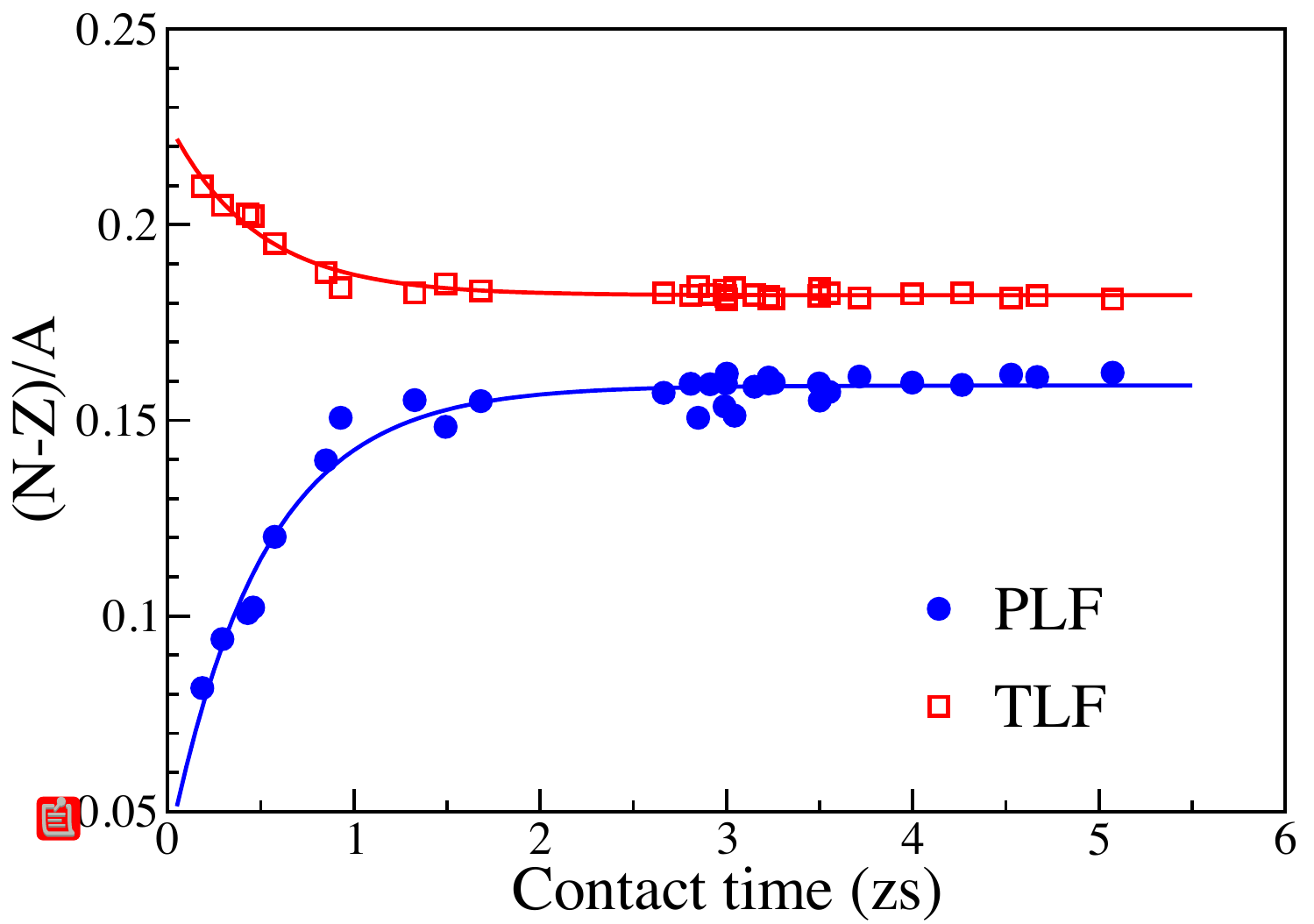}
\caption{ Evolution of $\frac{N-Z}{A}$ of the outgoing projectile (target)  like fragments PLF (TLF) as a function of contact times in $^{78}$Kr$+^{208}$Pb at $E=8.5$ MeV/nucleon. From Ref.  \cite{umar2017}. 
}
\label{fig:NZeq}
\end{center}
\end{figure}

Equilibration of initial asymmetry between proton and neutron numbers, quantified by $$I=(N_1-Z_1)-(N_2-Z_2)$$ is shown in Fig.~\ref{fig:Equil}(b).
Here, the equilibrium values $I_\infty$ are not necessarily zero, as illustrated in Fig.~\ref{fig:NZeq} and need to be determined for each system by taking $I(\tau)$ for large times. 
The neutron-proton equilibration is a fast process that occurs within a timescale of $\tau\sim1$~zs in good agreement with experiment~\cite{jedele2017}. 

Figure \ref{fig:Equil}(c) shows the timescale for dissipation of total kinetic energy of the fragments where TKE$_\infty$ is determined for each system and each energy.
This process occurs with similar timescale, $1-2$~zs, as in neutron-to-proton equilibration.  
The good reproduction of TKE-angle correlations shown in Fig.~\ref{fig:NiNi} \cite{williams2018} gives confidence on this extracted timescale for energy dissipation.

The dissipation of orbital angular momentum $L$ between the fragments is shown in Fig.~\ref{fig:Equil}(d).
 Here, $$L_{loss}(\tau) = L_0-L(\tau)$$ is defined as the difference between initial and final angular momentum, and  the equilibrium value ${L_{loss_\infty}}$ as the maximum value of $L_{loss}(\tau)$. 
Like neutron-proton equilibration and energy dissipation, angular momentum dissipation is a fast process occurring within a timescale of approximatively  $1$~zs. 

\begin{figure}
\begin{center}
\includegraphics[width=7cm]{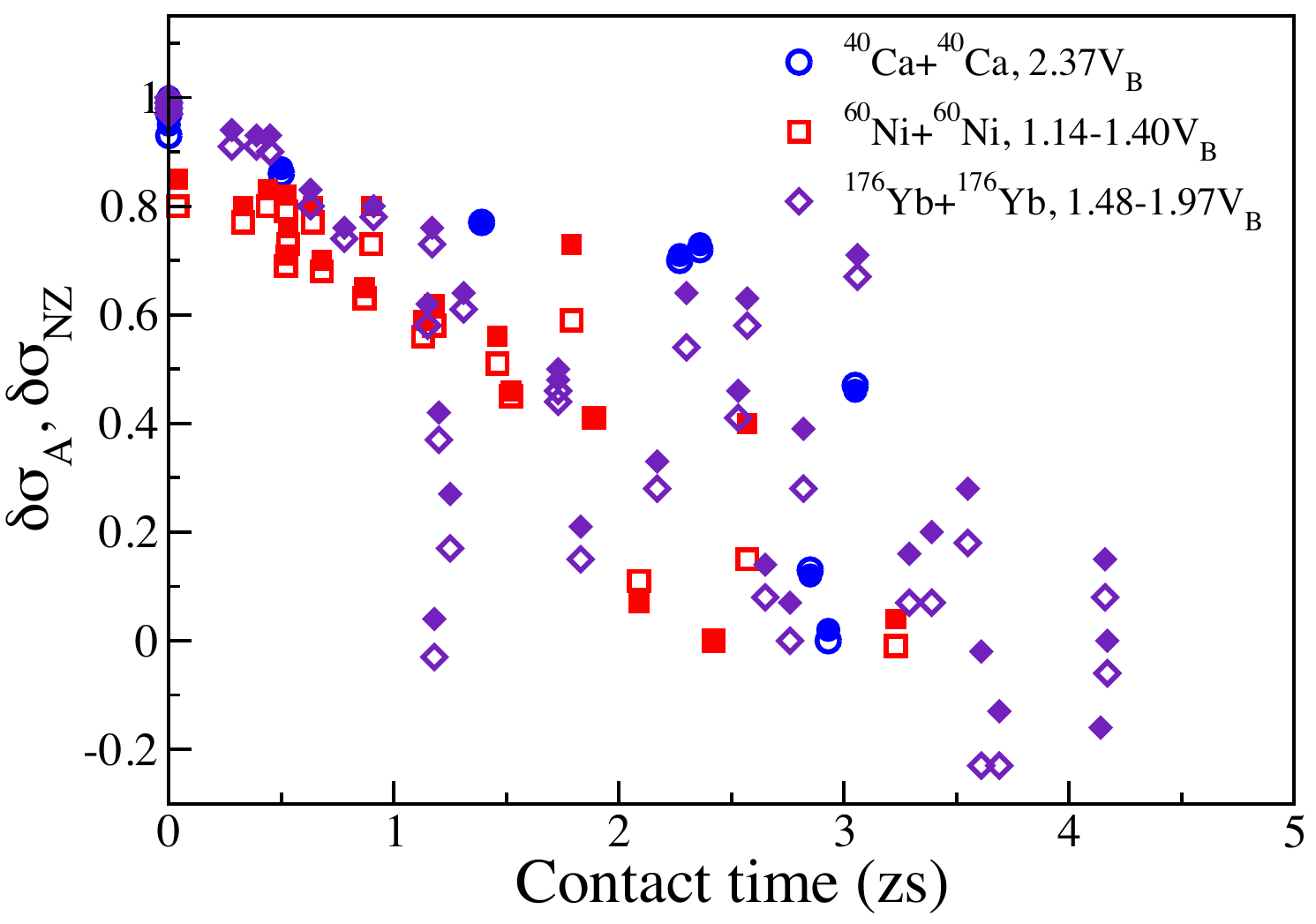}
\caption{Evolution of $\delta \sigma_A$ associated with mass fluctuations (open symbols) and $\delta \sigma_{NZ}$ associated with neutron-proton correlations (full symbols) as a function of contact times from TDRPA calculations \cite{sim11,williams2018,godbey2020b}. From Ref. \cite{simenel2020}. 
}
\label{fig:fluc_corr}
\end{center}
\end{figure}

Fluctuations of particle numbers in the fragments produced in symmetric deep inelastic collisions have been evaluated within the TDRPA formalism (see Sec.~\ref{sec:DIC}).
The evolutions of $\delta\sigma_A(\tau)$ and $\delta\sigma_{NZ}(\tau)$ with contact time are reported in Fig.~\ref{fig:fluc_corr}. 
The particle numbers in the incoming fragments are well defined, thus we have $\sigma_{A_0}=\sigma_{NZ_0}=0$. 
The values of $\sigma_{A,NZ\,_\infty}$ are taken from the most central DIC.
Fluctuations and correlations are found to build up within a timescale of about 3~zs. 

Several conclusions can be drawn from a comparison between these timescales for equilibration, dissipation and fluctuations: 
\begin{itemize}
\item The timescales are essentially independent on systems and energies, indicating that  the underlying mechanisms are universal.
\item Neutron-to-proton equilibration occurs on very similar timescales as angular momentum and kinetic energy dissipations, pointing toward a  correlation between these mechanisms, and indicating that nucleon transfer (as occurring in neutron-proton equilibration) is likely to be the main mechanism for dissipation \cite{randrup1978}.
\item Mass equilibration is much slower and is not expected to be a significant contributor to dissipation processes. 
\item Particle number fluctuations and correlations  build up within a few zeptoseconds, which is a bit slower than dissipation, yet much faster than mass equilibration, indicating that fluctuations are not correlated to the latter. 
\end{itemize}
It would be interesting to investigate timescales for particle number fluctuations in asymmetric collisions, e.g., with stochastic mean field calculations.

\subsection{Actinide collisions \label{sec:actinides}}

The collision of actinides form, during few zs, the heaviest nuclear systems available on Earth. 
In one hand, such systems are interesting to study the stability of the QED vacuum under strong electric fields~\cite{rei81,ack08,gol09}. 
In the other hand, they might be used to form neutron-rich heavy and super-heavy elements via multi-nucleon transfer reactions. 

\begin{figure}
\begin{center}
\includegraphics[width=7cm]{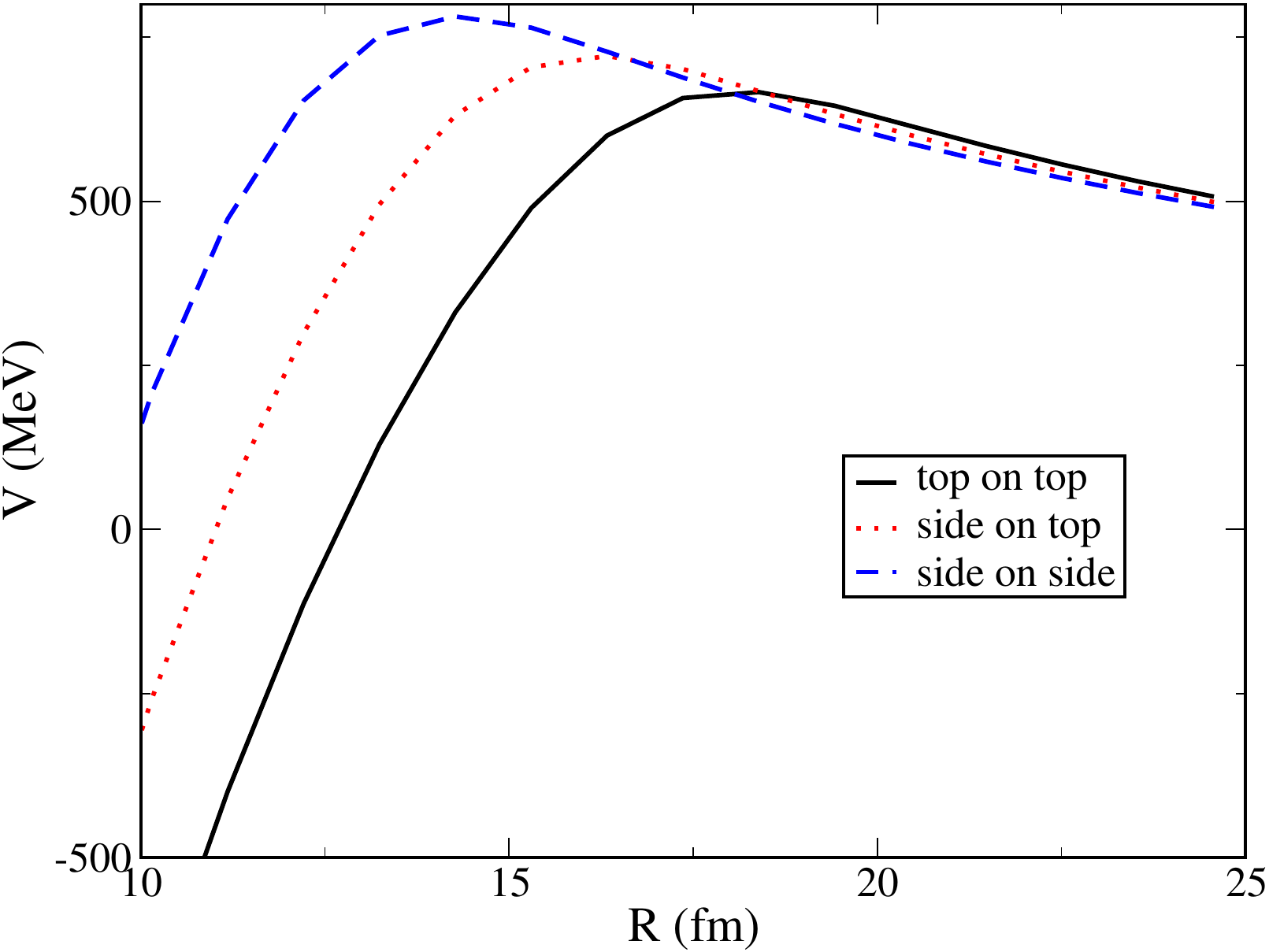}
\caption{Frozen-HF nucleus-nucleus potentials of the $^{238}$U+$^{238}$U system obtained for different orientations of the actinides. }
\label{fig:pot_frozen}
\end{center}
\end{figure}

\begin{figure}
\begin{center}
\includegraphics[width=7cm]{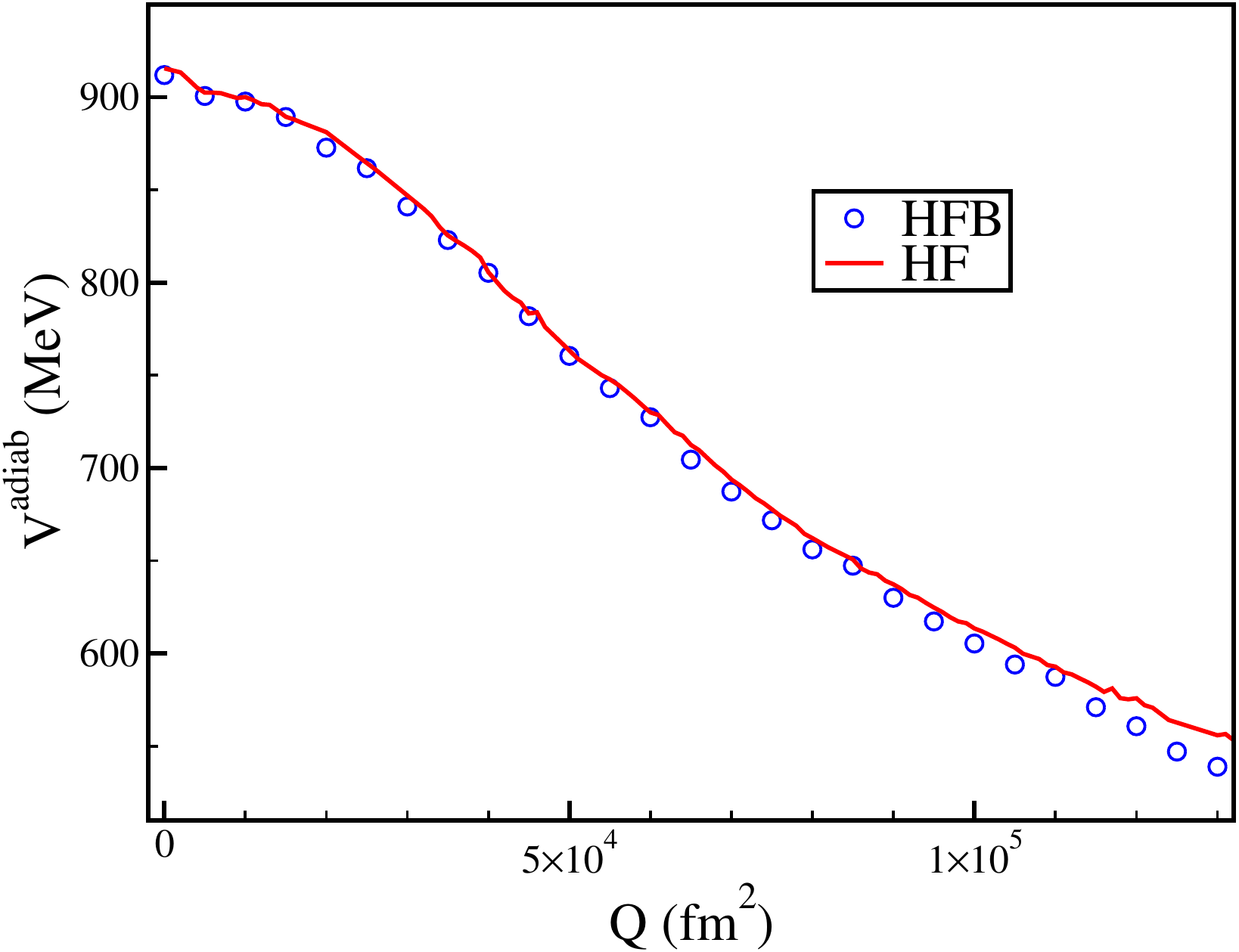}
\caption{The adiabatic potential of the $^{238}$U+$^{238}$U system is computed with the \textsc{ev8} code~\cite{bon05} as a function of the total quadrupole moment.  Calculations are performed with (HFB) and without (HF) pairing residual interaction.}
\label{fig:pot_adiab}
\end{center}
\end{figure}

As we saw in section~\ref{sec:QF}, there is a link between the collision time and the amount of transferred nucleons. 
It is then important to optimise collision times in order to favour the formation of heavy systems. 
It was initially believed that the potentials between two actinides have a barrier (and then a pocket)~\cite{sei85}, leading to possible long contact times at energies close to the barrier. 
Frozen Hartree-Fock potentials (see Sec. \ref{sec:FHF}) are shown in Fig.~\ref{fig:pot_frozen} where the  potentials have been computed with two $^{238}$U HF ground states with different orientations. 
However, Berger and collaborators  showed with constrained HFB calculations of the composite system of $^{238}$U $+^{238}$U that there is, in fact, no barrier in their (adiabatic) nucleus-nucleus potential~\cite{ber90}.
This result  is confirmed with the Skyrme functional in Fig.~\ref{fig:pot_adiab}.

\begin{figure}
\includegraphics[width=8.8cm]{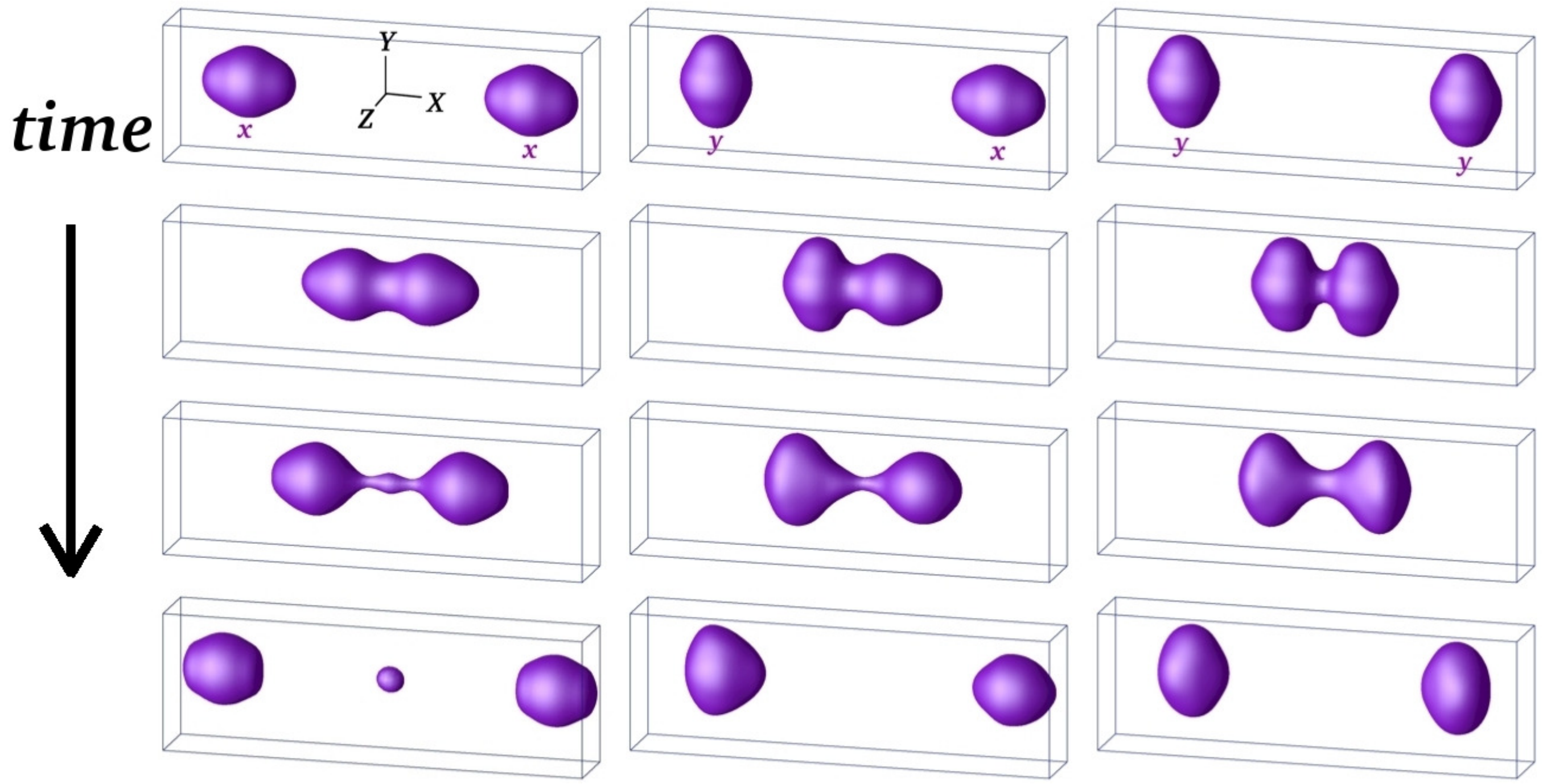}
\caption{Snapshots of the isodensity at half the saturation density in $^{238}$U+$^{238}$U central collisions at $E_{c.m.}=900$~MeV from TDHF calculations. Snapshots are given at times $t=0$, 1.5, 2.7, and $4.2$~zs from top to bottom. From Ref.~\cite{gol09}.}
\label{fig:density}
\end{figure}

The dynamics of actinide collisions have been studied with the TDHF approach~\cite{cus80,gol09,ked10,sim11b,ayik2017,umar2018,ayik2020,zhang2024b} and the quantum molecular dynamics (QMD) formalism~\cite{tia08,zha09}. 
These microscopic studies complement other works with a quantum master equation~\cite{sar09}, the dinuclear system (DNS) model \cite{ada05,fen09}, multidimensional Langevin equations~\cite{zag06}, and the constrained molecular dynamics model~\cite{mar02}. 

Figure~\ref{fig:density} shows zeptosecond snapshots of the density obtained in TDHF calculations of $^{238}$U+$^{238}$U central collisions at $E_{c.m.}=900$~MeV. 
We see that the initial orientation of the nuclei plays a crucial role on the reaction mechanism~\cite{gol09}. 
For instance, we observe the formation of a third fragment in the left column.
A net transfer is also obtained in the middle column.
Indeed, integration of  proton and neutron 
densities in each reactant indicates an average transfer of $\sim6$~protons and $\sim11$~neutrons from the right to the left nucleus. In this case, transfer occurs from the 
tip of the aligned nucleus to the side of the other. 
This configuration is then expected to favor the formation of nuclei heavier than~$^{238}$U.

\begin{figure}
\begin{center}
\includegraphics[width=6.5cm]{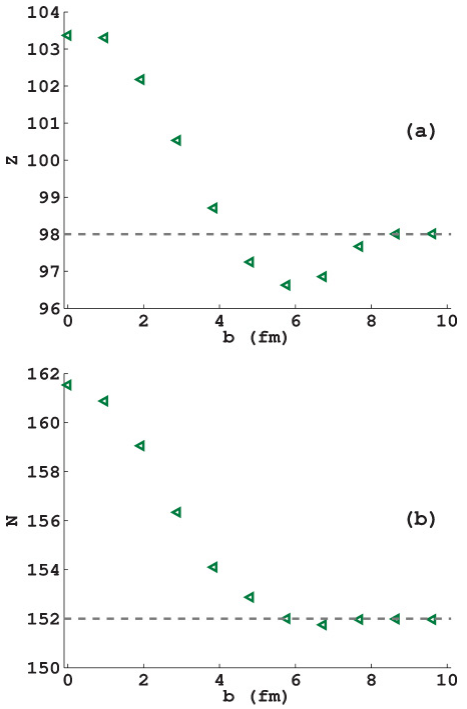}
\caption{(a) Proton and (b) neutron numbers in the heavy fragment of $^{232}$Th$+^{250}$Cf at $E_{c.m.}=915.8$~MeV as a function of the impact parameter $b$. The orientation is chosen so that the tip of $^{232}$Th hits the side of $^{250}$Cf at $b=0$.  The dashed lines represent the number of protons and neutrons in $^{250}$Cf. 
 From Ref.~\cite{ked10}.}
\label{fig:ThCf}
\end{center}
\end{figure}

Pursuing this idea, a new ``inverse quasi-fission'' mechanism has been identified~\cite{ked10}.
The inverse process of quasi-fission corresponds to a transfer of nucleons from the light collision partner to the heavy one. 
Such a mechanism may occur in actinide collisions due to shell effects in the $^{208}$Pb region~\cite{vol78,zag06}. 
The mechanism proposed in Ref.~\cite{ked10} is different (but may be complementary).
It occurs for specific orientations of the actinides, where the tip of the lighter one is in contact with the side of the heavier one. 
An example of inverse quasi-fission reaction is shown in  Fig.~\ref{fig:ThCf}, where a $^{232}$Th+$^{250}$Cf collision produces a $^{265}$Lr fragment\footnote{This fragment is not exactly a primary fragment as at the final time of the calculation, about 3 neutrons have been emitted in the entire system (see Fig.~4 of~\cite{ked10}). However, it is still excited  and may cool down by emitting additional neutrons or, of course, by fissioning.} at $b=0$.

\subsection{Conclusions and perspectives}

This section was devoted to the study of heavy-ion reaction mechanisms in order to investigate quantum dynamical effects in complex systems and to predict the formation of new nuclei.

The TDHF theory was shown to reproduce energy thresholds for fusion for systems spanning the entire  nuclear chart.
The effect of deformation and transfer on fusion barriers is well treated.
The competition with quasi-fission leading to a fusion hindrance in heavy systems is also included. 
The quasi-fission mechanism has been investigated and the role of shell effects in the fragment formation has been discussed. 
In some reactions, the latter could be strong enough to induce an inverse quasi-fission reaction. 

It was also shown that the charge equilibration process in $N/Z$ asymmetric collision affects the reaction mechanism: excitation of a preequilibrium GDR in the compound system (which can be used to study the path to fusion), enhancement of proton stripping and neutron pickup, and correlations between fragment $N$ and $Z$ distributions in DIC (which should be sensitive to the symmetry energy).

Actinide collisions have been investigated. 
Inverse quasi-fission mechanism associated with specific orientations was found in numerical simulations.
This mechanism might be investigated to produce and study  heavy nuclei.

Quantum effects at the single-particle level are well treated in the TDHF approach, allowing for realistic predictions of independent particle transfer probabilities using a particle-number projection technique. 
Particle number fluctuations and correlations require beyond TDHF approaches such as TDRPA, in particular in deep inelastic collisions.
However, application of TDRPA are limited to symmetric collisions. Other approaches, e.g., based on a stochastic mean-field description, should overcome this limitation.  

A strong limitation of the TDHF approach is that it does not allow for quantum tunnelling of the many-body wave function. 
As a result, sub-barrier fusion cannot be studied directly and requires the extraction of nucleus-nucleus potentials, e.g., with density constrained techniques. 
The latter have been used extensively, including in reactions relevant to astrophysical processes,  where changed quantum tunnelling rates can have drastic effects.
Alternatively, a possible approach is to use an extension of the TDHF theory based on imaginary time evolution \cite{neg82,mcglynn2020}.

\section{Conclusions}

Nuclei are ideal to investigate fundamental  aspects of the quantum many-body problem. 
They exhibit collective motions built from coherent superpositions of the states of their constituents.
Examples range from collective vibrations to the formation of a compound system in collisions.
These features are common to other composite systems (atomic clusters, molecules, Bose-Einstein condensates...).
Their study in nuclear systems is obviously part of a wider physics field. 

The Balian-V\'en\'eroni variational principle offers a powerful approach to the many-body quantum dynamics. 
Both the observable and the state of the system are considered as variational quantities. 
In the limit of independent particles, one obtains the time-dependent Hartree-Fock formalism for the expectation of one-body observables, and the time-dependent RPA for their fluctuations and correlations. 

Different studies of nuclear dynamics, from collective vibrations to heavy-ion collisions have been presented in this review. 
A particular attention was devoted to the interplay between collective motions and internal degrees of freedom within a unified theoretical description. 

Strongly interacting systems such as nuclei  exhibit collective vibrations in the continuum, and their direct decay could be used to infer their microscopic structure. 
We also questioned the harmonic nature of these vibrations.
In particular, a source of anharmonicity from the coupling between different vibrational modes was identified.
Nuclei are known to exhibit superfluidity due to pairing residual interaction and it is natural to wonder if the pairing field could also vibrate. 
The theoretical approach was then extended to study such pairing vibrations. 

Large amplitude collective motions were investigated in the framework of fission and heavy-ion collisions. 
The role of shell effects in the formation of fission fragments was studied. 
We discussed how fusion is affected by the internal structure of the collision partners, including superfluidity. 
Mechanisms in competition with fusion, i.e., transfer reactions, deep-inelastic collisions, and quasi-fission  were investigated. 
Finally, actinide collisions were studied.

\section*{Acknowledgements}

This work is dedicated to P. Bonche, who is the main author of the \textsc{tdhf3d} code and was one of the principal contributors to this field. 

The tremendous amount of work performed in this field since the first edition of this review is due to very active early and mid-career researchers, including Abhishek, S. Ebata, K. Godbey, Lu Guo, I. Iwata, P. McGlynn, G. Scamps, B. Schuetrumpf, K. Sekizawa, Y. Tanimura and K. Washiyama. The number of groups who work in the development of TDHF and its extensions considerably increased, now including groups in Australia, China, Croatia, France, Germany, Japan, Poland, UK, and USA. Numerous students and postdocs of these groups have contributed to work reported in this new edition. 

Computational resources were provided in part by the Centre de Calcul Recherche et Technologie of the Commissariat \`a l'\'Energie Atomique, France, and by the Australian Government through the National Computational Infrastructure (NCI) under the ANU Merit Allocation Scheme, the Adapter scheme, and the National Committee Merit Allocation Scheme.
This work has been partly supported by the Australian Research Council Discovery Project grants DP0879679,  
DP110102858, DP110102879, DP160101254 and DP190100256.

{
\appendix 
\section*{APPENDICES}
\addcontentsline{toc}{section}{APPENDICES}
\section{TDHF approach from the standard variational principle\label{annexe:standardTDHF}}

Consider a state of $N$ independent particles described by the Slater determinant $\kfi$.
The action defined in Eq.~(\ref{eq:SDirac}) reads
\oeq
S \equiv S_{t_0,t_1}[\phi] = \int_{t_0}^{t_1} \stb d t \stf \<\phi (t)| \( i\hb \frac{d }{{d} t} - \oH \) |\phi(t)\>
\ceq
where $\oH$ is the Hamiltonian of the system. 
In such a state, every $M$-body density matrix ($1\le M\le N$) is simply expressed as a function of the one-body density matrix $\ro$~(see appendix~C of Ref.~\cite{sim10a}).
Then, all the information on the system is contained in $\ro$.
The expectation value of the Hamiltonian on the state $\kfi$ may then be written as a functional of~$\ro$: $E[\ro] = \bfi \oH \kfi$.
In addition, we have
\oeq
\bfi \frac{d}{d t} \kfi = \sum_{i=1}^N \<\az_i|  \frac{d}{d t} |{\az}_i\>
\ceq
and, then,
\oeqn
S &=& \int_{t_0}^{t_1} \stb d t \stf \( i\hb\sum_{i=1}^N \<\az_i| \frac{ d}{ d t} |{\az}_i\>
-E[\ro(t)]\) \nonumber \\
&=& \int_{t_0}^{t_1} \stb  d t \stf \(i\hb \sum_{i=1}^N \sdf \int \stb  d x \stf \az_i^*(x,t) \sdf \frac{ d}{ d t} {\az}_i(x,t)
-E[\ro(t)]\) \nonumber \\
\ceqn
where $x\equiv (\vr s q)$ describes all the single-particle degrees of freedom (position $\vr$, spin $s$ and isospin $q$). 
We introduced the notation $\int \sdb d x = \sum_{q s} \int \sdb  d \vr$.

The variational principle reads $\del S = 0$. 
The variation must be done on each independent variable.
Here, these variables are the real part $\az^{\rm Re}_\al$ and the imaginary part $\az_\al^{\rm Im}$ of each occupied single particle state $\az_\al$.
We must then consider 
\oeq
\frac{\delta S}{\delta  \az^{\rm Re}_\al(x,t)} = 0 \mbox{\hspace{1cm} and \hspace{1cm}}
\frac{\delta S}{\delta  \az^{\rm Im}_\al(x,t)} = 0 
\label{eq:delRe}
\ceq
for each  $\al \in \{1...N\}$, for all $t$ such as $t_0\le t \le t_1$, and for all $x$. 

However, the calculation is more straightforward if we use $\az_\al$ and $\az^*_\al$ as {\it independent} variables instead of $\az^{\rm Re}_\al$ and $\az_\al^{\rm Im}$.
We can then consider the variations over $\az$ and over $\az^*$ independently. 
Note, however, that we loose the property ''$\az^*$ 
being complex conjugate of $\az$'' that we should restore later.

Equations (\ref{eq:delRe}) are then replaced by
\oeq
\frac{\delta S}{\delta  \az_\al(x,t)} = 0 \mbox{\hspace{1cm} and \hspace{1cm}}
\frac{\delta S}{\delta  \az^*_\al(x,t)} = 0 .
\label{eq:delphiphistar}
\ceq
The variation over $\az^*$  gives
\oeq
\frac{\delta S}{\delta  \az^*_\al(x,t)} = i\hb \sdf \frac{ d}{ d t} \az_\al(x,t) - 
 \int_{t_0}^{t_1} \stb  d t' \stf \frac{\delta E\[\ro(t')\]}{\delta  \az^*_\al(x,t)}.
\ceq
The functional derivative of $E$ can be re-written thanks to a change of variable
\oeq 
 \frac{\delta E\[\ro(t')\]}{\delta  \az^*_\al(x,t)}
 =  \int \stb  d y \,  d y' \stf \frac{\delta E\[\ro(t')\]}{\delta  \ro(y,y';t')}
\sdf  \frac{\delta  \ro(y,y';t')}{\delta  \az^*_\al(x,t)}.
\ceq
Using
\oeq 
  \frac{\delta  \ro(y,y';t')}{\delta  \az^*_\al(x,t)} =  \az_\al(y,t') \sdf \delta(y'-x) \sdf \delta(t-t')
\ceq
and noting the single-particle Hartree-Fock Hamiltonian $h$ with matrix elements
\oeq
h(x,y;t) = \frac{\delta E\[\ro(t)\]}{\del \ro(y,x;t)},
\label{eq:hEHF}
\ceq
we get the TDHF equation for the set of occupied states
\oeq
\fbox{$ \displaystyle i\hb \sdf \frac{ d }{ d t} \az_\al(x,t) = \int \stb  d y \stf h(x,y;t) \sdf \az_\al(y,t)$}\sdf. \label{eq:TDHF_az}
\ceq

The variation over $\az$ gives, after integrating by part the term with the time derivative, 
\oeqn
\frac{\delta S}{\delta  \az_\al(x,t)} &=& \frac{\delta }{\delta  \az_\al(x,t)}  \int_{t_0}^{t_1} \stb  d t' \nonumber\\
&&\stb\stb\stb\stb\stb\stb\stb \left[ i \hb \frac{ d}{ d t'}
 \(\sum_\be \int \stb  d y \stf \az^*_\be(y,t')\sdf \az_\be(y,t')\) \right. \nonumber \\
&&\stb\stb\stb\stb\stb\stb\stb\left.
-i\hb \sdf  \sum_\be \int \stb  d y \stf \(\frac{ d}{ d t'} \az^*_\be(y,t')\) \sdf \az_\be(y,t')-  E\[\ro(t')\]\right].\nonumber\\
\label{eq:var_az}
\ceqn
The first term in the r.h.s. cannot  be replaced {\it  a priori} by  $\Tr [\ro] = N$
because we considered $\az$ and $\az^*$ to be independent variables. 
Let us note the variation of $\az$ at a specific ''point'' of the Hilbert space at time~$t$ and affecting only the state $\az_\al$. This variation can be defined as 
\oeq
\delta_{\al x t} \, \az_\be(y,t') = \Delta \az \sdf f(t) \sdf \delta(t-t') \sdf \delta(x-y) \sdf \delta_{\al \be}.
\ceq
Using the usual definition of the functional derivative, we may write
\oeqn
&& \stb\frac{\delta }{\delta  \az_\al(x,t)}
 \int_{t_0}^{t_1} \stb  d t' \stf  \frac{ d}{ d t'}
 \(\sum_\be \int \stb  d y \stf \az^*_\be(y,t')\sdf \az_\be(y,t')\) \nonumber \\
&&\stb =\lim_{\Delta \az \rightarrow 0} 
 \int_{t_0}^{t_1} \stb  d t' \stf  \frac{ d}{ d t'}
\sdf \frac{1}{\Delta \az} \(\sum_\be \int \stb d y \stf \az^*_\be(y,t')\sdf \delta_{\al x t} \az_\be(y,t')\) \nonumber \\
&&\stb= \[ f(t) \sdf \delta(t-t') \sdf \az^*_\al(x,t') \]_{t'=t_0}^{t_1}.
\ceqn
We have to choose $f(t_0)=f(t_1)=0$ in oder to cancel this term at $t=t_0$ or $t=t_1$. 
It is equivalent to forbid variations of $\az$ at $t_0$ and $t_1$. 
As a result, Eq.~(\ref{eq:var_az}) leads to the complex conjugate of the TDHF equation~(\ref{eq:TDHF_az}), 
restoring the fact that $\az^*$ is the complex conjugate of $\az$. 
This last property is also necessary for energy and particle number conservations.

\section{Fluctuations and correlations of one-body observables with the Balian-V\'en\'eroni variational principle \label{annexe:BV}}

In this appendix, we detail the derivation of the Balian-V\'en\'eroni prescription (also referred to as TDRPA) for fluctuation and correlation of one-body observables, given in eqs.~(\ref{eq:CiiBV}) and~(\ref{eq:CijBV}), respectively. 
We consider the particular case where the independent particle state of the system is a pure state described by a unique Slater determinant at all time. 
The more general derivation with mean-field states of the form given in Eq.~(\ref{eq:defD}) can be found in Refs.~\cite{bal92,bro09}.

As mentioned in section~\ref{sec:fluccor}, fluctuations and correlations of the one-body observables 
\oeq
\oQ_i = \sum_{\al\be} Q_{i_{\al\be}} \oad_\al \oa_\be
\ceq
are obtained by evaluating 
\oeq
\oA_1=\exp \(-\sum_i \varepsilon_i \oQ_i\) 
\label{eq:boundA1}
\ceq
at the final time $t_1$, and using Eq.~(\ref{eq:lnA}) for small $\varepsilon_i$. 
The expectation value of $\oA_1$ at $t_1$ is obtained from
\oeq
\<\oA_1\>(t_1)=\Tr \[ \oA(t_1) \oD(t_1) \],
\ceq
where the observable is assumed to be a time-dependent operator $\oA(t)$ with the boundary condition $\oA(t_1) = \oA_1$. The state of the system is assumed to be known at the initial time $t_0$, with the boundary condition $\oD(t_0)=\oD_0$. 
As a result, the variations of the observable and the state obey 
\oeq
\delta \oA(t_1) = \delta \oD (t_0) = 0.
\ceq

\subsection*{Variational space and parametrisation of variational quantities}

The BV variational principle is solved by requiring the stationarity of the action-like quantity $J$ defined 
 in Eq.~(\ref{eq:JA}), or, equivalently, in Eq.~(\ref{eq:JD}). 
 In the present application, the variational space for the observable is restricted to exponential of one-body operators. The observable is then parametrised as 
\oeq
\oA(t)=\exp[-\oL(t)]
\ceq
with 
\oeq 
\oL(t)=\sum_{\al\be}L_{\al\be}(t) \oad_\al \oa_\be
\ceq
The mean-field density-matrix is parametrised as
\oeq
\oD(t)=\exp[-m(t)-\oM(t)]
\ceq
with 
\oeq 
\oM(t)=\sum_{\al\be}M_{\al\be}(t) \oad_\al \oa_\be.
\ceq
As discussed in section~\ref{sec:fluccor}, the case of a pure Slater determinant corresponds to the limit where the eigenvalues of the matrix $M(t)$ tend to $\pm \infty$ and $m(t)\rightarrow +\infty$ in such a way that $z(t)=\Tr\oD(t)=1$~\cite{bal85}. 

According to Eqs~(\ref{eq:rhoM}) and (\ref{eq:mz1}), we have the following relationships 
\oeqn
\rho(t)&=&\frac{1}{1+e^{M(t)}}\mbox{ and}\label{eq:roeM}\\
m(t)&=&\tr \ln (1+e^{-M(t)}).
\ceqn
where $\rho$ is the one-body density matrix with elements
\oeq
\rho_{\al\be}(t) = \Tr[\oD(t)\oad_\be\oa_\al].
\label{eq:rhoalbe}
\ceq 
The inverse relationships read
\oeqn
m(t)&=&-\tr [\ln (1-\rho(t))] \mbox{ and}\label{eq:m}\\
e^{-M(t)} &=& \frac{\rho(t)}{1-\rho(t)}.\label{eq:e-M}
\ceqn
As expected, we see that the Slater determinant can be entirely parametrised by $\rho(t)$.

By analogy, we introduce a similar parametrisation of the observable using eqs.~(\ref{eq:rhoM}) and~(\ref{eq:TrD}). We get 
\oeqn
\sigma(t)&=&\frac{1}{1+e^{L(t)}}\mbox{ and}\\
y(t)&=&\exp \[\tr \ln (1+e^{-L(t)})\], 
\ceqn
with the inverse relationships
\oeqn
\ln y(t)&=&-\tr[\ln(1-\sigma(t))] \mbox{ and}\label{eq:lny}\\
e^{-L(t)} &=& \frac{\sigma(t)}{1-\sigma(t)}.\label{eq:e-L}
\ceqn

Equations~(\ref{eq:JA}) and (\ref{eq:JD}) contain the products $\oA\oD$ and $\oD\oA$. 
It is convenient to define a similar parametrisation of these quantities, using the fact that the product of the exponential of one-body operators is also an exponential of a one-body operator
\oeqn
\oA(t)\oD(t)&=&e^{-\oL(t)}e^{-\oM(t)}=e^{-\oL'(t)}\mbox{ and}\\
\oD(t)\oA(t)&=&e^{-\oM(t)}e^{-\oL(t)}=e^{-\oM'(t)}.\label{eq:DA}
\ceqn
$\oL'$ and $\oM'$ are one-body operators which are parametrised as 
\oeqn
\oL'(t)&=&m(t)+\sum_{\al\be}L'_{\al\be}(t) \oad_\al\oa_\be \mbox{ and}\\
\oM'(t)&=&m(t)+\sum_{\al\be}M'_{\al\be}(t) \oad_\al\oa_\be,
\ceqn
with
\oeqn
e^{-L'(t)}&=&e^{-L(t)}e^{-M(t)} \mbox{ and}\\
e^{-M'(t)}&=&e^{-M(t)}e^{-L(t)}.\label{eq:defM'}
\ceqn
By analogy with Eq.~(\ref{eq:rhoalbe}), we introduce two new matrices, $\rho'$ and $\sigma'$, with elements
\oeqn
\rho'_{\al\be}(t)&=& \frac{\Tr[\oD(t)\oA(t)\oad_\be\oa_\al]}{\omega(t)} \mbox{ and}\\
\sigma'_{\al\be}(t)&=& \frac{\Tr[\oA(t)\oD(t)\oad_\be\oa_\al]}{\omega(t)},
\ceqn
and obeying the following relationships:
\oeqn
\rho'(t)&=&\frac{1}{1+e^{M'(t)}}=\frac{1}{1+e^{L(t)}e^{M(t)}} \mbox{ and}\label{eq:rho'}\\
\sigma'(t)&=&\frac{1}{1+e^{L'(t)}}=\frac{1}{1+e^{M(t)}e^{L(t)}}.
\ceqn
Using eqs.~(\ref{eq:DA}) and (\ref{eq:TrD}), the normalisation factor 
\oeq
\omega(t)=\Tr[\oD(t)\oA(t)]
\label{eq:defomega}
\ceq
becomes
\oeq
\omega(t)=\exp\[-m(t)+\tr\(\ln(1+e^{-M(t)}e^{-L(t)})\)\].
\label{eq:omegamML}
\ceq
Using eqs.~(\ref{eq:m}), (\ref{eq:e-M}), (\ref{eq:lny}) and (\ref{eq:e-L}), we get
\oeq
\omega(t)=y(t)\exp\[\tr\(\ln(1-\rho(t)-\sigma(t)+2\sigma(t)\rho(t))\)\].
\label{eq:omega}
\ceq

The action contains a term with a time derivative, like $\Tr[\oA\frac{d\oD}{dt}]$ in Eq.~(\ref{eq:JA}), which, starting from Eq.~(\ref{eq:omega}), becomes
\oeq
\Tr\[\oA\frac{d\oD}{dt}\] = \omega \sdf\tr\[ \( \frac{2\sigma -1}{1-\ro-\sigma+2\sigma\rho} \) \frac{d\rho}{dt}\]
\ceq
The other terms of the action contain the Hamiltonian $\oH$.
For mean-field states, all the information on the system is contained in the one-body density matrix $\rho$ and we can write the expectation value of the Hamiltonian as an energy density functional, i.e.,
\oeq
\Tr\[\oD(t)\oH\]= E[\rho(t)].
\ceq
Similarly, we have
\oeqn
\Tr\[ \oD(t)\oA(t)\oH \] &=& \omega(t)E[\rho'(t)]\mbox{ and}\\
\Tr\[ \oA(t)\oD(t)\oH \] &=& \omega(t)E[\sigma'(t)].
\ceqn
As a result, the action defined in Eq.~(\ref{eq:JA}) becomes
\oeqn
 \stb\stb\stb \stb J&=& \omega(t_1) - \int_{t_0}^{t_1} dt \, \omega(t) \[\frac{}{} iE[\rho'(t)] -iE[\sigma'(t)]\right. \nonumber\\
&+& \left.\tr \( \frac{2\sigma(t) -1}{1-\ro(t)-\sigma(t)+2\sigma(t)\rho(t)}  \frac{d\rho(t)}{dt}\)\].
\ceqn

\subsection*{Equations of motion for $\rho$, $\rho'$ and $\sigma'$}

We now seek for an equation of motion for $\rho(t)$. 
This is obtained by requiring the stationarity of $J$ when $y$ and $\sigma$ vary. 
$J$ depends on $y$ due to  $\omega$  which is linear in $y$, i.e., 
\oeq
\frac{d\omega}{dy}=\frac{\omega}{y}.
\ceq
Requiring the stationarity of the action when $y(t)$ varies implies
\oeq
0=iE[\rho'(t)] -iE[\sigma'(t)]+ \tr \( \eta(t) \frac{d\rho(t)}{dt}\)
\label{eq:vary}
\ceq
where we have introduced a new matrix 
\oeq
\eta(t) = \frac{2\sigma(t) -1}{1-\ro(t)-\sigma(t)+2\sigma(t)\rho(t)} .
\ceq
A variation of $\sigma$ implies a variation of $\eta$ according to 
\oeq
\delta \eta=\frac{d\eta}{d\sigma}\delta \sigma .
\ceq 
Then, instead of considering the variation of $\sigma$, we equivalently study the variation of $J$ with $\eta$.  
The $\eta$ matrix can be re-written in such a way that it contains an explicit dependence on $\rho'$ and $\sigma'$:
\oeqn
\eta &=& \rho^{-1}(\rho'-\rho)(1-\ro)^{-1}\\
&=&(1-\ro)^{-1}(\sigma'-\rho)\rho^{-1}.
\ceqn
This tells us how $\rho'$ and $\sigma'$ vary with a variation of $\eta$:
\oeqn
\delta \rho' &=& \rho \, \delta \eta  \, (1-\rho) \mbox{ and}\\
\delta \sigma' &=& (1-\rho) \, \delta \eta \, \rho.
\ceqn
These relations are used to determine the variation of the energies $E[\rho']$ and $E[\sigma']$. Note that the latter can be written as a function of the self-consistent HF Hamiltonian $h$ using Eq.~(\ref{eq:hEHF}):
\oeqn
E[\rho']&=& \tr\(h[\rho'] \rho' \)\mbox{ and}\\
E[\sigma']&=& \tr\(h[\sigma'] \sigma' \).
\ceqn
A variation of $\eta$ induces, then, a variation of Eq.~(\ref{eq:vary}) which can be written as
\oeq
\tr\[\delta \eta \( \frac{d\ro}{dt} +i(1-\ro)\,h[\rho']\,\ro-i\ro \,h[\sigma'] \,(1-\ro)\)\]=0.
\ceq
Requesting this equation to hold for any variation $\delta \eta$ gives the equation of motion for $\rho$:
\oeq
i\frac{d\ro}{dt} =(1-\ro)\,h[\rho']\,\ro-\ro \,h[\sigma'] \,(1-\ro).
\label{eq:deltaeta}
\ceq

To get the equations of motion for $\ro'$ and $\sigma'$, it is convenient to express the time derivative term with the $L$ and $M$ matrices.
From Eq.~(\ref{eq:e-M}), we have
\oeq
\frac{d}{dt} e^{-M} = \frac{1}{1-\ro} \, \frac{d\ro}{dt} \, \frac{1}{1-\ro}.
\ceq
Using, Eq.~(\ref{eq:deltaeta}), we get
\oeq
i\frac{d}{dt}e^{-M} = h[\ro']\, e^{-M} - e^{-M}\, h[\sigma'].
\ceq
Similarly, we have
\oeq
i\frac{d}{dt}e^{-L} = h[\sigma']\, e^{-L} - e^{-L}\, h[\rho'].
\ceq
We can then write
\oeqn
\frac{d }{dt} e^{-M'} &=& \frac{de^{-M}}{dt} e^{-L} + e^{-M} \frac{de^{-L}}{dt} \nonumber \\
&=& i \[e^{-M'},h[\ro']\].
\label{eq:de-Mdt}
\ceqn
From Eq.~(\ref{eq:rho'}), we have
\oeq
1-\ro'=\frac{1}{1+e^{-M'}}.\label{eq:1-ro'}
\ceq
This leads to
\oeq
\frac{d}{dt} \ro' = (1-\ro') \, \frac{de^{-M'}}{dt} \, (1-\ro').
\ceq
Using Eq.~(\ref{eq:de-Mdt}) we get the equation of motion for $\ro'$:
\oeq
i\frac{d\rho'}{dt}=\[h[\ro'],\ro'\].
\label{eq:drho'}
\ceq
Similarly, we have
\oeq
i\frac{d\sigma'}{dt}=\[h[\sigma'],\sigma'\].
\label{eq:dsigma'}
\ceq

\subsection*{Development in powers of $\varepsilon$}

From the definition of $\omega(t)$ in Eq.~(\ref{eq:defomega}), we see that the expectation of $\oA_1$ at the final time $t_1$ is equal to $\omega(t_1)$. 
We now show that, when the action is stationary, $\omega(t)$ is in fact constant in time.
The one-body density matrix associated with a Slater determinant has eigenvalues 1 for the occupied states and 0 for the others. As a result, we see from Eq.~(\ref{eq:trln1+A}) that $m(t)$ is equal to an infinite constant and, then, $dm/dt=0$. 
Using this property and eqs.~(\ref{eq:defM'}), (\ref{eq:rho'}), (\ref{eq:omegamML}), and (\ref{eq:de-Mdt}), we get
\oeqn
\frac{d}{dt}\omega &\propto& \frac{d}{dt} \tr\(\ln (1+e^{-M'})\)\nonumber \\
&\propto&\tr \frac{\frac{d}{dt}e^{-M'}}{1+e^{-M'}}\nonumber \\
&\propto& \tr\[ h[\ro'],\ro' \]=0.
\ceqn
$\omega(t)$ is then a constant we need to evaluate to get the fluctuations and correlations from Eq.~(\ref{eq:lnA}). 

$\ln \omega$ can be expressed as a function of $\rho$ and $L$ from Eq.~(\ref{eq:omegamML}) and using eqs.~(\ref{eq:m}) and (\ref{eq:e-M}):
\oeq
\ln  \omega = \tr \[ \ln \( 1-\ro+\ro\, e^{-L} \)\].
\label{eq:lnomega}
\ceq
Deriving $\ln \omega$ according to the parameter $\varepsilon_i$, we get
\oeqn
\frac{d}{d\varepsilon_i}\ln \omega &=& -\<\oQ_i\>+\sum_j \varepsilon_jC_{ij}+ \cdots \label{eq:epsilondel} \\
&=&\tr\[\ro'e^L\frac{de^{-L}}{d\varepsilon_i}\]\label{eq:dode}
\ceqn
where the first identity is obtained from Eq.~(\ref{eq:lnA}) and the second from eqs.~(\ref{eq:lnomega}), (\ref{eq:roeM}) and (\ref{eq:rho'}).
Let us expand $\rho$ and $L$ in powers of the $\varepsilon$:
\oeqn
\rho(t)&=& \rho^{(0)}(t) + \rho^{(1)}(t) +\cdots \mbox{ and}\\
L(t)&=&  L^{(1)}(t) +\cdots 
\ceqn
where the superscript denotes the power in $\varepsilon$. 
The fact that $L^{(0)}=0$ is due to the boundary condition in Eq.~(\ref{eq:boundA1}) implying
\oeq
L(t_1)=\sum_i \varepsilon_i Q_i.
\label{eq:bcL}
\ceq 
We now expand $\rho'$ up to first order in $\varepsilon$ using eqs.~(\ref{eq:rho'}) and  (\ref{eq:e-M}):
\oeq
\rho'=\rho^{(0)} + \rho^{(1)} - \rho^{(0)}L^{(1)} (1-\rho^{(0)}) +O(\varepsilon^2).
\ceq
Separating the zeroth and first order contributions, i.e.,  $\rho'\simeq\rho'^{(0)}+\rho'^{(1)}$, we get
\oeqn
\rho'^{(0)}&=&\rho^{(0)} \nonumber\\
\rho'^{(1)}&=&\rho^{(1)} - \rho^{(0)}L^{(1)} (1-\rho^{(0)}) \label{eq:rho'exp}
\ceqn

\subsection*{Expectation value of one-body observables}

At the zeroth order, Eq.~(\ref{eq:dode}) becomes
\oeqn
\<\oQ_i\>&=& \tr\[\rho^{(0)}(t)\frac{dL^{(1)}(t)}{d\varepsilon_i}\] \nonumber \\
&=& \tr\[\rho^{(0)}(t_1)Q_i\],
\label{eq:Qtdhf}
\ceqn
where we have used the boundary condition in Eq.~(\ref{eq:bcL}).
From eqs.~(\ref{eq:drho'}) and~(\ref{eq:rho'exp}) we see that $\rho^{(0)}$ obeys the TDHF equation. 
Equation~(\ref{eq:Qtdhf}) is then exactly the TDHF result for the expectation value of one-body observables. 

\subsection*{Fluctuations and correlations}

We now expand Eq.~(\ref{eq:dode}) up to first order in $\varepsilon$, using eqs.~(\ref{eq:rho'exp}) and (\ref{eq:bcL}):
\oeqn
&&\frac{d\ln \omega}{d\varepsilon_i} =\nonumber\\
&& \tr\[\(\ro^{(0)}(t_1)+\ro^{(1)}(t_1)-\ro^{(0)}(t_1)\sum_j\varepsilon_jQ_j(1-\ro^{(0)}(t_1))\)
\right.\nonumber\\
&&\left.  \( 1+\sum_n\varepsilon_nQ_n\)\(-Q_i+ \sum_m \varepsilon_mQ_iQ_m\) \],
\ceqn
where we assumed $[Q_i,Q_j]=0$ for simplicity.
Identifying with Eq.~(\ref{eq:epsilondel}), we get
\oeqn
\sum_j\varepsilon_jC_{ij} &=& \sum_j \varepsilon_j \tr\[Q_i\rho^{(0)}(t_1) Q_j \(1-\rho^{(0)}(t_1)\) \] \nonumber \\
&&- \tr\(\rho^{(1)}(t_1)Q_i\).\label{eq:epsCij}
\ceqn
The first term in the right hand side gives the usual fluctuations and correlations computed at the TDHF level [see Eq.~(\ref{eq:CiiTDHF})]. 
The present approach, which is more general as it uses a larger variational space for the evolution of the observable than in the TDHF formalism, is expected to improve the fluctuations and correlations thanks to the second term in the right hand side. 

We see in Eq.~(\ref{eq:epsCij}) that the time evolution of $\rho^{(1)}$ is needed in addition to $\rho^{(0)}$. 
Unfortunately, $\rho^{(1)}$ does not obey a simple equation and its determination is not trivial. 
It is however possible to find an equation for $C_{ij}$ which is easier to implement. 
In fact, although Eq.~(\ref{eq:epsCij}) has been obtained by evaluating the matrices at time $t_1$, it is possible to show that a similar equation holds at any time. 
Let us first introduce the quantity
\oeqn
F_{ij}(t) &=& \frac{1}{2} \tr \( \rho^{(0)} [L_i^{(1)},L_j^{(1)}]\)+ \tr \[ L_i^{(1)}\ro^{(0)} L_j^{(1)}(1-\rho^{(0)})\]\nonumber \\
&&-  \tr\(\rho^{(1)}L_i^{(1)}\),\label{eq:Fij}
\ceqn
where the matrices in the right hand side are expressed at time~$t$ and the $L_i^{(1)}$ obey the boundary condition
\oeq
L_i^{(1)}(t_1)=\varepsilon_iQ_i.
\ceq
We have also the boundary condition 
\oeq
\sum_j F_{ij}(t_1)=\sum_j\varepsilon_i\varepsilon_jC_{ij}.
\ceq 
Note that a similar term than the first term in the right hand side of Eq.~(\ref{eq:Fij}) appears in Eq.~(\ref{eq:epsCij}) when we consider the general case $[Q_i,Q_j]\ne0$. 
We now show that, in fact, $F_{ij}$ does not depend on time. 
We first re-write $F_{ij}$ as a function of the matrices $L$, $\sigma'$ and $\rho'$:
\oeq
F_{ij}=\frac{1}{2}\tr \[ L_i^{(1)} (\sigma_j'^{(1)}+\rho_j'^{(1)}) \],
\ceq
where we have used Eq.~(\ref{eq:rho'exp}) and its equivalent for $\sigma'$.
The matrices $\rho'^{(1)}$ and $\sigma'^{(1)}$ are obtained by taking the first order in $\varepsilon$ in the equations (\ref{eq:drho'}) and (\ref{eq:dsigma'}), respectively.
We obtain the time-dependent RPA equations
\oeqn
i\frac{d\rho'^{(1)}}{dt}&=&\[h[\rho^{(0)}] , \rho'^{(1)}\] + \[ \tr_2(\bar{v}  \rho'^{(1)}), \rho^{(0)}\]\label{eq:idrho'}\\
i\frac{d\sigma'^{(1)}}{dt}&=&\[h[\rho^{(0)}] , \sigma'^{(1)}\] + \[ \tr_2(\bar{v}  \sigma'^{(1)}), \rho^{(0)}\],\label{eq:idsigma'}
\ceqn
 where $\bar{v}$ is the antisymetrised two-body interaction defined in Eq.~(\ref{eq:vbar}). 
We also get a similar equation for $L$:
\oeq
i\frac{dL^{(1)}}{dt}=\[h[\rho^{(0)}] , L^{(1)}\] + \tr_2\[ \bar{v} [L^{(1)}, \rho^{(0)}]\].\label{eq:idL}
\ceq
The time derivative of $F_{ij}$ obeys
\oeqn
\frac{d}{dt}F_{ij} &=& -\frac{1}{2}\tr\[ \frac{dL_{i}^{(1)}}{dt}{\rho'}_j^{(1)} + L_{i}^{(1)} \frac{d{\rho'}_j^{(1)}}{dt} \right.\nonumber \\
&&+ \left.\frac{dL_{i}^{(1)}}{dt}{\sigma'}_j^{(1)} + L_{i}^{(1)} \frac{d{\sigma'}_j^{(1)}}{dt} \].
\ceqn
 Replacing with eqs.~(\ref{eq:idrho'}), (\ref{eq:idsigma'}) and (\ref{eq:idL}), we get 
\oeq 
\frac{d}{dt} F_{ij}(t) = 0.
\ceq
$F_{ij}$ is then constant with time. 

It is convenient to write it with the matrices at the time $t_0$ instead of $t_1$. Indeed, in this case, we have $\rho^{(1)}(t_0)=0$ due to the boundary condition at initial time. This simplifies the expression for the fluctuations and correlations:
\oeqn
\stb\stb C_{ij} &=& -\lim_{\varepsilon_i,\varepsilon_j\rightarrow0} \frac{1}{2\varepsilon_i\varepsilon_j} \nonumber \\
&&\tr \[ [L_i^{(1)}(t_0),\rho^{(0)}(t_0)] [L_j^{(1)}(t_0),\rho^{(0)}(t_0)]  \] \label{eq:CijLro}
\ceqn
Let us introduce a new matrix $\eta_i(t,\varepsilon_i)$ defined as
\oeqn
\eta_i(t,\varepsilon_i) &=& \rho^{(0)}(t) + i \[ L_i^{(1)}(t) , \rho^{(0)}(t) \].\label{eq:etafull}\\
&\approx&  e^{iL_i^{(1)}(t)}\rho^{(0)}(t) e^{-iL_i^{(1)}(t)},\label{eq:etaapp}
\ceqn
where the last equation holds for small $\varepsilon$. 
We can show that $\eta_i(t,\varepsilon_i)$ follows also a TDHF equation.
Note that the fact that $\rho^{(0)}$ is a Slater determinant implies that 
 $\eta_i$ is also a Slater determinant according to 
Eq.~(\ref{eq:etaapp}).

From eqs.~(\ref{eq:CijLro}) and~(\ref{eq:etafull}) we get our final result
\oeqn
C_{ij} = \lim_{\varepsilon_i,\varepsilon_j\rightarrow0} \frac{1}{2\varepsilon_i\varepsilon_j}
&\tr&\[\(\rho^{(0)}(t_0)-\eta_i(t_0,\varepsilon_i)\)\right.\nonumber\\
&&\stb \stb \stb\left.\(\rho^{(0)}(t_0)-\eta_j(t_0,\varepsilon_j)\)\],
\ceqn
with the boundary condition  at the final time $t_1$
\oeq
\eta_i(t_1,\varepsilon_i)=e^{i\varepsilon_iQ_i}\rho^{(0)}(t_1)e^{-i\varepsilon_iQ_i}.
\ceq

\section{From Fock to single-particle space with exponential of one-body operators \label{annexe:formula}}

Several relations are derived which are used in the derivation of the Balian-V\'en\'eroni prescription for fluctuation and correlation of one-body operators in appendix~\ref{annexe:BV}.
They relate operators or their associated matrices in Fock space to matrices expressed in the single-particle space.

\subsection*{Link between the density matrix of an independent-particle state and the one-body density matrix}

Our goal is to show the relation
\oeq
\boxed{\rho=\frac{1}{1+e^M}}
\label{eq:rhoM}
\ceq
where $\rho$ is the one-body density matrix with elements 
\oeq
\rho_{\al\be}=\<\oad_\be\oa_\al\>=\Tr\(\oD\oad_\be\oa_\al\),
\label{eq:rho_albe}
\ceq
and
\oeq
\oD=e^{-m-\oM}
\label{eq:D}
\ceq
is the density matrix associated with an independent-particle state described by a Slater determinant.
The later obeys
\oeq
\Tr\oD=1
\ceq
and $\oM$ is a one-body operator of the form 
\oeq
\oM=\sum_{\al\be}M_{\al\be}\oad_\al\oa_\be.
\label{eq:oM}
\ceq
We first show the following property: 
\oeq
\oad_\al\oD=\sum_\be\(e^M\)_{\be\al}\oD\oad_\be
\label{eq:oadD}
\ceq

{
Define $\oF(x)=e^{x\oA}\oB e^{-x\oA}.$
Its Taylor development reads
$$\oF(x)=\sum_{n=0}^\infty\frac{x^n}{n!}\of_n \stf \mbox{with}\stf \of_n=\left.\frac{\partial^n\oF}{\partial x^n}\right|_{x=0}$$
\oeqn
\oF(x)&=&\sum_{n=0}^\infty\frac{(x\oA)^n}{n!}\sdf\oB\sdf\sum_{m=0}^\infty\frac{(x\oA)^m}{m!}\nonumber\\
&=&\oB+x\[\oA,\oB\]+\frac{x^2}{2!}\[\oA,[\oA,\oB]\]+\cdots\nonumber
\ceqn
$\Rightarrow \of_0=\oB, \stf \of_{n+1}=[\oA,\of_n]$. 
Using this relation with $x=1$, we get
$$
\oD^{-1}\oad_\al\oD=\sum_{n=0}^\infty \frac{1}{n!} \of_n \stf \mbox{with}\stf 
\left\{\begin{array}{cl}
\of_0&=\oad_\al \\
\of_{n+1}&=\[\oM,\of_n\],
\end{array}\right.
$$
where the $\of_n$ have an implicit label $\al$.
Using Eq.~(\ref{eq:oM}) and the commutation rules for creators and annihilators of fermions, we find
$$\of_n=\sum_\be\(M^n\)_{\be\al} \oad_\be.$$
We finally get
$$\oD^{-1}\oad_\al\oD=\sum_\be\(e^M\)_{\be\al}\oad_\be,$$
leading to Eq.~(\ref{eq:oadD}).
}

Using $\oad_\al\oa_\be=\delta_{\al\be}-\oa_\be\oad_\al$ and eqs.~(\ref{eq:rho_albe}) and~(\ref{eq:oadD}) leads directly to
\oeq
\rho_{\be\al} = \delta_{\be\al} - \Tr\(\oad_\al\oD\oa_\be\) =  \delta_{\be\al} - \sum_\ga \(e^M\)_{\ga\al}\rho_{\be\ga}. 
\ceq
We get $\rho=1-\rho e^M$ and finally Eq.~(\ref{eq:rhoM}).

\subsection*{Trace of the exponential of a one-body operator}
Our goal is to show the relation
\oeq
\boxed{\Tr e^{-\oA}=\det(1+e^{-A})}
\label{eq:e-A}
\ceq
where $\oA$ is a one-body operator.

We first show that
\oeq
\Tr\oB=\int dz^*dz \sdf e^{-z^*z}\<z|\oB|-z\>
\label{eq:TrB}
\ceq
where $\oB$ contains only even powers of $\oad$ and $\oa$ (this is the case, for instance, when $\oB$ is the exponential of a one-body operator), and $z$ is an element of the Grassmann algebra (see appendix~\ref{annexe:Grassmann}). 

{
With $\{|\xi\>\}$ a basis of the Fock space, the trace of $\oB$ reads
$ \Tr\oB=\sum_\xi \<\xi|\oB|\xi\>.$
Inserting the closure relation from Eq.~(\ref{eq:zclosure}), we get
\oeq
\Tr\oB= \int dz^*dz e^{-\mathbf{z}^*\mathbf{z}}\sum_\xi \<\xi|\oB|z\>\<z|\xi\>.
\label{eq:TrB2}
\ceq
From Eq.~(\ref{eq:zAz'}), we have
$$\<\xi|\oB|z\>= B(\overleftarrow{\partial}_\mathbf{z},\mathbf{z})\<\xi|z\>.$$
$\{|\xi\>\}$ can be chosen to be an ensemble of Slater determinants with different particle numbers $N_\xi$:
$$|\xi\>=\(\prod_{i=1}^{N_\xi} \oad_{\xi_i}\) |-\>.$$ 
Using Eq.~(\ref{eq:az}) we get
$$\<\xi|z\>=\(\prod_{i=1}^{N_\xi} z_{\xi_i}\) \<-|z\>=\prod_{i=1}^{N_\xi} z_{\xi_i}.$$
Indeed, from Eq.~(\ref{eq:coh_state}), we have 
$$\<-|z\>=\<-|e^{\sum_\al z_\al\oad_\al}|-\>=1.$$
Then, using Eq.~(\ref{eq:z_alz_be}), we get
$$\<\xi|z\>\,\<z|\xi\> = (-1)^{N_\xi}\<z|\xi\> \<\xi|z\>=\<z|\xi\> \<\xi|-z\>.$$
Using the closure relation $\sum|\xi\>\<\xi|=\hat{1}$, Eq.~(\ref{eq:TrB2}) finally becomes
\oeqn
\Tr\oB&=& \int dz^*dz \sdf e^{-\mathbf{z}^*\mathbf{z}}\, B(\overleftarrow{\partial}_\mathbf{z},\mathbf{z})\,\<z|-z\>\nonumber\\
&=& \int dz^*dz \sdf e^{-\mathbf{z}^*\mathbf{z}}\, \<z|\oB|-z\>,\nonumber
\ceqn
}
i.e., Eq.~(\ref{eq:TrB}).

The second step is to show the relation
\oeq
e^{-\oA}|z\> = |e^{-A}z\>.
\label{eq:eA}
\ceq
{
Similarly to Eq.~(\ref{eq:oadD}), we have
$$\oa_\al e^{-\oA}= e^{-\oA} \sum_\be \(e^{-A}\)_{\al\be}\, \oa_\be $$
and then, using Eq.~(\ref{eq:az}), we get
\oeqn
\oa_\al e^{-\oA}|z\>&=& e^{-\oA} \sum_\be \(e^{-A}\)_{\al\be}\, z_\be |z\> \nonumber \\
&=&  \(e^{-A}\mathbf{z}\)_\al e^{-\oA}|z\> \equiv \oa_\al|e^{-A}z\>.\nonumber
\ceqn
}

Finally, we write Eq.~(\ref{eq:TrB}) with $\oB=e^{-\oA}$ and we use eqs.~(\ref{eq:eA}) and (\ref{eq:zz'}) to get
\oeqn 
\Tr e^{-\oA} &=& \int dz^*dz \sdf e^{-\mathbf{z}^*\mathbf{z}}\<z|-e^{-A}z\>\nonumber \\
&=&\int dz^*dz \sdf e^{-\mathbf{z}^* (1+e^{-A})\mathbf{z}}.\nonumber 
\ceqn
Using Eq.~(\ref{eq:detM}) we obtain our final result in Eq.~(\ref{eq:e-A}).

\subsection*{Trace of the density matrix}

Our goal is to show the relation
\oeq
\boxed{\Tr \oD = e^{-m+\tr [\ln (1+e^{-M})]}}
\label{eq:TrD}
\ceq
where $\Tr$ denotes the trace in the Fock space while $\tr$ denotes the trace of a single-particle matrix.
$\oD$ is the density matrix of an independent particle system, as defined by eqs.~(\ref{eq:D}) and~(\ref{eq:oM}). However we do not assume a normalised state, i.e., $\Tr\oD$ is not necessarily equal to 1. 

From Eq.~(\ref{eq:e-A}), we have
\oeq\Tr \oD= e^{-m} \det (1+e^{-M}) \label{eq:TrD2}.\ceq
We now show the property 
\oeq
\tr \ln (1+A) = \ln \det (1+A) 
\label{eq:trln}
\ceq
where $A$ is a  matrix which can be diagonalised.
{
In its diagonal form, $A_{ij} = a_i \delta_{ij}$, we have
$\tr A^n=\sum_i a_i^n.$
Then we get 
\oeqn
\tr \ln (1+A) &=& \tr (A -\frac{1}{2} A^2 + \frac{1}{3} A^3 - \cdots) \nonumber \\
&=& \sum_i (a_i -\frac{1}{2}  a_i^2 + \frac{1}{3} a_i^3 - \cdots )\nonumber \\
&=&  \sum_i \ln (1+a_i) \nonumber\\
&=& \ln \prod_i (1+a_i) \label{eq:trln1+A}\\
&=&\ln \det (1+A). 
\ceqn}
Taking the exponential of Eq.~(\ref{eq:trln}) with $A=e^{-M}$ and replacing in Eq.~(\ref{eq:TrD2})  gives Eq.~(\ref{eq:TrD}). 
 
 In the particular case where the system is described by a Slater determinant, the state is normalised ($\Tr \oD=1$) and one gets 
 \oeq
 m=\tr \ln (1+e^{-M}).
\label{eq:mz1}
 \ceq
 
 \section{Grassmann algebra \label{annexe:Grassmann}}

A brief introduction to Grassmann algebra can be found in p.~21-27 of Ref.~\cite{bla86}.
Here, we recall mostly the main relations which are used in appendix~\ref{annexe:formula}. 
Similarly to the imaginary number $i$ which has been introduced to satisfy $i^2=-1$, 
the Grassmann algebra has been developed such that its elements vanish when squared.
Elements $\{z_\al\}$ of the Grassmann algebra and their complex conjugate elements obey
\oeq
z_\al^2\equiv{z_\al^*}^2=0 
\label{eq:z_al}
\ceq
\oeq
z_\al z_\be + z_\be z_\al\equiv z_\al z_\be^* + z_\be^* z_\al=0.
\label{eq:z_alz_be}
\ceq
They are eigenvalues of annihilation and creation operators such that
\oeq
\oa_\al |z\>=z_\al|z\> \stf\mbox{ and }\stf  \<z|\oad_\al = \<z|z_\al^*
\label{eq:az}
\ceq
where $|z\>$ and $\<z|$ are coherent states of the form 
\oeq
|z\> = e^{\sum_\al z_\al\oad_\al}|-\> \stf \mbox{ and } \stf \<z|=\<-|e^{\sum_\al z_\al^*\oa_\al},
\label{eq:coh_state}
\ceq
and $|-\>$ is the particle vacuum. 

Elements of the Grassmann algebra are polynomial of second degree at most, as, for example, 
\oeq
P(z_\al,z_\al^*)=P_0+P_1z_\al+P_2z_\al^*+P_{12}z_\al z_\al^*.
\ceq
Derivation is defined  as
\oeqn
\overrightarrow{\partial}_{z_\al}P  &=& P_1+P_{12}z_\al^* \nonumber\\
\overrightarrow{\partial}_{z_\al^*}P  &=& P_2-P_{12}z_\al \nonumber\\
(\mbox{indeed }\overrightarrow{\partial}_{z_\al^*}z_\al z_\al^*&=&-(\overrightarrow{\partial}_{z_\al^*}z_\al^*) z_\al=-z_\al)\nonumber\\
\overleftarrow{\partial}_{z_\al}P  &=& P_1-P_{12}z_\al^* \nonumber\\
 \overleftarrow{\partial}_{z_\al^*}P  &=& P_2+P_{12}z_\al. 
\ceqn
We can show the following properties
\oeq 
\oad_\al|z\>=\overleftarrow{\partial}_{z_\al}|z\> \stf \mbox{ and } \stf \<z|\oa_\al=\overrightarrow{\partial}_{z_\al^*}\<z|.
\ceq
Integration is defined as
\oeq
\int dz_\al = \int dz_\al^* = 0
\label{eq:intdz}
\ceq
and
\oeq
\stf \int dz_\al z_\al= \int dz_\al^* z_\al^* = 1.
\ceq

The overlap of two coherent states reads
\oeq
\<z|z'\>=e^{\sum_\al z_\al^*z_\al'}=e^{\mathbf{z}^*\mathbf{z}'}.
\label{eq:zz'}
\ceq
{
Indeed, using  eqs.~(\ref{eq:z_al}) and (\ref{eq:coh_state}),  and the Wick theorem, we have
\oeqn
\<z|z'\>&=&\<-|\sum_n\frac{1}{n!}\(\sum_\al z_\al^*\oa_\al\)^n\sum_m\frac{1}{m!}\(\sum_\be {z'_\be}\oad_\be\)^m|-\>\nonumber\\
&=& \sum_n\frac{1}{n!^2}\<-|\(\sum_\al z_\al^*\oa_\al\)^n\( \sum_\be z_\be'\oad_\be\)^n|-\>\nonumber\\
&=&1+\sum_{\al\be} z_\al^*\<-|\oa_\al\oad_\be|-\>z_\be'\nonumber\\
&&+\frac{1}{2!^2}\sum_{\al\be\ga\delta}z_\al^*z_\be^*\<-|\oa_\al\oa_\be\oad_\ga\oad_\delta|-\>z'_\ga z'_\delta+\cdots\nonumber\\
&=& \!1\!+\!\sum_{\al} z_\al^*z_\al'+\frac{1}{2!^2}\sum_{\al\be}(z_\al^*z_\be^*z'_\be z'_\al\!-\!z_\al^*z_\be^*z'_\al z'_\be)\!+\!\cdots\nonumber\\
&=& 1+\sum_{\al} z_\al^*z_\al'+\frac{1}{2!}\sum_{\al}z_\al^*z'_\al\sum_\be z_\be^*z'_\be +\cdots\nonumber\\
&=&e^{\sum_\al z_\al^*z'_\al}.\nonumber
\ceqn
}
Equation~(\ref{eq:zz'}) can be used to define the metric entering the following closure relation
\oeq
\int dz^*dz \sdf e^{-\mathbf{z}^*\mathbf{z}} |z\>\, \<z| = \hat{1},
\label{eq:zclosure}
\ceq
where we introduced the notation $ \int dz^*dz  \equiv \int \prod_\al dz_\al^*dz_\al$.

An operator $\oA(\oad,\oa)\equiv \oA(\cdots,\oad_\al,\cdots,\oa_\be,\cdots)$ may be represented by the differential operator $A$ defined as
\oeq
\<z|\oA(\oad,\oa)|z'\>= A(\overleftarrow{\partial}_{\mathbf{z}'},\mathbf{z}')\,\<z|z'\>
\label{eq:zAz'}
\ceq
or
\oeq
\<z|\oA(\oad,\oa)|z'\>= A(\mathbf{z}^*,\overrightarrow{\partial}_{\mathbf{z}^*})\,\<z|z'\>
\ceq
where we used the notation $\overleftarrow{\partial}_\mathbf{z}\equiv\{\cdots,\overleftarrow{\partial}_{z_\al},\cdots\}$ and $\overrightarrow{\partial}_\mathbf{z}\equiv\{\cdots,\overrightarrow{\partial}_{z_\al},\cdots\}$.

We now show the relation
\oeq
\int  dz^*dz\sdf e^{-\mathbf{z}^*M\mathbf{z}} = \det M
\label{eq:detM}
\ceq
{
Starting from the Taylor development
$$
\int  dz^*dz\sdf e^{-\mathbf{z}^*M\mathbf{z}} = \int  dz^*dz \sum_{n=0}^\infty \frac{1}{n!}(-\mathbf{z}^*M\mathbf{z})^n ,
$$
and using Eq.~(\ref{eq:intdz}), we see that the only non-vanishing term of the Taylor development is for $n=N$ where $N$ is the dimension of the Fock space:
$$
\int  dz^*dz\sdf e^{-\mathbf{z}^*M\mathbf{z}} = \frac{(-1)^N}{N!} \int  dz^*dz \sdf (\mathbf{z}^*M\mathbf{z})^N 
$$
with
$$ (\mathbf{z}^*M\mathbf{z})^N=\sum_{\al_1\be_1}z_{\al_1}^*M_{\al_1\be_1}z_{\be_1}\cdots\sum_{\al_N\be_N}z_{\al_N}^*M_{\al_N\be_N}z_{\be_N}.
$$
From Eq.~(\ref{eq:z_al}), we see that all terms with $\al_i=\al_{j\neq i}$ or $\be_i=\be_{j\neq i}$ vanish. It leads to two sums over all possible permutations of the $\al$ and of the $\be$ indices:
\oeqn
 (\mathbf{z}^*M\mathbf{z})^N=\sum_{\al,\be\in Perm_N}&&z_{\al(1)}^*M_{\al(1)\be(1)}z_{\be(1)}\cdots \nonumber\\
&&z_{\al(N)}^*M_{\al(N)\be(N)}z_{\be(N)}.\nonumber
\ceqn
Noting the following relations:
\oeqn
\int dz^*dz&=&(-1)^{N(N-1)/2} \nonumber\\
&&\int dz_1^*\cdots dz_N^* dz_1 \cdots dz_N,\nonumber \\
z^*_{\al(1)} z_{\be(1)} \cdots z^*_{\al(N)} z_{\be(N)}
&=& (-1)^{N(N+1)/2}\nonumber\\
&&z_{\be(1)}\cdots z_{\be(N)}z_{\al(1)}^*\cdots z_{\al(N)}^*,\nonumber\\
z_{\be(1)}\cdots z_{\be(N)} &=& (-1)^{N(N-1)/2} \mbox{sign}(\be) z_N\cdots z_1,\nonumber
\ceqn
and the similar equation for the $z_\al^*$, we get
$$
\int \sdb dz^*dz\sdf {z_{\al(1)}}^*z_{\be(1)}\cdots {z_{\al(N)}}^*z_{\be(N)}=(-1)^N \mbox{sign}(\al) \mbox{sign}(\be) 
$$
 where sign$(\al)$ is the sign of the permutation $\al$, i.e., +1 for an even number of permutations and -1 for an odd one. This leads to Eq.~(\ref{eq:detM})
\oeqn
 \int  dz^*dz e^{-\mathbf{z}^*M\mathbf{z}} &=& \frac{1}{N!}\sum_{\al,\be\in Perm_N} \mbox{sign}(\al) \mbox{sign}(\be)  \nonumber\\
&& \stf\stf\stf\stf\stf\stf\stf\stf\stf\stf\stf\stf M_{\al(1)\be(1)}\cdots M_{\al(N)\be(N)} \nonumber \\
&=&  \sum_{\al\in Perm_N} \mbox{sign}(\al)  M_{\al(1)1}\cdots M_{\al(N)N} \nonumber \\
&=& \det M.\nonumber
  \ceqn
}

}

\bibliographystyle{epj}
\bibliography{biblio}

\begin{thebibliography}{553}

\bibitem{simenel2012}
C.~Simenel, Eur. Phys. J. A \textbf{48}, 152 (2012)

\bibitem{nakatsukasa2016}
T.~Nakatsukasa, K.~Matsuyanagi, M.~Matsuo, K.~Yabana, Rev. Mod. Phys.
  \textbf{88}, 045004 (2016)

\bibitem{simenel2018}
C.~Simenel, A.S. Umar, Prog. Part. Nucl. Phys. \textbf{103}, 19 (2018)

\bibitem{washiyama2020}
K.~Washiyama, K.~Sekizawa, Front. Phys. \textbf{8}, 93 (2020)

\bibitem{lacroix2015}
D.~Lacroix, Y.~Tanimura, G.~Scamps, C.~Simenel, Int. J. Mod. Phys. E
  \textbf{24}, 1541005 (2015)

\bibitem{bulgac2020}
A.~Bulgac, S.~Jin, I.~Stetcu, Front. Phys. \textbf{8}, 63 (2020)

\bibitem{lacroix2014}
D.~Lacroix, S.~Ayik, Eur. Phys. J. A \textbf{50}, 95 (2014)

\bibitem{tohyama2020}
M.~Tohyama, Front. Phys. \textbf{8}, 67 (2020)

\bibitem{sekizawa2019}
{Kazuyuki Sekizawa}, Front. Phys. \textbf{7}, 20 (2019)

\bibitem{godbey2020}
K.~Godbey, A.S. Umar, Front. Phys. \textbf{8}, 40 (2020)

\bibitem{stevenson2019}
P.D. Stevenson, M.C. Barton, Prog. Part. Nucl. Phys. \textbf{104}, 142 (2019)

\bibitem{sun2022c}
X.X. Sun, L.~Guo, Commun. Theor. Phys. \textbf{74}, 097302 (2022)

\bibitem{magierski2024}
{Magierski, Piotr}, {Barresi, Andrea}, {Makowski, Andrzej}, {Pcak, Daniel},
  {Wlazłowski, Gabriel}, Eur. Phys. J. A \textbf{60}, 186 (2024)

\bibitem{colonna2020}
M.~Colonna, Prog. Part. Nucl. Phys. \textbf{113}, 103775 (2020)

\bibitem{neg98}
J.W. Negele, H.~Orland, \emph{Quantum Many-Particle Systems (Advanced Books
  Classics)} (Westview Press, Boulder, CO, 1998), ISBN 0738200522

\bibitem{bla86}
J.P. Blaizot, G.~Ripka, \emph{Quantum Theory of Finite Systems} (MIT Press,
  1986)

\bibitem{sim10a}
C.~Simenel, D.~Lacroix, B.~Avez, \emph{Quantum Many-Body Dynamics: Applications
  to Nuclear Reactions} (VDM Verlag, Sarrebruck, Germany, 2010),
  \texttt{arXiv:0806.2714}

\bibitem{lac04}
D.~Lacroix, S.~Ayik, P.~Chomaz, Prog. Part. Nucl. Phys. \textbf{52}, 497 (2004)

\bibitem{dir30}
P.A.M. Dirac, Proc. Camb. Phil. Soc. \textbf{26}, 376 (1930)

\bibitem{lic76}
P.C. Lichtner, J.J. Griffin, Phys. Rev. Lett. \textbf{37}, 1521 (1976)

\bibitem{bla81}
J.~Blaizot, G.~Ripka, Phys. Lett. B \textbf{105}, 1 (1981)

\bibitem{bog46}
N.N. Bogoliubov, J. Phys. (URSS) \textbf{10}, 256 (1946)

\bibitem{bor46}
H.~Born, H.S. Green, Proc. Royl. Soc. \textbf{A188}, 10 (1946)

\bibitem{kir46}
J.G. Kirkwood, J. Chem. Phys. \textbf{14}, 180 (1946)

\bibitem{cas90}
W.~Cassing, U.~Mosel, Prog. Part. Nucl. Phys. \textbf{25}, 235 (1990)

\bibitem{bla92}
F.V. De~Blasio, W.~Cassing, M.~Tohyama, P.F. Bortignon, R.A. Broglia, Phys.
  Rev. Lett. \textbf{68}, 1663 (1992)

\bibitem{luo99}
H.G. Luo, W.~Cassing, S.J. Wang, Nucl. Phys. A \textbf{652}, 164  (1999)

\bibitem{toh01}
M.~Tohyama, Phys. Rev. C \textbf{64}, 067304 (2001)

\bibitem{toh02a}
M.~Tohyama, A.S. Umar, Phys. Lett. B \textbf{549}, 72 (2002)

\bibitem{toh02b}
M.~Tohyama, A.S. Umar, Phys. Rev. C \textbf{65}, 037601 (2002)

\bibitem{ass09}
M.~{Assie}, D.~{Lacroix}, Phys. Rev. Lett. \textbf{102}, 202501 (2009)

\bibitem{tohyama2016}
M.~Tohyama, A.S. Umar, Phys. Rev. C \textbf{93}, 034607 (2016)

\bibitem{lackner2015}
F.~Lackner, I.~B{\v{r}}ezinov{\'{a}}, T.~Sato, K.L. Ishikawa, J.~Burgd\"orfer,
  Phys. Rev. A \textbf{91}, 023412 (2015)

\bibitem{lackner2017}
F.~Lackner, I.~B{\v{r}}ezinov{\'{a}}, T.~Sato, K.L. Ishikawa, J.~Burgd\"orfer,
  Phys. Rev. A \textbf{95}, 033414 (2017)

\bibitem{wen2018}
K.~Wen, M.C. Barton, A.~Rios, P.D. Stevenson, Phys. Rev. C \textbf{98}, 014603
  (2018)

\bibitem{barton2021}
M.~Barton, P.~Stevenson, A.~Rios, Phys. Rev. C \textbf{103}, 064304 (2021)

\bibitem{kru85}
H.~Kruse, B.V. Jacak, J.J. Molitoris, G.D. Westfall, H.~St\"ocker, Phys. Rev. C
  \textbf{31}, 1770 (1985)

\bibitem{aic85}
J.~Aichelin, G.~Bertsch, Phys. Rev. C \textbf{31}, 1730 (1985)

\bibitem{dirac1930}
P.A.M. Dirac, Math. Proc. Camb. Phil. Soc. \textbf{26}, 376 (1930)

\bibitem{bal81}
R.~Balian, M.~V\'en\'eroni, Phys. Rev. Lett. \textbf{47}, 1353 (1981)

\bibitem{koo77}
S.E. Koonin, K.T.R. Davies, V.~Maruhn-Rezwani, H.~Feldmeier, S.J. Krieger, J.W.
  Negele, Phys. Rev. C \textbf{15}, 1359 (1977)

\bibitem{dav78}
K.T.R. Davies, V.~Maruhn-Rezwani, S.E. Koonin, J.W. Negele, Phys. Rev. Lett.
  \textbf{41}, 632 (1978)

\bibitem{das79}
C.H. Dasso, T.~D{\o}ssing, H.C. Pauli, Z. Phys. A \textbf{289}, 395 (1979)

\bibitem{bal84}
R.~Balian, M.~V\'en\'eroni, Phys. Lett. B \textbf{136}, 301 (1984)

\bibitem{bal92}
R.~Balian, M.~V\'en\'eroni, Ann. Phys. \textbf{216}, 351 (1992)

\bibitem{tro85}
T.~Troudet, D.~Vautherin, Phys. Rev. C \textbf{31}, 278 (1985)

\bibitem{mar85}
J.B. Marston, S.E. Koonin, Phys. Rev. Lett. \textbf{54}, 1139 (1985)

\bibitem{bon85}
P.~Bonche, H.~Flocard, Nucl. Phys. A \textbf{437}, 189 (1985)

\bibitem{zie88}
M.~Zielinska-Pfab\'e, C.~Gr\'egoire, Phys. Rev. C \textbf{37}, 2594 (1988)

\bibitem{bro08}
J.M.A. Broomfield, P.D. Stevenson, J. Phys. G \textbf{35}, 095102 (2008)

\bibitem{bro09}
J.M.A. Broomfield, Ph.D. thesis, University of Surrey (2009)

\bibitem{sim11}
C.~Simenel, Phys. Rev. Lett. \textbf{106}, 112502 (2011)

\bibitem{scamps2015a}
G.~Scamps, C.~Simenel, D.~Lacroix, Phys. Rev. C \textbf{92}, 011602({R}) (2015)

\bibitem{williams2018}
E.~Williams, K.~Sekizawa, D.J. Hinde, C.~Simenel, M.~Dasgupta, I.P. Carter,
  K.J. Cook, D.Y. Jeung, S.D. McNeil, C.S. Palshetkar et~al., Phys. Rev. Lett.
  \textbf{120}, 022501 (2018)

\bibitem{godbey2020b}
K.~Godbey, C.~Simenel, A.S. Umar, Phys. Rev. C \textbf{101}, 034602 (2020)

\bibitem{gao2025}
Z.~Gao, K.~Sekizawa, L.~Zhu, \emph{Time-dependent random phase approximation
  for particle-number fluctuations and correlations in deep-inelastic
  collisions of $^{144}$sm+$^{144}$sm and $^{154}$sm+$^{154}$sm} (2025),
  \texttt{https://arxiv.org/abs/2504.09436}

\bibitem{mar91}
C.~Martin, D.~Vautherin, Phys. Lett. B \textbf{260}, 1 (1991)

\bibitem{mar95}
C.~Martin, Phys. Rev. D \textbf{52}, 7121 (1995)

\bibitem{mar99}
C.~Martin, Ann. Phys. \textbf{278}, 202 (1999)

\bibitem{ben99}
M.~Benarous, H.~Flocard, Ann. Phys. \textbf{273}, 242 (1999)

\bibitem{bou10}
A.~Boudjem{\^a}a, M.~Benarous, Eur. Phys. J. D \textbf{59}, 427 (2010)

\bibitem{flo82}
H.~Flocard, in \emph{Time-Dependent \text{Hartree-Fock} and Beyond}, edited by
  K.~Goeke, P.G. Reinhard (Springer-Verlag, Berlin/New York, 1982)

\bibitem{bal85}
R.~Balian, M.~V\'en\'eroni, Ann. Phys. \textbf{164}, 334 (1985)

\bibitem{ayi08}
S.~Ayik, Phys. Lett. B \textbf{658}, 174 (2008)

\bibitem{ayi09}
S.~{Ayik}, K.~{Washiyama}, D.~{Lacroix}, Phys. Rev. C \textbf{79}, 054606
  (2009)

\bibitem{was09b}
K.~{Washiyama}, S.~{Ayik}, D.~{Lacroix}, Phys. Rev. C \textbf{80}, 031602
  (2009)

\bibitem{yil11}
B.~{Yilmaz}, S.~{Ayik}, D.~{Lacroix}, K.~{Washiyama}, Phys. Rev. C \textbf{83},
  064615 (2011)

\bibitem{ayik2015}
S.~Ayik, O.~Yilmaz, B.~Yilmaz, A.S. Umar, A.~Gokalp, G.~Turan, D.~Lacroix,
  Phys. Rev. C \textbf{91}, 054601 (2015)

\bibitem{ayik2016}
S.~Ayik, O.~Yilmaz, B.~Yilmaz, A.S. Umar, Phys. Rev. C \textbf{94}, 044624
  (2016)

\bibitem{tanimura2017}
Y.~Tanimura, D.~Lacroix, S.~Ayik, Phys. Rev. Lett. \textbf{118}, 152501 (2017)

\bibitem{sekizawa2020}
K.~Sekizawa, S.~Ayik, Phys. Rev. C \textbf{102}, 014620 (2020)

\bibitem{ayik2020}
S.~Ayik, B.~Yilmaz, O.~Yilmaz, A.S. Umar, Phys. Rev. C \textbf{102}, 024619
  (2020)

\bibitem{ayik2020b}
S.~Ayik, K.~Sekizawa, Phys. Rev. C \textbf{102}, 064619 (2020)

\bibitem{ayik2021}
S.~Ayik, M.~Arik, E.C. Karanfil, O.~Yilmaz, B.~Yilmaz, A.S. Umar, Phys. Rev. C
  \textbf{104}, 054614 (2021)

\bibitem{arik2023}
M.~Arik, S.~Ayik, O.~Yilmaz, A.S. Umar, Phys. Rev. C \textbf{108}, 064604
  (2023)

\bibitem{ayik2023a}
S.~Ayik, M.~Arik, O.~Yilmaz, B.~Yilmaz, A.S. Umar, Phys. Rev. C \textbf{107},
  014609 (2023)

\bibitem{ayik2023b}
S.~Ayik, M.~Arik, E.~Erbayri, O.~Yilmaz, A.S. Umar, Phys. Rev. C \textbf{108},
  054605 (2023)

\bibitem{lacroix2012}
D.~Lacroix, S.~Ayik, B.~Yilmaz, Phys. Rev. C \textbf{85}, 041602 (2012)

\bibitem{lacroix2013}
D.~Lacroix, D.~Gambacurta, S.~Ayik, Phys. Rev. C \textbf{87}, 061302 (2013)

\bibitem{lacroix2014b}
D.~Lacroix, S.~Hermanns, C.M. Hinz, M.~Bonitz, Phys. Rev. B \textbf{90}, 125112
  (2014)

\bibitem{yilmaz2014a}
B.~Yilmaz, D.~Lacroix, R.~Curebal, Phys. Rev. C \textbf{90}, 054617 (2014)

\bibitem{lacroix2016}
D.~Lacroix, Y.~Tanimura, S.~Ayik, B.~Yilmaz, Eur. Phys. J. A \textbf{52}, 94
  (2016)

\bibitem{ulgen2019}
I.~Ulgen, B.~Yilmaz, D.~Lacroix, Phys. Rev. C \textbf{100}, 054603 (2019)

\bibitem{lacroix2022}
D.~Lacroix, A.B. Balantekin, M.J. Cervia, A.V. Patwardhan, P.~Siwach, Phys.
  Rev. D \textbf{106}, 123006 (2022)

\bibitem{lacroix2024}
D.~Lacroix, A.~Bauge, B.~Yilmaz, M.~Mangin-Brinet, A.~Roggero, A.B. Balantekin,
  Phys. Rev. D \textbf{110}, 103027 (2024)

\bibitem{czuba2020}
{Czuba, Thomas}, {Lacroix, Denis}, {Regnier, David}, {Ulgen, Ibrahim}, {Yilmaz,
  Bulent}, Eur. Phys. J. A \textbf{56}, 111 (2020)

\bibitem{may50}
M.G. Mayer, Phys. Rev. \textbf{78}, 22 (1950)

\bibitem{dea03}
D.J. Dean, M.~Hjorth-Jensen, Rev. Mod. Phys. \textbf{75}, 607 (2003)

\bibitem{bar57}
J.~Bardeen, L.N. Cooper, J.R. Schrieffer, Phys. Rev. \textbf{108}, 1175 (1957)

\bibitem{rin80}
P.~Ring, P.~Schuck, \emph{The Nuclear Many-Body Problem} (Springer Verlag,
  1980)

\bibitem{blocki1976}
J.~Blocki, H.~Flocard, Nucl. Phys. A \textbf{273}, 45 (1976)

\bibitem{reinhard1997}
P.G. Reinhard, M.~Bender, K.~Rutz, J.A. Maruhn, Z. Phys. A \textbf{358}, 277
  (1997)

\bibitem{eba10}
S.~Ebata, T.~Nakatsukasa, T.~Inakura, K.~Yoshida, Y.~Hashimoto, K.~Yabana,
  Phys. Rev. C \textbf{82}, 034306 (2010)

\bibitem{sca12}
G.~Scamps, D.~Lacroix, G.~Bertsch, K.~Washiyama, Phys. Rev. C \textbf{85},
  034328 (2012)

\bibitem{scamps2013a}
G.~Scamps, D.~Lacroix, Phys. Rev. C \textbf{87}, 014605 (2013)

\bibitem{tanimura2015}
Y.~Tanimura, D.~Lacroix, G.~Scamps, Phys. Rev. C \textbf{92}, 034601 (2015)

\bibitem{bon76}
P.~Bonche, S.~Koonin, J.W. Negele, Phys. Rev. C \textbf{13}, 1226 (1976)

\bibitem{neg82}
J.W. Negele, Rev. Mod. Phys. \textbf{54}, 913 (1982)

\bibitem{sky56}
T.~Skyrme, Phil. Mag. \textbf{1}, 1043 (1956)

\bibitem{has12}
{Hashimoto, Yukio}, Eur. Phys. J. A \textbf{48}, 55 (2012)

\bibitem{hashimoto2013}
Y.~Hashimoto, Phys. Rev. C \textbf{88}, 034307 (2013)

\bibitem{hashimoto2016}
Y.~Hashimoto, G.~Scamps, Phys. Rev. C \textbf{94}, 014610 (2016)

\bibitem{scamps2017b}
G.~Scamps, Y.~Hashimoto, Phys. Rev. C \textbf{96}, 031602 (2017)

\bibitem{scamps2019b}
G.~Scamps, Y.~Hashimoto, Phys. Rev. C \textbf{100}, 024623 (2019)

\bibitem{ren2020a}
Z.~Ren, P.~Zhao, J.~Meng, Phys. Lett. B \textbf{801}, 135194 (2020)

\bibitem{ren2020b}
Z.X. Ren, P.W. Zhao, J.~Meng, Phys. Rev. C \textbf{102}, 044603 (2020)

\bibitem{ren2022}
Z.X. Ren, J.~Zhao, D.~Vretenar, T.~Nik\v{s}i\'{c}, P.W. Zhao, J.~Meng, Phys.
  Rev. C \textbf{105}, 044313 (2022)

\bibitem{ren2022b}
Z.X. Ren, D.~Vretenar, T.~Nik\v{s}i\'{c}, P.W. Zhao, J.~Zhao, J.~Meng, Phys.
  Rev. Lett. \textbf{128}, 172501 (2022)

\bibitem{li2023}
B.~Li, D.~Vretenar, Z.X. Ren, T.~Nik\ifmmode \check{s}\else
  \v{s}\fi{}i\ifmmode~\acute{c}\else \'{c}\fi{}, J.~Zhao, P.W. Zhao, J.~Meng,
  Phys. Rev. C \textbf{107}, 014303 (2023)

\bibitem{zhang2024a}
D.D. Zhang, D.~Vretenar, T.~Nik\ifmmode \check{s}\else
  \v{s}\fi{}i\ifmmode~\acute{c}\else \'{c}\fi{}, P.W. Zhao, J.~Meng, Phys. Rev.
  C \textbf{109}, 024614 (2024)

\bibitem{zhang2024b}
D.D. Zhang, B.~Li, D.~Vretenar, T.~Nik\ifmmode \check{s}\else
  \v{s}\fi{}i\ifmmode~\acute{c}\else \'{c}\fi{}, Z.X. Ren, P.W. Zhao, J.~Meng,
  Phys. Rev. C \textbf{109}, 024316 (2024)

\bibitem{li2024}
B.~Li, D.~Vretenar, T.~Nik{\v s}i{\'c}, J.~Zhao, P.W. Zhao, J.~Meng, Front.
  Phys. \textbf{19}, 44201 (2024)

\bibitem{cha98}
E.~Chabanat, P.~Bonche, P.~Haensel, J.~Meyer, R.~Schaeffer, Nucl. Phys. A
  \textbf{635}, 231 (1998)

\bibitem{eng75}
Y.~Engel, D.~Brink, K.~Goeke, S.~Krieger, D.~Vautherin, Nucl. Phys. A
  \textbf{249}, 215  (1975)

\bibitem{bon87}
P.~Bonche, H.~Flocard, P.~Heenen, Nucl. Phys. A \textbf{467}, 115 (1987)

\bibitem{uma06a}
A.S. Umar, V.E. Oberacker, Phys. Rev. C \textbf{73}, 054607 (2006)

\bibitem{mar06}
J.A. {Maruhn}, P.G. {Reinhard}, P.D. {Stevenson}, M.R. {Strayer}, Phys. Rev. C
  \textbf{74}, 027601 (2006)

\bibitem{loe12}
N.~Loebl, A.S. Umar, J.A. Maruhn, P.G. Reinhard, P.D. Stevenson, V.E.
  Oberacker, Phys. Rev. C \textbf{86}, 024608 (2012)

\bibitem{fracasso2012}
S.~Fracasso, E.B. Suckling, P.D. Stevenson, Phys. Rev. C \textbf{86}, 044303
  (2012)

\bibitem{dai2014a}
G.~Dai, L.~Guo, E.~Zhao, S.~Zhou, Sci. China Phys. \textbf{57}, 1618 (2014)

\bibitem{stevenson2016}
P.D. Stevenson, E.B. Suckling, S.~Fracasso, M.C. Barton, A.S. Umar, Phys. Rev.
  C \textbf{93}, 054617 (2016)

\bibitem{bender2003}
M.~Bender, P.H. Heenen, P.G. Reinhard, Rev. Mod. Phys. \textbf{75}, 121 (2003)

\bibitem{bulgac1999}
{Aurel Bulgac}, arXiv:nucl-th/9907088  (1999)

\bibitem{bulgac2002b}
{Aurel Bulgac}, {Yongle Yu}, Phys. Rev. Lett. \textbf{88}, 042504 (2002)

\bibitem{dobaczewski1996}
J.~Dobaczewski, W.~Nazarewicz, T.R. Werner, J.F. Berger, C.R. Chinn,
  J.~Decharge, Phys. Rev. C \textbf{53}, 2809 (1996)

\bibitem{kim97}
K.H. Kim, T.~Otsuka, P.~Bonche, J. Phys. G \textbf{23}, 1267 (1997)

\bibitem{mar05}
J.A. Maruhn, P.G. Reinhard, P.D. Stevenson, J.R. Stone, M.R. Strayer, Phys.
  Rev. C \textbf{71}, 064328 (2005)

\bibitem{maruhn2014}
J.A. Maruhn, P.G. Reinhard, P.D. Stevenson, A.S. Umar, Comput. Phys. Commun.
  \textbf{185}, 2195 (2014)

\bibitem{schuetrumpf2018}
B.~Schuetrumpf, P.G. Reinhard, P.D. Stevenson, A.S. Umar, J.A. Maruhn, Comput.
  Phys. Commun. \textbf{229}, 211 (2018)

\bibitem{abhishek2024}
Abhishek, P.~Stevenson, Y.~Shi, E.~Yüksel, A.~Umar, Comp. Phys. Commun.
  \textbf{301}, 109239 (2024)

\bibitem{shi2024}
Y.~Shi, P.D. Stevenson, N.~Hinohara, \emph{A program for 3d nuclear static and
  time-dependent density-functional theory with full skyrme energy density
  functional: Hit3d} (2024), \texttt{2403.12539}

\bibitem{giuliani2023}
S.A. Giuliani, L.M. Robledo, Eur. Phys. J. A \textbf{59}, 301 (2023)

\bibitem{solermiras2025}
D.~Soler~Miras, A.~Rios, PoS \textbf{QNP2024}, 228 (2025)

\bibitem{uma05}
A.S. Umar, V.E. Oberacker, Phys. Rev. C \textbf{71}, 034314 (2005)

\bibitem{nak05}
T.~Nakatsukasa, K.~Yabana, Phys. Rev. C \textbf{71}, 024301 (2005)

\bibitem{sekizawa2013}
{Kazuyuki Sekizawa}, {Kazuhiro Yabana}, Phys. Rev. C \textbf{88}, 014614 (2013)

\bibitem{jin2021}
S.~Jin, K.J. Roche, I.~Stetcu, I.~Abdurrahman, A.~Bulgac, Comp. Phys. Commun.
  \textbf{269}, 108130 (2021)

\bibitem{sim03}
C.~Simenel, P.~Chomaz, Phys. Rev. C \textbf{68}, 024302 (2003)

\bibitem{cho04}
P.~Chomaz, C.~Simenel, Nucl. Phys. A \textbf{731}, 188 (2004)

\bibitem{ste04}
P.D. Stevenson, M.R. Strayer, J.~Rikovska~Stone, W.G. Newton, Int. J. Mod.
  Phys. E \textbf{13}, 181 (2004)

\bibitem{rei07}
P.G. Reinhard, L.~Guo, J.A. Maruhn, Eur. Phys. J. A \textbf{32}, 19 (2007)

\bibitem{sim09}
C.~Simenel, P.~Chomaz, Phys. Rev. C \textbf{80}, 064309 (2009)

\bibitem{ste11}
I.~Stetcu, A.~Bulgac, P.~Magierski, K.J. Roche, Phys. Rev. C \textbf{84},
  051309 (2011)

\bibitem{avez2013}
B.~Avez, C.~Simenel, Eur. Phys. J. A \textbf{49}, 76 (2013)

\bibitem{scamps2013b}
G.~Scamps, D.~Lacroix, Phys. Rev. C \textbf{88}, 044310 (2013)

\bibitem{pardi2013}
C.I. Pardi, P.D. Stevenson, Phys. Rev. C \textbf{87}, 014330 (2013)

\bibitem{goddard2013}
P.M. Goddard, N.~Cooper, V.~Werner, G.~Rusev, P.D. Stevenson, A.~Rios,
  C.~Bernards, A.~Chakraborty, B.P. Crider, J.~Glorius et~al., Phys. Rev. C
  \textbf{88}, 064308 (2013)

\bibitem{scamps2014a}
G.~Scamps, D.~Lacroix, Phys. Rev. C \textbf{89}, 034314 (2014)

\bibitem{stetcu2015}
I.~Stetcu, C.A. Bertulani, A.~Bulgac, P.~Magierski, K.J. Roche, Phys. Rev.
  Lett. \textbf{114}, 012701 (2015)

\bibitem{burrello2019}
S.~Burrello, M.~Colonna, G.~Col\`o, D.~Lacroix, X.~Roca-Maza, G.~Scamps,
  H.~Zheng, Phys. Rev. C \textbf{99}, 054314 (2019)

\bibitem{shi2020}
Y.~Shi, N.~Hinohara, B.~Schuetrumpf, Phys. Rev. C \textbf{102}, 044325 (2020)

\bibitem{aitbenmennana2020}
A.A.B. Mennana, Y.E. Bassem, M.~Oulne, Physica Scripta \textbf{95}, 065301
  (2020)

\bibitem{aitbenmennana2021}
A.~Ait Ben~Mennana, M.~Oulne, The European Physical Journal Plus \textbf{136},
  85 (2021)

\bibitem{li2023b}
B.~Li, D.~Vretenar, T.~Nik\ifmmode \check{s}\else
  \v{s}\fi{}i\ifmmode~\acute{c}\else \'{c}\fi{}, P.W. Zhao, J.~Meng, Phys. Rev.
  C \textbf{108}, 014321 (2023)

\bibitem{aitbenmennana2023}
A.~{Ait Ben Mennana}, M.~Oulne, Nucl. Phys. A \textbf{1034}, 122644 (2023)

\bibitem{marevic2023}
P.~Marevi\ifmmode~\acute{c}\else \'{c}\fi{}, D.~Regnier, D.~Lacroix, Phys. Rev.
  C \textbf{108}, 014620 (2023)

\bibitem{shi2023}
Y.~Shi, P.D. Stevenson, Chin. Phys. C \textbf{47}, 034105 (2023)

\bibitem{simenel2013a}
C.~Simenel, R.~Keser, A.S. Umar, V.E. Oberacker, Phys. Rev. C \textbf{88},
  024617 (2013)

\bibitem{simenel2013b}
C.~Simenel, M.~Dasgupta, D.J. Hinde, E.~Williams, Phys. Rev. C \textbf{88},
  064604 (2013)

\bibitem{simenel2016b}
{Simenel, C.}, {Buete, J.}, {Vo-Phuoc, K.}, EPJ Web Conf. \textbf{123}, 01004
  (2016)

\bibitem{vophuoc2016}
K.~Vo-Phuoc, C.~Simenel, E.C. Simpson, Phys. Rev. C \textbf{94}, 024612 (2016)

\bibitem{simenel2014a}
C.~Simenel, A.S. Umar, Phys. Rev. C \textbf{89}, 031601(R) (2014)

\bibitem{goddard2015}
P.M. Goddard, P.D. Stevenson, A.~Rios, Phys. Rev. C \textbf{92}, 054610 (2015)

\bibitem{goddard2016}
P.M. Goddard, P.D. Stevenson, A.~Rios, Phys. Rev. C \textbf{93}, 014620 (2016)

\bibitem{bulgac2016}
{Aurel Bulgac}, {Piotr Magierski}, {Kenneth J. Roche}, {Ionel Stetcu}, Phys.
  Rev. Lett. \textbf{116}, 122504 (2016)

\bibitem{scamps2018}
G.~Scamps, C.~Simenel, Nature \textbf{564}, 382 (2018)

\bibitem{scamps2019}
G.~Scamps, C.~Simenel, Phys. Rev. C \textbf{100}, 041602({R}) (2019)

\bibitem{bulgac2019}
A.~Bulgac, S.~Jin, I.~Stetcu, Phys. Rev. C \textbf{100}, 014615 (2019)

\bibitem{bulgac2019b}
A.~Bulgac, S.~Jin, K.J. Roche, N.~Schunck, I.~Stetcu, Phys. Rev. C
  \textbf{100}, 034615 (2019)

\bibitem{pancic2020}
M.~Pancic, Y.~Qiang, J.~Pei, P.~Stevenson, Front. Phys. \textbf{8}, 351 (2020)

\bibitem{bulgac2021}
A.~Bulgac, I.~Abdurrahman, S.~Jin, K.~Godbey, N.~Schunck, I.~Stetcu, Phys. Rev.
  Lett. \textbf{126}, 142502 (2021)

\bibitem{qiang2021a}
Y.~Qiang, J.C. Pei, P.D. Stevenson, Phys. Rev. C \textbf{103}, L031304 (2021)

\bibitem{qiang2021b}
Y.~Qiang, J.C. Pei, Phys. Rev. C \textbf{104}, 054604 (2021)

\bibitem{bulgac2022}
A.~Bulgac, I.~Abdurrahman, K.~Godbey, I.~Stetcu, Phys. Rev. Lett. \textbf{128},
  022501 (2022)

\bibitem{scamps2022}
G.~Scamps, Phys. Rev. C \textbf{106}, 054614 (2022)

\bibitem{iwata2022}
Y.~Iwata, T.~Nishikawa, Phys. Rev. C \textbf{105}, 044603 (2022)

\bibitem{scamps2024a}
G.~Scamps, Phys. Rev. C \textbf{109}, L011602 (2024)

\bibitem{tong2022b}
L.~Tong, S.~Yan, Phys. Rev. C \textbf{106}, 044611 (2022)

\bibitem{abdurrahman2024}
I.~Abdurrahman, M.~Kafker, A.~Bulgac, I.~Stetcu, Phys. Rev. Lett. \textbf{132},
  242501 (2024)

\bibitem{huang2024a}
Y.~Huang, X.X. Sun, L.~Guo, Eur. Phys. J. A \textbf{60}, 100 (2024)

\bibitem{huang2024b}
Y.~Huang, X.X. Sun, L.~Guo, Phys. Rev. C \textbf{110}, 064318 (2024)

\bibitem{seb09}
F.~S\'ebille, S.~Figerou, V.~de~la Mota, Nucl. Phys. A \textbf{822}, 51 (2009)

\bibitem{seb11}
F.~S\'ebille, V.~de~la Mota, S.~Figerou, Phys. Rev. C \textbf{84}, 055801
  (2011)

\bibitem{schuetrumpf2013}
B.~Schuetrumpf, M.A. Klatt, K.~Iida, J.A. Maruhn, K.~Mecke, P.G. Reinhard,
  Phys. Rev. C \textbf{87}, 055805 (2013)

\bibitem{schuetrumpf2014}
B.~Schuetrumpf, K.~Iida, J.A. Maruhn, P.G. Reinhard, Phys. Rev. C \textbf{90},
  055802 (2014)

\bibitem{schuetrumpf2015a}
B.~Schuetrumpf, W.~Nazarewicz, Phys. Rev. C \textbf{92}, 045806 (2015)

\bibitem{schuetrumpf2015b}
B.~Schuetrumpf, M.A. Klatt, K.~Iida, G.E. Schr\"oder-Turk, J.A. Maruhn,
  K.~Mecke, P.G. Reinhard, Phys. Rev. C \textbf{91}, 025801 (2015)

\bibitem{wlazlowski2016}
G.~Wlaz\l{}owski, K.~Sekizawa, P.~Magierski, A.~Bulgac, M.M. Forbes, Phys. Rev.
  Lett. \textbf{117}, 232701 (2016)

\bibitem{sekizawa2022}
K.~Sekizawa, S.~Kobayashi, M.~Matsuo, Phys. Rev. C \textbf{105}, 045807 (2022)

\bibitem{yoshimura2024}
K.~Yoshimura, K.~Sekizawa, \emph{Superfluid extension of the self-consistent
  time-dependent band theory for neutron star matter: Anti-entrainment versus
  superfluid effects in the slab phase} (2024), \texttt{2306.03327}

\bibitem{sim01}
C.~Simenel, P.~Chomaz, G.~de~France, Phys. Rev. Lett. \textbf{86}, 2971 (2001)

\bibitem{sim04}
C.~Simenel, P.~Chomaz, G.~de~France, Phys. Rev. Lett. \textbf{93}, 102701
  (2004)

\bibitem{uma06b}
A.S. Umar, V.E. Oberacker, Phys. Rev. C \textbf{74}, 021601 (2006)

\bibitem{uma06c}
A.S. Umar, V.E. Oberacker, Phys. Rev. C \textbf{74}, 024606 (2006)

\bibitem{uma06d}
A.S. Umar, V.E. Oberacker, Phys. Rev. C \textbf{74}, 061601 (2006)

\bibitem{guo07}
L.~Guo, J.A. Maruhn, P.G. Reinhard, Phys. Rev. C \textbf{76}, 014601 (2007)

\bibitem{sim07}
C.~Simenel, P.~Chomaz, G.~de~France, Phys. Rev. C \textbf{76}, 024609 (2007)

\bibitem{uma07}
A.S. Umar, V.E. Oberacker, Phys. Rev. C \textbf{76}, 014614 (2007)

\bibitem{sim08}
C.~Simenel, B.~Avez, Int. J. Mod. Phys. E \textbf{17}, 31 (2008)

\bibitem{uma08b}
A.S. Umar, V.E. Oberacker, Phys. Rev. C \textbf{77}, 064605 (2008)

\bibitem{was08}
K.~{Washiyama}, D.~{Lacroix}, Phys. Rev. C \textbf{78}, 024610 (2008)

\bibitem{uma09a}
A.S. Umar, V.E. Oberacker, J.A. Maruhn, P.G. Reinhard, Phys. Rev. C
  \textbf{80}, 041601 (2009)

\bibitem{uma09b}
A.S. Umar, V.E. Oberacker, Eur. Phys. J. A \textbf{39}, 243 (2009)

\bibitem{uma09c}
A.S. Umar, V.E. Oberacker, J. Phys. G \textbf{36}, 025101 (2009)

\bibitem{was09a}
K.~{Washiyama}, D.~{Lacroix}, Int. J. Mod. Phys. E \textbf{18}, 2114 (2009)

\bibitem{uma10b}
A.S. {Umar}, V.E. {Oberacker}, J.A. {Maruhn}, P.G. {Reinhard}, Phys. Rev. C
  \textbf{81}, 064607 (2010)

\bibitem{obe10}
V.E. {Oberacker}, A.S. {Umar}, J.A. {Maruhn}, P.G. {Reinhard}, Phys. Rev. C
  \textbf{82}, 034603 (2010)

\bibitem{loe11}
N.~Loebl, J.A. Maruhn, P.G. Reinhard, Phys. Rev. C \textbf{84}, 034608 (2011)

\bibitem{iwa11}
Y.~Iwata, J.A. Maruhn, Phys. Rev. C \textbf{84}, 014616 (2011)

\bibitem{leb12}
D.~Lebhertz, S.~Courtin, F.~Haas, D.G. Jenkins, C.~Simenel, M.D. Salsac, D.A.
  Hutcheon, C.~Beck, J.~Cseh, J.~Darai et~al., Phys. Rev. C \textbf{85}, 034333
  (2012)

\bibitem{uma12a}
A.S. Umar, V.E. Oberacker, J.A. Maruhn, P.G. Reinhard, Phys. Rev. C
  \textbf{85}, 017602 (2012)

\bibitem{obe12}
V.E. Oberacker, A.S. Umar, J.A. Maruhn, P.G. Reinhard, Phys. Rev. C
  \textbf{85}, 034609 (2012)

\bibitem{kes12}
R.~Keser, A.S. Umar, V.E. Oberacker, Phys. Rev. C \textbf{85}, 044606 (2012)

\bibitem{uma12b}
A.S. Umar, V.E. Oberacker, C.J. Horowitz, Phys. Rev. C \textbf{85}, 055801
  (2012)

\bibitem{iwata2013}
Y.~Iwata, K.~Iida, N.~Itagaki, Phys. Rev. C \textbf{87}, 014609 (2013)

\bibitem{desouza2013}
R.T. deSouza, S.~Hudan, V.E. Oberacker, A.S. Umar, Phys. Rev. C \textbf{88},
  014602 (2013)

\bibitem{dai2014}
G.F. Dai, L.~Guo, E.G. Zhao, S.G. Zhou, Phys. Rev. C \textbf{90}, 044609 (2014)

\bibitem{umar2014a}
A.S. Umar, C.~Simenel, V.E. Oberacker, Phys. Rev. C \textbf{89}, 034611 (2014)

\bibitem{steinbach2014}
T.K. Steinbach, J.~Vadas, J.~Schmidt, C.~Haycraft, S.~Hudan, R.T. deSouza, L.T.
  Baby, S.A. Kuvin, I.~Wiedenh\"over, A.S. Umar et~al., Phys. Rev. C
  \textbf{90}, 041603 (2014)

\bibitem{bourgin2016}
D.~Bourgin, C.~Simenel, S.~Courtin, F.~Haas, Phys. Rev. C \textbf{93}, 034604
  (2016)

\bibitem{reinhard2016a}
P.G. Reinhard, A.S. Umar, P.D. Stevenson, J.~Piekarewicz, V.E. Oberacker, J.A.
  Maruhn, Phys. Rev. C \textbf{93}, 044618 (2016)

\bibitem{simenel2017}
C.~Simenel, A.S. Umar, K.~Godbey, M.~Dasgupta, D.J. Hinde, Phys. Rev. C
  \textbf{95}, 031601(R) (2017)

\bibitem{godbey2017}
K.~Godbey, A.S. Umar, C.~Simenel, Phys. Rev. C \textbf{95}, 011601(R) (2017)

\bibitem{magierski2017}
P.~Magierski, K.~Sekizawa, G.~Wlaz\l{}owski, Phys. Rev. Lett. \textbf{119},
  042501 (2017)

\bibitem{sekizawa2017b}
K.~Sekizawa, G.~Wlaz\l{}owski, P.~Magierski, {EPJ} {W}eb {C}onf. \textbf{163},
  00051 (2017)

\bibitem{guo2018}
L.~Guo, C.~Simenel, L.~Shi, C.~Yu, Phys. Lett. B \textbf{782}, 401 (2018)

\bibitem{guo2018b}
L.~Guo, K.~Godbey, A.S. Umar, Phys. Rev. C \textbf{98}, 064607 (2018)

\bibitem{sekizawa2019b}
K.~Sekizawa, K.~Hagino, Phys. Rev. C \textbf{99}, 051602 (2019)

\bibitem{godbey2019b}
K.~Godbey, C.~Simenel, A.S. Umar, Phys. Rev. C \textbf{100}, 024619 (2019)

\bibitem{godbey2019c}
K.~Godbey, L.~Guo, A.S. Umar, Phys. Rev. C \textbf{100}, 054612 (2019)

\bibitem{li2019}
X.~Li, Z.~Wu, L.~Guo, Sci. China-Phys. Mech. Astron. \textbf{62}, 122011 (2019)

\bibitem{umar2021}
A.S. Umar, C.~Simenel, K.~Godbey, Phys. Rev. C \textbf{104}, 034619 (2021)

\bibitem{tong2022a}
L.~Tong, S.~Yan, Phys. Rev. C \textbf{105}, 014613 (2022)

\bibitem{sun2022}
X.X. Sun, L.~Guo, A.S. Umar, Phys. Rev. C \textbf{105}, 034601 (2022)

\bibitem{sun2022b}
X.X. Sun, L.~Guo, Phys. Rev. C \textbf{105}, 054610 (2022)

\bibitem{godbey2022}
K.~Godbey, A.S. Umar, C.~Simenel, Phys. Rev. C \textbf{106}, L051602 (2022)

\bibitem{magierski2022}
P.~Magierski, A.~Makowski, M.C. Barton, K.~Sekizawa, G.~Wlaz\l{}owski, Phys.
  Rev. C \textbf{105}, 064602 (2022)

\bibitem{gumbel2023}
R.~Gumbel, C.~Ross, A.S. Umar, Phys. Rev. C \textbf{108}, L051602 (2023)

\bibitem{tong2023}
L.~Tong, S.~Yan, Phys. Rev. C \textbf{107}, 054615 (2023)

\bibitem{sun2023}
X.X. Sun, L.~Guo, Phys. Rev. C \textbf{107}, L011601 (2023)

\bibitem{sun2023b}
X.X. Sun, L.~Guo, Phys. Rev. C \textbf{107}, 064609 (2023)

\bibitem{yao2024}
H.~Yao, H.~Yang, N.~Wang, Phys. Rev. C \textbf{110}, 014602 (2024)

\bibitem{desouza2024}
R.T. deSouza, K.~Godbey, S.~Hudan, W.~Nazarewicz, Phys. Rev. C \textbf{109},
  L041601 (2024)

\bibitem{iwa09}
Y.~{Iwata}, T.~{Otsuka}, J.A. {Maruhn}, N.~{Itagaki}, Eur. Phys. J. A
  \textbf{42}, 613 (2009)

\bibitem{iwa10a}
Y.~Iwata, T.~Otsuka, J.A. Maruhn, N.~Itagaki, Phys. Rev. Lett. \textbf{104},
  252501 (2010)

\bibitem{iwa10b}
Y.~Iwata, T.~Otsuka, J.A. Maruhn, N.~Itagaki, Nucl. Phys. A \textbf{836}, 108
  (2010)

\bibitem{iwa12}
Y.~Iwata, J. Mod. Phys. \textbf{3}, 476 (2012)

\bibitem{sim12}
C.~Simenel, D.J. Hinde, R.~du~Rietz, M.~Dasgupta, M.~Evers, C.J. Lin, D.H.
  Luong, A.~Wakhle, Phys. Lett. B \textbf{710}, 607 (2012)

\bibitem{umar2017}
A.S. Umar, C.~Simenel, W.~Ye, Phys. Rev. C \textbf{96}, 024625 (2017)

\bibitem{stone2017}
J.R. Stone, P.~Danielewicz, Y.~Iwata, Phys. Rev. C \textbf{96}, 014612 (2017)

\bibitem{simenel2020}
C.~Simenel, K.~Godbey, A.S. Umar, Phys. Rev. Lett. \textbf{124}, 212504 (2020)

\bibitem{guo08}
L.~Guo, J.A. Maruhn, P.G. Reinhard, Y.~Hashimoto, Phys. Rev. C \textbf{77},
  041301 (2008)

\bibitem{uma08a}
A.S. Umar, V.E. Oberacker, J.A. Maruhn, Eur. Phys. J. A \textbf{37}, 245 (2008)

\bibitem{sim10b}
C.~Simenel, Phys. Rev. Lett. \textbf{105}, 192701 (2010)

\bibitem{eve11}
M.~Evers, M.~Dasgupta, D.J. Hinde, D.H. Luong, R.~Rafiei, R.~du~Rietz,
  C.~Simenel, Phys. Rev. C \textbf{84}, 054614 (2011)

\bibitem{sekizawa2014}
K.~Sekizawa, K.~Yabana, Phys. Rev. C \textbf{90}, 064614 (2014)

\bibitem{yilmaz2014}
B.~Yilmaz, S.~Ayik, D.~Lacroix, O.~Yilmaz, Phys. Rev. C \textbf{90}, 024613
  (2014)

\bibitem{sekizawa2015}
K.~Sekizawa, K.~Yabana, {EPJ} {W}eb {C}onf. \textbf{86}, 00043 (2015)

\bibitem{sekizawa2016}
K.~Sekizawa, K.~Yabana, Phys. Rev. C \textbf{93}, 054616 (2016)

\bibitem{scamps2017a}
G.~Scamps, C.~Rodr\'{\i}guez-Tajes, D.~Lacroix, F.~Farget, Phys. Rev. C
  \textbf{95}, 024613 (2017)

\bibitem{sekizawa2017}
K.~Sekizawa, Phys. Rev. C \textbf{96}, 014615 (2017)

\bibitem{sekizawa2017a}
K.~Sekizawa, Phys. Rev. C \textbf{96}, 041601(R) (2017)

\bibitem{regnier2018}
D.~Regnier, D.~Lacroix, G.~Scamps, Y.~Hashimoto, Phys. Rev. C \textbf{97},
  034627 (2018)

\bibitem{ayik2018}
S.~Ayik, B.~Yilmaz, O.~Yilmaz, A.S. Umar, Phys. Rev. C \textbf{97}, 054618
  (2018)

\bibitem{yilmaz2018}
B.~Yilmaz, S.~Ayik, O.~Yilmaz, A.S. Umar, Phys. Rev. C \textbf{98}, 034604
  (2018)

\bibitem{ayik2019}
S.~Ayik, B.~Yilmaz, O.~Yilmaz, A.S. Umar, Phys. Rev. C \textbf{100}, 014609
  (2019)

\bibitem{ayik2019b}
S.~Ayik, O.~Yilmaz, B.~Yilmaz, A.S. Umar, Phys. Rev. C \textbf{100}, 044614
  (2019)

\bibitem{wu2019}
Z.~Wu, L.~Guo, Phys. Rev. C \textbf{100}, 014612 (2019)

\bibitem{wu2020}
Z.~Wu, L.~Guo, Sci. China-Phys. Mech. Astron. \textbf{63}, 242021 (2020)

\bibitem{roy2022}
B.J. Roy, S.~Santra, A.~Pal, H.~Kumawat, S.K. Pandit, V.V. Parkar,
  K.~Ramachandran, K.~Mahata, K.~Sekizawa, Phys. Rev. C \textbf{105}, 044611
  (2022)

\bibitem{wu2022}
Z.~Wu, L.~Guo, Z.~Liu, G.~Peng, Phys. Lett. B \textbf{825}, 136886 (2022)

\bibitem{li2024b}
B.~Li, D.~Vretenar, T.~Nik\ifmmode \check{s}\else
  \v{s}\fi{}i\ifmmode~\acute{c}\else \'{c}\fi{}, D.D. Zhang, P.W. Zhao,
  J.~Meng, Phys. Rev. C \textbf{110}, 034611 (2024)

\bibitem{wakhle2014}
A.~Wakhle, C.~Simenel, D.J. Hinde, M.~Dasgupta, M.~Evers, D.H. Luong,
  R.~du~Rietz, E.~Williams, Phys. Rev. Lett. \textbf{113}, 182502 (2014)

\bibitem{oberacker2014}
V.E. Oberacker, A.S. Umar, C.~Simenel, Phys. Rev. C \textbf{90}, 054605 (2014)

\bibitem{ayik2015a}
S.~Ayik, B.~Yilmaz, O.~Yilmaz, Phys. Rev. C \textbf{92}, 064615 (2015)

\bibitem{umar2015a}
A.S. Umar, V.E. Oberacker, C.~Simenel, Phys. Rev. C \textbf{92}, 024621 (2015)

\bibitem{hammerton2015}
K.~Hammerton, Z.~Kohley, D.J. Hinde, M.~Dasgupta, A.~Wakhle, E.~Williams, V.E.
  Oberacker, A.S. Umar, I.P. Carter, K.J. Cook et~al., Phys. Rev. C
  \textbf{91}, 041602(R) (2015)

\bibitem{umar2016}
A.S. Umar, V.E. Oberacker, C.~Simenel, Phys. Rev. C \textbf{94}, 024605 (2016)

\bibitem{morjean2017}
M.~Morjean, D.J. Hinde, C.~Simenel, D.Y. Jeung, M.~Airiau, K.J. Cook,
  M.~Dasgupta, A.~Drouart, D.~Jacquet, S.~Kalkal et~al., Phys. Rev. Lett.
  \textbf{119}, 222502 (2017)

\bibitem{yu2017}
{Chong Yu}, {Lu Guo}, Sci. China Phys. \textbf{60}, 092011 (2017)

\bibitem{guo2018c}
L.~Guo, C.~Shen, C.~Yu, Z.~Wu, Phys. Rev. C \textbf{98}, 064609 (2018)

\bibitem{zheng2018}
H.~Zheng, S.~Burrello, M.~Colonna, D.~Lacroix, G.~Scamps, Phys. Rev. C
  \textbf{98}, 024622 (2018)

\bibitem{godbey2019}
K.~Godbey, A.S. Umar, C.~Simenel, Phys. Rev. C \textbf{100}, 024610 (2019)

\bibitem{simenel2021}
C.~Simenel, P.~McGlynn, A.S. Umar, K.~Godbey, Phys. Lett. B \textbf{822},
  136648 (2021)

\bibitem{li2022}
L.~Li, L.~Guo, K.~Godbey, A.S. Umar, Phys. Lett. B \textbf{833}, 137349 (2022)

\bibitem{stevenson2022}
P.D. Stevenson, Front. Phys. \textbf{10}, 1019285 (2022)

\bibitem{mcglynn2023}
P.~McGlynn, C.~Simenel, Phys. Rev. C \textbf{107}, 054614 (2023)

\bibitem{lee2024}
H.~Lee, P.~McGlynn, C.~Simenel, Phys. Rev. C \textbf{110}, 024606 (2024)

\bibitem{scamps2024b}
G.~Scamps, Phys. Rev. C \textbf{110}, 054605 (2024)

\bibitem{li2024c}
L.~Li, L.~Guo, K.~Godbey, A.S. Umar, Phys. Rev. C \textbf{110}, 064607 (2024)

\bibitem{wang2016}
N.~Wang, L.~Guo, Phys. Lett. B \textbf{760}, 236 (2016)

\bibitem{roy2018}
B.J. Roy, Y.~Sawant, P.~Patwari, S.~Santra, A.~Pal, A.~Kundu, D.~Chattopadhyay,
  V.~Jha, S.K. Pandit, V.V. Parkar et~al., Phys. Rev. C \textbf{97}, 034603
  (2018)

\bibitem{jiang2025}
X.~Jiang, N.~Wang, R.~An, Phys. Rev. C \textbf{111}, 044604 (2025)

\bibitem{besse2020}
G.~Besse, V.~de~la Mota, E.~Bonnet, P.~Eudes, P.~Napolitani, Z.~Basrak, Phys.
  Rev. C \textbf{101}, 054608 (2020)

\bibitem{uma10a}
A.S. {Umar}, J.A. {Maruhn}, N.~{Itagaki}, V.E. {Oberacker}, Phys. Rev. Lett.
  \textbf{104}, 212503 (2010)

\bibitem{iwata2015}
Y.~Iwata, T.~Ichikawa, N.~Itagaki, J.A. Maruhn, T.~Otsuka, Phys. Rev. C
  \textbf{92}, 011303 (2015)

\bibitem{schuetrumpf2017}
B.~Schuetrumpf, W.~Nazarewicz, Phys. Rev. C \textbf{96}, 064608 (2017)

\bibitem{stevenson2020b}
P.D. Stevenson, J.L. Willerton, SciPost Phys. Proc. \textbf{3}, 47 (2020)

\bibitem{gol09}
C.~Golabek, C.~Simenel, Phys. Rev. Lett. \textbf{103}, 042701 (2009)

\bibitem{ked10}
D.J. Kedziora, C.~Simenel, Phys. Rev. C \textbf{81}, 044613 (2010)

\bibitem{ayik2017}
S.~Ayik, B.~Yilmaz, O.~Yilmaz, A.S. Umar, G.~Turan, Phys. Rev. C \textbf{96},
  024611 (2017)

\bibitem{umar2018}
{Umar, A.S.}, {Simenel, C.}, Acta Phys. Pol. B \textbf{49}, 573 (2018)

\bibitem{blo79}
J.~B{\l}ocki, H.~Flocard, Phys. Lett. B \textbf{85}, 163 (1979)

\bibitem{kortelainen2012}
M.~Kortelainen, J.~{McDonnell}, W.~Nazarewicz, P.G. Reinhard, J.~Sarich,
  N.~Schunck, M.V. Stoitsov, S.M. Wild, Phys. Rev. C \textbf{85}, 024304 (2012)

\bibitem{degiovannini2012}
U.~{De Giovannini}, D.~Varsano, M.A.L. Marques, H.~Appel, E.K.U. Gross,
  A.~Rubio, Phys. Rev. A \textbf{85}, 062515 (2012)

\bibitem{boucke1997}
K.~Boucke, H.~Schmitz, H.J. Kull, Phys. Rev. A \textbf{56}, 763 (1997)

\bibitem{mangin-brinet1998}
M.~Mangin-Brinet, J.~Carbonell, C.~Gignoux, Phys. Rev. A \textbf{57}, 3245
  (1998)

\bibitem{pardi2014}
C.I. Pardi, P.D. Stevenson, K.~Xu, Phys. Rev. E \textbf{89}, 033312 (2014)

\bibitem{schuetrumpf2016}
B.~Schuetrumpf, W.~Nazarewicz, P.G. Reinhard, Phys. Rev. C \textbf{93}, 054304
  (2016)

\bibitem{rei06}
P.G. Reinhard, P.D. Stevenson, D.~Almehed, J.A. Maruhn, M.R. Strayer, Phys.
  Rev. E \textbf{73}, 036709 (2006)

\bibitem{ave08}
B.~Avez, C.~Simenel, P.~Chomaz, Phys. Rev. C \textbf{78}, 044318 (2008)

\bibitem{car01}
I.~Carusotto, Y.~Castin, J.~Dalibard, Phys. Rev. A \textbf{63}, 023606 (2001)

\bibitem{jui02}
O.~Juillet, P.~Chomaz, Phys. Rev. Lett. \textbf{88}, 142503 (2002)

\bibitem{won78}
C.Y. Wong, H.H.K. Tang, Phys. Rev. Lett. \textbf{40}, 1070 (1978)

\bibitem{won79}
C.Y. Wong, H.H.K. Tang, Phys. Rev. C \textbf{20}, 1419 (1979)

\bibitem{dan84}
P.~Danielewicz, Ann. Phys. \textbf{152}, 305 (1984)

\bibitem{bot90}
W.~Botermans, R.~Malfliet, Phys. Rep. \textbf{198}, 115 (1990)

\bibitem{ayi80}
S.~Ayik, Z. Phys. A \textbf{298}, 83 (1980)

\bibitem{lac99}
D.~Lacroix, P.~Chomaz, S.~Ayik, Nucl. Phys. A \textbf{651}, 369 (1999)

\bibitem{rei92}
P.G. Reinhard, E.~Suraud, Ann. Phys. \textbf{216}, 98 (1992)

\bibitem{ayi01}
S.~Ayik, Y.~Abe, Phys. Rev. C \textbf{64}, 024609 (2001)

\bibitem{lac01}
D.~Lacroix, S.~Ayik, P.~Chomaz, Phys. Rev. C \textbf{63}, 064305 (2001)

\bibitem{wu1997}
J.S. Wu, K.C. Wong, M.R. Strayer, M.~Baranger, Phys. Rev. C \textbf{56}, 857
  (1997)

\bibitem{abada1992}
A.~Abada, D.~Vautherin, Phys. Rev. C \textbf{45}, 2205 (1992)

\bibitem{mcglynn2020}
P.~McGlynn, C.~Simenel, Phys. Rev. C \textbf{102}, 064614 (2020)

\bibitem{hil53}
D.L. Hill, J.A. Wheeler, Phys. Rev. \textbf{89}, 1102 (1953)

\bibitem{gri57}
J.J. Griffin, J.A. Wheeler, Phys. Rev. \textbf{108}, 311 (1957)

\bibitem{rei83}
P.G. Reinhard, R.Y. Cusson, K.~Goeke, Nucl. Phys. A \textbf{398}, 141 (1983)

\bibitem{ben08}
M.~Bender, P.H. Heenen, Phys. Rev. C \textbf{78}, 024309 (2008)

\bibitem{del10}
J.P. Delaroche, M.~Girod, J.~Libert, H.~Goutte, S.~Hilaire, S.~P\'eru,
  N.~Pillet, G.F. Bertsch, Phys. Rev. C \textbf{81}, 014303 (2010)

\bibitem{rod10}
T.R. Rodr\'iguez, J.L. Egido, Phys. Rev. C \textbf{81}, 064323 (2010)

\bibitem{gou05}
H.~Goutte, J.F. Berger, P.~Casoli, D.~Gogny, Phys. Rev. C \textbf{71}, 024316
  (2005)

\bibitem{schunck2016}
N.~Schunck, L.M. Robledo, Rep. Prog. Phys. \textbf{79}, 116301 (2016)

\bibitem{schunck2022}
N.~Schunck, D.~Regnier, Prog Part Nucl Phys \textbf{125}, 103963 (2022)

\bibitem{regnier2018b}
D.~Regnier, N.~Dubray, M.~Verri{\`{e}}re, N.~Schunck, Comput. Phys. Commun.
  \textbf{225}, 180 (2018)

\bibitem{bender2020}
M.~Bender, R.~Bernard, G.~Bertsch, S.~Chiba, J.J. Dobaczewski, N.~Dubray,
  S.~Giuliani, K.~Hagino, D.~Lacroix, Z.~Li et~al., J. Phys. G: Nucl. Part.
  Phys. \textbf{47}, 113002 (2020)

\bibitem{verriere2020}
M.~Verriere, D.~Regnier, Front. Phys. \textbf{8}, 233 (2020)

\bibitem{lau2022}
N.W.T. Lau, R.N. Bernard, C.~Simenel, Phys. Rev. C \textbf{105}, 034617 (2022)

\bibitem{marevic2024}
P.~Marevi{\'c}, D.~Regnier, D.~Lacroix, Eur. Phys. J. A \textbf{60}, 10 (2024)

\bibitem{regnier2019a}
D.~Regnier, D.~Lacroix, Phys. Rev. C \textbf{99}, 064615 (2019)

\bibitem{hasegawa2020}
N.~Hasegawa, K.~Hagino, Y.~Tanimura, Phys. Lett. B \textbf{808}, 135693 (2020)

\bibitem{schunck2019}
\emph{{Energy Density Functional Methods for Atomic Nuclei}} (IOP Publishing
  Ltd., Bristol, UK, 2019), edited by N. Schunck

\bibitem{simenel2014b}
C.~{S}imenel, J. Phys. G: Nucl. Part. Phys. \textbf{41}, 094007 (2014)

\bibitem{boh75}
A.~Bohr, B.~Mottelson, \emph{Nuclear Structure} (2 vol., W.A. Benjamin, Inc.,
  1975)

\bibitem{har01}
M.N. Harakeh, A.~van~der Woude, \emph{Giant Resonances: Fundamental
  High-Frequency Modes of Nuclear Excitations} (Oxford University Press, New
  York, 2001)

\bibitem{bal47}
G.C. Baldwin, G.S. Klaiber, Phys. Rev. \textbf{71}, 3 (1947)

\bibitem{gol48}
M.~Goldhaber, E.~Teller, Phys. Rev. \textbf{74}, 1046 (1948)

\bibitem{fuk72}
S.~Fukuda, Y.~Torizuka, Phys. Rev. Lett. \textbf{29}, 1109 (1972)

\bibitem{mar76}
N.~Marty, A.~Willis, V.~Comparat, R.~Frascaria, M.~Morlet, Orsay report
  IPNO76-03  (1976)

\bibitem{har77}
M.N. Harakeh, K.~van~der Borg, T.~Ishimatsu, H.P. Morsch, A.~van~der Woude,
  F.E. Bertrand, Phys. Rev. Lett. \textbf{38}, 676 (1977)

\bibitem{you77}
D.H. Youngblood, C.M. Rozsa, J.M. Moss, D.R. Brown, J.D. Bronson, Phys. Rev.
  Lett. \textbf{39}, 1188 (1977)

\bibitem{pin66}
D.~Pines, P.~Nozi\`eres, \emph{The Theory of Quantum Liquids} (Benjamin, New
  York, 1966)

\bibitem{cho95}
P.~Chomaz, N.~Frascaria, Phys. Rep. \textbf{252}, 275 (1995)

\bibitem{aum98}
T.~Aumann, P.F. Bortignon, H.~Emling, Ann. Rev. Nucl. Part. Sci. \textbf{48},
  351 (1998)

\bibitem{sca04}
J.A. Scarpaci, Nucl. Phys. A \textbf{731}, 175 (2004)

\bibitem{vol95}
C.~Volpe, F.~Catara, P.~Chomaz, M.V. Andr{\'e}s, E.G. Lanza, Nucl. Phys. A
  \textbf{589}, 521 (1995)

\bibitem{bor97}
P.F. Bortignon, C.H. Dasso, Phys. Rev. C \textbf{56}, 574 (1997)

\bibitem{fal03}
M.~Fallot, P.~Chomaz, M.V. Andr\'es, F.~Catara, E.G. Lanza, J.A. Scarpaci,
  Nucl. Phys. A \textbf{729}, 699 (2003)

\bibitem{lan06}
E.G. Lanza, F.~Catara, M.V. Andr\'es, P.~Chomaz, M.~Fallot, J.A. Scarpaci,
  Phys. Rev. C \textbf{74}, 064614 (2006)

\bibitem{kre74}
S.~Krewald, J.~Birkholz, A.~Faessler, J.~Speth, Phys. Rev. Lett. \textbf{33},
  1386 (1974)

\bibitem{liu76}
K.~Liu, N.V. Giai, Phys. Lett. B \textbf{65}, 23 (1976)

\bibitem{kam98}
S.~Kamerdzhiev, R.J. Liotta, E.~Litvinova, V.~Tselyaev, Phys. Rev. C
  \textbf{58}, 172 (1998)

\bibitem{mat01}
M.~Matsuo, Nucl. Phys. A \textbf{696}, 371 (2001)

\bibitem{hag01}
K.~Hagino, H.~Sagawa, Nucl. Phys. A \textbf{695}, 82 (2001)

\bibitem{kha02}
E.~Khan, N.~Sandulescu, M.~Grasso, N.~Van~Giai, Phys. Rev. C \textbf{66},
  024309 (2002)

\bibitem{cho87}
P.~Chomaz, N.V. Giai, S.~Stringari, Phys. Lett. B \textbf{189}, 375 (1987)

\bibitem{pac88}
J.M. Pacheco, E.~Maglione, R.A. Broglia, Phys. Rev. C \textbf{37}, 2257 (1988)

\bibitem{str79}
S.~Stringari, D.~Vautherin, Phys. Lett. B \textbf{88}, 1 (1979)

\bibitem{chinn1996}
C.R. Chinn, A.S. Umar, M.~Valli\'eres, M.R. Strayer, Phys. Rep. \textbf{264},
  107 (1996)

\bibitem{ave09}
B.~Avez, Ph.D. thesis, University of Paris XI (2009)

\bibitem{geng2025}
J.~Geng, Z.H. Wang, P.W. Zhao, Y.F. Niu, H.~Liang, W.H. Long, Phys. Rev. C
  \textbf{111}, 024305 (2025)

\bibitem{bennaceur2005}
K.~Bennaceur, J.~Dobaczewski, Comp. Phys. Commun. \textbf{168}, 96 (2005)

\bibitem{you04}
D.H. Youngblood, Y.W. Lui, H.L. Clark, B.~John, Y.~Tokimoto, X.~Chen, Phys.
  Rev. C \textbf{69}, 034315 (2004)

\bibitem{die88}
S.S. Dietrich, B.L. Bermann, At. Data Nucl. Data Tables \textbf{39}, 199 (1988)

\bibitem{bertrand1981}
F.E. Bertrand, Nucl. Phys. A \textbf{354}, 129 (1981)

\bibitem{sim05}
C.~Simenel, P.~Chomaz, T.~Duguet, arXiv:nucl-th/0504050

\bibitem{sur89}
E.~Suraud, M.~Pi, P.~Schuck, Nucl. Phys. A \textbf{492}, 294 (1989)

\bibitem{patel2014}
D.~Patel, Ph.D. thesis, University of {N}otre {D}ame (2014)

\bibitem{per08}
S.~P\'eru, H.~Goutte, Phys. Rev. C \textbf{77}, 044313 (2008)

\bibitem{lan09}
E.G. Lanza, F.~Catara, D.~Gambacurta, M.V. Andr\'es, P.~Chomaz, Phys. Rev. C
  \textbf{79}, 054615 (2009)

\bibitem{eng99}
J.~Engel, M.~Bender, J.~Dobaczewski, W.~Nazarewicz, R.~Surman, Phys. Rev. C
  \textbf{60}, 014302 (1999)

\bibitem{fra05}
S.~Fracasso, G.~Col\`o, Phys. Rev. C \textbf{72}, 064310 (2005)

\bibitem{peru2014}
{Péru, S.}, {Martini, M.}, Eur. Phys. J. A \textbf{50}, 88 (2014)

\bibitem{bes66}
D.~B\`es, R.~Broglia, Nucl. Phys. \textbf{80}, 289 (1966)

\bibitem{rip69}
G.~Ripka, R.~Padjen, Nucl. Phys. A \textbf{132}, 489 (1969)

\bibitem{oer01}
W.~von Oertzen, A.~Vitturi, Rep. Prog. Phys. \textbf{64}, 1247 (2001)

\bibitem{kha04}
E.~Khan, N.~Sandulescu, N.~Van~Giai, M.~Grasso, Phys. Rev. C \textbf{69},
  014314 (2004)

\bibitem{pll11}
E.~Pllumbi, M.~Grasso, D.~Beaumel, E.~Khan, J.~Margueron, J.~van~de Wiele,
  Phys. Rev. C \textbf{83}, 034613 (2011)

\bibitem{shi11}
H.~Shimoyama, M.~Matsuo, Phys. Rev. C \textbf{84}, 044317 (2011)

\bibitem{potel2013}
G.~Potel, A.~Idini, F.~Barranco, E.~Vigezzi, R.A. Broglia, Rep. Prog. Phys.
  \textbf{76}, 106301 (2013)

\bibitem{gra12}
M.~Grasso, D.~Lacroix, A.~Vitturi, Phys. Rev. C \textbf{85}, 034317 (2012)

\bibitem{ave10}
B.~Avez, P.~Chomaz, T.~Duguet, C.~Simenel, Mod. Phys. Lett. A \textbf{25}, 1997
  (2010)

\bibitem{bor98}
P.F. Bortignon, A.~Bracco, R.A. Broglia, \emph{Giant Resonances: Nuclear
  Structure at Finite Temperature} (Harwood Academic Publishers, Amsterdam,
  1998)

\bibitem{vau87}
D.~Vautherin, J.~Treiner, M.~V\'en\'eroni, Phys. Lett. B \textbf{191}, 6 (1987)

\bibitem{bon84}
P.~Bonche, S.~Levit, D.~Vautherin, Nucl. Phys. A \textbf{428}, 95 (1984)

\bibitem{lac98}
D.~Lacroix, P.~Chomaz, S.~Ayik, Phys. Rev. C \textbf{58}, 2154 (1998)

\bibitem{dro90}
S.~Dro\.{z}d\.{z}, S.~Nishizaki, J.~Speth, J.~Wambach, Phys. Rep. \textbf{197},
  1 (1990)

\bibitem{lac00}
D.~Lacroix, A.~Mai, P.~von Neumann-Cosel, A.~Richter, J.~Wambach, Phys. Lett. B
  \textbf{479}, 15 (2000)

\bibitem{wan85}
S.J. Wang, W.~Cassing, Ann. Phys. \textbf{159}, 328 (1985)

\bibitem{qiang2025}
Y.~Qiang, J.~Pei, K.~Godbey, Phys. Lett. B \textbf{861}, 139248 (2025)

\bibitem{negele1978}
J.W. Negele, S.E. Koonin, P.~M\"oller, J.R. Nix, A.J. Sierk, Phys. Rev. C
  \textbf{17}, 1098 (1978)

\bibitem{caamano2015}
M.~Caama\~no, F.~Farget, O.~Delaune, K.H. Schmidt, C.~Schmitt, L.~Audouin, C.O.
  Bacri, J.~Benlliure, E.~Casarejos, X.~Derkx et~al., Phys. Rev. C \textbf{92},
  034606 (2015)

\bibitem{tanimura2017err}
Y.~Tanimura, D.~Lacroix, S.~Ayik, Phys. Rev. Lett. \textbf{121}, 059902 (2018)

\bibitem{hulet1986}
E.K. Hulet, J.F. Wild, R.J. Dougan, R.W. Lougheed, J.H. Landrum, A.D. Dougan,
  M.~Schadel, R.L. Hahn, P.A. Baisden, C.M. Henderson et~al., Phys. Rev. Lett.
  \textbf{56}, 313 (1986)

\bibitem{meitner1939}
L.~Meitner, O.R. Frisch, Nature (London) \textbf{143}, 239 (1939)

\bibitem{gustafsson1971}
C.~Gustafsson, P.~M\"oller, S.G. Nilsson, Phys. Lett. B \textbf{34}, 349 (1971)

\bibitem{bernard2023}
{Bernard, R. N.}, {Simenel, C.}, {Blanchon, G.}, Eur. Phys. J. A \textbf{59},
  51 (2023)

\bibitem{wilkins1976}
B.D. Wilkins, E.P. Steinberg, R.R. Chasman, Phys. Rev. C \textbf{14}, 1832
  (1976)

\bibitem{zhang2016}
C.L. Zhang, B.~Schuetrumpf, W.~Nazarewicz, Phys. Rev. C \textbf{94}, 064323
  (2016)

\bibitem{sadhukhan2016}
J.~Sadhukhan, W.~Nazarewicz, N.~Schunck, Phys. Rev. C \textbf{93}, 011304
  (2016)

\bibitem{andreyev2010}
A.N. Andreyev, J.~Elseviers, M.~Huyse, P.~Van~Duppen, S.~Antalic, A.~Barzakh,
  N.~Bree, T.E. Cocolios, V.F. Comas, J.~Diriken et~al., Phys. Rev. Lett.
  \textbf{105}, 252502 (2010)

\bibitem{prasad2020}
E.~Prasad, D.J. Hinde, M.~Dasgupta, D.Y. Jeung, A.C. Berriman, B.M.A.
  {Swinton-Bland}, C.~Simenel, E.C. Simpson, R.~Bernard, E.~Williams et~al.,
  Phys. Lett. B \textbf{811}, 135941 (2020)

\bibitem{swinton2020b}
B.M.A. Swinton-Bland, M.A. Stoyer, A.C. Berriman, D.J. Hinde, C.~Simenel,
  J.~Buete, T.~Tanaka, K.~Banerjee, L.T. Bezzina, I.P. Carter et~al., Phys.
  Rev. C \textbf{102}, 054611 (2020)

\bibitem{mahata2022}
K.~Mahata, C.~Schmitt, S.~Gupta, A.~Shrivastava, G.~Scamps, K.H. Schmidt, Phys.
  Lett. B \textbf{825}, 136859 (2022)

\bibitem{kaur2024}
P.~Kaur, M.~Maiti, R.~Kumar, A.~Singh, H.~Sharma, Y.~Arafat, N.~Saneesh,
  A.~Parihari, M.~Kumar, K.S. Golda et~al., Phys. Rev. C \textbf{110}, 034613
  (2024)

\bibitem{moller2012}
P.~M\"oller, J.~Randrup, A.J. Sierk, Phys. Rev. C \textbf{85}, 024306 (2012)

\bibitem{ichikawa2012}
T.~Ichikawa, A.~Iwamoto, P.~M\"oller, A.J. Sierk, Phys. Rev. C \textbf{86},
  024610 (2012)

\bibitem{andreev2012}
A.V. Andreev, G.G. Adamian, N.V. Antonenko, Phys. Rev. C \textbf{86}, 044315
  (2012)

\bibitem{panebianco2012}
S.~Panebianco, J.L. Sida, H.~Goutte, J.F. Lema{\^{\i}}tre, N.~Dubray,
  S.~Hilaire, Phys. Rev. C \textbf{86}, 064601 (2012)

\bibitem{warda2012}
M.~Warda, J.L. Egido, Phys. Rev. C \textbf{86}, 014322 (2012)

\bibitem{andreev2013}
A.V. Andreev, G.G. Adamian, N.V. Antonenko, A.N. Andreyev, Phys. Rev. C
  \textbf{88}, 047604 (2013)

\bibitem{mcdonnell2014}
J.D. {McDonnell}, W.~Nazarewicz, J.A. Sheikh, A.~Staszczak, M.~Warda, Phys.
  Rev. C \textbf{90}, 021302 (2014)

\bibitem{bernard2024}
{Bernard, Rémi N.}, {Simenel, Cédric}, {Blanchon, Guillaume}, {Lau, Ngee-Wein
  T.}, {McGlynn, Patrick}, Eur. Phys. J. A \textbf{60}, 192 (2024)

\bibitem{unik1974}
J.P. Unik, L.E. Glendenin, K.F. Flynn, A.~Gorski, R.K. Sjoblom, \emph{Fragment
  mass and kinetic energy distributions for fissioning systems ranging from
  mass 230 to 256} (IAEA, Vienna, 1974), Vol.~II, pp. 19--45

\bibitem{brosa1990}
U.~Brosa, S.~Grossmann, A.~M\"{u}ller, Phys. Rep. \textbf{197}, 167 (1990)

\bibitem{bockstiegel2008}
C.~B\"{o}ckstiegel, S.~Steinh\"{a}user, K.H. Schmidt, H.G. Clerc, A.~Grewe,
  A.~Heinz, M.~{de Jong}, A.R. Junghans, J.~M\"{u}ller, B.~Voss, Nucl. Phys. A
  \textbf{802}, 12 (2008)

\bibitem{mcglynn2024}
P.~McGlynn, C.~Simenel, Phys. Rev. C \textbf{111}, 034619 (2025)

\bibitem{becke1990}
A.D. Becke, K.E. Edgecombe, J. Chem. Phys. \textbf{92}, 5397 (1990)

\bibitem{reinhard2011}
P.G. Reinhard, J.A. Maruhn, A.S. Umar, V.E. Oberacker, Phys. Rev. C
  \textbf{83}, 034312 (2011)

\bibitem{wilson2021}
J.N. Wilson, D.~Thisse, M.~Lebois, N.~Jovan{\v{c}}evi{\'{c}}, D.~Gjestvang,
  R.~Canavan, M.~Rudigier, D.~{\'E}tasse, R.B. Gerst, L.~Gaudefroy et~al.,
  Nature \textbf{590}, 566 (2021)

\bibitem{scamps2023b}
G.~Scamps, I.~Abdurrahman, M.~Kafker, A.~Bulgac, I.~Stetcu, Phys. Rev. C
  \textbf{108}, L061602 (2023)

\bibitem{bertsch2019c}
G.F. Bertsch, T.~Kawano, L.M. Robledo, Phys. Rev. C \textbf{99}, 034603 (2019)

\bibitem{marevic2021}
P.~Marevi\'{c}, N.~Schunck, J.~Randrup, R.~Vogt, Phys. Rev. C \textbf{104},
  L021601 (2021)

\bibitem{bon78}
P.~Bonche, B.~Grammaticos, S.~Koonin, Phys. Rev. C \textbf{17}, 1700 (1978)

\bibitem{umar1986a}
A.S. Umar, M.R. Strayer, P.G. Reinhard, Phys. Rev. Lett. \textbf{56}, 2793
  (1986)

\bibitem{umar1989}
A.S. Umar, M.R. Strayer, P.G. Reinhard, K.T.R. Davies, S.J. Lee, Phys. Rev. C
  \textbf{40}, 706 (1989)

\bibitem{reinhard1988}
P.G. Reinhard, A.S. Umar, K.T.R. Davies, M.R. Strayer, S.J. Lee, Phys. Rev. C
  \textbf{37}, 1026 (1988)

\bibitem{mor99}
C.R. Morton, A.C. Berriman, M.~Dasgupta, D.J. Hinde, J.O. Newton, K.~Hagino,
  I.J. Thompson, Phys. Rev. C \textbf{60}, 044608 (1999)

\bibitem{bas77}
R.~Bass, Phys. Rev. Lett. \textbf{39}, 265 (1977)

\bibitem{guo2012}
L.~Guo, T.~Nakatsukasa, {EPJ} {W}eb {C}onf. \textbf{38}, 09003 (2012)

\bibitem{bla54}
J.S. Blair, Phys. Rev. \textbf{95}, 1218 (1954)

\bibitem{bas80}
R.~Bass, \emph{Nuclear Reactions with Heavy-Ions} (Springer-Verlag, 1980)

\bibitem{tserruya1978}
I.~Tserruya, Y.~Eisen, D.~Pelte, A.~Gavron, H.~Oeschler, D.~Berndt, H.L.
  Harney, Phys. Rev. C \textbf{18}, 1688 (1978)

\bibitem{fernandez1978}
B.~Fernandez, C.~Gaarde, J.S. Larsen, S.~Pontoppidan, F.~Videbaek, Nucl. Phys.
  A \textbf{306}, 259 (1978)

\bibitem{kolata1979}
J.J. Kolata, R.M. Freeman, F.~Haas, B.~Heusch, A.~Gallmann, Phys. Rev. C
  \textbf{19}, 2237 (1979)

\bibitem{wu1984}
S.C. Wu, C.A. Barnes, Nucl. Phys. A \textbf{422}, 373 (1984)

\bibitem{thomas1986}
J.~Thomas, Y.T. Chen, S.~Hinds, D.~Meredith, M.~Olson, Phys. Rev. C
  \textbf{33}, 1679 (1986)

\bibitem{esbensen2012}
H.~Esbensen, Phys. Rev. C \textbf{85}, 064611 (2012)

\bibitem{flambaum1999}
V.V. Flambaum, V.G. Zelevinsky, Phys. Rev. Lett. \textbf{83}, 3108 (1999)

\bibitem{zielinski2024b}
R.~Zielinski, P.~McGlynn, C.~Simenel, Eur. Phys. J. C \textbf{84}, 967 (2024)

\bibitem{deleo2009}
S.~De~Leo, P.~Rotelli, Eur. Phys. J. C \textbf{62}, 793 (2009)

\bibitem{zielinski2024a}
R.~Zielinski, C.~Simenel, P.~McGlynn, Eur. Phys. J. C \textbf{84}, 992 (2024)

\bibitem{lei95}
J.R. Leigh, M.~Dasgupta, D.J. Hinde, J.C. Mein, C.R. Morton, R.C. Lemmon, J.P.
  Lestone, J.O. Newton, H.~Timmers, J.X. Wei et~al., Phys. Rev. C \textbf{52},
  3151 (1995)

\bibitem{hin96}
D.J. Hinde, M.~Dasgupta, J.R. Leigh, J.C. Mein, C.R. Morton, J.O. Newton,
  H.~Timmers, Phys. Rev. C \textbf{53}, 1290 (1996)

\bibitem{das98}
M.~Dasgupta, D.J. Hinde, N.~Rowley, A.M. Stefanini, Ann. Rev. Nucl. Part. Sci.
  \textbf{48}, 401 (1998)

\bibitem{row91}
N.~Rowley, G.~Satchler, P.~Stelson, Phys. Lett. B \textbf{254}, 25 (1991)

\bibitem{hag99}
K.~Hagino, N.~Rowley, A.~Kruppa, Comp. Phys. Commun. \textbf{123}, 143 (1999)

\bibitem{nay07}
B.K. Nayak, R.K. Choudhury, A.~Saxena, P.K. Sahu, R.G. Thomas, D.C. Biswas,
  B.V. John, E.T. Mirgule, Y.K. Gupta, M.~Bhike et~al., Phys. Rev. C
  \textbf{75}, 054615 (2007)

\bibitem{bon81}
P.~Bonche, N.~Ng\^o, Phys. Lett. B \textbf{105}, 17 (1981)

\bibitem{parascandolo2016}
C.~Parascandolo, D.~Pierroutsakou, R.~Alba, A.~Del~Zoppo, C.~Maiolino,
  D.~Santonocito, C.~Agodi, V.~Baran, A.~Boiano, M.~Colonna et~al., Phys. Rev.
  C \textbf{93}, 044619 (2016)

\bibitem{parascandolo2022}
C.~Parascandolo, D.~Pierroutsakou, R.~Alba, A.~Del~Zoppo, C.~Maiolino,
  D.~Santonocito, C.~Agodi, V.~Baran, A.~Boiano, M.~Colonna et~al., Phys. Rev.
  C \textbf{105}, 064611 (2022)

\bibitem{fli96}
S.~Flibotte, P.~Chomaz, M.~Colonna, M.~Cromaz, J.~DeGraaf, T.E. Drake,
  A.~Galindo-Uribarri, V.P. Janzen, J.~Jonkman, S.W. Marshall et~al., Phys.
  Rev. Lett. \textbf{77}, 1448 (1996)

\bibitem{jac62}
J.D. Jackson, \emph{Classical Electrodynamics} (Wiley, New York, 1962)

\bibitem{bar01}
V.~Baran, D.M. Brink, M.~Colonna, M.~Di~Toro, Phys. Rev. Lett. \textbf{87},
  182501 (2001)

\bibitem{swi82}
W.~Swiatecki, Nucl. Phys. A \textbf{376}, 275 (1982)

\bibitem{itk11}
I.M. Itkis, E.M. Kozulin, M.G. Itkis, G.N. Knyazheva, A.A. Bogachev, E.V.
  Chernysheva, L.~Krupa, Y.T. Oganessian, V.I. Zagrebaev, A.Y. Rusanov et~al.,
  Phys. Rev. C \textbf{83}, 064613 (2011)

\bibitem{nas07}
A.~Nasirov, A.~Muminov, R.~Utamuratov, G.~Fazio, G.~Giardina, F.~Hanappe,
  G.~Mandaglio, M.~Manganaro, W.~Scheid, Eur. Phys. J. A \textbf{34}, 325
  (2007)

\bibitem{sim09b}
C.~Simenel, B.~Avez, C.~Golabek, proceeding of the KERNZ08 conference,
  arXiv:0904.2653

\bibitem{blo77}
J.~B{\l}ocki, J.~Randrup, W.~Swiatecki, C.~Tsang, Ann. Phys. \textbf{105}, 427
  (1977)

\bibitem{den02}
V.Y. Denisov, W.~N{\"o}renberg, Eur. Phys. J. A \textbf{15}, 375 (2002)

\bibitem{brueckner1968}
K.A. Brueckner, J.R. Buchler, M.M. Kelly, Phys. Rev. \textbf{173}, 944 (1968)

\bibitem{fliessbach1971}
T.~Fliessbach, Z. Phys. \textbf{247}, 117 (1971)

\bibitem{brink1975}
D.M. Brink, F.~Stancu, Nucl. Phys. A \textbf{243}, 175 (1975)

\bibitem{zint1975}
P.G. Zint, U.~Mosel, Phys. Lett. B \textbf{56}, 424 (1975)

\bibitem{fliessbach1975}
T.~Fliessbach, Z. Phys. A \textbf{272}, 39 (1975)

\bibitem{beck1978}
F.~Beck, K.H. M\"uller, H.S. K\"ohler, Phys. Rev. Lett. \textbf{40}, 837 (1978)

\bibitem{sinha1979}
B.~Sinha, S.A. Moszkowski, Phys. Lett. B \textbf{81}, 289 (1979)

\bibitem{den10a}
V.Y. Denisov, V.A. Nesterov, Phys. At. Nucl. \textbf{73}, 1142 (2010)

\bibitem{den10b}
V.Y. Denisov, O.I. Davidovskaya, Phys. At. Nucl. \textbf{73}, 404 (2010)

\bibitem{cusson1985}
R.Y. Cusson, P.G. Reinhard, M.R. Strayer, J.A. Maruhn, W.~Greiner, Z. Phys. A
  \textbf{320}, 475 (1985)

\bibitem{umar1985}
A.S. Umar, M.R. Strayer, R.Y. Cusson, P.G. Reinhard, D.A. Bromley, Phys. Rev. C
  \textbf{32}, 172 (1985)

\bibitem{liu2019b}
S.~Liu, D.~Zhao, C.~Rong, T.~Lu, S.~Liu, J. Chem. Phys. \textbf{150}, 204106
  (2019)

\bibitem{li2020}
T.~Li, M.Z. Chen, C.L. Zhang, W.~Nazarewicz, M.~Kortelainen, Phys. Rev. C
  \textbf{102}, 044305 (2020)

\bibitem{cus85}
R.Y. Cusson, P.G. Reinhard, M.R. Strayer, J.A. Maruhn, W.~Greiner, Z. Phys. A
  \textbf{320}, 475 (1985)

\bibitem{goeke1983}
K.~Goeke, F.~Gr\"ummer, P.G. Reinhard, Ann. Phys. \textbf{150}, 504 (1983)

\bibitem{washiyama2008}
{Kouhei Washiyama}, {Denis Lacroix}, Phys. Rev. C \textbf{78}, 024610 (2008)

\bibitem{dobaczewski1995}
J.~Dobaczewski, J.~Dudek, Phys. Rev. C \textbf{52}, 1827 (1995)

\bibitem{umar2012a}
A.S. Umar, V.E. Oberacker, C.J. Horowitz, Phys. Rev. C \textbf{85}, 055801
  (2012)

\bibitem{oberacker2013}
V.E. Oberacker, A.S. Umar, Phys. Rev. C \textbf{87}, 034611 (2013)

\bibitem{jiang2014}
X.~Jiang, J.A. Maruhn, S.~Yan, Phys. Rev. C \textbf{90}, 064618 (2014)

\bibitem{jiang2002}
C.L. Jiang, H.~Esbensen, K.E. Rehm, B.B. Back, R.V.F. Janssens, J.A. Caggiano,
  P.~Collon, J.~Greene, A.M. Heinz, D.J. Henderson et~al., Phys. Rev. Lett.
  \textbf{89}, 052701 (2002)

\bibitem{dasgupta2007}
M.~Dasgupta, D.J. Hinde, A.~Diaz-Torres, B.~Bouriquet, C.I. Low, G.J. Milburn,
  J.O. Newton, Phys. Rev. Lett. \textbf{99}, 192701 (2007)

\bibitem{stefanini2010}
A.M. Stefanini, G.~Montagnoli, L.~Corradi, S.~Courtin, E.~Fioretto,
  A.~Goasduff, F.~Haas, P.~Mason, R.~Silvestri, P.P. Singh et~al., Phys. Rev. C
  \textbf{82}, 014614 (2010)

\bibitem{back2014}
B.B. Back, H.~Esbensen, C.L. Jiang, K.E. Rehm, Rev. Mod. Phys. \textbf{86}, 317
  (2014)

\bibitem{hagino2022}
K.~Hagino, K.~Ogata, A.M. Moro, Progress in Particle and Nuclear Physics
  \textbf{125}, 103951 (2022)

\bibitem{aljuwair1984}
H.A. Aljuwair, R.J. Ledoux, M.~Beckerman, S.B. Gazes, J.~Wiggins, E.R. Cosman,
  R.R. Betts, S.~Saini, O.~Hansen, Phys. Rev. C \textbf{30}, 1223 (1984)

\bibitem{montagnoli2012}
G.~Montagnoli, A.M. Stefanini, C.L. Jiang, H.~Esbensen, L.~Corradi, S.~Courtin,
  E.~Fioretto, A.~Goasduff, F.~Haas, A.F. Kifle et~al., Phys. Rev. C
  \textbf{85}, 024607 (2012)

\bibitem{bourgin2014}
D.~Bourgin, S.~Courtin, F.~Haas, A.M. Stefanini, G.~Montagnoli, A.~Goasduff,
  D.~Montanari, L.~Corradi, E.~Fioretto, J.~Huiming et~al., Phys. Rev. C
  \textbf{90}, 044610 (2014)

\bibitem{horowitz2008}
C.J. Horowitz, H.~Dussan, D.K. Berry, Phys. Rev. C \textbf{77}, 045807 (2008)

\bibitem{jiang2007}
C.L. Jiang, K.E. Rehm, B.B. Back, R.V.F. Janssens, Phys. Rev. C \textbf{75},
  015803 (2007)

\bibitem{cumming2001}
{Andrew Cumming}, {Lars Bildsten}, Astrophys. J. \textbf{559}, L127 (2001)

\bibitem{strohmayer2002}
T.E. Strohmayer, E.F. Brown, Astrophys. J. \textbf{566}, 1045 (2002)

\bibitem{hoyle1954}
F.~Hoyle, Astrophys. J. Suppl. Ser. \textbf{1}, 121 (1954)

\bibitem{monpribat2022}
{Monpribat, E.}, {Martinet, S.}, {Courtin, S.}, {Heine, M.}, {Ekstr\"om, S.},
  {Jenkins, D. G.}, {Choplin, A.}, {Adsley, P.}, {Curien, D.}, {Moukaddam, M.}
  et~al., Astron. Astrophys. \textbf{660}, A47 (2022)

\bibitem{dumont2024}
{Dumont, T.}, {Monpribat, E.}, {Courtin, S.}, {Choplin, A.}, {Bonhomme, A.},
  {Ekström, S.}, {Heine, M.}, {Curien, D.}, {Nippert, J.}, {Meynet, G.},
  Astron. Astrophys. \textbf{688}, A115 (2024)

\bibitem{close2025}
G.~Close, P.~Stevenson, A.~Diaz-Torres, \emph{Quantum dynamical microscopic
  approach to stellar carbon burning} (2025),
  \texttt{https://arxiv.org/abs/2502.11240}

\bibitem{jiang2013}
C.L. Jiang, B.B. Back, H.~Esbensen, R.V.F. Janssens, K.E. Rehm, R.J. Charity,
  Phys. Rev. Lett. \textbf{110}, 072701 (2013)

\bibitem{umar2023}
A.S. Umar, K.~Godbey, C.~Simenel, Phys. Rev. C \textbf{107}, 064605 (2023)

\bibitem{stefanini2009}
A.M. Stefanini, G.~Montagnoli, R.~Silvestri, L.~Corradi, S.~Courtin,
  E.~Fioretto, B.~Guiot, F.~Haas, D.~Lebhertz, P.~Mason et~al., Phys. Lett. B
  \textbf{679}, 95 (2009)

\bibitem{morton1999}
C.R. Morton, A.C. Berriman, M.~Dasgupta, D.J. Hinde, J.O. Newton, K.~Hagino,
  I.J. Thompson, Phys. Rev. C \textbf{60}, 044608 (1999)

\bibitem{vid77}
F.~Videb\ae{}k, R.B. Goldstein, L.~Grodzins, S.G. Steadman, T.A. Belote, J.D.
  Garrett, Phys. Rev. C \textbf{15}, 954 (1977)

\bibitem{rafferty2016}
D.C. Rafferty, M.~Dasgupta, D.J. Hinde, C.~Simenel, E.C. Simpson, E.~Williams,
  I.P. Carter, K.J. Cook, D.H. Luong, S.D. {McNeil} et~al., Phys. Rev. C
  \textbf{94}, 024607 (2016)

\bibitem{lacroix2019}
D.~Lacroix, S.~Ayik, Phys. Rev. C \textbf{101}, 014310 (2020)

\bibitem{bulgac2019c}
A.~Bulgac, Phys. Rev. C \textbf{100}, 034612 (2019)

\bibitem{bender03}
M.~Bender, P.H. Heenen, P.G. Reinhard, Rev. Mod. Phys. \textbf{75}, 121 (2003)

\bibitem{cor09}
L.~Corradi, G.~Pollarolo, S.~Szilner, J. Phys. G \textbf{36}, 113101 (2009)

\bibitem{bro91}
R.A. Broglia, A.~Winther, \emph{Heavy Ion Reactions: Lecture Notes the
  Elementary Processes (Frontiers in Physics)} (Addison Wesley Publishing
  Company, New-York, 1991), ISBN 0201513927

\bibitem{corradi1997}
L.~Corradi, A.M. Stefanini, J.H. He, S.~Beghini, G.~Montagnoli, F.~Scarlassara,
  G.F. Segato, G.~Pollarolo, C.H. Dasso, Phys. Rev. C \textbf{56}, 938 (1997)

\bibitem{sim11b}
C.~Simenel, C.~Golabek, D.J. Kedziora, EPJ Web of Conferences \textbf{17},
  09002 (2011)

\bibitem{vio85}
V.E. Viola, K.~Kwiatkowski, M.~Walker, Phys. Rev. C \textbf{31}, 1550 (1985)

\bibitem{roy77}
J.C. Roynette, H.~Doubre, N.~Frascaria, J.C. Jacmart, N.~Poff\'e, M.~Riou,
  Phys. Lett. B \textbf{67}, 395 (1977)

\bibitem{boc82}
R.~Bock, Y.T. Chu, M.~Dakowski, A.~Gobbi, E.~Grosse, A.~Olmi, H.~Sann,
  D.~Schwalm, U.~Lynen, W.~M{\"u}ller et~al., Nucl. Phys. A \textbf{388}, 334
  (1982)

\bibitem{tok85}
J.~T{\"o}ke, R.~Bock, G.~Dai, A.~Gobbi, S.~Gralla, K.~Hildenbrand,
  J.~Kuzminski, W.~M{\"u}ller, A.~Olmi, H.~Stelzer et~al., Nucl. Phys. A
  \textbf{440}, 327  (1985)

\bibitem{she87}
W.Q. Shen, J.~Albinski, A.~Gobbi, S.~Gralla, K.D. Hildenbrand, N.~Herrmann,
  J.~Kuzminski, W.F.J. M{\"u}ller, H.~Stelzer, J.~T{\"o}ke et~al., Phys. Rev. C
  \textbf{36}, 115 (1987)

\bibitem{hinde2021}
D.J. Hinde, M.~Dasgupta, E.C. Simpson, Prog. Part. Nucl. Phys. \textbf{118},
  103856 (2021)

\bibitem{sdvision}
D.~Pomarede, B.~Thooris{\it~et al.},
  \texttt{http://irfu.cea.fr/Projets/COAST/visu.htm}

\bibitem{itkis2004}
M.G. Itkis, J.~\"Ayst\"o, S.~Beghini, A.A. Bogachev, L.~Corradi, O.~Dorvaux,
  A.~Gadea, G.~Giardina, F.~Hanappe, I.M. Itkis et~al., Nucl. Phys. A
  \textbf{734}, 136 (2004)

\bibitem{nishio2008}
K.~Nishio, H.~Ikezoe, S.~Mitsuoka, I.~Nishinaka, Y.~Nagame, Y.~Watanabe,
  T.~Ohtsuki, K.~Hirose, S.~Hofmann, Phys. Rev. C \textbf{77}, 064607 (2008)

\bibitem{hinde2018}
D.J. Hinde, D.Y. Jeung, E.~Prasad, A.~Wakhle, M.~Dasgupta, M.~Evers, D.H.
  Luong, R.~du~Rietz, C.~Simenel, E.C. Simpson et~al., Phys. Rev. C
  \textbf{97}, 024616 (2018)

\bibitem{jeung2022}
D.Y. Jeung, D.J. Hinde, M.~Dasgupta, C.~Simenel, E.C. Simpson, K.J. Cook, H.M.
  Albers, J.~Buete, I.P. Carter, C.~D\"ullmann et~al., Phys. Lett. B
  \textbf{837}, 137641 (2022)

\bibitem{pal2024}
A.~Pal, S.~Santra, P.C. Rout, A.~Kundu, D.~Chattopadhyay, R.~Gandhi, P.N.
  Patil, R.~Tripathi, B.J. Roy, Y.~Sawant et~al., Phys. Rev. C \textbf{110},
  034601 (2024)

\bibitem{chizhov2003}
{\relax A. Yu}.~Chizhov, M.G. Itkis, I.M. Itkis, G.N. Kniajeva, E.M. Kozulin,
  N.A. Kondratiev, I.V. Pokrovsky, R.N. Sagaidak, V.M. Voskressensky, A.V.
  Yeremin et~al., Phys. Rev. C \textbf{67}, 011603 (2003)

\bibitem{hinde2022}
D.J. Hinde, R.d. Rietz, D.Y. Jeung, K.J. Cook, M.~Dasgupta, E.C. Simpson, R.G.
  Thomas, M.~Evers, C.J. Lin, D.H. Luong et~al., Phys. Rev. C \textbf{106},
  064614 (2022)

\bibitem{reinhard2021}
P.G. Reinhard, B.~Schuetrumpf, J.A. Maruhn, Comput. Phys. Commun. \textbf{258},
  107603 (2021)

\bibitem{zag06}
V.I. Zagrebaev, Y.T. Oganessian, M.G. Itkis, W.~Greiner, Phys. Rev. C
  \textbf{73}, 031602 (2006)

\bibitem{randrup1982}
J.~Randrup, Nucl. Phys. A \textbf{383}, 468 (1982)

\bibitem{randrup1978}
J.~Randrup, Nucl. Phys. A \textbf{307}, 319 (1978)

\bibitem{swiatecki2005}
W.J. \'{S}wi\c{a}tecki, K.~Siwek-Wilczy\'nska, J.~Wilczy\'nski, Phys. Rev. C
  \textbf{71}, 014602 (2005)

\bibitem{prasad2016}
E.~Prasad, A.~Wakhle, D.J. Hinde, E.~Williams, M.~Dasgupta, M.~Evers, D.H.
  Luong, G.~Mohanto, C.~Simenel, K.~Vo-Phuoc, Phys. Rev. C \textbf{93}, 024607
  (2016)

\bibitem{jedele2017}
A.~Jedele, A.B. {McIntosh}, K.~Hagel, M.~Huang, L.~Heilborn, Z.~Kohley, L.W.
  May, E.~{McCleskey}, M.~Youngs, A.~Zarrella et~al., Phys. Rev. Lett.
  \textbf{118}, 062501 (2017)

\bibitem{rei81}
J.~Reinhardt, B.~M\"uller, W.~Greiner, Phys. Rev. A \textbf{24}, 103 (1981)

\bibitem{ack08}
E.~Ackad, M.~Horbatsch, Phys. Rev. A \textbf{78}, 062711 (2008)

\bibitem{bon05}
P.~Bonche, H.~Flocard, P.H. Heenen, Comp. Phys. Commun. \textbf{171}, 49 (2005)

\bibitem{sei85}
M.~Seiwert, W.~Greiner, W.T. Pinkston, J. Phys. G \textbf{11}, L21 (1985)

\bibitem{ber90}
J.F. Berger, J.D. Anderson, P.~Bonche, M.S. Weiss, Phys. Rev. C \textbf{41},
  R2483 (1990)

\bibitem{cus80}
R.Y. Cusson, J.A. Maruhn, H.~St\"ocker, Z. Phys. A \textbf{294}, 257 (1980)

\bibitem{tia08}
J.~Tian, X.~Wu, K.~Zhao, Y.~Zhang, Z.~Li, Phys. Rev. C \textbf{77}, 064603
  (2008)

\bibitem{zha09}
K.~Zhao, X.~Wu, Z.~Li, Phys. Rev. C \textbf{80}, 054607 (2009)

\bibitem{sar09}
V.V. Sargsyan, Z.~Kanokov, G.G. Adamian, N.V. Antonenko, W.~Scheid, Phys. Rev.
  C \textbf{80}, 047603 (2009)

\bibitem{ada05}
G.G. Adamian, N.V. Antonenko, A.S. Zubov, Phys. Rev. C \textbf{71}, 034603
  (2005)

\bibitem{fen09}
Z.Q. Feng, G.M. Jin, J.Q. Li, Phys. Rev. C \textbf{80}, 067601 (2009)

\bibitem{mar02}
T.~Maruyama, A.~Bonasera, M.~Papa, S.~Chiba, Eur. Phys. J. A \textbf{14}, 191
  (2002)

\bibitem{vol78}
V.V. Volkov, Phys. Rep. \textbf{44}, 93 (1978)

\end{thebibliography}

\end{document}